\documentclass[a4paper]{JHEP3}
\usepackage[centertags]{amsmath}
\usepackage{amssymb}
\usepackage{fixltx2e}
\MakeRobust{\eqref} 
\usepackage{graphicx}
\usepackage{cite}
\usepackage{subfigure}
\preprint{ TIFR/TH/14-12\\
HRI/ST/1405 \\
ICTS-2014-04}

\newcommand{\Tr}{\text{Tr}} 
\newcommand{\ra}{\rangle}
\newcommand{\la}{\langle}

\def \beal#1 {\begin{align}#1\end{align}}

\def\Tr{\mathrm{Tr}}

\def\nn{\nonumber\\}

\newcommand{\wt}{\widetilde}

\def\={\stackrel{\bullet}{=}}
\def\nn{\notag\\}

\def\Tr{\mathrm{Tr}}

\def\[{\left[}
\def\]{\right]}
\def\({\left(}
\def\){\right)}

\def \be {\begin{equation}}
\def \ee {\end{equation}}
\def \bes {\begin{split}}
\def \ees {\end{split}}
\def \bea {\begin{eqnarray}}
\def \eea {\end{eqnarray}}
\def \beal#1 {\begin{align}#1\end{align}}
\def \nn {\notag\\}

\def\aver#1{\left\langle #1 \right\rangle}

\newcommand{\al}{\alpha}
\newcommand{\ga}{\gamma}
\newcommand{\de}{\delta}

\def \be {\begin{equation}}
\def \ee {\end{equation}}
\def \bea {\begin{eqnarray}}
\def \eea {\end{eqnarray}}
\def \beal#1 {\begin{align}#1\end{align}}
\def \nn {\notag\\}

\def\aver#1{\left\langle #1 \right\rangle}

\usepackage{enumerate}

\title{
Unitarity, Crossing Symmetry and Duality 
of the S-matrix in Large N Chern-Simons Theories with Fundamental Matter}

\author{
Sachin Jain$^{a),1}$, 
Mangesh Mandlik$^{a),2}$, 
Shiraz Minwalla$^{a), b) 3}$, 
Tomohisa Takimi$^{c),4}$, 
Spenta R. Wadia$^{a),d),5}$, 
Shuichi Yokoyama$^{a),6}$
\\
$^{a)}$Department of Theoretical Physics, Tata Institute of Fundamental Research,
Homi Bhabha Road, Mumbai 400005, India\\
$^{b)}$School of Natural Sciences, Institute for Advanced Study, Princeton, NJ08540, USA\\
$^{c)}$Harish-Chandra Research Institute, Chhatnag Road, Jhusi,Allahabad 211019 ,India\\
$^{d)}$International Centre for Theoretical Sciences, Tata Institute of Fundamental Research,
TIFR Centre Building, Indian Institute of Science, Bangalore 560004, India \\
{\small \tt E-mail: $^1$sachin@theory.tifr.res.in,${}^2$mangesh@theory.tifr.res.in, ${}^3$minwalla@theory.tifr.res.in,
 ${}^4$takimitomohisa@hri.res.in, ${}^5$wadia@theory.tifr.res.in, ${}^6$yokoyama@theory.tifr.res.in}
}

\abstract{We present explicit computations and conjectures for $2 \to 2$ scattering matrices in large $N$ {\it $U(N)$} Chern-Simons
 theories coupled to fundamental bosonic or fermionic matter to all orders in the 
't Hooft coupling expansion. The bosonic and fermionic S-matrices map to each other 
under the recently conjectured Bose-Fermi duality after a level-rank transposition. 
The S-matrices presented in this paper may be regarded as relativistic generalization of Aharonov-Bohm scattering. 
They have unusual structural features: they include a non- 
analytic piece localized on forward scattering, and obey modified crossing symmetry rules. We conjecture that 
these unusual features are properties of S-matrices in all Chern-Simons matter theories. 
The S-matrix in one of the exchange channels in our paper has an anyonic character; 
the parameter map of the conjectured Bose-Fermi duality may be derived by equating the anyonic phase in the 
bosonic and fermionic theories.}

\begin{document}

\section{Introduction}

It has recently been conjectured that $U(N)$ Chern-Simons theories coupled to a multiplet
of fundamental Wilson-Fisher bosons at level $k$ are dual to $U(|k|-N)$ Chern-Simons theories 
coupled to fundamental fermions at level $-k$. \footnote{Our notation is as follows.  $k$ is the coefficient 
of the Chern-Simons term in the bulk Lagrangian in the dimensional reduction scheme
utilized throughout in this paper. It is useful to define $\kappa={\rm sgn} (k)( |k|-N)$. $|\kappa|$ is the level of the WZW theory dual to the pure Chern-Simons theory. Note that $|k| >N$.  In terms of $\kappa$ and $N$ the duality map takes the level-rank form $N'= |\kappa|,~~ \kappa'= -{\rm sgn}(\kappa) N$.  }
The evidence for this conjecture is threefold. First the spectrum of `single trace'
operators and the three point functions of these operators have also been computed exactly in
the 't Hooft limit, and have been found to match\cite{Giombi:2011kc,Aharony:2011jz,Maldacena:2011jn,Maldacena:2012sf,Aharony:2012nh,GurAri:2012is}. 
Second the thermal partition functions of these theories have also been computed in the
't Hooft large $N$ limit and have been shown to match 
\cite{Aharony:2011jz,
Giombi:2011kc,Jain:2012qi,
Aharony:2012ns,Jain:2013py,Takimi:2013zca,Aharony:2012nh}.
Finally the duality described above has been demonstrated to follow from an extreme deformation of the known Giveon-Kutasov type duality 
\cite{Giveon:2008zn,Benini:2011mf} 
between supersymmetric theories~\cite{Jain:2013gza}.

Assuming the duality described above does indeed hold, it is interesting to better understand
the map that transforms bosons into fermions. Morally, we would like an explicit construction 
of the fundamental fermionic field $\psi_a(x)$ as a function of the fundamental bosonic fields
\footnote{The template here is the formula $\psi =e^{i \phi}$ of two dimensional bosonization. 
In some respects the already well known map between the gauge invariant higher spin currents on the two sides of the duality 
is the $2+1$ dimensional analogues of the $1+1$ dimensional 
 relation between global $U(1)$ symmetry currents $\partial \phi \sim {\tilde \psi} \psi$.};
such a formula cannot, however, be given precise meaning in the current context as $\psi_a(x)$ is not
 gauge invariant and its offshell correlators  are ill-defined. 

The  on shell limit of correlators of the elementary bosonic and fermionic fields, however, 
are physical as they compute the S-matrix for the scattering of bosonic or fermionic quanta.  As Chern-Simons theory has no propagating gluonic states, the S-matrix is free of soft gluon infrared divergences when the fundamental fields (bosons and fermions) are taken to be massive. 
An identity relating well-defined bosonic and fermionic S-matrices appears to be the
closest we can come to a precise bosonization map. Motivated by this observation, in this 
paper we present a detailed study of $2 \rightarrow 2$ S-matrices in Chern-Simons theories 
with fundamental bosonic and fermionic fields. 

Even independent of the Bose-Fermi duality, it is interesting that it is possible to determine 
exact results for the S-matrix of these theories as a function of the  't Hooft coupling 
constant $\lambda=\frac{N}{k}$. Exact results for scattering amplitudes as a function of a gauge coupling 
constant are rare, and should be studied when available for qualitative lessons. 
As we will see below, the explicit formulae for S-matrices presented in this paper turn out to possess several unfamiliar and unusual  structural features. Some of these unusual features appear to have a simple physical interpretation; we anticipate that they are general properties of S-matrices 
in all matter Chern-Simons theories. \footnote{These features include the presence of an non-analytic $\delta$ function piece in the S-matrix localized on forward scattering, and modified crossing symmetry relations as we describe below.}

As we have mentioned above, it is possible to determine (or conjecture) explicit results for the 
$2 \rightarrow 2$ scattering amplitudes for large $N$ fundamental matter Chern-Simons theories. 
In this paper we present explicit formulae for all these scattering amplitudes. In the rest of this 
introduction we will describe the most important qualitative features of our results. We first briefly review some kinematics in order to set terminology.

Consider the $2 \rightarrow 2$ scattering of particles in representations $R_1$ and $R_2$ of $U(N)$.  Let the tensor product of these two representations decompose as 
\begin{equation}\label{irreps}
 R_1 \times R_2 = \sum_m R_m.
\end{equation}
It follows from $U(N)$ invariance that the  S-matrix for the process takes the schematic form 
\begin{equation} \label{sms}
S= \sum_m P_m S_m ,
\end{equation}
where $P_m$ is the projector onto the $m^{th}$ representation, and $S_m$ is the scattering matrix
in the `$m^{th}$' channel.

In this paper we study the $2 \rightarrow 2$ scattering matrices of the elementary quanta of theories 
with only fundamental matter.  In this situation $R_1$ and $R_2$,  are either both fundamentals, or one fundamental and one antifundamental.~\footnote{The scattering of two antifundamentals is simply 
related to the scattering of two fundamentals, and will not be considered separately in this paper.}
In the case of fundamental - fundamental scattering, $R_m$ is either the `symmetric' representation 
with two boxes in the first row (and no boxes in any other row) of the Young Tableaux, 
or the `antisymmetric' representation with two boxes in the first column and no boxes 
in any other column. In the case of fundamental - antifundamental 
scattering, $R_m$ is either the singlet or the adjoint representation. In this paper we will 
present computations or conjectures for the all orders S-matrices in all the four channels 
mentioned above (symmetric, antisymmetric, singlet and adjoint) in both the bosonic 
and the fermionic theories. 

The scattering matrices of interest to us in this paper are already well known 
in the non-relativistic limit (i.e. in the limit in which the masses of the scattering particles 
and the center of mass energy are both taken to infinity at fixed momentum transfer) as we now 
very briefly review. The Chern-Simons equation of motion  ensures that 
each particle traps magnetic flux. The Aharonov-Bohm effect then ensures that the particle 
$R_1$ picks up the phase $2 \pi \nu_m$ as it circles around\footnote{Readers familiar with the relationship between Chern-Simons theory and WZW theory may recognize this formula in another guise. $\frac{C_2(R)}{k}$ is the holomorphic scaling dimension 
of a primary operator in the integrable representation $R$, and $e^{2 \pi i \nu_m}$ is the monodromy 
of the four point function $<R_1, ~R_2,  ~{\bar R}_1, ~ {\bar R}_2>$ in the conformal block 
corresponding to the OPE $ R_1 R_2 \rightarrow R_m$ .}  the particle $R_2$, where 
\begin{equation}\label{flux}
2 \pi \nu_m =\frac{4 \pi T_1^a T_2^a}{k}= 2 \pi \frac{ C_2(R_m)-C_2(R_1) - C_2(R_2)}{k},
\end{equation}
(where $T_{1/2}^a$ are the representation matrices for the group generators in representations 
$R_{1}$ and $R_2$  and $C_2(A)$ is the quadratic Casimir in representation $A$).
It follows as a consequence~\cite{Bak:1994dj} that the non-relativistic scattering amplitude in the $R_m$ exchange channel is given by the Aharonov-Bohm scattering amplitude of a $U(1)$ particle of unit charge of a point like 
magnetic flux of strength $2 \pi \nu_m$.  

It is easily verified that $\nu_m={\cal O}(\frac{1}{N})$ or smaller in the symmetric, antisymmetric or 
adjoint channels. In the singlet channel, however, it turns out that to leading order in the large 
$N$ limit  $\nu_m=\frac{N}{k}= \lambda$. 
It follows that the rotation by $\pi$ which interchanges the two scattering particles is accompanied by 
a phase $e^{-i \pi \lambda_B}$ in the bosonic theory and $(-1) e^{-i \pi \lambda_F}=e^{-i \pi (- {\rm sgn}(\lambda_F) + \lambda_F)}$ 
in the fermionic theory. \footnote{ The additional -1 in the fermionic theory 
comes from Fermi statistics. We have used  $-1=e^{ \pm i \pi}
= e^{-i \pi {\rm sgn}(\lambda_F)}$.} Note that these phases are identical when 
\begin{equation}\label{dm}
\lambda_B= \lambda_F -{\rm sgn}(\lambda_F).
\end{equation}
However \eqref{dm} is precisely the map between $\lambda_B$ and $\lambda_F$ 
\cite{Aharony:2012nh} induced by the level-rank duality 
transformation described at the beginning of this introduction. In the singlet channel, in  other words the bosons and conjecturally dual fermions are both effectively anyonic, with the same anyonic phase. This 
observation provides a partial  physical explanation for the duality map \eqref{dm}.

We note in passing that the anyonic phase $\pi \lambda_B$ is precisely twice the phase 
of the bulk interaction term in the conjectured Vasiliev duals to these theories \cite{Giombi:2011kc,Chang:2012kt}. Indeed 
the first speculation of the  bosonization duality for matter Chern-Simons theories \cite{Giombi:2011kc}
was  motivated by argument very similar to that presented in the previous paragraph but in the 
context of Vasiliev theories (deformations of the bosonic and fermionic theory that lead to the 
same interaction phase ought to be the same theory).  It would certainly be very interesting to find
 a logical link between the phase of interactions in Vasiliev theory and the anyonic phase of the previous paragraph, but we will 
not peruse this thread in this paper. 

Moving away from the non-relativistic limit, in this paper we have (following the lead of \cite{Aharony:2012nh}) summed all planar graphs to determine the exact relativistic S-matrix for both the bosonic as well as the fermionic theories in the symmetric, antisymmetric and adjoint channels.  Even though our completely explicit solutions are quite simple, they possess a rich analytic structure (see section \ref{res} for a detailed listing of results). It is a simple matter to compare the explicit results for the S-matrices in the bosonic and fermionic theories that are conjecturally dual to each 
other. We find that the bosonic and fermionic S-matrices agree perfectly in the adjoint channel. 
On the other hand the bosonic S-matrix in the symmetric/antisymmetric channels matches the 
fermionic S-matrix in the antisymmetric/symmetric channels. Our results are all consistent with the following rule: the bosonic S-matrix in the exchange channel $R_m$ is identical with the fermionic 
S-matrix in the exchange channel $R_m^T$, where $R_m^T$ is the dual representation
under level-rank duality. \footnote{In the large $N$ and large $k$ limit, the dual of a 
representation with a finite number of boxes plus a finite number of anti-boxes in the Young 
Tableaux is given by the following rule: we simply transpose the boxes and the anti-boxes 
in the Young Tableaux (i.e. exchange rows and columns independently for boxes and 
anti-boxes). According to this rule the fundamental, antifundamental, singlet and adjoint
representations are self-dual, while the symmetric and antisymmetric representations map 
to each other.}

 The match of S-matrices upto transposition appears to make perfect sense from 
 several points of view. Let us focus attention on the particle - particle scattering and 
 consider a multi-particle asymptotic state. As the Aharonov-Bohm phases 
 $\nu_m$ vanish in the large $N$ limit considered in this paper, the multi-particle 
 state in question is effectively a collection of non interacting bosonic particles, and so 
 must obey Bose statistics. As an example, consider a multi-particle state that is 
 completely antisymmetric under the interchange of its momenta. 
 In order to meet the requirement of Bose statistics, this state must also be completely 
 antisymmetric under the interchange of color indices. The corresponding dual 
 asymptotic state in the fermionic theory is also completely antisymmetric under the 
 interchange of momenta. In order to meet the requirement of fermionic statistics, 
 this state must thus be completely {\it symmetric} under the interchange of color 
 indices. In other words the map between bosonic and fermionic asymptotic states 
 must involve a transposition of color representations; this transposition is part of the 
 duality map between asymptotic states of the two theories, and is a reflection 
 of the bose -fermi nature of the duality. \footnote{It is not difficult to see how 
 the transposition of S-matrices emerges out of the difference between Bose and 
 Fermi statistics at the diagrammatic level. Scattering processes involving 
identical particles (both fundamentals or both antifundamentals) receive contributions both from 
`direct' scattering processes as well as `exchange' scattering process. The usual rules tell us that 
direct and exchange processes must be added together with a positive sign in the bosonic theory
but with a negative sign in the fermionic theory. The difference in relative signs implies that  
S-matrix in the symmetric channel (the sum of the exchange and direct S-matrices) 
in the bosonic theory is interchanged with 
the antisymmetric S-matrix (the difference between exchange and direct processes) 
the fermionic theory. }  See section \ref{disc} for further discussion of the map between the 
multi-particle states of this theory induced by duality. 
  
The transposition of exchange representations above might also have been anticipated from another 
point of view. In the 
pure gauge sector (i.e. upon decoupling the fundamental bosonic and fermionic fields by making 
them very massive), the conjectured duality between the bosonic and fermionic theories 
reduces to the level-rank duality between two distinct pure Chern-Simons theories. It is well known 
that, under level-rank duality, a Wilson line in representation $R$ maps to a Wilson line in the 
representation $R^T$. As a Wilson line in representation $R$ represents the trajectory of a particle 
in representation $R$, it seems very natural that the exchange channels in a dynamical scattering 
process also map to each other only after a transposition.

Before proceeding we pause to address an issue of possible confusion. 
We have asserted above that scalar and spinor S-matrices map to each other under duality. 
The reader whose intuition is built from the study of four dimensional scattering processes may 
find this confusing.  Scalar and spinor S-matrices cannot be equated in four or higher dimensions as they are functions of different variables. Scalar S-matrices are labelled by the momenta of the participating particles. On the other hand spinor S-matrices are labelled by both the momentum and the `polarization spinor' of the participating particles. In precisely three
dimensions, however, the Dirac equation uniquely determines the polarization spinor
of particles and antiparticles as a function of of their momenta \footnote{A related fact: the  little group for massive particles in 2+1 dimensions is $SO(2)$, which admits nontrivial one dimensional representations.}. 
It follows that three dimensional spinorial and scalar S-matrices are both functions  only of the 
momenta of the scattering particles, so these S-matrices can be sensibly identified.

For a technical reason we explain below we are unable to directly compute the S-matrix in the 
singlet exchange channel by summing graphs; given this technical limitation we are constrained 
to simply conjecture a result for this S-matrix. The reader familiar with the usual lore on 
scattering matrices may think this is an easy task. According to traditional wisdom, the  
S-matrices in a relativistic quantum field theory enjoy crossing symmetry.  Particle-antiparticle 
scattering in both channels should be determined from the results of particle-particle 
scattering; given the scattering amplitudes in the  symmetric and antisymmetric exchange 
channels, we should be able to obtain the results of scattering in the  singlet and adjoint exchange 
channels by analytic continuation. This principle yields a conjecture for the S-matrix 
in the singlet channel which, however, fails every consistency check: it has the wrong non 
relativistic limit and does not obey the constraints of unitarity. For this reason we propose that 
the usual rules of crossing symmetry are modified in the study of S-matrices in matter Chern 
Simons theories. 

 A hint that crossing symmetry might be complicated in these theories is present already in the non-relativistic limit as the Aharonov-Bohm scattering amplitude has an unusual $\delta$ function 
contribution localized about forward scattering \cite{Ruijsenaars:1981fp}. This contribution to the S-matrix has a simple 
physical origin: a wave packet of one particle that passes through another is diluted by the 
factor $\cos (\pi \nu_m)$ compared to the usual expectations because of destructive interference 
from Aharonov-Bohm phases; as a consequence the S-matrix includes a term proportional 
to $(\cos (\pi \nu_m) -1) I$ ($I$ is the identity S 
 matrix; see subsection \ref{ik} for more details).    The non-analyticity of this term
makes it difficult to imagine it can be obtained from a procedure involving analytic continuation. 
 
In addition to the singular  $\delta$ function piece, the scattering amplitude has an 
analytic part. In this paper (and in the large $N$ limit studied here) we conjecture that 
this analytic piece is given by the naive analytic continuation from the particle-particle 
sector, multiplied by the factor 
$$f(\lambda)=\frac{\sin (\pi \lambda)}{\pi \lambda}.$$
This conjecture passes several consistency checks; it yields a result consistent with 
the expectations of unitarity, and has the right non-relativistic limit, and yields
$S$-channel S-matrices that transform into each other under Bose-Fermi duality.

The factor $f(\lambda)$ is familiar in the study of pure Chern-Simons theory; $N$ times this factor is the expectation value of a circular Wilson loop on $S^3$ in the large $N$ limit. In section 
\ref{expmd} below we present a tentative explanation for why one should have {\it expected} S-matrices in matter 
Chern-Simons theories to obey the modified analyticity relation with precisely the  factor 
$f(\lambda)$. Our tentative explanation has its roots in the fact that the fully gauge invariant object that obeys crossing symmetry is the `S-matrix' computed in this paper dressed with external Wilson 
lines linking the scattering particles. The presence of the Wilson lines leads to an additional 
contribution (in addition to those considered in this paper) that we argue to be channel dependent; 
in fact we argue that the ratio of the additional contributions in the two channels is precisely the 
given by the factor above, explaining why the `bare' S-matrix computed in this paper has 
`renormalized' crossing symmetry properties. If our tentative explanation of this feature is 
along the right tracks, then it should be possible to find a refined argument that predicts the 
analytic structure and crossing properties of the S-matrix at finite values of $N$ and 
$k$.  We leave this exciting task for  the future. 

We note also that the factor $f(\lambda)$ appears also in the normalization of two point functions 
of, for instance, two stress tensors (see \cite{Aharony:2012nh}). The appearance of this factor
in the two point functions of gauge invariant operators seems tightly tied to the appearance of the 
same factor in scattering in the singlet channel, as the diagrams that contribute to these processes 
are very similar. It would be interesting to understand this relationship better.

This paper is organized as follows. In section \ref{back} below we describe the theories we 
study in this paper, review the conjectured level-rank dualities between the 
bosonic and fermionic theories, set up the notation and conventions for the scattering process we study, 
review the constraints of unitarity on scattering and review the known non-relativistic limits of 
the scattering matrices. In section \ref{res} below we briefly summarize the method we use to 
 compute S-matrices, and  provide a detailed listing of the principal results and conjectures of 
our paper. We then turn to a systematic presentation of our results. In section \ref{boson} we 
compute the S-matrices of the bosonic theories by solving the relevant Schwinger-Dyson 
equations. In section \ref{landauonshell} we verify the results of section \ref{boson} at one loop 
by a direct diagrammatic evaluation of the S-matrix in the covariant Landau gauge. 
In section \ref{fermion} we compute the S-matrix of the  fermionic theories by solving
a Schwinger-Dyson equation and verify 
the equivalence of our bosonic and fermionic results under duality. In section \ref{sing} we present 
our conjecture for the $S$-channel scattering amplitudes (in the bosonic and fermionic systems) of our theory, and provide a heuristic explanation for the unusual transformation properties under crossing symmetry 
obeyed by our conjecture. In section \ref{disc} we end with a discussion of our results and
of promising future directions of research. Several appendices contain technical details of the 
computations presented in this paper.

\section{Statement of the problem and review of  background material}\label{back}

This section is organized as follows. In subsection \ref{th} we 
describe the theories we study.  In subsection \ref{id} we review the conjectured duality between the bosonic and fermionic theories. In subsection \ref{ik} we review relevant aspects of 
the kinematics of $2 \rightarrow 2$ scattering in 3 dimensions, with particular emphasis on
the structure of the `identity' scattering amplitude, which will turn out to be 
renormalized in matter Chern-Simons theories. In subsection \ref{chan} 
 describe the precise scattering processes we study in this paper. 
In subsection \ref{abs} we review the known non-relativistic limits of these scattering amplitudes. 
In subsection \ref{uni} we describe the constraints on these amplitudes from the requirement of 
unitarity.

\subsection{Theories}\label{th}

As we have explained above, in this paper we study two classes of large $N$ Chern-Simons theories coupled to matter fields in the fundamental representation. The first family of theories we study involves  a single complex bosonic field, in the fundamental representation of $U(N)$, minimally coupled to a 
Chern-Simons coupled gauge field. In the rest of this paper we refer to this class of theories as 
`bosonic theories'. The second family of theories we study involves a single complex fermionic field 
in the fundamental representation of $U(N)$, minimally coupled to a 
Chern-Simons coupled gauge field. In the rest of this paper we refer to this class of theories as 
`fermionic theories'.

\begin{figure}[tbp] 
   \begin{center}
     \subfigure[]{\includegraphics[scale=0.75]{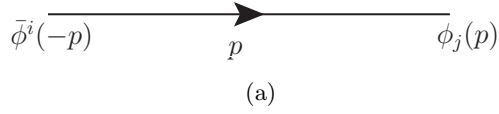}
 }
  
   \end{center}
   \vspace{-0.5cm}
   \caption{Propagator of bosonic particles.}
\label{B-pro}
   \end{figure}

 The bosonic system we study is described by the Euclidean Lagrangian 
\beal{
S  &= \int d^3 x  \biggl[i \varepsilon^{\mu\nu\rho} \frac{k_{B}}{4 \pi}
\Tr( A_\mu\partial_\nu A_\rho -\frac{2 i}{3}  A_\mu A_\nu A_\rho)
+  D_\mu \bar \phi  D^\mu\phi 
+m_B^2 \bar\phi \phi+\frac{1}{2 N_B} b_4(\bar{\phi}\phi)^2\biggl]
\label{sclag}} 
with $\lambda_{B}=\frac{N_B}{k_B}$.  
Throughout this paper we employ the dimensional regularization 
scheme and light cone gauge employed in the original study of \cite{Giombi:2011kc}.
The theory \eqref{sclag} has been studied  intensively in the recent literatures~\cite{Aharony:2012nh,Aharony:2011jz,Aharony:2012ns,GurAri:2012is,Jain:2012qi,Jain:2013gza,Jain:2013py,Takimi:2013zca,Yokoyama:2013pxa,Maldacena:2012sf}. It has in particular been 
demonstrated that in the regulation scheme and gauge employed in this paper, the 
bosonic propagator is given, at all orders in $\lambda_B$,  by the extremely simple form  
\begin{equation} \label{propB}
\begin{split}
\langle { \phi_j(p) \bar \phi}^i(-q) \rangle &= \frac{(2 \pi)^3 \delta^i_j \delta^3(-p+q)}
{ p^2+ c_B^2 } \\
\end{split}
\end{equation}
where the pole mass, $c_B$ is a function 
of $m_B, b_4$ and $\lambda_B$, given by 
\begin{equation}\label{cbf}
c_B^2= \frac{\lambda_B^2}{4}c_B^2- \frac{b_4}{4 \pi} |c_B|+m_B^2.
\end{equation}
 (see e.g. 
Eqn $1.5$ of \cite{Jain:2013gza} setting $x_4=0$ setting temperature $T$ to zero).
In all the Feynman diagrams presented in this paper, we adopt the following convention. 
The propagator \eqref{propB} is denoted by a line with an arrow from 
${\bar \phi}$ to $\phi$, 
with moment $p$ in the direction of the arrow (see Fig.~\ref{B-pro}). 

\begin{figure}[tbp] 
   \begin{center}
     \subfigure[]{\includegraphics[scale=0.75]{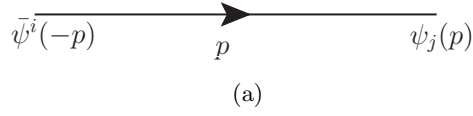}
 }
  
   \end{center}
   \vspace{-0.5cm}
   \caption{Propagator of fermionic particles.}
\label{F-pro}
   \end{figure}

The fermionic system we study is described by the   Lagrangian
\beal{
S  &= \int d^3 x  \biggl[i \varepsilon^{\mu\nu\rho} \frac{k_F}{4 \pi}
\Tr( A_\mu\partial_\nu A_\rho -\frac{2 i}{3}  A_\mu A_\nu A_\rho)
+  \bar\psi \gamma^\mu  D_\mu \psi
+m_F \bar\psi \psi \biggl]\label{ffcs}}
with $\lambda_{F}=\frac{N_F}{k_F}$. This theory has also been studied intensively in the 
recent literatures~\cite{Aharony:2012ns,GurAri:2012is,
Jain:2012qi,Jain:2013gza,Jain:2013py,Takimi:2013zca,
Yokoyama:2013pxa,Giombi:2011kc,
Maldacena:2012sf}. In particular it has been demonstrated that the fermionic propagator 
 is given (in the light cone gauge and dimensional regulation scheme of this paper), to all orders in $\lambda_F$, by \cite{Giombi:2011kc,
Aharony:2012ns,GurAri:2012is,Jain:2013gza} 
\begin{equation} \label{propF}
\begin{split}
\aver{ \psi_j(p) \bar\psi^i(-q)} &= \frac{ \delta^i_j  (2 \pi)^3 \delta^3(-p+q)}{i\gamma^\mu  p_\mu +  \Sigma_F(p)},
\end{split}
\end{equation}
where
\beal{ 
{\Sigma_F}(p)=& i\gamma^{\mu}\Sigma_{\mu}(p)+\Sigma_{I}(p) I,  \nn
\Sigma_{I} ( p) =& m_F+\lambda_F \sqrt{c_F^2+p_s^2}, \nn
\Sigma_{\mu}(p) =&\delta_{+\mu}\frac{p_{+}}{p_s^2}\left(c_F^2- \Sigma_{I}^2(p) \right),\nn
c_{F}^2=&\left(\frac{{m_F}}{{\rm sgn}(m_F)-\lambda_F} \right)^2.
\label{sigmafg}
}
Here  $\gamma^{\mu}$ compose the Euclidean Clifford algebra, 
$$\{\gamma^{\mu}, \gamma^{\nu}\} = 2\delta^{\mu\nu}, 
\quad [\gamma^{\mu}, \gamma^{\nu}]= 2i\epsilon^{\mu\nu\rho}\gamma_{\rho}.
$$
The fermionic propagator presented above has a pole at $p^2=c_F^2$; 
so the quantity $c_F$ is the pole mass  - or true mass - of the fermionic quanta .
In all the Feynman diagrams presented in this paper, we adopt the following convention. 
The propagator \eqref{propF} is denoted by a line with an arrow from 
$\bar{\psi}$ to $\psi$, 
with momentum $p$ in the direction of the arrow (see Fig.~\ref{F-pro}). 

\subsection{Conjectured Bose-Fermi duality} \label{id}

The bosonic theory \eqref{sclag} may be rewritten as 
\begin{equation}\label{sclagm}
 \begin{split}
S  =& \int d^3 x  \biggl[i \varepsilon^{\mu\nu\rho} \frac{k_{B}}{4 \pi}
\Tr( A_\mu\partial_\nu A_\rho -\frac{2 i}{3}  A_\mu A_\nu A_\rho)
+  D_\mu \bar \phi  D^\mu\phi 
+m_B^2 \bar\phi \phi+\frac{1}{2 N_B} b_4(\bar{\phi}\phi)^2 \\ 
&- \frac{N_B}{2 b_4} \left( \sigma - \frac{b_4}{N_B} {\bar \phi \phi} 
 -m_B^2 \right)^2 \biggl].
\end{split}
\end{equation}
We have introduced a new field $\sigma$ in \eqref{sclagm}; upon integrating 
$\sigma$ out \eqref{sclagm} trivially reduces to \eqref{sclag}. Expanding out the 
last bracket in \eqref{sclagm} and ignoring the constant term, we find that 
\eqref{sclagm} may be rewritten as 
\begin{equation}\label{sclagmr}
S  = \int d^3 x  \biggl[i \varepsilon^{\mu\nu\rho} \frac{k_{B}}{4 \pi}
\Tr( A_\mu\partial_\nu A_\rho -\frac{2 i}{3}  A_\mu A_\nu A_\rho)
+  D_\mu \bar \phi  D^\mu\phi 
+ \sigma {\bar \phi} \phi + N_B \frac{m_B^2}{b_4} \sigma
- N_B \frac{\sigma^2}{2 b_4}  \biggl].
\end{equation}
The so called Wilson-Fisher limit of the bosonic theory is obtained by taking the limit 
\begin{equation}\label{crilim}
b_4 \to \infty, ~~~ m_B \to \infty , ~~~ \frac{4 \pi m_B^2}{b_4}= m_B^{ {\rm cri }} = {\rm fixed}.
\end{equation}
In this limit the last term in \eqref{sclagmr} may be omitted; moreover 
 it follows from \eqref{cbf} that in this limit 
$$|c_B|=m_B^{ {\rm cri}}.$$
Note, of course, that this equation has no solution for negative $m_B^{ {\rm cri}}$. 
As was explained in \cite{Aharony:2012nh, Jain:2013gza} this is plausibly 
a reflection of the fact that \eqref{cbf} is the saddle point equation for an uncondensed 
solution, whereas the scalar in the theory wants to condense when $m_B^{ {\rm cri}} <0$. 
The determination of the condensed saddle point is a fascinating but unsolved problem, 
and in this paper we restrict our attention to the case $m_B^{ {\rm cri}}>0$.

As we have mentioned in the introduction, it has been has been conjectured 
that the scalar theory in the Wilson-Fisher limit described above is dual 
to the theory \eqref{ffcs},
\footnote{A preliminary suggestion for this duality may be found in 
\cite{Giombi:2011kc} . The  conjecture was first clearly stated, for the massless theories in 
\cite{Aharony:2012ns}, making heavy use of the results of \cite{Maldacena:2011jn,Maldacena:2012sf}. 
The conjecture was generalized to the massive theories in \cite{Aharony:2012nh} and further 
generalized in \cite{Jain:2013gza}. Additional evidence for this conjecture is presented in 
\cite{GurAri:2012is, Jain:2013py,Takimi:2013zca}. } once we identify 
parameters according to 
\begin{equation} \label{dualitymap}
\begin{split}
k_F&= -k_B, \\
N_F&=|k_B|-N_B, \\
\lambda_B&=\lambda_F -{\rm sgn}(\lambda_F), \\
m_F&=-m_B^{ {\rm cri}} \lambda_B.\\
\end{split}
\end{equation}
As we have explained above, we will restrict our attention to bosonic theories with 
$m_B^{{\rm cri}} >0$. It follows from \eqref{dualitymap} that, for the purpose of studying the 
bose-fermi duality, \footnote{We emphasize that all results obtained directly in the fermionic 
theory are valid irrespective of whether or not \eqref{fti} is obeyed. However we do not have 
a corresponding bosonic results to compare with when this inequality is not obeyed.} we should restrict attention to fermionic theories that obey the inequality 
\begin{equation} \label{fti}
\lambda_F m_F >0 .
\end{equation}

It is easily verified that \eqref{dualitymap} implies that 
\begin{equation}\label{pme}
|c_F|=|c_B|.
\end{equation}
In other words the bosonic and fermionic fields have equal pole masses under duality. 
This observation already makes it seem likely that the duality map should involve
some sort of identification of elementary bosonic and fermionic quanta. \footnote{Note that this 
is very different from sine-Gordon-Thirring duality, in which elementary fermionic 
quanta are identified with solitons in the bosonic theory.} The relationship between
bosonic and fermionic S-matrices, proposed in this paper, helps to flesh this identification
out.

\subsection{Scattering kinematics} \label{ik}

In this paper we study $2 \rightarrow 2$ particle scattering; for this purpose we work in 
Minkowski space.  Let the 3 momenta of the initial particles be denoted 
by $p_1$ and $p_2$ and let the momenta of the final particles be denoted by  $-p_3$ and $-p_4$. Momentum conservation ensures $p_1+p_2+ p_3+p_4=0$. We use the mostly positive 
sign convention, and define the Lorentz invariants $s, t, u$ in the usual manner
\begin{equation}\label{stu1}
s=-(p_1+p_2)^2, ~~t=-(p_1+p_3)^2, ~~u=-(p_1+p_4)^2, ~~s+t+u=4c_{B}^2
\end{equation}
where $c_{B}$ is the pole mass of the scattering particles (the scattering particles have equal mass). 

The $S$ matrix for the scattering processes is given by (see below for slight modifications to deal with 
bosonic or fermionic statistics)
\begin{equation}\label{smatrix} \begin{split}
{\bf S}(p_1, p_2,- p_3, -p_4)=&(2 E_{ {\vec p}_1}) (2 \pi)^2 \delta^2( {\vec p}_1 + {\vec p}_3)
(2 E_{{\vec p}_2}) (2 \pi)^2 \delta^2( {\vec p}_2 + {\vec p}_4) \\
&+
i (2 \pi)^3 \delta^3(p_1+p_2+p_3+p_4)  T(s, t, u, E(p_1, p_2, p_3)),\\
 E_{{\vec p}}=&\sqrt{c_{B}^2+ {\vec p}^2},\\
E(p_1, p_2, p_3)=&\pm 1 = {\rm sgn} \left (\epsilon_{\mu\nu\rho} p_1^\mu p_2^\nu p_3^\rho \right),~~~
\epsilon_{012}=- \epsilon^{012}= 1
\end{split}
\end{equation}

The fact that $2 \rightarrow 2$ scattering can depend on the $Z_2$ valued variable 
$E(p_1, p_2, p_3)$ rather than just $s, t, u$ is a kinematical peculiarity of 
3-dimenensions. Note that $E(p_1, p_2, p_3)$ 
measures the `handedness' of the triad of vectors $p_1, p_2, p_3$. The symbol ${\vec p}$ that appears in \eqref{smatrix} denotes the spatial part of the 3-vector $p$. It might seem to be 
strange that ${\vec p}$ makes 
any appearance in the formula for a Lorentz covariant S-matrix. 
Note, however, that the various 3-vectors we deal with are always on-shell, so the knowledge of ${\vec p}$ is sufficient 
to permit the reconstruction of the full 3-vector $p$. Using the 
on-shell condition it is not difficult to verify that $(2 E_{\vec p}) (2\pi)^2\delta^2( {\vec p} + {\vec r})$ is Lorentz invariant, even though this is not completely manifest.

The manifestly Lorentz invariant rule for the multiplication of two S-matrices is 
\begin{equation}\label{smatmult}  \begin{split}
& [{\bf S_1 S_2 }](p_1, p_2,- p_3, -p_4) \\
=&\int \frac{d^3 r_1  (2 \pi) \theta(r_1^0)\delta(r_1^2+c_{B}^2) }{(2 \pi)^3}  \frac{d^3 r_2 ( 2 \pi) 
\theta(r_2^0)\delta(r_2^2+c_{B}^2)}{(2 \pi)^3} 
\\
&~~ \times {\bf S_1}(p_1, p_2, -r_1,- r_2)  {\bf S_2}(r_1, r_2, -p_3,- p_4)\\
=& \int \frac{d^2 {\vec r}_1}{2 E_{ {\vec r}_1} (2 \pi)^2 } \frac{d^2 {\vec r}_2}{2 E_{ {\vec r}_2} (2 \pi)^2 }
 {\bf S_1}(p_1, p_2, -r_1,- r_2)  {\bf S_2}(r_1, r_2,- p_3,- p_4).\\
\end{split}
\end{equation}
The quantity 
\begin{equation}\label{identsm} 
I(p_1, p_2,- p_3,- p_4)=(2 E_{ {\vec p}_1}) (2 \pi)^2 \delta^2( {\vec p}_1 + {\vec p}_3)
(2 E_{{\vec p}_2}) (2 \pi)^2 \delta^2( {\vec p}_2 + {\vec p}_4) 
\end{equation}
that appears in the first line of \eqref{smatrix} is clearly the identity matrix for this multiplication rule. 

The identity matrix may be rewritten in a manifestly Lorentz invariant form (see Appendix \ref{idt} )  
\begin{equation}\label{idm}
I(p_1, p_2,- p_3,- p_4)= \lim_{\epsilon \to 0} 4 \pi \sqrt{s} \delta \left( \sqrt{ \frac{4 t}{t+u} } - \epsilon \right) (2 \pi)^3 
\delta^3(p_1+p_2+p_3+p_4).
\end{equation} 

It is sometimes convenient to study $2 \rightarrow 2$ scattering in the center of mass frame. In this frame the scattering momenta may be taken to be 
\begin{equation}\label{scatmom} \begin{split}
& p_1=(\sqrt{k^2+c_{B}^2}, k, 0), ~~~p_2=(\sqrt{k^2+c_{B}^2}, -k, 0) \\
&p_3=(-\sqrt{k^2+c_{B}^2}, -k \cos (\theta), 
-k \sin (\theta)), ~~~p_4=(-\sqrt{k^2+c_{B}^2}, 
k \cos (\theta),  k\sin (\theta)). \\
\end{split}
\end{equation}
The kinematical invariants are given by 
\begin{equation}\label{stucm}
s= 4(c_{B}^2 +k^2), ~~~t=-2 k^2\left(1-\cos(\theta) \right), ~~~u=-2 k^2\left(1 + \cos (\theta) \right),
\end{equation}
and the S-matrix 
takes the form 
\begin{equation}\label{sform}
{\bf S}= (2 \pi)^3 \delta(p_1+p_2+p_3+p_4) S(\sqrt{s}, \theta),
\end{equation}
where $\theta$ is the scattering angle - the angle between $-{\vec p}_3$ and ${\vec p}_1$. More precisely, 
let ${\vec p}_1$ point along the positive $x$ axis so that ${\vec p}_2$ points along the negative $x$ axis. 
$\theta \in (-\pi, \pi)$ is defined as the rotation in the clockwise direction (here clockwise is defined w.r.t. the orientation of the 
usual $x, y$ axis system) that is needed to rotate ${\vec p}_1$ into $-{\vec p}_3$. Note that parity transformations,
that take $\theta$ to $-\theta$, are generically not symmetries of our theory.  In the center of mass system $E(p_1, p_2, p_3)$ defined in \eqref{smatrix} is given by 
\begin{equation}\label{eppp}
E(p_1, p_2, p_3)={\rm sgn}(\theta).
\end{equation}
For later use we note the following center of mass reduction formulae
\begin{equation}\begin{split}\label{redform}
E(p_1, p_2, p_3) \sqrt{ \frac{su}{t} }& \rightarrow \sqrt{s} \cot \left(\frac{\theta}{2}\right),\\
E(p_1, p_2, p_3) \sqrt{ \frac{st}{u} }& \rightarrow \sqrt{s} \tan \left(\frac{\theta}{2}\right),\\
E(p_1, p_2, p_3) \sqrt{ \frac{tu}{s} }& \rightarrow  \frac{2 k^2}{\sqrt{s} } \sin (\theta).\\
\end{split}
\end{equation}

The rule \eqref{smatmult}
induces the following multiplication rule for the functions $S(\sqrt{s}, \theta)$:
\begin{equation} \label{cmprod}
[S_1 S_2](\sqrt{s}, \theta) = \int \frac{d \alpha}{8 \pi \sqrt{s}} S_1(\sqrt{s}, \alpha) S_2(\sqrt{s}, \theta - \alpha),
\end{equation}
The identity matrix for this multiplication rule is clearly given by 
\begin{equation}\label{cmf}
S_I(\sqrt{s}, \theta)=8 \pi \sqrt{s} \delta(\theta) = \lim_{\epsilon \to 0} 4 \pi \sqrt{s} 
\left[ \delta ( \theta + \epsilon) + \delta(\theta -\epsilon) \right],
\end{equation}
in agreement with \eqref{idm} recast in center of mass coordinates. 

The Hermitian conjugate of an S-matrix functions for $S^\dagger$ are given by 
\begin{equation}\label{dsdag}\begin{split}
&[{\bf S}^\dagger](p_1, p_2, -p_3,- p_4)= {\bf S}^*(p_3, p_4, -p_1, -p_2),\\
&[ S^\dagger](\sqrt{s}, \theta)= S^*(\sqrt{s}, -\theta).\\
\end{split}
\end{equation}
The S-matrix must be unitary, i.e. must obey the equation  $S^\dagger S=1$. This implies 
\begin{equation}\label{unitarity}
-i (T-T^\dagger)=T^\dagger T.
\end{equation}
Written out as an explicit equation for the $T$ functions this boils down to 
\begin{equation}\begin{split} \label{consteq}
 &-i \left(T(p_1,p_2,-p_3,-p_4)-T^{\ast}(p_3,p_4,-p_1,-p_2)\right)\delta^3(p_1+p_2+p_3+p_4)\\
&=\int \frac{d^3 l}{(2\pi)^3}\frac{d^3 r}{(2\pi)^3} 
\biggl[
\theta(-l_0)\theta(-r_0)\delta^3(p_1+p_2+p_3+p_4)\delta^3(p_1+p_2+l+r) 
\\
&~~~~~~~~~~~\times (2 \pi) \delta(r^2+c_{B}^2)
(2 \pi) \delta(l^2+c_{B}^2)
T^*(-p_1,-p_2,l,r)T(-p_3,-p_4,l,r)\biggr] + \ldots
\end{split}
\end{equation}
where the $\ldots$ denotes the contribution of intermediate states with more than two particles. 
We will return to this formula below 

\subsection{Channels of scattering} \label{chan}

A theory of a fundamental field has two kinds of elementary quanta: those that transform in the fundamental
of $U(N)$ and those that transform in the antifundamental of that gauge group.  
In this paper we refer to quanta in the fundamental of $U(N)$ as particles; we refer to quanta in the
antifundamental of $U(N)$ as antiparticle. We use the symbol $P_i(p)$ to denote a particle with color
index $i$ and three momentum $p$, while $A^i(p)$ denotes an antiparticle with color index $i$ and
three momentum $p$. We employ this notation for both the bosonic and the fermionic theories 
described in the previous subsection.

In this paper we study $2 \rightarrow 2 $ scattering. There are essentially two distinct $2 \rightarrow 2$
scattering process; particle-particle scattering and Particle-antiparticle scattering \footnote{The case of 
antiparticle-antiparticle scattering is related to that of particle-particle scattering by CPT, and 
so needn't be considered separately.}

\subsubsection{Particle - antiparticle scattering} \label{pa}

The tensor product of a fundamental and an antifundamental consists of the 
adjoint and the singlet representations. It follows that Particle-antiparticle scattering 
is characterized by two scattering functions. We adopt the following terminology: we 
refer to scattering in the singlet channel as scattering in the $S$-channel. Scattering 
in the adjoint channel is referred to as scattering in the $T$-channel. 

It follows from $U(N)$ invariance that the S-matrix for the process
\begin{equation}\label{pas}
P_i(p_1) + A^j(p_2) \rightarrow P_m(-p_3) + A^n(-p_4)
\end{equation}
is given by 
\begin{equation}
 {\bf S}=\delta_{i}^{m}\delta^{j}_{n}I(p_1, p_2, -p_3,- p_4)+ i T_{in}^{jm}(p_1,p_2,-p_3,-p_4) (2 \pi)^3\delta^3(p_1+p_2+p_3+p_4).
\end{equation}
(see the previous subsection for the definition of $I$). The S-matrix may be decomposed into adjoint and singlet scattering matrices
\begin{equation}\label{sde}
 {\bf S}=\left( \delta_{i}^{m}\delta^{j}_{n} - \frac{\delta_i^j \delta^m_n}{N} \right) S_T
+ \frac{\delta_i^j \delta^m_n}{N} S_S
\end{equation}
where 
\begin{equation} \label{st} \begin{split}
&{\bf S}_T= I(p_1, p_2, -p_3,- p_4)+ i T_T(p_1,p_2,-p_3,-p_4) (2 \pi)^3 \delta^3(p_1+p_2+p_3+p_4)\\
&{\bf S}_S= I(p_1, p_2,- p_3, -p_4)+ 
i T_S(p_1,p_2,-p_3,-p_4)(2 \pi)^3\delta^3(p_1+p_2+p_3+p_4)
\end{split}
\end{equation}
and
\begin{equation}\label{Tijmn}
T_{in}^{jm}(p_1,p_2,-p_3,-p_4) = \left( \delta_{i}^{m}\delta^{j}_{n} - \frac{\delta_i^j \delta^m_n}{N} \right) T_T(p_1,p_2,-p_3,-p_4)
+ \frac{\delta_i^j \delta^m_n}{N} T_S(p_1,p_2,-p_3,-p_4)
\end{equation}

\subsubsection{Particle - particle scattering} \label{pp}

The tensor product of two fundamentals consists of the representation with two 
boxes in the first row of the Young Tableaux, and another representation with two 
boxes in the first column of the Young Tableaux. We refer to these two representations 
as the symmetric $U$-channel and the antisymmetric $U$-channel respectively. 
It follows that particle- particle scattering is characterized by the scattering functions 
in these two channels. 
 
 More quantitatively, the S-matrix for the process 
\begin{equation} \label{pps}
P_i(p_1) + P_j(p_2) \rightarrow P_m(-p_3) + P_n(-p_4) 
\end{equation} takes the form
\begin{equation}
\begin{split}
 {\bf S}&=\pm \delta_{i}^{m}\delta_{j}^{n} I(p_1, p_2, p_3, p_4) +\delta_{i}^{n}\delta_{j}^{m} I(p_1, p_2, p_4, p_3)\\
&+ i T_{ij}^{mn}(p_1,p_2,p_3,p_4) (2 \pi)^3 \delta^3(p_1+p_2+p_3+p_4)               
\end{split}
\end{equation} where the $\pm$ in the first line is for bosons/fermions. The S-matrix may be 
decomposed into the symmetric and antisymmetric channels 
\begin{equation}\label{sdecomp}
{\bf S}= \frac{ \delta_{i}^{n}\delta_{j}^{m} + \delta_{i}^{m}\delta_{j}^{n} }{2} {\bf S}_{U_s}
+ \frac{ \delta_{i}^{n}\delta_{j}^{m} - \delta_{i}^{m}\delta_{j}^{n} }{2} {\bf S}_{U_a}
\end{equation}
where 
\begin{equation}\label{stpp} \begin{split}
{\bf S}_{U_s} &= \pm I(p_1, p_2, p_3, p_4)   +  I(p_1, p_2, p_4, p_3)  \\ 
& +i T_{U_s}(p_1, p_2, p_3, p_4) (2 \pi)^3 \delta(p_1+p_2+p_3+p_4)\\
{\bf S}_{U_a}&=-(\pm) I(p_1, p_2, p_3, p_4)   +  I(p_1, p_2, p_4, p_3) \\
&+ i T_{U_a}(p_1, p_2, p_3, p_4) (2 \pi)^3 \delta(p_1+p_2+p_3+p_4).\\
\end{split}
\end{equation}
We will sometimes need to work with the direct 
and exchange scattering amplitudes (${\bf S}_{U_d}$ and  ${\bf S}_{U_e}$) by 
\begin{equation}\label{sdecomp1}
{\bf S}= \delta_{i}^{m}\delta_{j}^{n} {\bf S}_{U_d} + \delta_{i}^{n}\delta_{j}^{m}  {\bf S}_{U_e}
\end{equation}
where 
\begin{equation}\label{stpp1} \begin{split}
{\bf S}_{U_d} &= \pm  I(p_1, p_2, p_3, p_4)  +i T_{U_d}(p_1, p_2, p_3, p_4) (2 \pi)^3 \delta(p_1+p_2+p_3+p_4)\\
{\bf S}_{U_e}&= I(p_1, p_2, p_4, p_3)+ i T_{U_a}(p_1, p_2, p_3, p_4) (2 \pi)^3 \delta(p_1+p_2+p_3+p_4)\\
\end{split}
\end{equation}
where
\begin{equation}\label{oac} \begin{split}
{\bf S}_{U_s} ={\bf S}_{U_d} +{\bf S}_{U_e}, ~~~{\bf S}_{U_a} ={\bf S}_{U_e} -{\bf S}_{U_d}, ~~~T_{U_s} =T_{U_d} +T_{U_e}, ~~~T_{U_a} =T_{U_e} -T_{U_d}.
\end{split}
\end{equation}
And
\begin{equation}\label{TijmnP}
T_{ij}^{mn}(p_1,p_2,p_3,p_4) = \delta_{i}^{m}\delta_{j}^{n} T_{U_d}(p_1,p_2,p_3,p_4)
+ \delta_i^n \delta^m_j T_{U_e}(p_1,p_2,p_3,p_4).
\end{equation}
We refer to  $S_{U_d}$ as the `direct S-matrix' in the $U$-channel. $S_{U_e}$, on the other 
hand is the `exchange S-matrix in the $U$-channel.

In this paper we study scattering in both the bosonic as well as fermionic theories described in the 
previous subsection. We use the superscript $B/F$ to denote the corresponding functions in the
 bosonic/fermionic theories. For example $S_T^B$ is the $T$-channel scattering matrix for bosons, 
while $S^F_S$ denotes the $S$-channel scattering matrix for fermions.

\subsection{Tree level scattering amplitudes in the bosonic and fermionic theories}\label{treel}

The evaluation of full S-matrix of the bosonic and fermionic theories of subsection \ref{th} is the main 
subject of this paper. The evaluation of the all loop amplitudes will require summing all planar diagrams
in lightcone gauge, together with some educated guesswork. However the tree level scattering amplitudes 
in these theories are, of course, easily evaluaed in a covariante Landau gauge. In this section we simply 
present the results for these tree level scattering amplitudes, in all scattering channels, in both the bosonic and the fermionic theories. In every case we present the results for the full $S$ matrix (rather than the 
$T$ matrix) to emphasize the relative sign between the identity piece and the scattering terms. 
In the scalar theories we work for simplicity at $b_4=0$. Our results 
in the fermionic theory are presented upto a physically irrelevant overall phase. The results presented in this 
subsection are all derived in Appendix \ref{treeap}.

At tree level we find
\be \label{treeanswers}
\begin{split}
{\bf S}_{B,U_d} &=   I(p_1, p_2, p_3, p_4)  -\frac{4\pi}{k_B}  \frac{\epsilon_{\mu\nu\rho}p_1^\mu p_2^{\nu}p_3^{\rho}}{(p_2+p_3)^2} (2 \pi)^3 \delta(p_1+p_2+p_3+p_4)\\
{\bf S}_{B,U_e}&= I(p_1, p_2, p_4, p_3)+ \frac{4\pi}{k_B}  \frac{\epsilon_{\mu\nu\rho}p_1^\mu p_2^{\nu}p_3^{\rho}}{(p_2+p_4)^2} (2 \pi)^3 \delta(p_1+p_2+p_3+p_4)\\   
{\bf S}_{B,T} &=   I(p_1, p_2, p_3, p_4)  +\frac{4\pi}{k_B}  \frac{\epsilon_{\mu\nu\rho}p_1^\mu p_2^{\nu}p_3^{\rho}}{(p_4+p_3)^2} (2 \pi)^3 \delta(p_1+p_2+p_3+p_4)\\
{\bf S}_{B,S}&=I(p_1, p_2, p_3, p_4)- 4\pi\lambda_B  \frac{\epsilon_{\mu\nu\rho}p_1^\mu p_2^{\nu}p_3^{\rho}}{(p_2+p_4)^2} (2 \pi)^3 \delta(p_1+p_2+p_3+p_4)\\   
{\bf S}_{F,U_d} &=  I(p_1, p_2, p_3, p_4) +\frac{4\pi}{k_F} \left( \frac{\epsilon_{\mu\nu\rho}p_1^\mu p_2^{\nu}p_3^{\rho}}{(p_2+p_3)^2}-2 i m_F\right) (2 \pi)^3 \delta(p_1+p_2+p_3+p_4)\\
{\bf S}_{F,U_e}&= I(p_1, p_2, p_4, p_3)- \frac{4\pi}{k_F} \left( \frac{\epsilon_{\mu\nu\rho}p_1^\mu p_2^{\nu}p_3^{\rho}}{(p_2+p_4)^2}+2 i m_F\right) (2 \pi)^3 \delta(p_1+p_2+p_3+p_4)\\   
{\bf S}_{F,T} &=   I(p_1, p_2, p_3, p_4)  -\frac{4\pi}{k_F}\left(  \frac{\epsilon_{\mu\nu\rho}p_1^\mu p_2^{\nu}p_3^{\rho}}{(p_4+p_3)^2}+2im_F\right) (2 \pi)^3 \delta(p_1+p_2+p_3+p_4)\\
{\bf S}_{F,S}&= -I(p_1, p_2, p_3, p_4)+ 4\pi\lambda_F \left( \frac{\epsilon_{\mu\nu\rho}p_1^\mu p_2^{\nu}p_3^{\rho}}{(p_2+p_4)^2}-2 i m_F\right) (2 \pi)^3 \delta(p_1+p_2+p_3+p_4).\\   
\end{split}
\ee

\subsection{The non-relativistic limit and Aharonov-Bohm scattering} \label{abs}

As we have explained above, in this paper we wish to compute the $2 \rightarrow 2$ scattering matrix 
of fundamental matter coupled to Chern-Simons theory. The result of this computation is already 
well known in the non-relativistic limit, i.e. the limit in which 
\begin{equation} \label{nrlims}
 \frac{s-4 c_{B}^2}{4 c_{B}^2} \to 0.
\end{equation}
\footnote{In the limit \eqref{nrlims} $t/4m^2$ and $u/4m^2$ also tend to zero, as is most easily 
seen in the center of mass frame.} 
In this limit the S-matrix is obtained from the scattering of  two non-relativistic particles interacting with a Chern-Simons gauge field.  The quantum description of this system may be obtained by 
first eliminating the non dynamical gauge field in a suitable gauge and then writing down 
the effective two particle Schrodinger equation see e.g. \cite{Bak:1994dj}). Moving to center of mass and relative coordinates further simplifies the problem to the study of the quantum mechanics of 
a single particle interacting with a point like flux tube located 
at the origin. The S-matrix may then be read off from the scattering solution of 
Aharonov-Bohm \cite{Aharonov:1959fk} with one interesting twist; the effective value of the flux depends on the scattering channel.

Let the scattering particles transform in the representations $R_1$ and $R_2$ of $U(N)$. As 
we have reviewed in the introduction, if 
\begin{equation}\label{cgd}
R_1 \times R_2 = \sum_m R_m
\end{equation}
then 
$$S= \sum_m S_m  P_m$$
where $P_m$ is the projector onto the representation $R_m$. It turns out that the 
scattering matrix in the $m^{th}$ channel $S_m$ is simply the Aharonov-Bohm scattering amplitude of  a unit charge  $U(1)$ particle scattering off a thin flux tube with integrated flux   $2 \pi \nu_m$ where 
\begin{equation}\label{effab}
\nu_m= \frac{C_2(R_m)-C_2(R_1)- C_2(R_2)}{k}.
\end{equation}

Let  $F$ denote the fundamental representation, $A$ the antifundamental
representation, $S$ the `symmetric' representation (with two boxes in the first row of the 
Young Tableaux, and no boxes in any other row), $AS$ the antisymmetric representation (with 
two boxes in the first column of the Young Tableaux, and no boxes in any other column), $Adj$
the adjoint representation and $I$ the  and the singlet. The Casimirs of these representations are 
\begin{equation}\label{qcas} \begin{split}
&C_2(F)=C_2(A)
=\frac{N^2-1}{2 N}, ~~~C_2(S)=\frac{N^2+N-2}{N} , ~~~C_2(AS)=\frac{N^2 -N-2}{N} \\
&C_2(Adj)=N, ~~~C_2(I)=0.
\end{split}
\end{equation}
In the symmetric and antisymmetric exchange channels respectively (for particle-particle scattering) 
\begin{equation}\label{effpp}
\nu_S= \frac{1}{k} -\frac{1}{Nk}, ~~~\nu_{AS}=-\frac{1}{k} -\frac{1}{Nk}.
\end{equation}
In the singlet and adjoint exchange channels respectively (for particle - antiparticle scattering)
\begin{equation}\label{effpa}
\nu_I= -\lambda_{B} +\frac{1}{Nk}, ~~~\nu_{AdJ}= \frac{1}{Nk}.
\end{equation}
Note that in the large $N$ limit, $\nu_I$ is of order unity,  $\nu_S$ and $\nu_{AS}$ 
are  both of order ${\cal O}(1/N)$ and $\nu_{Adj}$ is  of order  ${\cal O}(1/N^2)$.

In the rest of this subsection we specialize to scattering in the scalar theory. As we have reviewed
in great detail in Appendix \ref{ab}, the quantum mechanics of a non-relativistic scalar scattering of
a point like flux tube with integrated flux $2 \pi \nu$ admits a `scattering' solution (the Aharonov 
Bohm solution), whose large radius asymptotics is given by 
\begin{equation}\label{absol}
\psi(r) = e^{ikx} + e^{-i \frac{\pi}{4}} h(\theta) e^{ikr} \sqrt{ \frac{2 \pi}{kr} }
\end{equation}
where
\begin{equation}\label{ht}
h(\theta)= 2 \pi \left( \cos (\pi \nu) -1 \right) \delta(\theta) +\sin (\pi \nu)   \left( {\rm Pv}  \cot \left(\frac{\theta}{2}\right)
- i {\rm sgn}(\nu) \right)
\end{equation}
where $Pv$ denotes the principal value. In the non-relativistic limit and in the center of mass frame the scattering amplitude $T$ is proportional to $h(\theta)$; more precisely
\begin{equation} \label{nrred}
T(s, \theta) = -4 i  h(\theta) \sqrt{s}.
\end{equation}  
Using  \eqref{stucm} 
\eqref{nrred} and \eqref{ht} together imply the covariant prediction 
\begin{equation}\label{htg} \begin{split}
T_m^{NR}(p_1, p_2, p_3, p_4, \lambda_{B}, b_4)=& -4 i \sqrt{s}  \sin (\pi \nu_m) 
\left(  E(p_1, p_2, p_3) \sqrt{ \frac{ t}{u} } - i {\rm sgn}(\nu_m)  \right) 
\\
&- i (\cos (\pi \nu_m) -1 ) I(p_1, p_2, p_3, p_4) 
\end{split}
\end{equation}
(see \eqref{identsm} \eqref{idm}, \eqref{cmf} for a definition of $I$)
where $T_m^{NR}$ is the non-relativistic limit of scattering in the $m^{th}$ channel, 
$\nu_m$ is the corresponding value of $\nu$ as described above.  

\eqref{htg} applies when the scattering particles are distinguishable (as in the case of particle - antiparticle 
scattering in the situation of interest to our paper). When the scattering particles are identical - as in the case 
of particle - particle scattering in our paper, $R_1=R_2=R$ and we have to add the contribution of exchange scattering. \eqref{htg}
is modified to 
\begin{equation}\label{htgident} \begin{split}
T_m^{NR}(p_1, p_2, p_3, p_4, \lambda_{B}, b_4)= & -4 i \sqrt{s}  \sin (\pi \nu_m) 
\left(  E(p_1, p_2, p_3) \sqrt{ \frac{ t}{u} } - i {\rm sgn}(\nu_m)  \right) 
\\
&- i (\cos (\pi \nu_m) -1 ) I(p_1, p_2, p_3, p_4) \\
&+ a \biggl[-4 i \sqrt{s}  \sin (\pi \nu_m) 
\left( - E(p_1, p_2, p_3) \sqrt{ \frac{ u}{t} } - i {\rm sgn}(\nu_m)  \right) 
\\
&~~~~~~~~- i (\cos (\pi \nu_m) -1 ) I(p_2, p_1, p_3, p_4)     \biggr]
\end{split}
\end{equation}
where the sign $a=1$ if the $R_3$ is symmetric in the $R s$ while $a=-1$ if $R_3$ is antisymmetric 
product of 2 $R$s (in the case that the scattering particles are fermionic, $a$ has an additional overall -1). 
In writing \eqref{htgident}  we have used the fact that $E(p_2, p_1, p_3)=-E(p_1, p_2, p_3)$.

\subsubsection{Non-relativistic limit of $S$-channel scattering}

In the $S$-channel $\nu_m=\lambda_{B}$ in the large $N$ limit  so  the $S$-channel S-matrix must reduce, in the 
limit \eqref{nrlims}, to  
\begin{equation}\label{Tsbnr} \begin{split}
(T^S_B)^{NR}(p_1, p_2, p_3, p_4, \lambda_{B}, b_4)=  
&4 i  \sqrt{s} \sin (\pi \lambda_{B}) 
\left(  E(p_1, p_2, p_3) \sqrt{ \frac{t}{u} } + i   {\rm sgn}(\lambda_{B}) \right) 
\\
&- i (\cos (\pi \lambda_{B}) - 1)  I(p_1, p_2, p_3, p_4).
\end{split}
\end{equation}
This prediction for the non-relativistic limit of the S-matrix in the $S$-channel has several striking 
features.
\begin{itemize}
\item $T_S^B$ is not an analytic function of kinematic variables. The term proportional to the 
$\delta$ function in that expression is singular, and is infact proportional to the identity 
scattering matrix (see subsection \ref{ik}).
\item $T_S^B$ is not an analytic function of  $\lambda_{B}$ at $\lambda_{B}=0$ (because of the term 
proportional to ${\rm sgn}(\lambda_{B})$. ) 
\item $T_S^B$ is universal, in the sense that it is independent of $b_4$ in this limit.
\end{itemize}

As we will see below, the last two features are artifacts of the non-relativistic limit.  On the other 
hand we will now argue that the last the term in \eqref{ht} $\propto \delta(\theta)$ is an exact feature of the 
S-matrix at all energy scales. 

The term proportional to $\delta(\theta)$ in \eqref{ht} was infact missed in the original analysis by 
Aharonov and Bohm. The presence of this term was discovered much later by Ruijsenaars 
\cite{Ruijsenaars:1981fp} (see also the later papers \cite{Jackiw:1989qp,Bak:1994dj,Bak:1994zz,AmelinoCamelia:1994we}
for further elaboration) where it was also pointed out that this contact term is necessary to unitarize 
Aharonov-Bohm scattering (see the next subsection for a review of this fact). In the rest of this 
subsection we will present a simple physical interpretation for this part of the Aharonov-Bohm S-matrix.

As we have reviewed extensively in~\eqref{ik}, 
the scattering matrix is postulated the form $S=I+iT$
where the factor $I$ accounts for the unscattered part of the wave packet. 
In the context of Aharonov-Bohm scattering, however, 
half of this unscattered wave packet passes above the scatterer and so picks up the phase $e^{i \pi \nu_m}$
while the other half passes below and so picks up the phase $e^{-i \pi \nu_m}$. The symmetry between up and 
down ensures that the part of the unscattered part of the S-matrix is modulated by a factor $\cos (\pi \nu_m)$ as 
it passes by the scatterer. In the current context, consequently, we should expect
$$S=\cos (\pi \nu_m) + i T'$$
where $T'$ is an analytic function of momentum.  
If we insist nonetheless on using the usual split $S=I+iT$ then we will find 
$$T=-i (\cos (\pi \nu_m) -1) I  + T'$$
(where $T'$ is an analytic function of the scattering angle) in perfect agreement with \eqref{htg}. 
As our physical explanation of the last term on the RHS of \eqref{htg} makes no reference to the non-relativistic limit, 
we expect this term to be an exact feature of the S-matrix in every channel, even away from the non-relativistic limit. 

All our comments about the term proportional to $I$ in the S-matrix hold also for the $T$ and the 
$U$-channels; the last term in \eqref{htg} is expected to be exact in these channels as well. 
As we have noted above, however, in these channels $\nu_m \leq {\cal O}(\frac{1}{N})$ so that 
$\cos (\pi\nu_m)-1 \leq {\cal O}( \frac{1}{N^2})$. 
It follows that the ${\cal O}(\frac{1}{N})$ computations
of these scattering matrices presented in the current paper will be insensitive to these terms.

\subsubsection{Non-relativistic limit of scattering in the other channels} \label{nrtc}

As we have seen above,  $\nu_m$ is of order $\frac{1}{N}$ or smaller in the other three scattering channels. 
All the calculations in this paper are done to leading order in the $\frac{1}{N}$, and so capture the first 
term in the Taylor expansion in $\nu_m$ of the scattering amplitude. In this subsubsection we 
merely emphasize the simple but confusing fact that the non-relativistic limit of this term need not agree with 
the first term in the Taylor expansion of the non-relativistic limit \eqref{htg} (this is an order of limits issue).

Let us consider a simple example for how this might work. Define 
$y= \frac{\nu_m (4 m^2)}{s-4m^2}$, and consider the function 
$$f=\frac{e^y - e^{-y} }{e^y+ e^{-y} }.$$
Taylor expanding this function to first order in $\nu_m$, we find 
$$f= 2 y +{\cal O} (\nu_m^2).$$
The non-relativistic limit the first term in this expansion diverges like $y$. 
On the other hand if we first take the non-relativistic limit \eqref{nrlims} 
$$f = {\rm sgn}(\nu_m).$$ 

Conservatively, therefore, we should conclude that the results of this subsection make no 
sharp prediction for the non-relativistic limit of the scattering amplitudes in the 
$U$ and $T$-channels. This is certainly the case for the term independent of $\theta$ in 
\eqref{ht}; as in the toy example above, this term is non-analytic in $\nu_m$, and so cannot
be Taylor expanded in $\nu_m$, and so makes no prediction for the non-relativistic limit 
of the Taylor expansion.

On the other hand the term in \eqref{ht} proportional to $\cot \left(\frac{\theta}{2}\right)$ and 
$\delta(\theta)$ are both analytic in $\theta$, and one might optimistically hope that 
the Taylor expansion of these terms in $\nu_m$ will accurately capture the non-relativistic
limits of the scattering amplitudes in the $U$ and the $T$-channels. Below we will see 
that this is indeed the case, though it works in a rather trivial way.

\subsection{Constraints from unitarity} \label{uni}

As we have already remarked above, the $S$ matrix in any quantum theory obeys the equation $S^\dagger S=1$. In subsection \ref{ik} we expanded this equation out in terms of the 
T-matrix to obtain \eqref{consteq}. 

In a general quantum field theory \eqref{consteq} does not constitute a closed equation for 
$2 \rightarrow 2$ scattering because of the terms indicated with the $\ldots$ - the contributions 
from $2 \times n$ scattering -  in the RHS of  \eqref{consteq}. 
It is easily verified, however, that at leading order in the large $N$ limit in the theories under consideration
the contribution of $ 2 \rightarrow n$ processes to the RHS of \eqref{unitarity} is suppressed, compared to the LHS, by a factor of $\frac{1}{N^{\frac{n-2}{2}}}$. In the large $N$ limit 
of interest to this paper,  it follows that we can drop the $\ldots$ on the RHS of
 \eqref{consteq}, which then turns into a 
powerful nonlinear closed constraint on $2 \rightarrow 2$ scattering matrix elements. 

\subsubsection{Constraints from unitarity in the various channels}

Let us work out the specific form of this constraint in the special case of particle - antiparticle scattering.
Using \eqref{st}, we find 
\begin{equation}\label{tsunit}
\begin{split}
& 
-i \biggl[\left(T_{T}(p_1,p_2,-p_3, -p_4)-T_{T}^{\ast}(p_3,p_4,p_1,p_2)\right)
\\
&~~~~~~\times
(2 \pi)^3 \left(\delta_{im}\delta_{jn}-\frac{1}{N}\delta_{ij}\delta_{mn}\right) (2 \pi)^3 \delta^3(p_1+p_2 -p_3-p_4)\biggr] \\
& -i \biggl[\left(T_{S}(p_1,p_2,-p_3,-p_4)-T_{S}^{\ast}(p_3,p_4,-p_1,-p_2)\right)\delta^3(p_1+p_2-p_3-p_4)\frac{1}{N}\delta_{ij}\delta_{mn}\biggr]\\
=&\int \frac{d^3 l}{(2\pi)^3}\frac{d^3 r}{(2\pi)^3} 
\Biggl[
(2\pi)^2\theta(l_0)\theta(r_0)\delta(r^2+c_{B}^2)\delta(l^2+c_{B}^2) 
\\
&~~~~~~~~~~~~~~~~~~~~\times (2 \pi)^6 \delta^3(p_1+p_2-p_3-p_4)\delta^3(p_1+p_2-l-r)
\\
&~~~~~~~~~~~~~~~~~~~~~\times
\biggl(\left(\delta_{im}\delta_{jn}-\frac{1}{N}\delta_{ij}\delta_{mn}\right)
T_{T}(p_1,p_2,-l,-r)T_{T}^{\ast}(p_3,p_4,-l-,r)
\\
&~~~~~~~~~~~~~~~~~~~~~~~~~~~+\frac{1}{N}T_{S}(p_1,p_2,-l,-r)T_{S}^{\ast}(p_3,p_4,-l,-r)\delta_{ij}\delta_{mn}\biggr)\Biggr].
                \end{split}
\end{equation}
 Equating the coefficients of the different index structures on the LHS and RHS we conclude that
\begin{equation} \label{tunit}
\begin{split}
&-i \left(T_{T}(p_1,p_2,-p_3,-p_4)-T_{T}^{\ast}(p_3,p_4,-p_1,-p_2)\right)\delta^3(p_1+p_2-p_3-p_4)\\
=&\int \frac{d^3 l}{(2\pi)^3}\frac{d^3 r}{(2\pi)^3} 
\biggl[(2\pi i)^2\theta(l_0)\theta(r_0)\delta(r^2+c_{B}^2)\delta(l^2+c_{B}^2)
\\
&~~~~~~~~~~~~~~~~~~~~~~
\times \delta^3(p_1+p_2-p_3-p_4)(2\pi)^3\delta^3(p_1+p_2-l-r)\\
&~~~~~~~~~~~~~~~~~~~~~~\times T_{T}(p_1,p_2,-l,-r)T_{T}^{\ast}(p_3,p_4, -l,-r)
\biggr],
                \end{split}
\end{equation}
and that
\begin{equation} \label{sunit}
\begin{split}
&-i \left(T_{S}(p_1,p_2,-p_3,-p_4)-T_{S}^{\ast}(p_3,p_4, -p_1,-p_2)\right)\delta^3(p_1+p_2-p_3-p_4)\\
=&\int \frac{d^3 l}{(2\pi)^3}\frac{d^3 r}{(2\pi)^3} \biggl[(2\pi i)^2\theta(l_0)\theta(r_0)\delta(r^2+c_{B}^2)\delta(l^2+c_{B}^2)
\\
 &~~~~~~~~~~~~~~~~~~~~~\times
\delta^3(p_1+p_2-p_3-p_4)(2\pi)^3\delta^3(p_1+p_2-l-r)\\
&~~~~~~~~~~~~~~~~~~~~~\times 
T_{S}(p_1,p_2,-l,-r)T_{S}^{\ast}(p_3,p_4,-l,-r)\biggr].
                \end{split}
\end{equation}

Now recall that the scattering matrix $T_T$ is ${\cal O}(\frac{1}{N})$. It follows that the RHS of 
\eqref{tunit} is subleading in $\frac{1}{N}$ compared to the LHS. In the large $N$ limit, consequently, 
\eqref{tunit} may be rewritten as 
\begin{equation} \label{tunitm}
\left(T_{T}(p_1,p_2,-p_3,-p_4)-T_{T}^{\ast}(p_3,p_4,-p_1,-p_2)\right)=0.
\end{equation}
Applying the same reasoning to particle-particle scattering, we reach the identical 
conclusion for $U$-channel scattering. It is easily verified that the slightly trivial, 
linear equations \eqref{tunitm} (and the analogous equation for $U$-channel scattering) are infact obeyed
 by the exact solutions for $T_T$ and $T_U$ presented below\footnote{This is related to the fact 
that these scattering amplitudes have no branch cuts in the physical domain for $T$ and $U$-channel 
scattering.}

On the other hand the $S$-channel scattering matrix $T_S^B$ is ${\cal O}(1)$ in the large 
$N$ limit. Consequently, the nonlinear equation \eqref{sunit} is a rather nontrivial 
constraint on $S$-channel scattering. 

\subsubsection{$S$-channel unitarity constraints in the center of mass frame}

The constraint on the $S$-channel S-matrix is most conveniently worked out in the center of mass frame. 
We choose the scattering momenta to take the form 
\begin{equation}\label{cmmoms}
\begin{split}
&p_1=\left(\sqrt{p^2+c_{B}^2},p,0\right),~~p_2=\left(\sqrt{p^2+c_{B}^2},-p,0\right),
\\
&p_3=\left(-\sqrt{p^2+c_{B}^2},-p \cos(\alpha),-p\sin(\alpha)\right),
~~p_4=\left(-\sqrt{p^2+c_{B}^2},p \cos(\alpha),p\sin(\alpha)\right),
\end{split}
\end{equation}  
In this frame $T_S= T_S(p, \alpha)$
or $T=T(s, \alpha)$ (recall $s=4 (p^2 +c_{B}^2)$ )  and the 
constraint from unitarity is simply a constraint on this function 
of two variables. 

In order to work out the precise form of this constraint we first process the delta functions inside the integrals.
\begin{equation}\label{cutint}
\begin{split}
&\int \frac{d^3 l}{(2\pi)^3}\frac{d^3 r}{(2\pi)^3} (2\pi)^2\theta(l_0)\theta(r_0)\delta(r^2+c_{B}^2)\delta(l^2+c_{B}^2)(2\pi)^3\delta^3(p_1+p_2-l-r)\\
=& \int \frac{d^3 l}{(2\pi)^3}(2\pi )^2\theta(l_0)
\theta(-l_0+(p_1)_0+(p_2)_0)\delta(l^2+c_{B}^2)
\delta((p_1+p_2)^2-2 (p_1+p_2)\cdot l)\\
=&\frac{1}{8\pi \sqrt{s}}\int d\theta dl_0 d\ell_{s}  \delta(l_0-\sqrt{p^2+c_{B}^2})\delta(\ell_{s}-p^2)\\
=&\frac{1}{8\pi \sqrt{s}}\int d\theta 
               \end{split}
\end{equation}
where $E_p=\sqrt{p^2+m^2}= \frac{\sqrt{s}}{2}$ and
$\ell_{s} = l^2 + l_{0}^2$.
It follows that the unitarity constraint is given by 
\begin{equation} \label{punit}
\begin{split}
&-i\left( T_{S}(s,\alpha)-T_{S}^{\ast}(s,-\alpha) \right) =\frac{1}{8\pi \sqrt{s} }\int d\theta  T_{S}(s,\theta)T_{S}^{\ast}(s, -(\alpha-\theta))
                \end{split}
\end{equation}
(this is essentially identical to the manipulation that produced the product rule \eqref{cmprod}).

\subsubsection{Unitarity of the non-relativistic limit}

As an example for how this works, we will now demonstrate that the non-relativistic limit of the 
$S$-channel S-matrix, \eqref{htg},  obeys the constraints of unitarity. In the center of mass frame 
\eqref{htg} takes the form
\begin{equation} \label{form}
 T_S(\sqrt{s},\alpha)=H(\sqrt{s}) T(\alpha)+W_1(\sqrt{s})- i W_2(\sqrt{s}) \delta(\alpha),
\end{equation}
where
$$T(\alpha)= i \cot\left( \frac{\alpha}{2}\right),$$ 
and 
\begin{equation}\label{nrtww} \begin{split}
& H(\sqrt{s})= 4 \sqrt{s} \sin (\pi \lambda_{B}), \\
& W_1(\sqrt{s})= -4 \sqrt{s} \sin (\pi \lambda_{B})  {\rm sgn}(\lambda_{B}),\\
& W_2(\sqrt{s})= 8 \pi  \sqrt{s} \left( \cos (\pi \lambda_{B}) - 1 \right).  \\
\end{split}
\end{equation}
With an eye to application later in the paper, we will first work out the unitarity constraint for 
arbitrary $H(\sqrt{s})$, $W_1(\sqrt{s})$ and $W_2(\sqrt{s})$, specializing to the specific forms \eqref{nrtww} 
only at the end. 

Using the formula \begin{equation}\label{cotcot}
 \int d\theta {\rm Pv} \cot \left(\frac{\theta}{2}\right)  {\rm Pv} \cot \left(\frac{\alpha-\theta}{2}\right)= 2\pi  -4 \pi^2 \delta(\alpha),
\end{equation}
(see footnote\footnote{
We can check the \eqref{cotcot}
by calculating the Fourier coefficients,
\begin{equation}
\begin{split}
&\int \frac{d \alpha}{2\pi} e^{-i n\alpha }
 \int d\theta {\rm Pv} \cot \left(\frac{\theta}{2}\right)  
{\rm Pv} \cot \left(\frac{\alpha-\theta}{2}\right)
\\
=&\oint \frac{d \omega}{2\pi \omega} \omega^{-n}
 \oint \frac{dz}{z} {\rm Pv} 
\left(\frac{z+1}{z-1}\right)
{\rm Pv} \left(\frac{z+\omega}{\omega-z}\right)
\\
=& \begin{cases}
 -i\oint dz \,{\rm Pv} \left(\frac{z+1}{z-1}\right)
z^{-n-1}
= -2\pi
& (n >0)
\\
 0 &   (n = 0)
\\
 i\oint dz \,{\rm Pv} \left(\frac{z+1}{z-1}\right)
z^{-n-1}
= -2\pi
& (n < 0)
   \end{cases}
\end{split}
\label{Fo-cotcot}
\end{equation}
where $z = e^{i \theta}$ and $\omega = e^{i \alpha}$.
By comparing 
\eqref{Fo-cotcot}
with Fourier coefficients of delta function,
\begin{equation}
\delta(\alpha)  = \frac{1}{2\pi}\sum_{n=-\infty}^{\infty} 
e^{i n \alpha},
\label{eq:delta-coef}
\end{equation}
we can immediately check \eqref{cotcot}. } for a check of \eqref{cotcot}), 
\eqref{punit} reduces to  
\begin{equation} \label{unitcond}
\begin{split}
 H -H^{\ast} &= \frac{1}{8\pi \sqrt{s}} \left( W_2 H^{\ast}-H W_2^{\ast}
\right),\\
W_2+W_2^{\ast} 
&= -\frac{1}{8\pi \sqrt{s}}\left( W_2W_2^{\ast} +4  \pi^2 H H^{\ast}
\right),\\
W_1-W_1^{\ast}&
=\frac{1}{8\pi \sqrt{s}}\left(W_2 W_1^{\ast}
-W_2^{\ast} W_1 \right)
-\frac{i}{4 \sqrt{s} }\left( HH^{\ast}-W_1 W_1^{\ast}\right).                 
\end{split}
\end{equation}

It is easily verified that the specific assignments \eqref{nrtww} 
obey the equation~\eqref{unitcond}. 
The first equation in \eqref{unitcond} is obeyed because $H$ and $W_2$, in \eqref{unitcond}, are both 
real. The third equation in \eqref{unitcond} is obeyed because $W_1$ is also real and $|H|^2 = |W_1|^2$. 
The second equation in \eqref{nrtww} reduces to the true trigonometric identity 
$$ 2 \left( 1-\cos (\pi \lambda_{B}) 
\right)=\left( 1 - \cos (\pi \lambda_{B}) \right)^2 + \sin^2 (\pi \lambda_{B}). $$
We conclude that the Aharonov-Bohm scattering amplitude obeys the equations of unitarity, though in a slightly trivial 
fashion as the coefficient of $\delta (\theta)$ was the only part of the 
S-matrix that had an imaginary 
piece. 

\subsubsection{Unitarity constraints on general S-matrices of the form \eqref{form}} \label{ug}

As we have seen in the last subsubsection, the functions ${\rm Pv} \cot \left(\frac{\theta}{2}\right)$, $1$ and 
$\delta(\theta)$ form a closed algebra under convolution (i.e the convolution of any two linear combinations
of these functions is, once again, a  linear combination of the same three functions). This nontrivial fact allowed us in the last subsection to find a simple solution of the unitarity equation 
of the form \eqref{form} (this was simply the Aharonov-Bohm solution).

Given the closure of \eqref{form} under convolution, it is tempting to conjecture that the scattering 
matrix in the $S$-channel takes the form \eqref{form} even outside the non-relativistic limit (we will 
find independent evidence below that this is indeed the case). With this conjecture in mind, in this 
subsection we will inquire to what extent the requirement of unitarity \eqref{unitcond} determines 
S-matrices of the form \eqref{form}.

Let us first do some counting. The data in S-matrices of the form \eqref{form} is three complex or 
six real functions of $s$ and $\lambda_{B}$. Unitarity provides 3 real equations. It follows that if we 
impose no more than the condition of unitarity, the general S-matrix is given in terms of three 
unknown real functions. 

In order to make further progress we need more information. In the previous subsection we have 
already argued that, on physical grounds, we expect the form of $W_2$ in \eqref{nrtww} to be exact 
even away from the non-relativistic limit. If we make this assumption, unitarity gives us 3 real 
equations for the remaining 4 unknown functions, and so the S-matrix is determined in 
terms of one unknown function. Let us see how this works in more detail. The first 
equation in \eqref{unitcond} forces the function $H$ to be real. The second equation 
in \eqref{unitcond} then forces $H$ to be given exactly by the expression in \eqref{nrtww}. 
We are left with a single unknown complex function $W_1$ subject to a single real equation; 
the third of \eqref{unitcond}.

Let us summarize. If we assume that the S-matrix takes the form \eqref{form} and further 
assume that the expression for $W_2$ in \eqref{nrtww} is exact, then unitarity also forces 
the expression for $H$ in \eqref{nrtww} to be exact, and constrains $W_1$ to obey the 
third of \eqref{nrtww}, which is one real equation for the unknown complex function $W_1$.

\section{Summary: method, results and conjectures} \label{res}

In this section we summarize the method we use to compute S-matrices and list our principal results 
and conjectures. 

\subsection{Method}\label{ime}

In this paper we compute the functions $T_T$, $T_{U_d}$ and $T_{U_{e}}$ for both the bosonic and the 
fermionic theories. We also present a conjecture for the functions $S_S$. We 
then study the transformation of our results under 
Bose-Fermi duality. The method we employ to 
compute the S-matrices  is completely straightforward; 
we sum all the off shell planar graphs with four external legs, and then obtain the S-matrices  by taking the appropriate on shell limits.

\begin{figure}[tbp] 
   \begin{center}
     \subfigure[]{\includegraphics[scale=0.75]{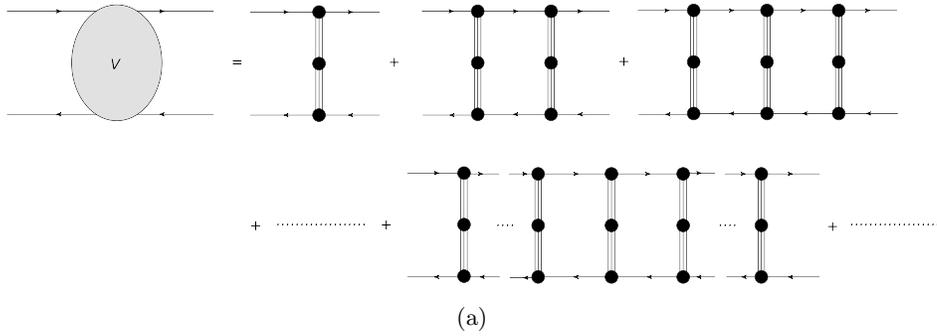}
 }
  
   \end{center}
   \vspace{-0.5cm}
   \caption{This diagram would contain a diagrammatic representation of the exact amplitude 
$V$ as a sum over ladders, where the `rungs' in the ladder are the triple line propagators.}
\label{ladder}
   \end{figure}

   \begin{figure}[tbp] 
   \begin{center}
     \subfigure[]{\includegraphics[scale=0.50]{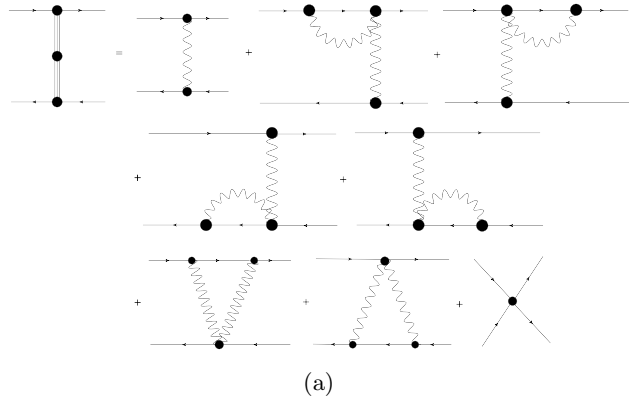}
 }
  
   \end{center}
   \vspace{-0.5cm}
   \caption{A diagrammatic representation of the effective single particle exchange four point amplitude 
for bosons. This amplitude is give by the sum of the tree level exchange of a gluon, dressed tree level 
exchanges of the gluon and the point interaction controlled by the parameter $b_4$}
\label{Unit}
   \end{figure}

\begin{figure}[tbp] 
   \begin{center}
     \subfigure[]{\includegraphics[scale=0.75]{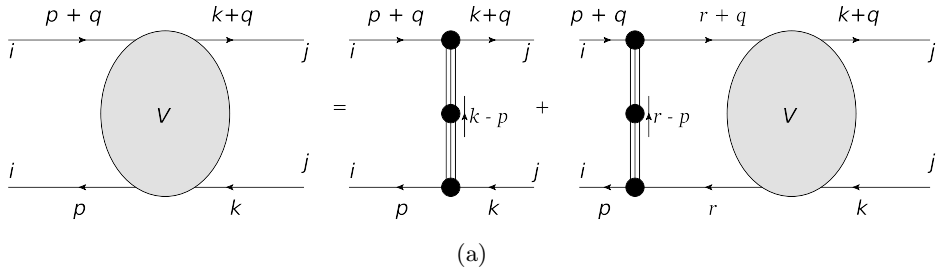}
 }
   \end{center}
   \vspace{-0.5cm}
   \caption{A diagrammatic depiction of the integral equation obeyed by offshell four point scattering
amplitudes. The blob here represents the all orders scattering amplitude while the triple line 
represents the effective single particle exchange four point interaction between quanta. Here, 
and in every Feynman diagram in this paper, all momenta flow in the direction 
of the arrows of the propagators.}
\label{IntB}
 \end{figure}

Following   \cite{Giombi:2011kc} and
several subsequent papers, we work in the lightcone gauge $A_-=0$. \footnote{Our notation is as follows. $x^+$, $x^-$ and $x^3$ are a set of coordinates on Minkowski space. $x^+$ and $x^-$ are lightcone coordinates while $x^3$ is a spatial coordinate.} The off shell four point amplitude receives contributions from an infinite number of Feynman graphs. The graphs that contribute may be enumerated 
very simply; they are simply the sum of all ladder graphs Fig \ref{ladder}, where the triple line is the 
 effective exchange interaction between fundamental particles. In the case of the bosonic theory, for instance, 
the triple line is given diagrammatically by Fig. \ref{Unit}. It is easy to convince oneself that the 
all orders amplitude depicted in Fig. \ref{ladder} obeys the integral equation depicted in  Fig \ref{IntB}
\cite{Giombi:2011kc, Aharony:2012nh}.

According to the labeling of momenta in Fig. \ref{IntB}, $q^\mu$ is the three momentum that flows, from left to right in graphs of Fig. \ref{ladder}.  $q^\mu$ is 
a `constant of motion' in the sense that if a given ladder diagram has a particular value of $q^\mu$ 
then every sub ladder within the original ladder also has the same value of $q^\mu$ (this is not true 
of the momenta $p$ and $k$ in Fig. \ref{ladder}).   This implies that different values of $q^\mu$ do not
`mix'  in the integral equation of Fig. \ref{IntB}. In other words Fig. \ref{IntB} represents an 
 infinite set of decoupled integral equations; one for every value of $q^\mu$. It was pointed out  in \cite{Aharony:2012nh} that the integral equations in Fig. \ref{IntB}  simplifies dramatically when 
 $q^{\pm}=0$. The authors of \cite{Aharony:2012nh} infact solved the relevant integral equations for the bosonic theory in massless limit. 
 In this paper to find exact formulae for the 
 sum over planar graphs with four external lines with  $q^\pm=0$ by explicitly solving the 
integral equations relevant to that case.  
 In the case of the bosonic theory our results are a generalization of those 
 of \cite{Aharony:2012nh} to nonzero mass\footnote{\cite{Aharony:2012nh} performed this summation in order 
 to evaluate three point functions of gauge invariant operators in special kinematical 
 configurations.} The integral equation turns out to be more complicated to solve in the 
case of the fermionic theory, but we are able to find the exact solution in this case as well.
 
 With exact off shell results in hand, we proceed to evaluate the S-matrices for our problem 
 by taking the appropriate on shell  limits.
The on shell condition determines the energy of each of the participating particles (in terms of 
their momenta) upto a sign. Energy and momentum conservation require that
two of the external lines have positive energy while the other two have negative energy, leaving 
a total of six distinct cases. 
\footnote{We say an external line has positive energy if 
$p_0$ is positive (or $p^0$ is negative) going into the graph. An external line with 
an ingoing arrow and positive energy represents an initial particle. An external line 
with an outgoing arrow and positive energy into the graph (or negative energy in the 
direction of the arrow) is an ingoing antiparticle. An external line with an arrow going 
into the graph and negative energy going into the graph is an outgoing antiparticle. An 
external line whose arrow points out of the graph and whose energy is negative going 
into the graph (or positive in the direction of the arrow) is an outgoing particle. }  Recalling that external lines with positive energy represent initial states while external lines with 
negative energy represent final states, it is not difficult to 
convince oneself that  one of these six cases determines the function $T_S$, another 
determines $T_T$,  two others determine $T_{U_d}$, $T_{U_e}$ respectively, while the
last two processes compute the CPT conjugates of scattering in the $U$-channel.  In other 
words the four different scattering functions introduced, in the previous subsection, 
are all different limits of the single four point amplitudes determined by the integral 
equation of Fig. \ref{IntB}.

As we have emphasized above, we have been able to evaluate the off shell four point 
amplitude only in the special case $q^{\pm}=0$. This technical limitation has different 
implications for our ability to compute the $S$ matrices in the different channels. 

$q^\mu$ turns out to be the center of mass 3 momentum for $S$-channel scattering.  The condition  $q^{\pm}=0$ ensures that the center of mass energy is spacelike; this is impossible for an onshell scattering process.  It follows that the technical limitations
which restricted us to $q^\pm=0$ forbid us from directly computing $S$-channel scattering, 
a fact that will force us to resort to conjecture in this channel. 

In the $T$ and $U$-channels, on the other hand, $q$ represents the 3 momentum transfer between
an initial and final particle. As all participating particles have the same mass, the 3 momentum transfer is always spacelike (this is most easily seen in the center of mass frame), there is no barrier
to setting $q^{\pm}=0$ in these processes. For an  arbitrary
$T$ or $U$-channel process, it is always possible to find an inertial frame in which $q^{\pm}=0$.
In these channels, in other words, the restriction to $q^{\pm}=0$ is simply a choice of frame.
Assuming that the S-matrix for our process is Lorentz invariant, the on shell limits of our 
off shell four point amplitude completely fix the S-matrix in these channels.  We are thus able 
to report definite results  for the scattering matrices in these channels. 
 
\subsection{Results in the $U$ and $T$ channels}\label{isg}

In this subsection we simply present our final results for $U$ and $T$-channel scattering, 
separately for the bosonic and the fermionic theories.  We first report our results for the  bosonic theory. 
In the $T$-channel (adjoint exchange) we find 
\begin{equation}\label{tchansg}\begin{split}
&T^B_{T}(p_1, p_2, p_3, p_4,  k_B,  \lambda_{B}, {\tilde b}_4, c_B)\\
=&E(p_1,p_2,p_3)\frac{4i\pi }{k_B}  \sqrt{\frac{u~t}{s}}
\\
&-\frac{4~i\pi}{k_B}  \sqrt{-t}~~\frac{({\tilde b}_4-4\pi i \lambda_{B}\sqrt{-t})e^{i\pi\lambda_B}+({\tilde b}_4+4\pi i \lambda_{B}\sqrt{-t})e^{2i\lambda_B \tan^{-1}\left(\frac{2|c_B|}{\sqrt{-t}}\right)}}{-({\tilde b}_4-4\pi i \lambda_{B}\sqrt{-t})e^{i\pi\lambda_B}+({\tilde b}_4+4\pi i \lambda_{B}\sqrt{-t})e^{2i\lambda_B  \tan^{-1}\left(\frac{2|c_B|}{\sqrt{-t}}\right)}}~~\\
=&E(p_1,p_2,p_3)\frac{4i\pi }{k_B}  \sqrt{\frac{u~t}{s}}
\\
&-\frac{4~i\pi}{k_B}  \sqrt{-t}~~\frac{({\tilde b}_4-4\pi i \lambda_{B}\sqrt{-t})+({\tilde b}_4+4\pi i \lambda_{B}\sqrt{-t})e^{-2i\lambda_B \tan^{-1}\left(\frac{\sqrt{-t}}{2|c_B|}\right)}}{-({\tilde b}_4-4\pi i \lambda_{B}\sqrt{-t})+({\tilde b}_4+4\pi i \lambda_{B}\sqrt{-t})e^{-2i\lambda_B \tan^{-1}\left(\frac{\sqrt{-t}}{2|c_B|}\right)}}~~\\
\end{split}\end{equation} where we have used 
$$\tan^{-1}(x)+\tan^{-1}(\frac{1}{x})=\frac{\pi}{2},~~\rm{for}~~x>0 $$ 
and  $\tilde b_4 = -b_4+2\pi\lambda_{B}^2 |c_B|.$ 
Here form of the $\tan^{-1}(x)$
is 
\begin{equation}
 \tan^{-1}x= \frac{1}{2 i} \ln \left( \frac{1+i x}{1-i x} \right)
\label{eq:tan-form}
\end{equation}
and the domain and the branch cut structure of the function 
$\tan^{-1}(x)$
are depicted in Fig.~\ref{tan}.

In the special case $b_4 \to \infty$, $T_T$ reduces to 
\begin{equation}\label{tchanss}
\begin{split}
T^{B \infty}_{T}(p_1, p_2, p_3, p_4, k_B,  \lambda_{B}, c_B)=&
E(p_1,p_2,p_3)\frac{4i\pi }{k_B}  \sqrt{\frac{u~t}{s}}
\\
&-\frac{4~i\pi}{k_B}  \sqrt{-t}~~\frac{1+e^{-2i\lambda_B \tan^{-1}\left(\frac{\sqrt{-t}}{2|c_B|}\right)}}{1-e^{-2i\lambda_B \tan^{-1}\left(\frac{\sqrt{-t}}{2|c_B|}\right)}}~~.
\end{split}
\end{equation}
In the $U$-channel we find 
\begin{equation}\label{uchansg}\begin{split}
&T^B_{U_d}(p_1, p_2, p_3, p_4,  k_B,  \lambda_{B},{\tilde b}_4, c_B) \\
=&E(p_1,p_2,p_3)\frac{4i\pi }{k_B}  \sqrt{\frac{s~t}{u}}\\
&-\frac{4~i\pi}{k_B}  \sqrt{-t}~~\frac{({\tilde b}_4-4\pi i \lambda_{B}\sqrt{-t})+({\tilde b}_4+4\pi i \lambda_{B}\sqrt{-t})e^{-2i\lambda_B \tan^{-1}\left(\frac{\sqrt{-t}}{2|c_B|}\right)}}{-({\tilde b}_4-4\pi i \lambda_{B}\sqrt{-t})+({\tilde b}_4+4\pi i \lambda_{B}\sqrt{-t})e^{-2i\lambda_B \tan^{-1}\left(\frac{\sqrt{-t}}{2|c_B|}\right)}}.~~\\
\end{split}\end{equation}
In the limit $b_4 \to \infty$ we have 
\begin{equation}\label{uchanss}
\begin{split}
T^{B \infty}_{U_d}(p_1, p_2, p_3, p_4,  k_B,  \lambda_{B}, c_B)
=&E(p_1,p_2,p_3)\frac{4i\pi }{k_B}  \sqrt{\frac{s~t}{u}}
\\
&-\frac{4~i\pi}{k_B}  \sqrt{-t}~~\frac{1+e^{-2i\lambda_B \tan^{-1}\left(\frac{\sqrt{-t}}{2|c_B|}\right)}}{1-e^{-2i\lambda_B \tan^{-1}\left(\frac{\sqrt{-t}}{2|c_B|}\right)}}.~~
\end{split}
\end{equation}
Finally, the amplitude $T^B_{U_e}$ is obtained from $T^B_{U_d}$ simply by interchanging the two initial momenta.
The usual symmetry of bosonic amplitudes immediately implies
\begin{equation}\label{uchansym}
T^B_{U_e}(p_1, p_2, p_3, p_4,  k_B,  \lambda_{B}, b_4, c_B)=
T^B_{U_d}(p_2, p_1, p_3, p_4,  k_B,  \lambda_{B}, b_4, c_B)
\end{equation}
with a similar formula for 
$S^\infty_{U_e}(p_1, p_2, p_3, p_4,  k_B,  \lambda_{B}, c_B).$

We now report our results for the fermionic theory. In this case S-matrix in the $T$-channel is given by
\begin{equation}\label{uchanf}\begin{split}
&T^F_T(p_1, p_2, p_3, p_4,  k_F,  \lambda_{F}, c_F)\\
=&-E(p_1,p_2,p_3)\frac{4i\pi }{k_F}  \sqrt{\frac{u~t}{s}}
\\
&+\frac{4~i\pi}{k_F}  \sqrt{-t}~~\frac{e^{i\pi\left(\lambda_F-{\rm sgn}(m_F)\right)}+e^{2i\left(\lambda_F-{\rm sgn}(m_F)\right) \tan^{-1}\left(\frac{2|c_F|}{\sqrt{-t}}\right)}}{e^{i\pi\left(\lambda_F-{\rm sgn}(m_F)\right)}-e^{2i\left(\lambda_F-{\rm sgn}(m_F)\right) \tan^{-1}\left(\frac{2|c_F|}{\sqrt{-t}}\right)}}~~\\
=&-E(p_1,p_2,p_3)\frac{4i\pi }{k_F}  \sqrt{\frac{u~t}{s}}
+\frac{4~i\pi}{k_F}  \sqrt{-t}~~\frac{1+e^{-2i\left(\lambda_F-{\rm sgn}(m_F)\right) \tan^{-1}\left(\frac{\sqrt{-t}}{2|c_F|}\right)}}{1-e^{-2i\left(\lambda_F-{\rm sgn}(m_F)\right) \tan^{-1}\left(\frac{\sqrt{-t}}{2|c_F|}\right)}}.~~\\
\end{split}\end{equation}
In the $U$-channel we find
\begin{equation}\label{uchans}\begin{split}
&T_{U_d}^F(p_1, p_2, p_3, p_4,  k_F,  \lambda_{F}, c_F)\\
=& -\Bigg(- E(p_1,p_2,p_3)\frac{4i\pi }{k_F}  \sqrt{\frac{s~t}{u}}
\\
&\qquad+\frac{4~i\pi}{k_F}  \sqrt{-t}~~\frac{e^{i\pi\left(\lambda_F-{\rm sgn}(m_F)\right)}
+e^{2i\left(\lambda_F-{\rm sgn}(m_F)\right) \tan^{-1}\left(\frac{2|c_F|}{\sqrt{-t}}\right)}}{e^{i\pi\left(\lambda_F-{\rm sgn}(m_F)\right)}-e^{2i\left(\lambda_F-{\rm sgn}(m_F)\right) \tan^{-1}\left(\frac{2|c_F|}{\sqrt{-t}}\right)}}\Bigg)~~\\
&=-\Bigg(- E(p_1,p_2,p_3)\frac{4i\pi }{k_F}  \sqrt{\frac{s~t}{u}}+\frac{4~i\pi}{k_F}  \sqrt{-t}~~\frac{1+e^{-2i\left(\lambda_F-{\rm sgn}(m_F)\right) \tan^{-1}\left(\frac{\sqrt{-t}}{2|c_F|}\right)}}{1-e^{-2i\left(\lambda_F-{\rm sgn}(m_F)\right) \tan^{-1}\left(\frac{\sqrt{-t}}{2|c_F|}\right)}} \Bigg).
\end{split}\end{equation}
Finally, the usual symmetry for fermionic amplitudes immediately implies that 
\begin{equation}\label{uchanfsym}
T^F_{U_e}(p_1, p_2, p_3, p_4,  k_F,  \lambda_{F},  c_F)=
-T^F_{U_d}(p_2, p_1, p_3, p_4,  k_F,  \lambda_{F}, c_F).
\end{equation}

As we have mentioned earlier in this introduction, in the limit $b_4 \to \infty$, the  bosonic theory studied in this paper has been conjectured to be dual to the fermionic theory, when the parameters of the two 
theories are related by \eqref{dualitymap}.
Our results for the  scattering amplitudes reported above are in perfect agreement with this 
conjecture. In particular it may be verified that, provided the inequality  \eqref{fti} is obeyed, 
the bosonic and fermionic S-matrices (including the identity pieces, see subsections \ref{ik} and \ref{chan}) 
\begin{equation}\label{smud} \begin{split}
&{\bf S}^{B\infty}_T(p_1, p_2, p_3, p_4,  -k_F,  \lambda_{F}-{\rm sgn}(\lambda_F), c_F) = {\bf S}^F_T (p_1, p_2, p_3, p_4, k_F,  \lambda_{F}, c_F),\\
&{\bf S}^{B\infty}_{U_d}(p_1, p_2, p_3, p_4,  -k_F,  \lambda_{F}-{\rm sgn}(\lambda_F), c_F)=-{\bf S}^F_{U_d}(p_1, p_2, p_3, p_4, k_F,  \lambda_{F}, c_F),\\
&{\bf S}^{B\infty}_{U_e}(p_1, p_2, p_3, p_4,  -k_F,  \lambda_{F}-{\rm sgn}(\lambda_F), c_F)={\bf S}^F_{U_e}(p_1, p_2, p_3, p_4, k_F,  \lambda_{F}, c_F),\\
&{\bf S}^{B\infty}_{U_s}(p_1, p_2, p_3, p_4,  -k_F,  \lambda_{F}-{\rm sgn}(\lambda_F), c_F)={\bf S}^F_{U_a}(p_1, p_2, p_3, p_4, k_F,  \lambda_{F}, c_F),\\
&{\bf S}^{B\infty}_{U_a}(p_1, p_2, p_3, p_4,  -k_F,  \lambda_{F}-{\rm sgn}(\lambda_F), c_F)={\bf S}^F_{U_s}(p_1, p_2, p_3, p_4, k_F,  \lambda_{F}, c_F).\\
\end{split}
\end{equation}

\subsection{A conjecture for identity exchange and modified crossing symmetry} \label{ip}

In the case of the bosonic theory we conjecture that   $S$ matrix in the $S$-channel is given by
\begin{equation} \label{conject}
{\bf S}_S^B=\cos (\pi \lambda_{B}) I(p_1, p_2, p_3, p_4) + i\frac{\sin (\pi \lambda_{B})}{\pi \lambda_{B}} T_S^{trial}  
\end{equation}
where 
$T_S^{trial}$ is the $S$-channel S-matrix obtained from analytic continuation of the $T$ or 
$U$-channel results using the usual rules of `naive' crossing symmetry, and is given by
\begin{equation} \label{ttrial} \begin{split}
T_S^{trial} =& 
\left( \pi \lambda_{B} \right)  4 ~i\sqrt{s}   
E(p_1, p_2, p_3) \sqrt{\frac{u}{t}}  
\\
&+
\left( \pi \lambda_{B} \right)  4 \sqrt{s}   
\left( \frac{ \left(4 \pi  \lambda_{B} \sqrt{s} +\wt b_4 \right)
  +  e^{i \pi \lambda_{B}} \left(-4 \pi  \lambda_{B} \sqrt{s} + \wt b_4 \right)  \left(  \frac{ \frac{1}{2} + \frac{c_B}{\sqrt{s} } }
{\frac{1}{2} - \frac{c_B}{\sqrt{s}} } \right)^{\lambda_{B}}
 } {\left(4 \pi  \lambda_{B} \sqrt{s}  + \wt b_4 \right) 
 -  e^{i \pi \lambda_{B}} \left(-4 \pi  \lambda_{B} \sqrt{s} + \wt b_4 \right)    \left(  \frac{ \frac{1}{2} + \frac{c_B}{\sqrt{s} } }
{\frac{1}{2} - \frac{c_B}{\sqrt{s}} } \right)^{\lambda_{B} }
} \right).  \\
\end{split}
\end{equation}
In the limit $b_4 \to \infty,$ $T_S^{trial}$ simplifies to 
\begin{equation} \label{ttriali}
T_S^{trial}
= \left( \pi \lambda_{B} \right)  4~i \sqrt{s}   \left( E(p_1, p_2, p_3) \sqrt{\frac{u}{t}}  +
\left( \frac{ 1
  +  e^{i \pi \lambda_{B}}   \left(  \frac{ \frac{1}{2} + \frac{c_B}{\sqrt{s} } }
{\frac{1}{2} - \frac{c_B}{\sqrt{s}} } \right)^{\lambda_{B}}
 } {1
 -  e^{i \pi \lambda_{B}}    \left(  \frac{ \frac{1}{2} + \frac{c_B}{\sqrt{s} } }
{\frac{1}{2} - \frac{c_B}{\sqrt{s}} } \right)^{\lambda_{B} }
} \right)  \right).  \\  
\end{equation}
In a similar manner we expect that the fermionic S-matrix is given by 
\begin{equation} \label{conjectf}\begin{split}
S_S^F&=\cos (\pi \lambda_F) I(p_1, p_2, p_3, p_4) + i\frac{\sin (\pi \lambda _F) }{\pi \lambda_F} T_F^{trial}\\
&= \sin (\pi \lambda_F)\Bigg(4 E(p_1,p_2,p_3)   \sqrt{\frac{s~t}{u}}+4 \sqrt{s}~~\frac{1+e^{-2i\left(\lambda_F-{\rm sgn}(m_F)\right) \tan^{-1}\left(\frac{\sqrt{s}}{2|c_F|}\right)}}{1-e^{-2i\left(\lambda_F-{\rm sgn}(m_F)\right) \tan^{-1}\left(\frac{\sqrt{s}}{2|c_F|}\right)}}~~
\Bigg)\\
&+ 
\cos (\pi \lambda_F )  I(p_1, p_2,p_3, p_4) .
\end{split}\end{equation}
It follows from \eqref{conject}, \eqref{conjectf} and the results of the previous subsection the fermionic and bosonic $S$-channel $S$ matrices map to each 
other under duality upto an overall minus sign (recall that overall phases in an S-matrix are unobservable and so unimportant).

\section{Scattering in the scalar theory}\label{boson}

In this section we compute the four point scattering amplitude in the theory of fundamental bosons 
coupled to Chern-Simons theory.  Very briefly we integrate out 
the gauge boson to obtain an offshell effective four boson term in the quantum effective action for our 
theory, given by 
\begin{equation} \label{defV}
  \frac{1}{2} \int \frac{d^3p}{(2 \pi)^3}  \frac{d^3 k}{(2 \pi)^3} \frac{d^3 q}{(2\pi)^3} V(p,k,q)
  {\phi}_{i}(p+q){\bar\phi}^{j}(-(k+q)){\bar\phi}^{i}(-p)\phi_{j}(k).
 \end{equation}
 We then take an appropriate on shell limit to evaluate the S-matrix.

\subsection{Integral equation for off shell four point amplitude}

 As explained in the previous section, $V(p, k, q)$ obeys the integral equation depicted in Fig \ref{IntB}. In formulas
\begin{equation}\begin{split}\label{sdeV}
V(p, k, q) &= V_0(p, k,  q) - i\int \frac{d^3 r}{(2 \pi)^3} V(p, r, q_3) 
\frac{N V_0(r, k, q_3)}{ \left(r^2 + c_B^2 -i \epsilon
\right)
\left( (r+q)^2+ c_B^2 - i \epsilon \right) }, \\
V(p, k, q) &= V_0(p, k,  q) -i \int \frac{d^3 r}{(2 \pi)^3} V_0(p, r, q_3) 
\frac{NV(r, k, q_3)}{ \left(r^2 + c_B^2 -i \epsilon
\right) \left( (r+q)^2+ c_B^2 - i \epsilon \right) } ,
\end{split}
\end{equation}
where the `one particle' amplitude $V_0$ is given by the sum of graphs in Fig. \ref{Unit}. 
Summing these graphs (see Appendix \ref{die} for details) we find\footnote{
If we include other multi-trace terms such as ${\lambda_p\over N^{p-1}}(\bar\phi\phi)^p$ in the action \eqref{sclag}, 
this effect only reflects a shift of $\tilde b_4$ by a linear term of $c_B^{p-2}\lambda_p$ with a suitable coefficient. 
The rest of calcuation of $2\to2$ scattering is the same as presented in this paper. }  
\begin{equation}\begin{split} \label{tle}
NV_0(p, k, q_3) & = -4 \pi i \lambda_{B} q_3 \frac{(k+p)_-}{(k-p)_-}+ \wt b_4,\\
\wt b_4& = 2 \pi \lambda_{B}^2  c_B - b_4.
\end{split} 
\end{equation}
Here 
\begin{equation} \label{convs}
d^3r= dr^0 dr^1 dr^3, ~~~k_{\pm}=\frac{\pm k_0+k_1}{\sqrt{2}}. ~~~
\end{equation}
\footnote{Note in that our definition of $k_-$ is the negative of the definition usually adopted in 
studies of Minkowskian physics. We adopt this definition because it will prove convenient once we 
continue to Euclidean space. }
\eqref{tle} is actually ambiguous as stated.  The first term on the RHS of \eqref{tle} 
is proportional to $\frac{1}{(k-p)_-} $: the gauge boson propagator in lightcone gauge. This 
term is ill defined when  $k_-=p_-$, a point that lies on the integration contour on the 
RHS of \eqref{sde}.

 The reason that the gauge boson has a codimension two singularity in momentum space 
is that the choice of lightcone gauge, $A_-=0$, leaves unfixed the 
residual gauge transformations that depend only on $x^+$ and $x^3$.  In this paper we 
resolve this ambiguity of the propagator at $p_-=0$ with the `Feynman' prescription
\begin{equation}\label{defsing}
\frac{1}{p_-} \rightarrow \frac{p_+}{p_+ p_- -  i \epsilon} .
\end{equation}
We adopt this prescription for several reasons. 
\begin{itemize}
\item{1.} It is the only resolution of the singularity of the gauge propagator 
that  permits continuation to Euclidean space. It therefore appears to be the only resolution of the 
singularity that can make contact with all the beautiful Euclidean results 
of \cite{Giombi:2011kc,Aharony:2011jz,Jain:2012qi,Maldacena:2011jn,Maldacena:2012sf, Aharony:2012ns,Jain:2013py,Takimi:2013zca,Aharony:2012nh}.
\item{2.} Its use leads to sensible results with no unphysical divergences.
~\footnote{Other potential 
resolutions of this singularity appear to lead to pathological results. For instance the replacement of 
$\frac{1}{p_-}$ by its principal value leads to unacceptable divergences in propagators.}
\item{3.} In special cases, results obtained by use of this prescription turn out to agree with 
results in the covariant Landau gauge (see subsection \ref{landauonshell} below).
\end{itemize}

Of course the pragmatic reasons spelt out above are ultimately unsatisfactory; we would like eventually to have a justification of this prescription on physical grounds (such a justification would presumably involve a careful accounting for the unfixed gauge symmetry of the problem). However we leave this potentially subtle exercise to future work.

\subsection{Euclidean continuation}

In order to solve the integral equation \eqref{sdeV} we will find it convenient to use a standard  maneuver to `continue this equation to Euclidean space'. Operationally, the procedure is to define 
a Euclidean amplitude via $V^E(p^0, k^0)=V(ip^0, ik^0)$. 
\footnote{In this paragraph we are interested only in the dependence of all quantities on $p^0$ and $k^0$
so we suppress the dependence of $V$ on other components of the momenta.}  Once the amplitude $V^E$ has been solved for, the amplitude of real physical 
interest, $V$, is obtained by the inverse relation 
$$V(p^0, k^0)= V^E(-ip^0, -ik^0).$$
Even though the  method of Euclidean continuation is standard in the study of scattering amplitudes, 
for completeness we recall the justification of this method, in the context of our problem, in 
Appendix \ref{er}. We emphasize that this procedure is valid only when the singularities of all propagators 
in the Lorentzian problem are resolved by the Feynman $i\epsilon$ prescription. This is one of the 
main reasons we adopted the $i \epsilon$ prescription of \eqref{defsing} above.  

The Euclidean continuation of the scattering amplitude obeys the integral equation
\begin{equation}\begin{split}\label{sdee} 
V^E(p, k, q) &= V^E_0(p, k,  q) + \int \frac{d^3 r}{(2 \pi)^3} V_0^E(p, r, q_3) \frac{N V^E(r, k, q_3)}{ \left(r^2 + c_B^2 
\right)
\left( (r+q)^2+ c_B^2 \right) } \\
V^E(p, k, q) &= V^E_0(p, k,  q) + \int \frac{d^3 r}{(2 \pi)^3} V^E(p, r, q_3) \frac{N V^E_0(r, k, q_3)}{ \left(r^2 + c_B^2 
\right)
\left( (r+q)^2+ c_B^2  \right) } \\
NV^E_0(p, k, q_3) & = -4 \pi i \lambda_{B} q_3 \frac{(k+p)_-}{(k-p)_-}+ \wt b_4\\
\end{split} 
\end{equation}
where
\begin{equation} \label{mconvs}
d^3r= dr^0 dr^1 dr^3, ~~~k_{\pm}=\frac{k_1 \pm i k_0}{\sqrt{2}}. ~~~
\end{equation}
Note, in particular, that $k_\pm$ are now complex conjugates of each other.  
Below we will sometimes use the notation 
\begin{equation}\label{ks}
k_s^2= 2 k_+k_-= k_1^2+k_0^2.
\end{equation}

\subsection{Solution of the Euclidean integral equation}
 
The integral equation \eqref{sdee} may be solved in a completely systematic manner. We have 
presented a detailed derivation of our solution of this equation in Appendix \ref{seie}. In this subsection 
we simply quote our final results. 

Our solution takes the form 
\begin{equation}\label{anonm}
N V=  e^{-2 i \lambda_{B}   \left( \tan ^{-1}
\left( \frac{2 (a(k))}{q_3} \right) - 
 \tan ^{-1}\left( \frac{2 (a(p))}{q_3} \right)\right)
   }  \left( 4 \pi i \lambda_{B}  q_3  \frac{p_- + k_-}{p_- - k_-}
+ j(q_3, \lambda_{B})\right)
\end{equation}
where
\begin{equation}\label{ap}
a(p)=\sqrt{2p_+p_- +c_{B}^2}
\end{equation}
and
\begin{equation}\label{anssefm}
 j(q_3, \lambda_{B})=  {4 \pi i \lambda_{B} q_3} \left( \frac{ \left(4 \pi i \lambda_{B} q_3 +\wt b_4 \right) e^{2 i \lambda_{B}  \tan^{-1}\left( \frac{2c_B}{q_3} \right)}
  +  \left(-4 \pi i \lambda_{B} q_3 + \wt b_4 \right)  e^{\pi i \lambda_{B} {\rm sgn} (q_3)}
 } {\left(4 \pi i \lambda_{B} q_3 + \wt b_4 \right) e^{2 i \lambda_{B}  \tan^{-1}\left( \frac{2c_B}{q_3} \right)}
 -  \left(-4 \pi i \lambda_{B} q_3 + \wt b_4 \right)  e^{\pi i \lambda_{B} {\rm sgn} (q_3)}
} \right).
 \end{equation}
It is not difficult to verify that 
\begin{equation}\label{jtr}
j(q_3, \lambda_{B})= j(-q_3, \lambda_{B})= j(q_3, -\lambda_{B})= j(-q_3, -\lambda_{B})= 
j(|q_3|, |\lambda_{B}|).
\end{equation}
In other words, $j$ is an even function of $q_3$ and $\lambda_{B}$ separately. It follows in particular that 
\begin{equation}\label{anssefmm}
 j(q, \lambda_{B})=  {4 \pi i \lambda_{B} |q|} \left( \frac{ \left(4 \pi i \lambda_{B} |q| +\wt b_4 \right) e^{2 i \lambda_{B}  \tan^{-1}\left( \frac{2c_B}{|q|} \right)}
  +  \left(-4 \pi i \lambda_{B} |q| + \wt b_4 \right)  e^{\pi i \lambda_{B} }
 } {\left(4 \pi i \lambda_{B} |q| + \wt b_4 \right) e^{2 i \lambda_{B}  \tan^{-1}\left( \frac{2c_B}{|q_3|} \right)}
 -  \left(-4 \pi i \lambda_{B} |q| + \wt b_4 \right)  e^{\pi i \lambda_{B}}
} \right).
\end{equation}
This formula may be rewritten as follows. Let us define 
\begin{equation}\begin{split} \label{hdeef}
H(q)&=\int \frac{d^3r}{(2 \pi)^3 }\frac{1}{\left(r^2+c_B^2\right) \left( (r+q)^2+ c_B^2 \right)}
=\left( -\frac{ \tan^{-1} \left( \frac{2c_B}{|q_3|} \right)}{4 \pi |q_3| } +\frac{1}{8 |q_3|}  \right)
\\
&=\frac{ \tan^{-1} \left( \frac{|q_3|}{2c_B} \right)}{4 \pi |q_3| }\\
&= \frac{1}{8 \pi  i |q|} \ln \left( \frac{ \frac{1}{2} + \frac{c_B}{iq } }
{-\frac{1}{2} + \frac{c_B}{ iq }  }   \right) .
\end{split}\end{equation}
Here to get the last line, we have used the formula \eqref{eq:tan-form}.
$H(q)$ is simply the one loop four boson scattering amplitude in $\phi^4$ theory. 
In terms of this function we have 
\begin{equation}\label{jfd}
j(q)= {4 \pi i \lambda_{B} |q|} \left( \frac{ \left(4 \pi i \lambda_{B} |q| +\wt b_4 \right)
  +  \left(-4 \pi i \lambda_{B} |q| + \wt b_4 \right)  e^{8 i\pi  \lambda_{B} |q|H(q) }
 } {\left(4 \pi i \lambda_{B} |q| + \wt b_4 \right) 
 -  \left(-4 \pi i \lambda_{B} |q| + \wt b_4 \right)  e^{8 i\pi  \lambda_{B} |q|H(q) }
} \right).
 \end{equation}
Using the last line in \eqref{hdeef} $j(q)$ may also be rewritten as 
\begin{equation}\label{jfe}
j(q)= 4 \pi i q \lambda_{B} 
\left( \frac{ \left(4 \pi  i q \lambda_{B} +\wt b_4 \right)
  +  \left(-4 \pi i q  \lambda_{B}  + \wt b_4 \right)  \left(  \frac{ \frac{1}{2} + \frac{c_B}{iq } }
{-\frac{1}{2} + \frac{c_B}{ iq }  }\right)^{\lambda_{B}}
 } {\left(4 \pi i q \lambda_{B}   + \wt b_4 \right) 
 -  \left(-4 \pi i q   \lambda_{B}  + \wt b_4 \right)    \left(  \frac{ \frac{1}{2} + \frac{c_B}{i q  } }
{-\frac{1}{2} + \frac{c_B}{ i q} } \right)^{\lambda_{B}}  } \right).
\end{equation}  

\subsubsection{Transformation under parity}

While parity transformations are not a symmetry of the bosonic theory, the simultaneous 
action of a parity transformation and the flip in the sign of $k_B$ 
(or $\lambda_B$) is 
symmetry of this theory.  Every physical quantity in this theory must, therefore, transform 
in a suitably `nice' way under the combined action of these two transformations.

The off shell Greens function computed in the previous subsection is not physical as it is 
not gauge invariant, and so need not transform `nicely' under parity operations. Indeed 
it is easily verified by inspection that the amplitude $V$ is left invariant by a reflection 
in the $3$ direction accompanied by a flip in the sign of $\lambda_{B}$. However the combined 
operation of a flip in the sign of $\lambda_{B}$ and a reflection in either the $0$ or $1$ directions
is not an invariance of this amplitude. The reason for this asymmetry is that  reflections in the 3 direction are the only parity transformations that commute with the choice of light cone 
gauge (for instance a reflection in the $1$ direction changes the gauge $A_-=0$ to $A_+=0$. ). 

As we will see below, the physical $S$ matrix indeed enjoys the full parity symmetry expected
of this theory.

\subsection{Analytic continuation of $j(q)$} \label{acj}

In our study of $S$-channel scattering later in this paper we will need to 
continue the function $j(q)$ to $q^2=-s$.  This analytic continuation is achieved by 
setting $q= \lim_{\alpha  \to \frac{\pi}{2} }e^{-i \alpha \frac{\pi}{2}} \sqrt{s}$ or equivalently 
by setting 
$$q \rightarrow -i \left( \sqrt{s} + i \epsilon \right)$$

\begin{figure}[tbp]
  \begin{center}
  \subfigure[]{\includegraphics[scale=.6]{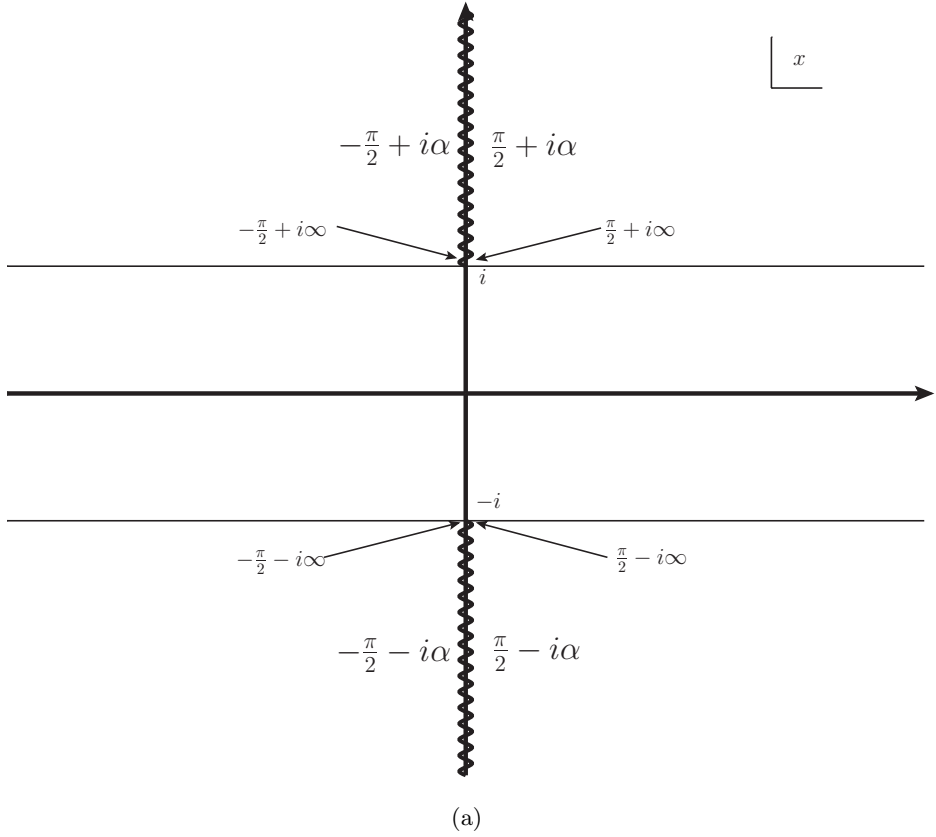}
  }\\
   \end{center}
  \vspace{-0.5cm}
  \caption{
Branch cut structure of the function $\tan^{-1}x$. $\alpha$ is a real function of x along the branch cut which vanishes at infinities
and becomes $\infty$ at $|Im(x)|=1$.
}
\label{tan}
  \end{figure}

The precise analytic continuation we will use is the following. We will take the function $j(q)$ to 
be defined by \eqref{jfd}, where $H(q)$ is defined by \eqref{hdeef}. The function $\tan^{-1}(x)$ 
that appears in some versions of the definition of $H(q)$ is taken to have the analytic structure 
depicted in Fig. \ref{tan}  \footnote{This analytic structure follows from the formula 
\begin{equation}
 \tan^{-1}x=\frac{1}{2i}\ln\left(\frac{1+i x}{1-i x}\right).
\end{equation}
if we define the logarithmic to be the usual log for positive real values, but to  have a branch cut along the negative real axis.}

The function $H(q)$ (see \eqref{hdeef} ) 
analytically continues to $H^M(\sqrt{s})$ 
\begin{equation}\label{acH} 
\begin{split}
H^M(\sqrt{s})  =&  
-i \int \frac{d^3r}{(2 \pi)^3 }\frac{1}{\left(r^2+c_B^2 - i \epsilon \right) 
\left( (r+q)^2+ c_B^2  - i \epsilon \right)}
\\
=& \frac{1}{8 \pi \sqrt{s} } \ln \left( \frac{ \frac{1}{2} + \frac{c_B}{\sqrt{s + i \epsilon }}}
{-\frac{1}{2} + \frac{c_B}{\sqrt{s + i \epsilon}} } \right).
\end{split}
\end{equation}
For  $\sqrt{s}< 2 c_B$, 
the factors of $i \epsilon$  make no difference in the formula  \eqref{acH}  
and may simply be dropped. When $\sqrt{s} >2 c_{B}$, 
the factors of $i \epsilon$ choose out the 
branch of logarithmic function and we have 
\begin{equation}\label{acHsl} 
 H^M(\sqrt{s})  =  
\begin{cases}
\frac{1}{8 \pi \sqrt{s} } \ln  \left(  \frac{ \frac{1}{2} + \frac{c_B}{\sqrt{s} } }
{-\frac{1}{2} + \frac{c_B}{\sqrt{s}} } \right)  &(\sqrt{s} < 2 c_{B}) \\
 \frac{1}{8 \pi \sqrt{s} }  \left(  \ln  \left( \frac{ \frac{1}{2} + \frac{c_B}{\sqrt{s}}}
{\frac{1}{2} - \frac{c_B}{\sqrt{s}} }  \right)+ i \pi \right) &
(\sqrt{s} > 2 c_{B})
\end{cases}.
\end{equation}
It follows, in particular,  that 
\begin{equation}\label{hc}
- i \left( H(\sqrt{s}- H^*(\sqrt{s}) \right) = \frac{\theta( \sqrt{s} -2 c_B)}{4 \pi \sqrt{s}} .
\end{equation}
Let $j^M$ denote the analytic continuation of  $j(q)$.  It follows that 
\begin{equation}\label{jm} \begin{split}
j^M(\sqrt{s})=&  (\pi \lambda_{B})  (4 \sqrt{s}) 
\left( \frac{ \left(4 \pi  \lambda_{B} \sqrt{s} +\wt b_4 \right) 
  +  \left(-4 \pi  \lambda_{B} \sqrt{s} + \wt b_4 \right)  e^{8 \pi  \lambda_{B}  \sqrt{s} H^M(\sqrt{s}) }
 } {\left(4 \pi  \lambda_{B} \sqrt{s}  + \wt b_4 \right) 
 -  \left(-4 \pi  \lambda_{B} \sqrt{s} + \wt b_4 \right)  e^{8 \pi  \lambda_{B}  \sqrt{s} H^M(\sqrt{s}) }
} \right),\\
j^M(\sqrt{s})=&
\begin{cases}
(\pi \lambda_{B})  (4 \sqrt{s}) 
\left( \frac{ \left(4 \pi  \lambda_{B} \sqrt{s} +\wt b_4 \right)
  +  \left(-4 \pi  \lambda_{B} \sqrt{s} + \wt b_4 \right)  \left(  \frac{ \frac{1}{2} + \frac{c_B}{\sqrt{s} } }
{-\frac{1}{2} + \frac{c_B}{\sqrt{s}} } \right)^{\lambda_{B}}
 } {\left(4 \pi  \lambda_{B} \sqrt{s}  + \wt b_4 \right) 
 -  \left(-4 \pi  \lambda_{B} \sqrt{s} + \wt b_4 \right)    \left(  \frac{ \frac{1}{2} + \frac{c_B}{\sqrt{s} } }
{-\frac{1}{2} + \frac{c_B}{\sqrt{s}} } \right)^{\lambda_{B}} 
} \right),  &(\sqrt{s} < 2 c_B)\\  
(\pi \lambda_{B})  (4 \sqrt{s}) 
\left( \frac{ \left(4 \pi  \lambda_{B} \sqrt{s} +\wt b_4 \right)
  +  e^{i \pi \lambda_{B}} \left(-4 \pi  \lambda_{B} \sqrt{s} + \wt b_4 \right)  \left(  \frac{ \frac{1}{2} + \frac{c_B}{\sqrt{s} } }
{\frac{1}{2} - \frac{c_B}{\sqrt{s}} } \right)^{\lambda_{B}}
 } {\left(4 \pi  \lambda_{B} \sqrt{s}  + \wt b_4 \right) 
 -  e^{i \pi \lambda_{B}} \left(-4 \pi  \lambda_{B} \sqrt{s} + \wt b_4 \right)    \left(  \frac{ \frac{1}{2} + \frac{c_B}{\sqrt{s} } }
{\frac{1}{2} - \frac{c_B}{\sqrt{s}} } \right)^{\lambda_{B}} 
} \right),  &(\sqrt{s} > 2 c_B)
\end{cases}
\end{split}
\end{equation}

\subsection{Poles of the functions $j(q)$ and $j^M(\sqrt{s})$ }

In this subsection we will analyze the conditions under which the functions $j(q)$ and 
$j^M(\sqrt{s})$ have poles for real values of their arguments. The conditions are most conveniently
presented in terms of inequalities on $b_4$ for fixed values of all other parameters.

Substituting $\wt b_4 = 2 \pi \lambda_{B}^2  c_B - b_4$ in the formulas \eqref{jm} and \eqref{jfe} we can see that for $b_4 >- 2\pi \lambda_{B} c_B(4-\lambda_{B})$ neither of the functions above has a pole at real values of its 
argument. When   $ -2 \pi \lambda_{B} c_B(4-\lambda_{B})\geq b_4 \geq -2 \pi c_B(4 - \lambda_{B}^2) $ the function $j^M$ 
has a pole, but $j$ has no pole. At the upper end of this interval the pole occurs at $\sqrt{s}=2 c_B$. 
 At the lower end of this interval the pole value is $\sqrt{s}=0$. For $b_4 \leq  -2 \pi c_B(4 - \lambda_{B}^2)$,
$j^M(\sqrt{s})$ has no real poles, but the function $j(q)$ develops a pole. This pole starts out at $q =0$
and migrates to $q=\infty$ as $b_4 \to- \infty$. 

A pole in the function $j^M(\sqrt{s})$ at $s=s_B$ signals the presence of a particle - antiparticle 
bound state in the singlet channel. As we have seen above, bound states exist only for $b_4$ less
than a certain minimum value. We will now explain how this result fits with physical intuition; let us 
first focus on the special case $\lambda_{B}=0$. In this case poles exist for $b_4 \leq 0$. In the 
non-relativistic limit a term $+\int b_4 \frac{ ({\bar \phi} \phi)^2}{2 N}$ in the  Minkowskian action 
represents a negative (attractive) delta function interaction between particles and antiparticles when $b_4 >0$.
It seems plausible that such an attractive potential could support a bound state, as appears to be the case. 
Clearly the binding energy of this system is proportional to $b_4$, and so goes to zero in the limit $b_4 \to 0$. 
In other words we should expect the mass of the bound state to be given precisely by $2c_B$ at $b_4=0$, 
exactly as we find. As $b_4$ decreases we should expect the binding energy to increase, i.e. for the 
bound state energy to decrease,  exactly as we find. Above a critical value of $c_B$ we find above that 
the binding energy is so large that the bound state energy vanishes. At even lower values of $b_4$
the vacuum is unstable as it is energetically favorable for particle - antiparticle pairs to spontaneously 
bubble out of the vacuum.
This instability is, presumably, signalled by the appearance of the tachyonic pole in $b_4$. The instability 
of the vacuum also seems reasonable from the viewpoint of quantum field theory; a large negative value of 
$b_4$ the classical scalar potential is unbounded from below; plausibly the same is true of the 
exact potential in the quantum effective action in this regime. 

The pattern is very similar at nonzero $\lambda_{B}$; though the precise values of the critical values for $b_4$ shift 
around. Apparently the anyonic interaction in the singlet channel renormalizes the effective interaction 
of the theory. 

Note that bound states do not exist in the limit $b_4 \to  \infty$, the limit in which the bosonic theory is dual to the fermionic theory.

It would be interesting to flesh out the qualitative discussion presented in this subsection. Near the threshold of bound state formation
the interacting particles are approximately non-relativistic, so it may be possible to reproduce the pole mass in this regime 
by solving a Schrodinger equation. We leave this to future work.

\subsection{Various limits of the function $j(q)$.} \label{eljq}

The explicit form of the function $j(q)$ (here $q=\sqrt{|q_3|^2}$) is one of the principal computational 
results of this section. $j(q)$ has the dimensions of mass. It is a function of one dimensionless variable 
$\lambda_{B}$, and three quantities of mass dimension 1; 
$q$, $c_B$ and $b_4$. It follows that 
$j$ takes the form $j = q h(x, y, \lambda_{B})$ where 
\begin{equation}\label{xyd}
x= \frac{q}{2 c_B}, ~~~y=\frac{q}{b_4}.
\end{equation} 
In this subsection we study the behavior of the function $j$ at extreme values of its three dimensionless 
arguments. 

\subsubsection{Large $b_4$ limit}

When $|b|_4 \gg \lambda_{B} q$ 
(i.e. when $ \lambda_{B} y \ll 1$) the function $j(q)$ simplifies to 
 \begin{equation} \label{anssc}
 j(q_3, \lambda_{B})={4 \pi i \lambda_{B} |q_3|} \left( \frac{ 1
  +   e^{8 i\pi  \lambda_{B} |q_3|H(q) }
 } {1 
 -    e^{8 i\pi  \lambda_{B} |q_3|H(q) }
} \right) = -4 \pi i \lambda_{B}  |q_3| \left( \frac{ 1  + e^{-2 i \lambda_{B}  \tan^{-1}\left( \frac{|q_3|}{2c_B} \right)} } { 1 - e^{-2 i \lambda_{B}  \tan^{-1}\left( \frac{|q_3|}{2c_B} \right)} }  
 \right) .
\end{equation}

\subsubsection{Small $\lambda_{B}$}

The function $j$ may be expanded in a Taylor series in $\lambda_{B}$ at fixed values of $x$ and $y$.  We 
find
\begin{equation}\label{sl}\begin{split}
j&= \frac{-b_4}{1+b_4 H(q)} -
\frac{16 \pi ^2 \lambda_{B} ^2 q_3^2}{3 b_4} \left(\frac{1}{\left(b_4 H(q)+1\right){}^2}-b_4 H(q)-1\right) + {\cal O}(\lambda_{B}^4).
\end{split}\end{equation}
The limits $\lambda_{B} \to 0$ and $b_4 \to \infty$ (i.e $\lambda_{B} \to 0$ and $y \to 0$ at fixed $x$ ) commute, 
so one may obtain 
the small $\lambda_{B}$ 
expansion of \eqref{anssc} by simply setting $b_4 \to \infty$
in \eqref{sl}.

Note that, in the strict $\lambda_{B} \to 0$ limit,
\begin{equation}\label{limlo} \begin{split}
&  \lim_{\lambda_{B} \to 0}N V(p, k, q^3)= \lim_{\lambda_{B} \to 0} j(q_3)= 
\frac{-b_4}{1+b_4 \frac{ \tan^{-1} \left( \frac{|q_3|} {2c_B}\right)}{4 \pi |q_3| }  }= \frac{-b_4}{1+H(q_3) b_4}.
\end{split}
\end{equation}
\eqref{limlo} is the well known result for the off shell amplitude in large $N$ $\phi^4$ theory. 
It is easily verified by directly solving the integral equation \eqref{sdee} at $\lambda_{B} =0$.

\subsubsection{The limit $|\lambda_{B}| \to 1$}

The expression for $j(q)$ simplifies somewhat in the limit $\lambda_{B} \to 1$.  The simplification 
is especially dramatic if we also take the limit $b_4 \to \infty$.  In the combined limit $y \to 0$ 
and $\lambda_{B} \to 1$ (the order of limits does not matter) we have 
\begin{equation}\label{lmto}
\begin{split}
j(q)=&{4 \pi i  q} \frac{ e^{i \tan^{-1}\left( \frac{2c_{B}}{q} \right)  } - e^{-i \tan^{-1}\left( \frac{2c_{B}}{q} \right)  } } { e^{i \tan^{-1}\left( \frac{2c_{B}}{q} \right)  } +e^{-i \tan^{-1}\left( \frac{2c_{B}}{q} \right)  } } 
\\
=&
 {4 \pi i q}
 (i)\tan \left( \tan^{-1}\left( \frac{2c_{B}}{q} \right)  \right) 
\\
=& -8 \pi  c_{B}.
\end{split}
\end{equation}

\subsubsection{The ultra-relativistic limit}

If $c_B$ and $b_4$ are held fixed while $\sqrt{-t}$ is taken to infinity (this is the case, 
for instance, in fixed angle high energy scattering in the $U$ and $T$-channels, see below) , 
we take $x$ and $y$ to infinity at fixed $\lambda_{B}$ and $j$ simplifies to 
\begin{equation}\label{ansit}
 j(q)=  4 \pi q \lambda_{B} \tan \frac{\pi \lambda_{B}}{2} .
\end{equation}

The ultra relativistic limit does not commute with the limit $b_4 \to \infty$. If $b_4$ is taken to 
$\infty$ first and $q \to \infty$ next then we work with $ y \to 0$, $ x \to \infty$ at fixed $\lambda_{B}$ 
and find
 \begin{equation}\label{ansith}
 j(q)= - 4 \pi  \lambda_{B} q \cot \left(\frac{\pi \lambda_{B}}{2}\right).
\end{equation}

The ultra relativistic limit also does not commute with the limit  $\lambda_{B} \to 0$. At $\lambda_{B}=0$ 
the function $j(q)$ tends to a constant proportional to $b_4$.  Physically this is 
we have a dimensionless coupling constant at nonzero $\lambda_{B}$, but only a dimensionful coupling 
constant at any finite $\lambda_{B}$; at zero lambda the theory is very weakly coupled at high energies, 
and receives contributions only from tree level graphs. 

\subsubsection{The massless limit}

If $c_B$ is taken to zero at fixed $b_4, \lambda_{B}$ 
and $q$ (i.e. if $x$ is taken to infinity at fixed $y$ and $\lambda_{B}$) 
then $j$ simplifies to the rational function
\begin{equation}\label{asio}
 j(q)=  {4 \pi  \lambda_{B} q} \left(  \frac{ 4 \pi \lambda_{B} \sin \left(\frac{\pi \lambda_{B}}{2}\right) 
q + {\tilde b}_4 \cos \left(\frac{\pi \lambda_{B}}{2}\right)  } {4 \pi  \lambda_{B}  \cos \left(\frac{\pi \lambda_{B}}{2}\right) q
-{\tilde b_4} \sin \left(\frac{\pi \lambda_{B}}{2}\right) } \right).
\end{equation}

The massless limit commutes with the limit $b_4 \to \infty$. In this limit \eqref{asio} reduces to 
\eqref{ansith}.

\subsubsection{The non-relativistic limit in the $U$ and $T$-channels}

As we will see below, the non-relativistic limit in the $U$ and $T$-channels is obtained by taking 
$c_B$ to infinity at fixed $q$. In other words, this limit is obtained by taking $x$ to zero at fixed 
$\lambda_{B}$ and $y$. In this limit 
$2 i \lambda_{B} \tan^{-1}\left( \frac{2 c_{B}}{q_3} \right)$ 
in \eqref{anssefmm} reduces to 
$\pi i \lambda_{B}$ and we have 
\begin{equation}\label{jnr}
j(\sqrt{-t})={\tilde b_4}.
\end{equation}

In this limit, in other words, the function $j$ receives contributions only from tree level 
scattering with the effective four point coupling ${\tilde b}_4$ in this limit. No genuine loop diagrams 
contribute to $T$ and $U$-channel scattering in this limit. 

If we first take $b_4 \to \infty$ and then take the non-relativistic limit we find 
\begin{equation}\label{jnrbi}
j(q)= -8 \pi c_B
\end{equation}
As  \eqref{jnr} and \eqref{jnrbi} both tend to infinity in the combined 
non-relativistic and $b_4 \to \infty$ limit, the reader may find herself tempted to conclude that 
the non-relativistic and $b_4 \to \infty$ commute. This conclusion is, infact, slightly misplaced. 
As we have emphasized in section \ref{abs}, the true dynamical information in the non-relativistic 
limit lies in the function 
$$h= -\frac{j}{8 i c_B}$$
which is derived from \eqref{nrred}.
The correct interpretation of the results of this subsection 
are that the function $h$ vanishes in the non-relativistic limit at fixed $b_4$, but reduces to 
a $\lambda_{B}$ independent numerical constant if $b_4$ is first taken to infinity.

\subsection{The non-relativistic limit in the $S$-channel}

As we will see below, the function relevant for scattering in the $S$-channel is the analytically 
continued function $j^M(\sqrt{s})$, see \eqref{jm}. The non-relativistic limit of 
$S$-channel scattering is obtained in the limit 
$\sqrt{s} \to 2 c_B$ where the limit is taken from 
above with all other parameters held fixed. It is easily seen from \eqref{jm} that in this limit 
\begin{equation}\label{jM}
j^M(\sqrt{s}) = - (\pi \lambda_{B}) (4 \sqrt{s}) {\rm sgn}( \lambda_{B}).
\end{equation}
Note that $j^M(\sqrt{s})$ 
is a non-analytic function of $\lambda_{B}$ as $\lambda_{B} \to 0$ in this 
limit. The non-analyticity is precisely of the form expected from the non-relativistic limit; 
infact, in this limit 
\begin{equation} \label{wo}
W_1(\sqrt{s})= \frac{\sin (\pi \lambda_{B})}{\pi \lambda_{B}} j^M(\sqrt{s}) .
 \end{equation}
 We will suggest an interpretation of this fact in section \ref{sing} below.

 \subsection{The onshell limit}

 In order to compute the physical S-matrix we analytically continue the amplitude $V$ to Minkowski space. 
It follows from \eqref{defV} that the onshell value of this analytically continued $V$ may directly be identified
with the scattering amplitude $T$ (see subsection \ref{ik}) once all momenta are taken onshell. 

As the 3 vectors $p$ and $p+q$ are simultaneously onshell, it follows that $p_3=-\frac{q_3}{2}$. Similarly $k_3=-\frac{q_3}{2}$. As  $p$ and $k$ are themselves onshell it follows that  
\footnote{The sign in the last two equations follows from the fact that $a(p)$ is defined with a square root with a branch cut on the negative real axis coupled with the fact that the rotation from Euclidean to Minkowski space proceeds in the clockwise direction.}
$$ a(p)^2=-\frac{q_3^2}{4}, ~~~~a(k)^2=-\frac{q_3^2}{4}, ~~~a(p)=a(k)=- i \frac{q_3}{2}.  $$

\subsubsection{An infrared `ambiguity' and its resolution}

The offshell amplitude \eqref{anon} takes the form 
\begin{equation}\label{ossf} 
\begin{split} 
&N V= P T,\\
T&=\left( 4 \pi i \lambda_{B}  q_3  \frac{p_- + k_-}{p_- - k_-}
+ j(q_3)\right),\\
P&=e^{-2 i \lambda_{B}   \left( \tan ^{-1}\left( \frac{2 (a(k)}{q_3} \right) - 
 \tan ^{-1}\left( \frac{2 (a(p)}{q_3} \right)\right).  }
\end{split}
\end{equation}
The expression $T$ defined above has a perfectly smooth on shell limit that we will study below. 
The onshell limit of $P$ is more singular,
\begin{equation}\label{pamb}
P= e^{-2 i \lambda_{B}   \left( \tan ^{-1}(-i) - \tan ^{-1}(-i) \right)   }, 
\end{equation}
recall that $\tan^{-1}(i)$ diverges, $P$ thus takes the schematic form 
$$P=e^{ i \lambda_{B} \left(\infty  -\infty \right)}$$
and is ambiguous.

The ambiguity in the expression for $P$ has its origins in ladder graphs in which the scalars 
interact via the exchange of a very soft gauge boson. The integration over very small 
gauge boson momenta is divergent; however we encounter two classes of divergences which 
could potentially cancel,  leading to the ambiguous result for $P$. 

In a theory with physical gluonic states, the IR divergence obtained upon integrating out soft 
gluons is a real effect in scattering amplitudes (even though it cancels out in physical IR safe
observables). However Chern-Simons theory has no physical gluons. On physical 
grounds, therefore, we do not expect the scattering amplitude to be divergent or ambiguous 
in any way. We will now explain  that the correct on shell value for $P$ is infact unity. 

We first note that the $\lambda_{B}$ dependence of the ambiguity is extremely 
simple; it follows that if we can accurately establish the on shell value of $P$ at one loop, we 
know its correct value at all loops. In order to determine $P$ at one loop, in Appendix \ref{olm} we have performed a careful computation of the one loop amplitude directly in Minkowski space. 
Offshell our result agrees perfectly with the analytic continuation of \eqref{anonm}, as we 
would expect. On being careful about all factors of $i \epsilon$ however, we find that the 
on shell result is unambiguous, and we find that the two terms in \eqref{pamb} actually 
cancel. It follows that the correct on shell continuation of $P$ above is simply unity. 
In the next subsection we present a completely  independent verification of this result 
from a rather different point of view. 

In this subsubsection 
we have already encountered an unusual phenomenon: the analytic 
continuation of the Euclidean answer is ambiguous or incomplete due to potential 
IR on shell singularities, and this ambiguity is resolved by performing a computation 
directly in Minkowski space. In the case at hand the ambiguity had a relatively simple 
and straightforward resolution. A similar issue will come back to haunt us in a more 
virulent form in our study of $S$-channel scattering below. 

\subsubsection{Covariantization of the amplitude}

We now turn to the onshell limit of $T$ in \eqref{ossf}. In this limit the 
expression for $T$ may equally well be written in the manifestly 
covariant form
\begin{equation}\label{onshells}
T=4 \pi i \lambda_{B}  \epsilon_{\mu \nu \rho}  \frac{q^\mu (p-k)^\nu (p+k)^\rho}{(p-k)^2}
+ j(\sqrt{q^2}).
\end{equation}
\footnote{The equivalence between \eqref{onshells} and 
\eqref{ossf} follows from the observation that, in
onshell, 
\begin{equation}
q \cdot (p-k) = 0 
\Rightarrow 
p_{3} - k_{3} = 0 
\Rightarrow 
(p_{-} - k_{-})
(p_{+} - k_{+}) = \frac{1}{2}(p-k)^2 \nonumber
\end{equation}
and the observation (see the previous subsection)
that $j(q_3)=j(-q_3)$.}
The manifestly covariant expression \eqref{onshells} also enjoys invariance under the simultaneous
operation of an arbitrary parity flip together with a flip in the sign of $\lambda_{B}$. The first 
term in \eqref{onshells} is odd under parity flips as well as under a flip in the sign of $\lambda_{B}$. 
The second term in \eqref{onshells} is even under both operations.

As we will explain in more detail below, the magnitude of the expression $\epsilon_{\mu \nu \rho}  \frac{q^\mu (p-k)^\nu (p+k)^\rho}{(p-k)^2}$ can be written in terms of the standard kinematical invariants
$s, t, u$. However the sign of this expression is not a function of these invariants. This is a peculiar  kinematical feature of 2-2 scattering in $2+1$ dimensions. The most general amplitude in this dimension 
is a function of $s, t$ and the $Z_2$ valued variable 
$$E(q, p-k, p+k)= {\rm sgn} 
\left(\epsilon_{\mu \nu \rho}  q^\mu (p-k)^\nu (p+k)^\rho \right).$$
The quantity $E(a, b, c)$ measures the `handedness' of the triad of three vectors $a,b, c$. 
Note that it is odd under parity as well as under the interchange of any two vectors. 

In order to obtain the onshell amplitude from the offshell one, one can utilize LSZ formula.
By making different choices for the signs of the energies of the four external particles, the 
single master expression \eqref{onshells} determines the T-matrix for particle-particle scattering 
in both channels, as well as the T-matrix for particle  antiparticle scattering in the adjoint channel; 
this observation also makes clear that these three T-matrices are related as usual by crossing symmetry. 
In the rest of this section we explicitly evaluate the T-matrix in each of these channels and 
comment on our results. 

\subsection{The S-matrix in the adjoint channel}

In order to determine the scattering function $T^B_T$ (particle - antiparticle scattering 
in the adjoint channel) we study the scattering process 
\begin{equation}\label{pft}
P_i(p_1) + A^j(p_2) \rightarrow P_i(p_3) + A^j(p_4)
\end{equation}
for $i \neq j$. It follows from the definitions \eqref{Tijmn}  that the scattering amplitude for this process is precisely the function $T^B_T$.

The S-matrix for the scattering process \eqref{pft} is evaluated by the exact onshell amplitude
\eqref{onshells}, once we make the identifications 
$$p_1=p+q, ~~~p_2=-(k+q), ~~~p_3=-p, ~~~p_4=k.$$
It follows that 
$$s=-(p-k)^2, ~~~t=-q^2, ~~~u=-(p+q+k)^2$$
which implies
\begin{align*}
&p_1^2=p_2^2=p_3^2=-c_{B}^2,
~~p_1\cdot p_2=\frac{-s+2c_{B}^2}{2},
\\
&p_1\cdot p_3=\frac{-t+2c_{B}^2}{2},
~~p_2\cdot p_3=\frac{-u+2c_{B}^2}{2}. 
\end{align*}
Note also that \footnote{In our notation $\epsilon_{012}=-\epsilon^{012}=1.$
} 
\begin{equation} \label{manip} \begin{split} 
&|\epsilon_{\mu\nu\rho} q^\mu (p-k)^\nu (p+k)^\rho |^2
=4 |\epsilon_{\mu\nu\rho} (p+q)^\mu (k+q)^\nu p^\rho |^2
=4 |\epsilon_{\mu\nu\rho} p_1^\mu p_2^\nu p_3^\rho |^2 \\
=&-4 \biggl( p_1^2 p_2^2 p_3^2 + 2 (p_1\cdot p_2) 
(p_2\cdot p_3) (p_3\cdot p_1)  
\\
&\qquad \quad -p_3^2 (p_1\cdot p_2)^2 
-p_2^2 (p_1\cdot p_3)^2 -p_1^2 (p_3\cdot p_2)^2 \biggr)\\
=& -\left(16 c_{B}^6-8 c_{B}^4 (s+t+u)+c_{B}^2 (s+t+u)^2-s~ t~ u \right) \\
=&s~t~u.
\end{split}
\end{equation}
It follows that 
\begin{equation} \label{smuc}
T_T^B(p_1, p_2, p_3, p_4, \lambda_{B}, b_4, c_B)= \frac{4i\pi}{k_B} E(p_1,p_2,p_3) \sqrt{ \frac{tu}{s}} 
+ \frac{1}{N}j(\sqrt{-t}),
\end{equation}
where the field renormalization factor is trivial in the leading order in $1/N$ expansion. 
In the center of mass frame, this S-matrix is given by 
\begin{equation}\label{smadjcm}
T_T^B(s,\theta, \lambda_{B}, b_4, c_B)= \frac{4i\pi}{k_B}\frac{s-4c_B^2}{2\sqrt{s}}\sin(\theta) 
+ \frac{1}{N}j\left(\sqrt{s-4c_B^2}\left|\sin\left(\frac{\theta}{2}\right)\right|\right).
\end{equation}
Notice that the scattering amplitude is completely regular at $\theta=0$; in particular 
In the non-relativistic limit we find that the scattering function $h(\theta)$ is given by 
\begin{equation}\label{nradjbnz}
h_T^B(\theta) = 0
\end{equation}
at finite $b_4$. If $b_4$ is taken to infinity first, on the other hand, in the non-relativistic limit we find
\begin{equation}\label{nradjbz}
h_T^B(\theta) = -i\pi.
\end{equation}
Notice that in neither case does $h(\theta)$ have a term proportional either to $\cot \left(\frac{\theta}{2}\right)$ 
or to $\delta(\theta)$ ) as anticipated in our discussion of the non-relativistic limit in 
 subsection \ref{nrtc}.

\subsection{The S-matrix for particle- particle scattering}

In order to determine the scattering function $T^B_{U_d}$ we study the scattering process 
\begin{equation}\label{pftpp}
P_i(p_1) + P_j(p_2) \rightarrow P_i(p_3) + P_j(p_4).
\end{equation}
It follows from the definitions \eqref{TijmnP}  that the scattering amplitude for this process is precisely 
the function $T^B_{U_d}$, provided $i \neq j$. 

The S-matrix for the scattering process \eqref{pft} is evaluated by the exact onshell amplitude
\eqref{onshells}, once we make the identifications 
$$p_1= p+q, ~~~p_2=k, ~~~p_3=-p, ~~~p_4=-(k+q).$$
It follows that 
$$s=-(p+q+k)^2, ~~~t= -q^2, ~~~u= - (p-k)^2$$
\begin{equation} \label{smucpp}
T^B_{U_d}(p_1, p_2, p_3, p_4, \lambda_{B}, b_4, c_B)= \frac{4i\pi}{k_B} E(p_1,p_2,p_3)\sqrt{ \frac{ts}{u}} + \frac{1}{N}j(\sqrt{-t})
\end{equation}
where $E(p_1,p_2,p_3)$ was defined in \eqref{smatrix}. Notice that, upto the issues involving the sign $E$, 
$T^B_{U_d}$ is obtained from $T^B_T$ by the interchange $s \leftrightarrow u$. 

In the bosonic theory under study, $T^B_{U_e}$ is obtained from $T^B_{U_d}$
by the interchange $p_1 \leftrightarrow p_2$. This interchange flips the sign of $E$ and also interchanges
$u$ and $t$, so we find  
\begin{equation} \label{smucppe}
T^B_{U_e}(p_1, p_2, p_3, p_4, \lambda_{B}, b_4, c_B)
= -\frac{4i\pi}{k_B} E(p_1,p_2,p_3)\sqrt{ \frac{us}{t}} 
+ \frac{1}{N}j(\sqrt{-u}).
\end{equation}

If the non-relativistic limit is taken at nonzero $b_4$ we have using \eqref{nrred}
\begin{equation}\label{nrlpp}
\begin{split}
h_{U_d}^B(\theta) &= -\frac{\pi}{k_B}\tan\left(\frac{\theta}{2}\right),\\
h_{U_e}^B(\theta)&= \frac{\pi}{k_B}\cot\left(\frac{\theta}{2}\right)  .  
\end{split}
\end{equation}
If $b_4$ is first taken to infinity, on the other hand, we have
\begin{equation}\label{nrlpp1}
\begin{split}
h_{U_d}^B(\theta) &= -\frac{\pi}{k_B}\tan\left(\frac{\theta}{2}\right)-i\pi,\\
h_{U_e}^B(\theta)&= \frac{\pi}{k_B}\cot\left(\frac{\theta}{2}\right)-i\pi   ,   
\end{split}
\end{equation}
in good agreement with the predictions of 
subsection \ref{nrtc}.

\section{The onshell one loop amplitude in Landau Gauge}\label{landauonshell}

In this section we present a consistency check of 
\eqref{onshells} and \eqref{anssefm}, the main 
results of the previous section. Our check proceeds  by independently evaluating the onshell 4 point 
function at one loop in the covariant Landau gauge. As we describe below, the results of our computation
are in perfect agreement with the expansion of \eqref{onshells} and \eqref{anssefm} to ${\cal O}(\lambda_{B}^2)$. 

We believe that the check performed in this subsection has value for several reasons. First, the lightcone 
gauge employed in this paper is nonstandard in several respects. It is not manifestly covariant. It leads
to a gauge boson propagator that is singular  when $p_-=0$: as we have emphasized above, in order 
to make progress in our computation we were forced to simply postulate an $i \epsilon$ prescription 
that resolves this singularity in an appealing manner. And finally the offshell result of this 
computation appears, at first sight, to be ambiguous when continued onshell. 

The computation we describe in this subsection, on the other hand, suffers from none of these deficiencies. 
It is manifestly covariant; it is an entirely standard computation, following rules that have been 
developed and repeatedly utilized over several decades, and it will turn out to have no confusing 
IR ambiguities. \footnote{Of course the weakness of the Landau gauge is that, unlike in the lightcone 
gauge,  it is very difficult to perform explicit computations in this gauge beyond low loop order, as 
the gauge condition does not remove all gauge boson self interactions.} 
 For this reason, the match between our results of the previous subsection and 
those that we report in this subsection may be regarded as rather nontrivial evidence that we have correctly 
dealt with all the tricky aspects of the computation in the lightcone gauge. 

\begin{figure}[tbp]
  \begin{center}
 \subfigure[]{\includegraphics[scale=.4]{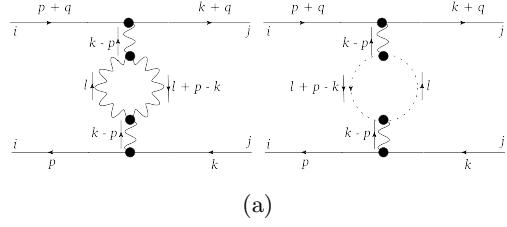}
\label{g}
  }
  \end{center}
  \vspace{-0.5cm}
  \caption{Gauge loop in gauge field propagator is cancelled by the ghost loop.}\label{Cancel}
  \end{figure}

We now turn to a brief description of the Landau gauge computation, relegating most details to 
Appendix \ref{landau}. For simplicity we work with the scalar theory in 
special case $b_4=0$. 
In the Landau Gauge, the gauge boson propagator receives two 
corrections at one loop:  from a gauge boson loop and from a ghost loop. It is easily verified that these two diagrams cancel each other (see Fig \ref{Cancel}).
It is also easily seen that the ghosts make no appearance in any other 
diagram that contributes to one loop scattering of four gauge bosons. It follows that, at the one loop 
level, we may ignore both renormalizations of the gauge boson propagator as well as the ghosts: These
two complications cancel each other out. 

\begin{figure}[tbp]
  \begin{center}
  \subfigure[]{\includegraphics[scale=.4]{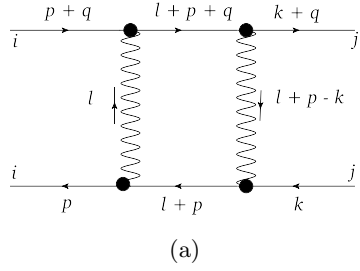}
  }\\
\end{center}
  \vspace{-0.5cm}
  \caption{The box diagram in Landau Gauge}
\label{LgBox}
  \end{figure}

\begin{figure}[tbp]
  \begin{center}
  \subfigure[]{\includegraphics[scale=.4]{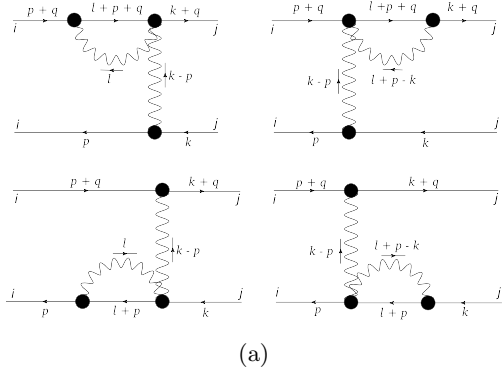}
  }\\
\end{center}
  \vspace{-0.5cm}
  \caption{H diagrams in the lightcone gauge. }
\label{LgH}
  \end{figure}
\begin{figure}[tbp]
  \begin{center}
  \subfigure[]{\includegraphics[scale=.4]{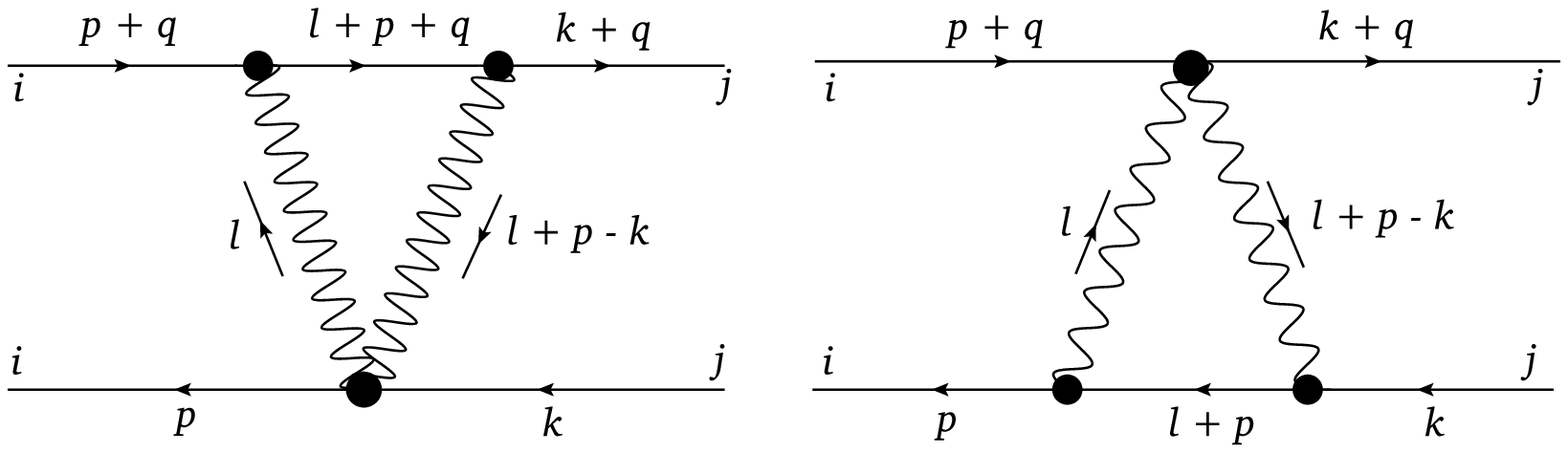}
  }\\
\end{center}
  \vspace{-0.5cm}
  \caption{V diagrams in the Landau Gauge.}
\label{LgV}
  \end{figure}
\begin{figure}[tbp]
  \begin{center}
  \subfigure[]{\includegraphics[scale=.4]{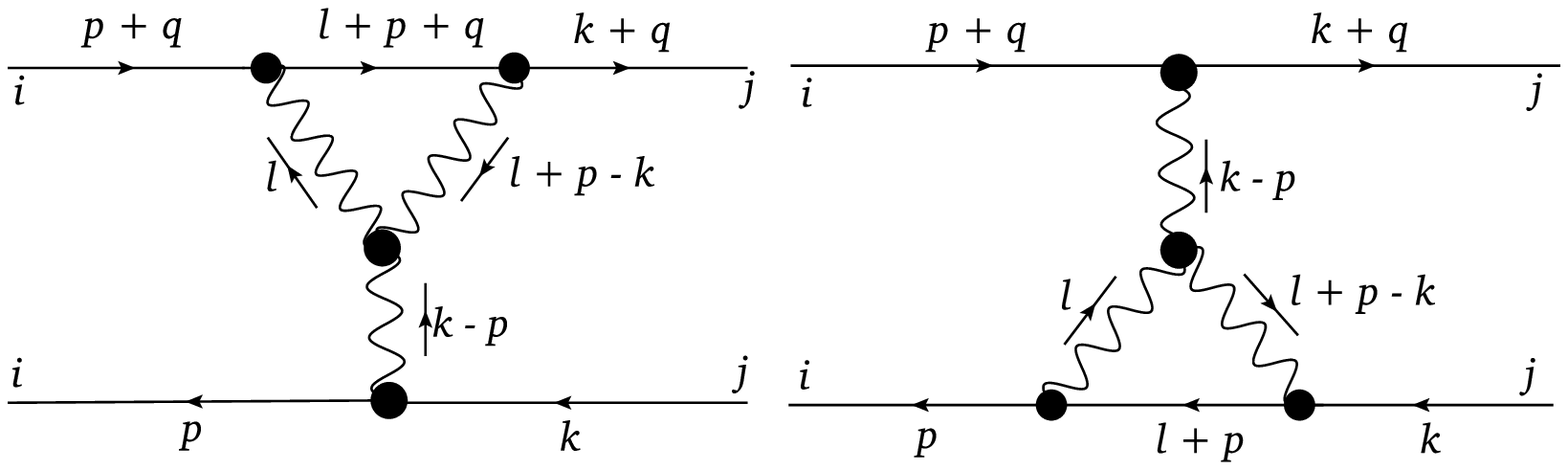}
  }\\
\end{center}
  \vspace{-0.5cm}
  \caption{Y diagram in the Landau gauge.}
\label{LgY}
  \end{figure}
\begin{figure}[tbp]
  \begin{center}
  \subfigure[]{\includegraphics[scale=.4]{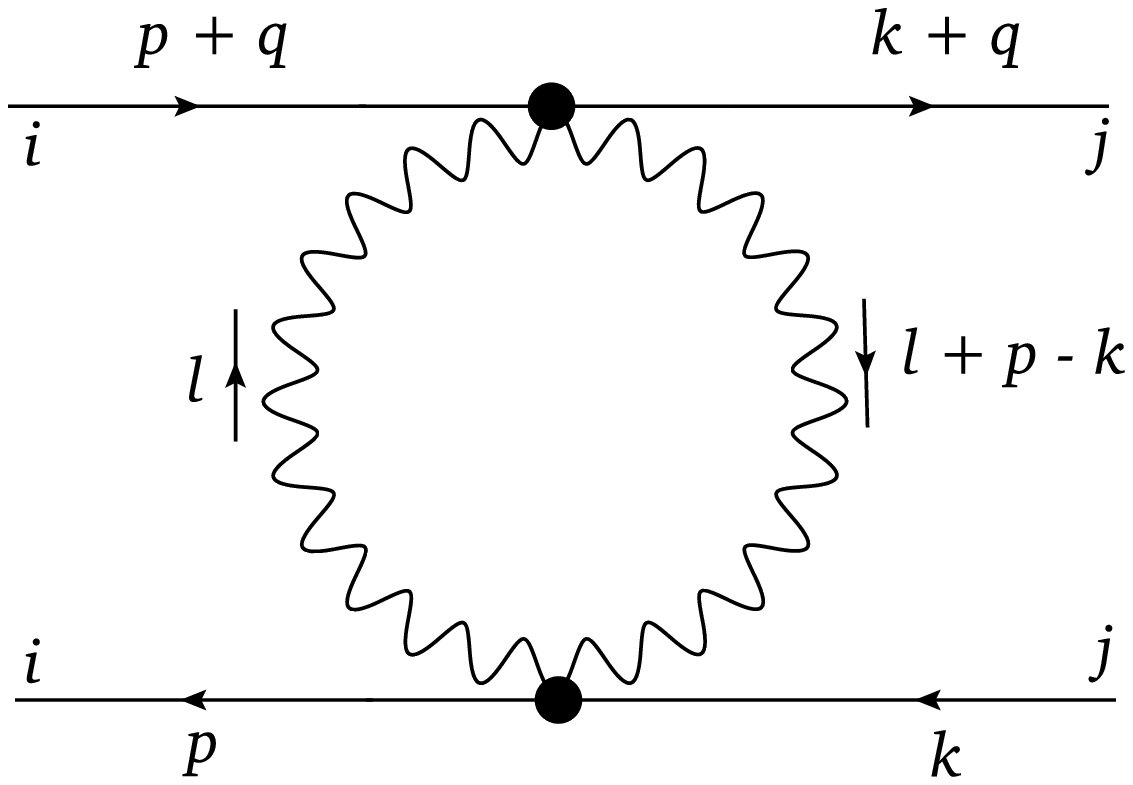}
  }\\
\end{center}
  \vspace{-0.5cm}
  \caption{Eye diagram in the Landau gauge.}
\label{LgEye}
  \end{figure}
\begin{figure}[tbp]
  \begin{center}
 \subfigure[]{\includegraphics[scale=.4]{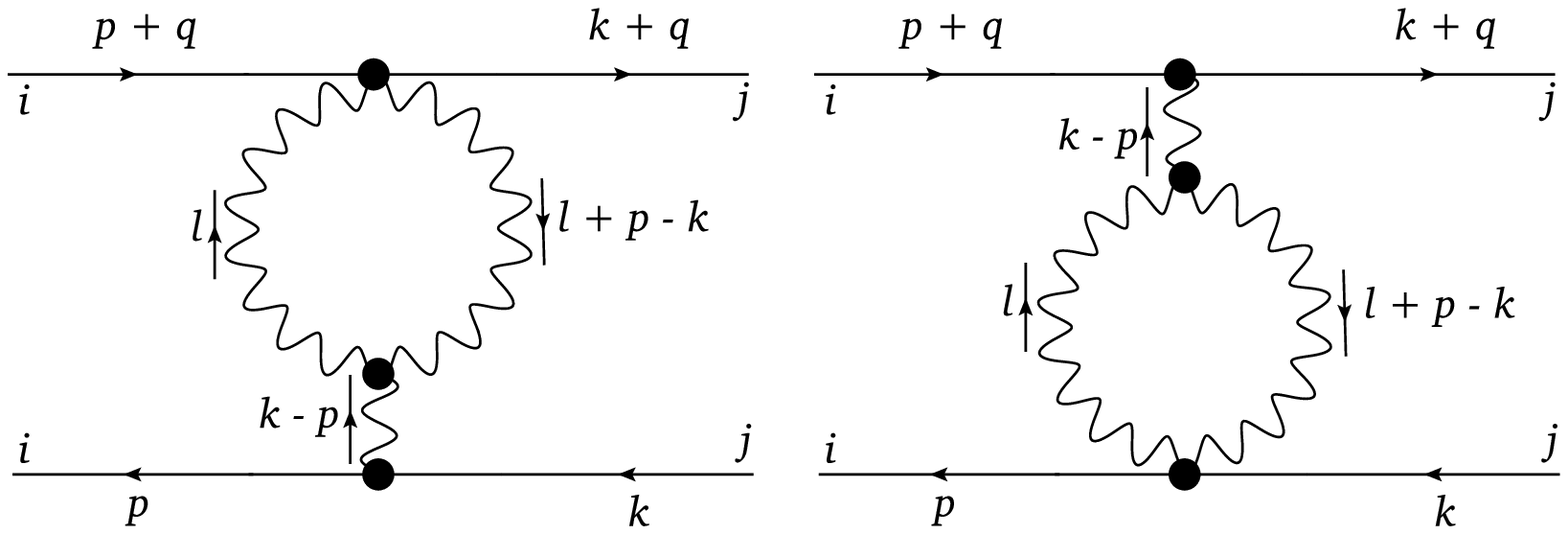}
  }\\
\end{center}
  \vspace{-0.5cm}
  \caption{Lollipop diagram in the Landau Gauge.}
\label{LgLol}
  \end{figure}

With this understanding it is easily verified that the one loop scattering amplitude of four scalar bosons 
receives contributions from six classes of diagrams, 
(see six figures, Figs.~\ref{LgBox}$\sim$\ref{LgLol}). 
These are the box diagrams 
of Fig.~\ref{LgBox}, 
the h diagrams of 
Fig.~\ref{LgH}, the V diagrams of Fig. \ref{LgV}, 
the Y diagrams of Fig.~\ref{LgY}, the Eye diagram of Fig.~\ref{LgEye},
and the Lollipop diagram of Fig.~\ref{LgLol}.
In order to evaluate the one loop contribution to four scalar scattering, we need to evaluate the 
sum of these six classes of diagrams.  
It is well known, however, that in the study of planar diagrams 
there is a canonical way to sum the integrands of these diagrams before performing the integral. 
We choose a uniform definition of the loop momentum across all the six sets of graphs; the loop 
momentum $l$ is the momentum that flows clockwise between the external line with momentum $p$ and the 
external line with momentum $p+q$ (see Fig. \ref{LgBox}).  Adopting this definition, we then evaluate 
the integrand for each class of diagrams, and sum the integrands. 

It turns out that the process of summing integrands leads to several cancellations and simplifications. 
 In order to see the cancellations between integrands, it is important that each integrand be expressed in a canonical form.
 There is, of course, a standard way to achieve this. It is a 
well known result that an arbitrary one loop integrand in $d$ dimensions may be reduced, under the 
integral sign \footnote{i.e. upto terms that integrate to zero.} 
to a linear sum over scalar integrals 
\footnote{A scalar integral, by definition, 
is the loop integral over a product of propagators in the loop, but
with numerator unity.} with at most $d$ propagators. 
The coefficients in this decomposition are 
rational functions of the external momenta. There also exists a rather simple algorithmic procedure 
for decomposing an arbitrary integrand into this canonical form. Finally the scalar integrals are not all 
independent. The canonical form of the integrand is obtained by decomposing the integrand into 
a linear combination of  linearly independent scalar integrands. 

Implementing this procedure (see Appendix \ref{landau} for several details) we find that the full one 
loop integrand for 4 scalar 
boson scattering turns out to be given by the remarkably simple expression
\begin{equation}\label{finintm}
\begin{split}
I_{full}=4\pi^2\lambda_{B}^2\biggl(&-\frac{2}{c_{B}^2+(l+p)^2}-\frac{2}{(l+p-k)^2}
\\
&-\frac{8 k \cdot q}{\left( c_{B}^2+(l+p)^2\right) 
\left( c_{B}^2+(p+q+l)^2 \right)}\biggr).
\end{split}
\end{equation}
In the dimensional regulation scheme that we employ, the integral of the first term in 
\eqref{finintm} is $4 \pi^2 \lambda_{B}^2  \times \frac{c_{B}}{2 \pi}$. 
The integral of the second 
term simply vanishes. 
The integral of the third term is $32 \pi^2 (k\cdot q) 
\lambda_{B}^2 H(q)$ 
where 
$H(q)$, the one loop amplitude for four boson scattering, was defined in \eqref{limlo}. It follows 
that the full one loop onshell scattering amplitude is given by 
\begin{equation}\label{vol}
\int \frac{d^3 l}{(2\pi)^3} I_{full} = 
V_{\rm one~loop}= 2 \pi c_{B} \lambda_{B}^2+ 32 \pi^2 (k\cdot q) \lambda_{B}^2 H(q)
\end{equation}
in perfect agreement with \eqref{sl} at $b_4=0$.

We end this brief subsection with two further comments. We first note that the one loop amplitude 
in the Landau gauge was manifestly infrared safe. While integrands that would have given rise 
to infrared divergences (associated with the exchange of arbitrarily soft gluons in loop) appear 
at intermediate stages in the computation, they all cancel already at the level of the integrand 
(i.e. before performing any integrals). This is the analogue of the slightly more subtle cancellation 
of IR divergences in lightcone gauge 
mentioned above and described in more detail in Appendix \ref{olm}.

The second comment is that the derivation integrand reported in \eqref{finintm}
uses a reduction formula that is valid only at generic values of external momenta. Our derivation of this 
formula fails, for instance, when two of the external momenta are collinear. In more familiar quantum 
field theories this caveat would be of little consequence; the analyticity of the amplitude as a function 
of external momenta would guarantee that the result applied at all values of the momenta. As we will 
see below, however, this amplitudes in Chern-Simons theories sometimes appear to have non 
analytic singularities, so the caveat spelt out in this paragraph may turn out to be more than a 
pedantic technicality.
\section{Scattering in the fermionic theory} \label{fermion}
\begin{figure}[tbp] 
   \begin{center}
     \subfigure[]{\includegraphics[scale=0.5]{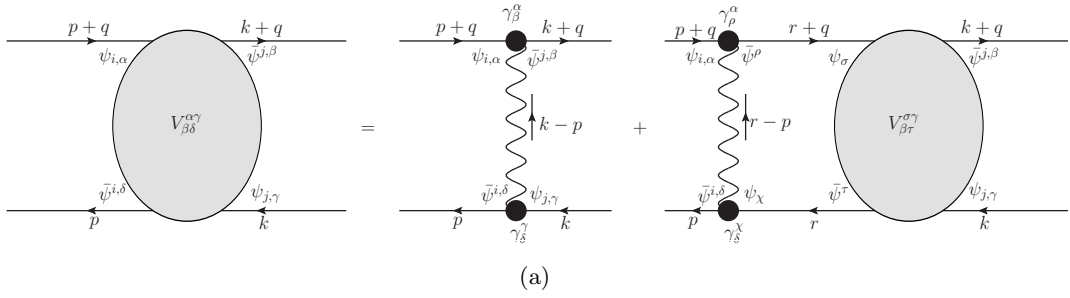}
 }
   \end{center}
   \vspace{-0.5cm}
   \caption{A diagrammatic depiction of the integral equation obeyed by offshell four point scattering
amplitudes in the fermionic theory. The blob here represents the all orders scattering amplitude.}
\label{IntF}
 \end{figure}
\begin{figure}[tbp] 
   \begin{center}
     \subfigure[]{\includegraphics[scale=0.6]{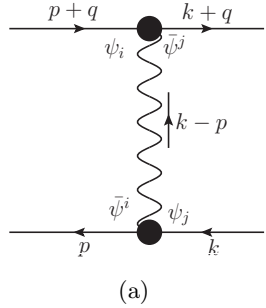}
 }
   \end{center}
   \vspace{-0.5cm}
   \caption{Fermionic tree level diagram}
\label{TF}
 \end{figure}
 
\begin{figure}[tbp] 
   \begin{center}
     \subfigure[]{\includegraphics[scale=0.6]{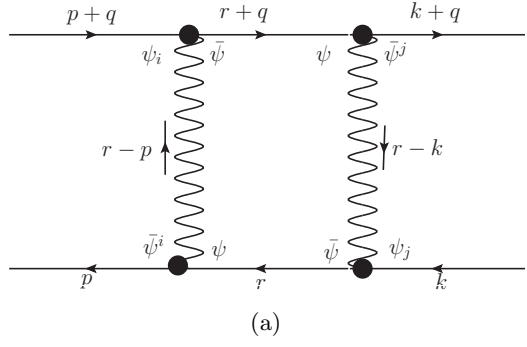}
 }
   \end{center}
   \vspace{-0.5cm}
   \caption{Fermionic 1 loop diagram}
\label{BoxF}
 \end{figure}
In this section we compute the four point scattering amplitude in the theory of fundamental fermions 
coupled to Chern-Simons theory. As in the bosonic theory, we integrate out 
the gauge boson to obtain an offshell effective four fermi term in the quantum effective action for our 
theory, given by 
\begin{equation} \label{defVf}
\frac{1}{2} \int \frac{d^3p}{(2 \pi)^3}  \frac{d^3 k}{(2 \pi)^3} \frac{d^3 q}{(2\pi)^3} 
   V^{\alpha\gamma}_{\beta\delta}(p,k,q){\psi}_{i,\alpha}(p+q){\bar\psi}^{j,\beta}
(-(k+q)){\bar\psi}^{i,\delta}(-p)\psi_{j,\gamma}(k).
 \end{equation}
 We then take an appropriate onshell limit to evaluate the S-matrix.
\subsection{The offshell four point amplitude}
As in the case of the bosonic theory, the offshell four point amplitude $V^{\alpha\gamma}_{\beta\delta}(p,k,q)$ obeys a closed Schwinger-Dyson equation. As for the bosonic theory, we work with 
the special case $q^\pm=0$. As above we first set up this Schwinger-Dyson equation for the Lorentzian theory, 
but find it more convenient, technically, to work with the Euclidean rotated amplitude. The Euclidean 
rotated amplitude is defined in a manner very similar to the bosonic theory (see below for a few more 
details), and may be shown to obey the Schwinger-Dyson equation
\begin{equation}\label{Masteq1}\begin{split}
 V^{\alpha\gamma}_{\beta\delta}(p,k,q)&=\frac{1}{2}(\ga^{\mu})^{\alpha}_{ \beta}G_{\mu\nu}(p-k)(\gamma^{\nu})^{\gamma}_{\de}\\
&+\frac{1}{2}\int \frac{d^{3}r}{(2\pi^3)}[\gamma^{\mu}G(r+q)]^{\alpha}_{\sigma}V^{\sigma\gamma}_{\beta\tau}(r,k,q)[G(r)\gamma^{\nu}]^{\tau}_{\delta}G_{\mu\nu}(p-r).
\end{split}\end{equation}
Here $G(p)_{\alpha \sigma}$ is the exact fermionic propagators determined in 
\eqref{propF} (see also 
\cite{Giombi:2011kc}),
while $G_{\mu\nu}$ is the gauge boson propagator defined by 
\begin{equation}\label{gaugepr}
\la A_\mu^a(-p) A^b_\nu(q) \ra 
= (2\pi)^3 \delta^3(p-q) G_{\mu \nu}(q)
\end{equation}
where $A_\mu= A^a T^a$ and we work with generators normalized so that 
\begin{equation}
 \sum_{a}(T^{a})^{i}_{j}(T^{a})^{k}_{l}=\frac{1}{2}\delta^{i}_{l} 
\delta^{k}_{j}.\label{normaln}
\end{equation}
And here $\gamma^{\mu}$ compose the Euclidean Clifford algebra, 
$$\{\gamma^{\mu}, \gamma^{\nu}\} = 2\delta^{\mu\nu}, 
\quad [\gamma^{\mu}, \gamma^{\nu}]= 2i\epsilon^{\mu\nu\rho}\gamma_{\rho},
\qquad 
(\epsilon^{103} = \epsilon_{103} = 1).
$$

In the lightcone gauge in which we work the only nonzero components of $G_{\mu\nu}$ are 
\begin{equation}\label{gcomp}
G_{+3}(p)=-G_{3+}(p)=\frac{4\pi ~i}{\kappa~p^{+}}.
\end{equation}

Now noting the fact that only non zero component is $G_{+3}(p)=-G_{3+}(p)$ and using rearrangement with $\gamma^{+} = \frac{i \gamma^0 + \gamma^1}{\sqrt{2}}$,
\begin{equation}\label{gamid}
 (\ga^{+})^{\alpha}_{ \beta}(\ga^{3})_{\delta}^{\gamma }-(\gamma^{3})^{\alpha}_{\beta}(\gamma^{+})_{\delta}^{\gamma}
=-\left( \delta^{\ga}_{\beta}(\gamma^{+})^{\alpha}_{\delta}-(\gamma^{+})^{\gamma}_{\beta} \delta^{\alpha}_{\delta}\right),
\end{equation}
as well as 
\begin{equation}\label{gamxid}
 \ga^{+} X \ga^{3}-\ga^{3} X \ga^{+}=-2(X_{I}\ga^{+}-X_{-}I),
\end{equation}
(here $X=X_i \gamma^i + X_I I$ is an arbitrary $2 \rightarrow 2$ matrix), 
we conclude that in the $\al,\de$ indices of R.H.S of the Eq.(\ref{Masteq1}) 
- and therefore 
the LHS , and so $V$ takes the form 
\begin{equation}\label{formv}
 V^{\alpha\gamma}_{\beta\delta}(p,k,q)=g(p,k,q) \delta^{\alpha}_{\delta}\delta^{\gamma}_{\beta}+
f(p,k,q) (\gamma^{+})^{\alpha}_{\delta}\delta^{\gamma}_{\beta}+g_1(p,k,q) \delta^{\alpha}_{\delta}(\gamma^{+})^{\gamma}_{\beta}+
f_1(p,k,q) (\gamma^{+})^{\alpha}_{\delta}(\gamma^{+})^{\gamma}_{\beta}.
\end{equation}

Plugging this form $V$ into \eqref{Masteq1} yields a set of four integral equations for the four component
functions in \eqref{formv}. We have succeeded in finding the exact solution to these equations.
We present the derivation of our solution in Appendix~\ref{AP:fermion}.  The final result for this 
offshell amplitude is extremely complicated. The result, which takes multiple pages to write, is given in \eqref{ansatz}, \eqref{fsw} and \eqref{fsb} of the Appendix. We see no benefit in reproducing 
this extremely complicated final result in the main text.

\subsection{The onshell limit}

As we have seen above, the offshell four point function defined in \eqref{defVf} is quite a complicated 
object. In this section we will argue that the onshell S-matrix is, however, rather simple. 

In order to study the S-matrix it is first convenient to continue our result for $V$ in \eqref{defVf} 
to Minkowski space. This is achieved by making the substitution 
$$ p^0  \rightarrow -i p^0, ~~~k^0 \rightarrow -i k^0, ~~~\gamma^0 \rightarrow -i \gamma^0,$$
on the Euclidean result of the previous subsection. This substitution yields the four fermi term in the 
effective action \eqref{defV} in Lorentzian space. 

In order to convert this four point vertex to a scattering amplitude, we must now go onshell. We now 
pause to carefully explain how this is achieved. 

In free field theory (i.e. in the absence of the four point function interaction) the fermion field operators 
may be expanded in creation and annihilation modes in the standard fashion
\begin{equation}\label{psift}
\begin{split}
\psi(x)&=\int \frac{d^3p}{(2\pi)^3}\psi(p)e^{ip\cdot x}  \\
&=\int \frac{d^2p}{(2\pi)^2}\frac{1}{\sqrt{2~E_p}}
\left(u(\vec{p})a_{\vec{p}} 
e^{ip\cdot x}+v(\vec{p})b_{\vec{p}}^{\dagger}e^{-ip \cdot x}\right),\\
 {\bar\psi}(x)&=\int \frac{d^3p}{(2\pi)^3}{\bar\psi}(p)e^{ip \cdot x}  \\     
&= \int \frac{d^2p}{(2\pi)^2} 
\frac{1}{\sqrt{2~E_p}}\left({\bar u}(\vec{p})a_{\vec{p}}^{\dagger} 
e^{-ip \cdot x}+{\bar v}(\vec{p})b_{\vec{p}}e^{ip \cdot x}\right),
\end{split}
\end{equation} 
where $p^0=\omega= \sqrt{ c_F^2 + p_1^2 + p_3^2}$. As always we use the mostly positive convention, 
so $e^{i p.x}$ has negative `frequency' in time, while $e^{-i p.x}$ has positive frequency in time. As is 
usual, the coefficients of negative frequency wave functions are annihilation operators, while the coefficients
of positive frequency wave functions are creation operators. We refer to $a$ and $a^\dagger$ as 
particle destruction and creation operators, while $b$ and $b^\dagger$ are antiparticle destruction and 
creation operators. The wave functions $u(p) e^{ip.x}$ and $v(p) e^{-ip.x}$ are solutions to the 
Dirac equation
$$ \left(i (p_\mu +\Sigma_\mu) \gamma^\mu + \Sigma_I \right) \psi(p)=0$$
and, as usual, ${\bar \psi} =i \psi^\dagger \gamma^0$,
where $\Sigma$ is defined in \eqref{sigmafg}. For later convenience, we introduce the following notation,
$$\Sigma_{I}(p_s)=f(p_s) p_s,~~~\Sigma_{+}=g(p_s) p_s.$$

The Dirac equation uniquely determines $u(p)$ and $v(p)$ upto multiplicative constants. 
We fix the normalization ambiguity by demanding 
\begin{equation}\label{normal}
\bar{u}(\vec{p})u(\vec{p}) = 2f(p_s)p_s, ~~~\bar{v}(\vec{p})v(\vec{p}) = -2f(p_s)p_s .
\end{equation}
\footnote{This normalization convention may be justified by performing a double analytic continuation, 
so that $x^0$ becomes a spatial direction and $x^3$ a temporal direction. Once this is done, the 
free Lagrangian is of first order in time, and so may be canonically quantized in the usual manner. 
The normalization described above are chosen to ensure that the usual anticommutation relations 
for the field operators $\psi$ translate to standard anticommutation relations for the creation and 
annihilation operators $a, a^\dagger, b, b^\dagger$. } These requirements leave the phase of the 
functions $u(p)$ and $v(p)$ undetermined: we will make an arbitrary choice for this phase below. 

We will find it useful to have explicit expressions for $u$ and $v$. In order to obtain these expressions, 
it is useful to fix a particular convention for $\gamma$ matrices.  In Euclidean space we make the 
choice 
$\gamma^{+}=\left(
\begin{array}{cc}
 0 & \sqrt{2} \\
 0 & 0
\end{array}
\right)$,
$\gamma^{-}=\left(
\begin{array}{cc}
 0 & 0 \\
 \sqrt{2} & 0
\end{array}
\right)$ and 
$\gamma^{3}=\left(
\begin{array}{cc}
 1 & 0 \\
 0 & -1
\end{array}
\right).$
This choice determines the Lorentzian $\gamma$ matrices to be 
\begin{align}\label{gams}
\begin{split}
\gamma^3 = \left(\begin{array}{cc}
                 1 & 0 \\ 0 & -1
                 \end{array} \right) ,
\qquad
\gamma^+ = \left(\begin{array}{cc} 0 & \sqrt{2} \\ 0 & 0 \end{array}\right) ,
\\
\gamma^- = \left(\begin{array}{cc} 0 & 0 \\\sqrt{2} & 0 \end{array}\right) ,
\qquad
\gamma^0 = \left(\begin{array}{cc} 0 & 1 \\ -1 & 0 \end{array}\right).
\end{split}
\end{align}
The quadratic Dirac Lagrangian consequently takes the explicit form 
\begin{equation}\label{diraceq}
\int \frac{d^3p}{(2\pi)^3}\bar{\psi}(-p)\left(\begin{array}{cc} i p_2 + f(p_s)p_s & i\sqrt{2}p_+( 1 + g(p_s)) \\ i\sqrt{2}p_- 
& -i p_2 + f(p_s)p_s 
		    \end{array}\right)\psi(p) .
\end{equation}
The equations of motion for $u$ and ${\bar u}$ are 
\begin{equation}\label{ueq}
\begin{split}
\left(\begin{array}{cc} i p_2 + f(p_s)p_s & -i(E_{\vec{p}}-p_1)( 1 + g(p_s)) \\ i(E_{\vec{p}}+p_1) & -i p_2 + f(p_s)p_s 
		    \end{array}\right)u(\vec{p}) &= 0,\\
\bar{u}(\vec{p})\left(\begin{array}{cc} i p_2 + f(p_s)p_s & -i(E_{\vec{p}}-p_1)( 1 + g(p_s)) \\ i(E_{\vec{p}}+p_1) & -i p_2 + f(p_s)p_s 
		    \end{array}\right) &= 0,
\end{split}
\end{equation}
while those for $v$ and ${\bar v}$ are 
\begin{equation}\label{veq}
\begin{split}
\left(\begin{array}{cc} i p_2 - f(p_s)p_s & -i(E_{\vec{p}}-p_1)( 1 + g(p_s)) \\ i(E_{\vec{p}}+p_1) & -i p_2 - f(p_s)p_s 
		    \end{array}\right)v(\vec{p}) &= 0,\\
\bar{v}(\vec{p})\left(\begin{array}{cc} i p_2 - f(p_s)p_s & -i(E_{\vec{p}}-p_1)( 1 + g(p_s)) \\ i(E_{\vec{p}}+p_1) & -i p_2 - f(p_s)p_s 
		    \end{array}\right) &= 0.
\end{split}\end{equation}
Note that, \eqref{ueq} and \eqref{veq} admits solution only when, determinant of the matrix appearing in those equations are zero. This 
gives onshell condition $p^2+c_F^2=0,$
equivalently
$$
p_2^2 + f(p_s)^2p_s^2 - \left(
E_{\vec{p}}^2 - p_{1}^2
\right)\left(1+ g(p_s)\right) = 0.
$$
Solving these equations subject to the normalization conventions described above (plus an arbitrary 
choice of phase) we find 
\begin{equation}\label{usol}
\begin{split}
u(\vec{p}) &= \frac{1}{\sqrt{E_{\vec{p}}+p_1}}
\begin{pmatrix} ip_2 - f(p_s)p_s 
\\ i(E_{\vec{p}}+p_1) \end{pmatrix},\\
\bar{u}(\vec{p}) &= \frac{1}{\sqrt{E_{\vec{p}}+p_1}}\left(\begin{array}{cc} -(E_{\vec{p}}+p_1) & ~~~p_2 - if(p_s)p_s \end{array}\right),
\end{split}
\end{equation}
and
\begin{equation}\label{vsol}
\begin{split}
v(\vec{p}) &= \frac{1}{\sqrt{E_{\vec{p}}+p_1}}\left(\begin{array}{c} ip_2 + f(p_s)p_s \\ i(E_{\vec{p}}+p_1) \end{array}\right),\\
\bar{v}(\vec{p}) &= \frac{1}{\sqrt{E_{\vec{p}}+p_1}}\left(\begin{array}{cc} -(E_{\vec{p}}+p_1) & ~~~p_2 + if(p_s)p_s \end{array}\right).
\end{split}
\end{equation}
\subsection{S-matrices}
With explicit expressions for $u({\vec p})$ and $v({\vec p})$ in hand, it might seem like an easy 
task to take the onshell limit of the ofshell 4 Fermi correlators. Infact that is not the case. As in the 
bosonic theory the onshell limit of these correlators is apparently ambiguous, and must be taken very 
carefully. The reader will recall that we discussed this issue at great detail in the bosonic theory, 
came to the conclusion that the correct final prescription is simply to first set $|{\vec k}|$ to $|{\vec p}|$
before taking either of these momenta individuallyonshell. We adopt a similar prescription for the bosonic theories. 
We first replace the quantities $E_p$ and $E_k$ that appear in our solutions for $u({\vec p})$ and 
$v({\vec p})$ with $\pm p^0$ and $ \pm k^0$ respectively. We then evaluate the offshell amplitude
with $|{\vec k}|=|{\vec p}|$ and only then take the momenta to individually be onshell. This process 
yields unambiguous answers which we present below. As in the bosonic case, it should be possible 
to justify this order of limits with a careful evaluation of the amplitude directly in Minkoski space
keeping careful track of the factors of $i \epsilon$ but we have not persued this thought.
\subsubsection{S-matrix for adjoint exchange in particle - antiparticle scattering}\label{TchanFF}
As we have explained above, the offshell four fermion scattering amplitude is extremely complicated. 
Quite remarkably, however, the onshell limit displays remarkable simplifications. 
In the $T$-channel the onshell S-matrix is given by  
\begin{equation}\begin{split} \label{tcf}
 T_T^F&=V^{\alpha\gamma}_{\beta\delta}(p,k,q){u}_{i,\alpha}(p+q){\bar v}^{j,\beta}
(-(k+q)){\bar u}^{i,\delta}(p)v_{j,\gamma}(-k)\\
&=-\frac{4 \pi i }{k_F}   q_3  \frac{p_- + k_-}{p_- - k_-}\\
&-\frac{4i\pi}{k_F}  q_3~~\frac{\left(q_3-2 i ~{\rm sgn}(m_F)|c_F|\right) e^{2 i \lambda_F  \tan ^{-1}\left(\frac{2 |c_F|}{q_3}\right)}
-e^{i \pi {\rm sgn}(q_3) \lambda_F } \left(q_3+2 i ~{\rm sgn}(m_F)|c_F|\right)}{e^{i
   \pi {\rm sgn}(q_3) \lambda_F } \left(q_3+2 i ~{\rm sgn}(m_F)|c_F|\right)+\left(q_3-2 i ~{\rm sgn}(m_F)|c_F|\right) e^{2 i \lambda_F  \tan ^{-1}\left(\frac{2 |c_F|}{q_3}\right)}}\\
&=-E(p_1,p_2,p_3)\frac{4i\pi }{k_F}  \sqrt{\frac{u~t}{s}}\\
&-\frac{4i\pi}{k_F}  \sqrt{-t}~~\frac{\left(\sqrt{-t}-2 i ~{\rm sgn}(m_F)|c_F|\right) e^{2 i \lambda_F  \tan ^{-1}\left(\frac{2 |c_F|}{\sqrt{-t}}\right)}
-e^{i \pi  \lambda_F } \left(\sqrt{-t}+2 i ~{\rm sgn}(m_F)|c_F|\right)}{e^{i
   \pi  \lambda_F } \left(\sqrt{-t}+2 i ~{\rm sgn}(m_F)|c_F|\right)+\left(\sqrt{-t}-2 i ~{\rm sgn}(m_F)|c_F|\right) e^{2 i \lambda_F  \tan ^{-1}\left(\frac{2 |c_F|}{\sqrt{-t}}\right)}}\\
&=-E(p_1,p_2,p_3)\frac{4i\pi }{k_F}  \sqrt{\frac{u~t}{s}}\\
&-\frac{4i\pi}{k_F}  \sqrt{-t}~~\frac{\left(\sqrt{-t}-2 i ~{\rm sgn}(m_F)|c_F|\right) e^{2 i \lambda_F  \tan ^{-1}\left(\frac{2 |c_F|}{\sqrt{-t}}\right)}
-e^{i \pi  \lambda_F } \left(\sqrt{-t}+2 i ~{\rm sgn}(m_F)|c_F|\right)}{e^{i
   \pi  \lambda_F } \left(\sqrt{-t}+2 i ~{\rm sgn}(m_F)|c_F|\right)+\left(\sqrt{-t}-2 i ~{\rm sgn}(m_F)|c_F|\right) e^{2 i \lambda_F  \tan ^{-1}\left(\frac{2 |c_F|}{\sqrt{-t}}\right)}}\\
&=-E(p_1,p_2,p_3)\frac{4i\pi }{k_F}  \sqrt{\frac{u~t}{s}}+\frac{4~i\pi}{k_F}  \sqrt{-t}~~\frac{e^{i\pi\left(\lambda_F-{\rm sgn}(m_F)\right)}+e^{2i\left(\lambda_F-{\rm sgn}(m_F)\right) \tan^{-1}\left(\frac{2|c_F|}{\sqrt{-t}}\right)}}{e^{i\pi\left(\lambda_F-{\rm sgn}(m_F)\right)}-e^{2i\left(\lambda_F-{\rm sgn}(m_F)\right) \tan^{-1}\left(\frac{2|c_F|}{\sqrt{-t}}\right)}}~~\\
&=-E(p_1,p_2,p_3)\frac{4i\pi }{k_F}  \sqrt{\frac{u~t}{s}}+\frac{4~i\pi}{k_F}  \sqrt{-t}~~\frac{1+e^{-2i\left(\lambda_F-{\rm sgn}(m_F)\right) \tan^{-1}\left(\frac{\sqrt{-t}}{2|c_F|}\right)}}{1-e^{-2i\left(\lambda_F-{\rm sgn}(m_F)\right) \tan^{-1}\left(\frac{\sqrt{-t}}{2|c_F|}\right)}}.~~\\
\end{split}
\end{equation}
As we have emphasized above, we have obtained this result only after taking the onshell limit in a particular 
manner. 
In particular, in the solution in \eqref{usol},\eqref{vsol} we 
treated $E_p$ as a free symbol to start with; we set  $p_s=k_s$ first and  then set  $E_p^2={\overrightarrow{p}}^2+c_{F}^2.$
\subsubsection{S-matrix for particle - particle scattering}\label{UchanFF}
In the $U$-channel
\begin{equation}\begin{split}\label{ucf}
 T_{U_d}^F&=V^{\alpha\gamma}_{\beta\delta}(p,k,q){u}_{i,\alpha}(p+q){\bar u}^{j,\beta}
(k+q){\bar u}^{i,\delta}(p)u_{j,\gamma}(k)\\
&=\frac{4 \pi i }{k_F}   q_3  \frac{p_- + k_-}{p_- - k_-}\\
&+\frac{4i\pi}{k_F}  q_3~~\frac{\left(q_3-2 i ~{\rm sgn}(m_F)|c_F|\right) e^{2 i \lambda_F  \tan ^{-1}\left(\frac{2 |c_F|}{q_3}\right)}
-e^{i \pi {\rm sgn}(q_3) \lambda_F } \left(q_3+2 i ~{\rm sgn}(m_F)|c_F|\right)}{e^{i
   \pi {\rm sgn}(q_3) \lambda_F } \left(q_3+2 i ~{\rm sgn}(m_F)|c_F|\right)+\left(q_3-2 i ~{\rm sgn}(m_F)|c_F|\right) e^{2 i \lambda_F  \tan ^{-1}\left(\frac{2 |c_F|}{q_3}\right)}}\\
&=-\Bigg( E(p_1,p_2,p_3)\frac{4i\pi }{k_B}  \sqrt{\frac{s~t}{u}}\\
&-\frac{4i\pi}{k_F}  \sqrt{-t}~~\frac{\left(\sqrt{-t}-2 i ~{\rm sgn}(m_F)|c_F|\right) e^{2 i \lambda_F  \tan ^{-1}\left(\frac{2 |c_F|}{\sqrt{-t}}\right)}
-e^{i \pi  \lambda_F } \left(\sqrt{-t}+2 i ~{\rm sgn}(m_F)|c_F|\right)}{e^{i
   \pi  \lambda_F } \left(\sqrt{-t}+2 i ~{\rm sgn}(m_F)|c_F|\right)+\left(\sqrt{-t}-2 i ~{\rm sgn}(m_F)|c_F|\right) e^{2 i \lambda_F  \tan ^{-1}\left(\frac{2 |c_F|}{\sqrt{-t}}\right)}}\Bigg)\\
&=-\Bigg(- E(p_1,p_2,p_3)\frac{4i\pi }{k_F}  \sqrt{\frac{s~t}{u}}+\frac{4~i\pi}{k_F}  \sqrt{-t}~~\frac{e^{i\pi\left(\lambda_F-{\rm sgn}(m_F)\right)}+e^{2i\left(\lambda_F-{\rm sgn}(m_F)\right) \tan^{-1}\left(\frac{2|c_F|}{\sqrt{-t}}\right)}}{e^{i\pi\left(\lambda_F-{\rm sgn}(m_F)\right)}-e^{2i\left(\lambda_F-{\rm sgn}(m_F)\right) \tan^{-1}\left(\frac{2|c_F|}{\sqrt{-t}}\right)}}~~
\Bigg)\\
&=-\Bigg(- E(p_1,p_2,p_3)\frac{4i\pi }{k_F}  \sqrt{\frac{s~t}{u}}+\frac{4~i\pi}{k_F}  \sqrt{-t}~~\frac{1+e^{-2i\left(\lambda_F-{\rm sgn}(m_F)\right) \tan^{-1}\left(\frac{\sqrt{-t}}{2|c_F|}\right)}}{1-e^{-2i\left(\lambda_F-{\rm sgn}(m_F)\right) \tan^{-1}\left(\frac{\sqrt{-t}}{2|c_F|}\right)}}~~
\Bigg).\\
\end{split}
\end{equation}

As in the previous subsubsection, we have obtained this resultafter taking the onshell limit in a particular 
manner. 
In particular, in the solution in \eqref{usol},\eqref{vsol} we 
treated $E_p$ as a free symbol to start with; we set  $p_s=k_s$ first and  then set  $E_p^2={\overrightarrow{p}}^2+c_{F}^2.$


\section{Scattering in the identity channel and crossing symmetry} \label{sing}

\subsection{Crossing symmetry} \label{ncd}

It is sometimes asserted  that the S-matrix for particle - antiparticle
scattering, in any quantum field theory,  may be obtained from the S-matrix for particle - particle scattering.
This claim goes by the name of crossing symmetry. In the context of the $2 \rightarrow 2$ scattering studied in this paper, the formulae asserted with the claim are (we work with the bosonic theory for 
definiteness)
\begin{equation}\label{pascat} 
T_S(s, t, u)=N T_{U_d}(t, u, s), ~~~~T_{T}(s, t, u)=      T_{U_d}(u, t, s).
\end{equation}
These equations assert that the formulae for particle - antiparticle scattering may 
be read off from the analytic continuation of the physical particle - particle scattering amplitude
\footnote{Analytic continuation is needed because physical scattering processes 
in the different channels utilize non overlapping domains of the (allegedly) single analytic 
`master' scattering formula. Consider, for instance, the first 
of \eqref{pascat}. Physical particle- particle scattering process are captured by the function 
$T_{U_d}(x, y, z)$ for $y, z <0$ ; given that $x+y+z=4m^2$, this implies $x>4m^2$. On the other hand 
on the RHS of the first of \eqref{pascat} we need the same function at $x, y <0$ and so $z>4 m^2$. 
It is clear that there is no overlap between these different domains. } .

In the case of an ungauged field theory - or in the case of the scattering of gauge invariant 
particles in a gauge theory, there is a rather straightforward intuitive 
 argument for crossing symmetry of amplitudes.  The LSZ formula relates S-matrices to 
onshell limits of well-defined offshell correlators. The offshell correlators  are expected to 
be analytic functions of their insertion positions.  The on shell limit of these correlators 
is the `master function' referred to in the footnote above which plausibly inherits analytic properties 
from those of the underlying correlators. 

This intuitive argument does not work for the scattering of non gauge singlet particles 
in a gauge theory, as the relevant scattering amplitudes 
cannot be obtained from the onshell limit of an offshell correlator (the putative offshell 
correlators are not gauge invariant and so are ill defined).

While the argument for crossing symmetry presented in this subsection does not apply to, 
for instance, the scattering of gluons in ${\cal N}=4$ Yang Mills theory, the final result 
(i.e. that scattering amplitudes obey crossing symmetry) is widely expected to hold true for these amlitudes,
 at least with a suitable definition of the scattering amplitudes (a definition is needed
to deal with IR ambiguities having to do with soft gluons and other soft particles). In this 
context we expect that the failure of the argument outlined in this subsection is just a technicality; 
other arguments (perhaps based on diagrammatics) guarantee the final result. 

As in the previous paragraph, current paper we are also interested in the scattering of non  singlet excitations. 
Unlike the case of gluonic scattering
in ${\cal N}=4$ Yang Mills, however, we will argue below that the failure of the argument for 
crossing symmetry  is more than a technicality. The crossing relations are 
{\it actually} modified in our theories. We suspect that the underlying reason for the modification
 is that the  Chern-Simons action, which controls the dynamics of our gauge fields, effectively turns 
our scattering particles into anyons. Apparently, the usual crossing relations are true 
for the scattering of bosons and fermions, but are modified in  the scattering of 
anyons. 

\subsection{A conjecture for the S-matrix in the singlet channel} \label{cs}

As we have explained above, a naive application of crossing symmetry predicts that, 
the $S$-channel scattering amplitude is given by $T_S^B(s,t,u)= N T^B_{U_d}(t,u,s)$. 
We have performed the analytic continuations needed to make sense of this formula 
in subsection \ref{acj}. Utilizing the results of that subsection, the naive prediction 
of crossing symmetry is 
\begin{equation} \label{ttrials} \begin{split}
&T_S^{trial}= \left( \pi \lambda_{B} \right)  4 i\sqrt{s}  E(p_1, p_2, p_3) \sqrt{\frac{u}{t}} + 
j^M(\sqrt{s}) \\
&= \left( \pi \lambda_{B} \right)  4 \sqrt{s}   \left( i~E(p_1, p_2, p_3) \sqrt{\frac{u}{t}}
+ \left( \frac{ \left(4 \pi  \lambda_{B} \sqrt{s} +\wt b_4 \right) 
  +  \left(-4 \pi  \lambda_{B} \sqrt{s} + \wt b_4 \right)  e^{8 \pi  \lambda_{B}  \sqrt{s} H^M(\sqrt{s}) }
 } {\left(4 \pi  \lambda_{B} \sqrt{s}  + \wt b_4 \right) 
 -  \left(-4 \pi  \lambda_{B} \sqrt{s} + \wt b_4 \right)  e^{8 \pi  \lambda_{B}  \sqrt{s} H^M(\sqrt{s}) }
} \right)  \right) \\
& = \left( \pi \lambda_{B} \right)  4 \sqrt{s}   \left( i~E(p_1, p_2, p_3)\sqrt{\frac{u}{t}}  +
\left( \frac{ \left(4 \pi  \lambda_{B} \sqrt{s} +\wt b_4 \right)
  +  e^{i \pi \lambda_{B}} \left(-4 \pi  \lambda_{B} \sqrt{s} + \wt b_4 \right)  \left(  \frac{ \frac{1}{2} + \frac{c_B}{\sqrt{s} } }
{\frac{1}{2} - \frac{c_B}{\sqrt{s}} } \right)^{\lambda_{B}}
 } {\left(4 \pi  \lambda_{B} \sqrt{s}  + \wt b_4 \right) 
 -  e^{i \pi \lambda_{B}} \left(-4 \pi  \lambda_{B} \sqrt{s} + \wt b_4 \right)    \left(  \frac{ \frac{1}{2} + \frac{c_B}{\sqrt{s} } }
{\frac{1}{2} - \frac{c_B}{\sqrt{s}} } \right)^{\lambda_{B}} 
} \right)  \right)  \\  
\end{split}
\end{equation}
(in the last line we have specialized to the physical domain $s \geq 4 c_B^2$). 

The function $T_S^{trial}$ cannot be the true scattering matrix in the $S$-channel for 
three related reasons. 
\begin{itemize}
\item $T_S^{trial}$ does not include the last term in \eqref{ht}; a term 
delta function localized on forward scattering with a  coefficient proportional to 
$(\cos (\pi \lambda_{B}) -1 )$. This term is certainly present in the scattering amplitude 
at least in the non-relativistic limit. 
\item Even ignoring the term localized at forward scattering, the non-relativistic limit of 
$T_S^{trial}$ does not agree with \eqref{ht}.
\item $T_S^{trial}$ does not obey the unitarity relation \eqref{punit}.
\end{itemize}

In the rest of this subsection we will demonstrate that all these problems are simultaneously 
cured if we conjecture that the scattering matrix in the $S$-channel is given by a rescaled 
$T_S^{trial}$ plus a contact term added by hand. We conjecture that the bosonic scattering 
matrix in the $S$-channel is given by 
\begin{equation} \label{conject1}
T_S^B= \frac{\sin (\pi \lambda_{B})}{\pi \lambda_{B}} T_S^{trial}  
-i (\cos (\pi \lambda_{B}) -1) I(p_1, p_2, p_3, p_4)
\end{equation}
(see subsection \ref{ik} for a definition of the Identity matrix).
In subsection \ref{expmd} we will present a tentative justification for the modification
of the usual rules of crossing symmetry implicit in \eqref{conject}. In the rest of this subsection 
we will demonstrate that the conjectured scattering amplitude $T_S^B$
passes various consistency checks. 

In the center of mass frame our conjectured scattering amplitude \eqref{conject1} takes the 
form \eqref{form} with 
\begin{equation}\label{nrtwc} \begin{split}
& H(\sqrt{s})= 4 \sqrt{s} \sin (\pi \lambda_{B}), \\
& W_1(\sqrt{s})= 4 \sqrt{s} \sin (\pi \lambda_{B}) G,  \\
& W_2(\sqrt{s})= 8 \pi  \sqrt{s} \left( \cos (\pi \lambda_{B}) - 1 \right),  \\
&  G=\left( \frac{ \left(4 \pi  \lambda_{B} \sqrt{s} +\wt b_4 \right)
  +  e^{i \pi {\lambda_{B}}} \left(-4 \pi  {\lambda_{B}} \sqrt{s} + \wt b_4 \right)  \left(  \frac{ \frac{1}{2} + \frac{c_B}{\sqrt{s} } }
{\frac{1}{2} - \frac{c_B}{\sqrt{s}} } \right)^{\lambda_{B}}
 } {\left(4 \pi  {\lambda_{B}} \sqrt{s}  + \wt b_4 \right) 
 -  e^{i \pi {\lambda_{B}}} \left(-4 \pi  {\lambda_{B}} \sqrt{s} + \wt b_4 \right)    \left(  \frac{ \frac{1}{2} + \frac{c_B}{\sqrt{s} } }
{\frac{1}{2} - \frac{c_B}{\sqrt{s}} } \right)^{\lambda_{B}} 
} \right).
\end{split}
\end{equation}

Let us first demonstrate that our conjectured expressions \eqref{nrtwc} have the correct 
non-relativistic limit. The functions $H$ and $W_2$ in  \eqref{nrtww} are independent of 
the energy $s$ and already agree perfectly with the same functions in \eqref{nrtww}.  Moreover
\begin{equation}\label{nrlpim}
\lim_{\sqrt{s} \rightarrow 2 c_B}G   = - {\rm sgn}({\lambda_{B}})
\end{equation}
it follows that 
\begin{equation}\label{nrwo}
\lim_{\sqrt{s} \to 2 c_B} W_1(\sqrt{s})= -4 \sqrt{s} |\sin (\pi {\lambda_{B}})|
\end{equation}
in agreement with \eqref{nrtww}. We conclude that our 
conjectured scattering amplitude \eqref{nrtwc} reduces precisely to the expected 
Aharonov-Bohm scattering amplitude in the non-relativistic limit. 

We next demonstrate that our conjecture for the $S$-channel S-matrix obeys 
the constraints of unitarity, i.e. that \eqref{nrtwc} obeys the equations \eqref{unitcond}.
As we have explained in subsection \ref{ug}, the fact that $H$ and $W_2$ in \eqref{nrtwc}
agree with the corresponding functions in \eqref{nrtww} immediately implies that the first 
two equations in \eqref{unitcond} are obeyed. We will now demonstrate that the 
functions in \eqref{nrtwc} also obey the third equation in \eqref{unitcond}. 
\footnote{A point here requires explanation. In our study of unitarity in section \ref{ug}, the function $H$
multiplies an S-matrix proportional to ${\rm Pv} \cot \frac{\theta}{2}$. Feynmam diagrams produce a
scattering amplitude in which the function $H$ multiplies $\frac{\sin \frac{\theta}{2}\cos\frac{\theta}{2} 
} {\sin^2 \frac{\theta}{2} -i \epsilon}$. These two expressions clearly coincide at nonzero $\theta$; 
interestingly enough they also coincide at $\theta=0$. Indeed it is not difficult to demonstrate that 
$$ {\rm Pv} \frac{1}{\theta}= \frac{\theta}{\theta^2 -i \epsilon}.$$ 
The key point here is that the second expression above has two poles; one of these lies above the 
real $\theta$ axis while the second one lies below it. The residue of each of these two poles is 
precisely half what it would have been for the simple pole $\frac{1}{\theta}$, demonstrating that  
the expression on the RHS is identical to the principal value. }

The third equation in \eqref{unitcond} may be rewritten, in terms of the function $G$, as 
$$G-G^*= (1- \cos (\pi {\lambda_{B}}))(G-G^*)-i  \sin (\pi {\lambda_{B}}) (1-G G^*)$$
This equation is holds if
\begin{equation}\label{ghold} 
G-G^*= -i \tan (\pi {\lambda_{B}}) (1-G G^*).
\end{equation}
Now 
$$G= \frac{1+e^{i \pi {\lambda_{B}}} y}{1- e^{i \pi {\lambda_{B}}} y},$$
where 
$$y=   \frac{ \left(-4 \pi  {\lambda_{B}} \sqrt{s} + \wt b_4 \right)}{ \left(4 \pi  {\lambda_{B}} \sqrt{s} +\wt b_4 \right)}\left(  \frac{ \frac{1}{2} + \frac{c_B}{\sqrt{s} } }
{\frac{1}{2} - \frac{c_B}{\sqrt{s}} } \right)^{\lambda_{B}}.$$
Note in particular that $y$ is real (its detailed form is irrelevant for what follows).  It follows that 
$$G- G^*= \frac{4 i y \sin (\pi {\lambda_{B}}) }{|1-e^{i \pi {\lambda_{B}}}y|^2}, 
~~~(1-G G^*)= \frac{-4 y\cos (\pi {\lambda_{B}})}{|1-e^{i \pi {\lambda_{B}}}y|^2}. $$
It follows 
that \eqref{ghold} is satisfied so that our proposal \eqref{conject} defines 
a unitary S-matrix. 

Finally,  in the limit ${\lambda_{B}} \to 0$, 
our conjecture reduces to (see the second line of \eqref{ttrial})
$$T_S^B=\frac{-b_4}{1+b_4 H^M(\sqrt{s})}.$$
It is easily independently verified that this is the correct formula for the scattering amplitude of the 
large $N$ $\phi^4$ theory that \eqref{sclag} reduces to in the small ${\lambda_{B}}$ limit. 
In other words our conjectured scattering amplitude has the correct small ${\lambda_{B}}$ limit.

\subsection{Bose-Fermi duality in the $S$-channel}

We have conjectured above that, in the $S$-channel, the bosonic S-matrix is given by 
\begin{equation}\label{bsm} 
T_S^B(s, t, u, \lambda_B)= \frac{k_B \sin (\pi \lambda_B)}{\pi}  T_{U_d}^B(t, u, s, \lambda_B) 
-i\left( \cos (\pi \lambda_B) -1 \right) I(p_1, p_2, p_3, p_4),
\end{equation}
This implies that the S-matrix in the $S$-channel is given by 
\begin{equation}\label{bsmf} 
S_S^B(s, t, u, \lambda_B)= i\frac{k_B \sin (\pi \lambda_B)}{\pi}  T_{U_d}^B(t, u, s, \lambda_B)+ 
\cos (\pi \lambda_B )  I(p_1, p_2, p_3, p_4),
\end{equation}
where $I$ is the identity S-matrix, see subsection \ref{ik}. 

In this section we have, so far, presented our conjecture for the $S$-channel S-matrix in the  bosonic theory.
It is natural to conjecture a similar formula in the fermionic theory. In analogy with our conjecture 
for the bosonic theory we conjecture that 
\begin{equation}\label{bsmf1} 
T_S^F(s, t, u, \lambda_F)= \frac{k_F \sin (\pi \lambda_F)}{\pi}  T_{U_d}^F(t, u, s, \lambda_F) -i\left( \cos (\pi \lambda_F) -1 \right) I(p_1, p_2, p_3, p_4)
\end{equation}
so that 
\begin{equation}\label{bsmfff} 
S_S^F(s, t, u, \lambda_F)= i \frac{k_F \sin (\pi \lambda_F)}{\pi}  T_{U_d}^F(t, u, s, \lambda_F)+ 
\cos (\pi \lambda_F )  I(p_1, p_2,p_3, p_4) .
\end{equation}

We will now demonstrate that these two conjectures map to each other under duality.
\begin{equation}\label{tf}\begin{split}
& \frac{k_B \sin (\pi \lambda_B)}{\pi}
=  \frac{k_F \sin (\pi \lambda_F)}{\pi}, \\
& T_{U_d}^B(t, u, s, \lambda_B)= -T_{U_d}^F(t, u, s, \lambda_F),\\
& \cos (\pi \lambda_B)= - \cos (\pi \lambda_F), \\
\end{split}
\end{equation}
(through this subsection we specialize to the limit $b_4 \to \infty$ in the bosonic theory). 
it follows that 
\begin{equation}\label{bsmff} 
S_S^B(s, t, u, \lambda_B)= - S_S^F(s,t,u, \lambda_F),
\end{equation}
which implies that 
\be\label{FSchn}\begin{split}
S_S^F(s,t,u, \lambda_F)=& \sin (\pi \lambda_F)\Bigg(4 E(p_1,p_2,p_3)   \sqrt{\frac{s~t}{u}}+4 \sqrt{s}~~\frac{1+e^{-2i\left(\lambda_F-{\rm sgn}(m_F)\right) \tan^{-1}\left(\frac{\sqrt{s}}{2|c_F|}\right)}}{1-e^{-2i\left(\lambda_F-{\rm sgn}(m_F)\right) \tan^{-1}\left(\frac{\sqrt{s}}{2|c_F|}\right)}}~~
\Bigg)\\
&+ 
\cos (\pi \lambda_F )  I(p_1, p_2,p_3, p_4).
\end{split}\ee Note that, $S_S^F(s,t,u, \lambda_F)$ reduces to correct tree level S-matrix presented in section \ref{treel}.
The overall minus sign on the RHS of \eqref{bsmff} has no physical significance, as the sign of 
fermionic scattering amplitudes is largely a matter of convention.
\footnote{ Indeed there does not even exist a particularly natural convention for the sign of a 
fermionic S-matrix. A fermionic transition amplitude 
could be defined either by $<a_4 a_3 | a_2^\dagger a_1^\dagger >$ or by the amplitude 
$<a_3 a_4|a_2^\dagger a_1^\dagger >$; both conventions are equally natural and yield 
S-matrices that differ by a minus sign. Note that the sign of all components of the S-matrix, including 
the identity term is flipped by this maneuver, just as in \eqref{bsmff}. }
\eqref{bsmff} demonstrates the unitarity singlet fermionic S-matrix obtained from the conjecture \eqref{bsmfff}, as we have already checked the unitarity of the bosonic S-matrix.

In summary, our conjecture for the $S$-channel S-matrices is consistent with Bose-Fermi duality. 
This observation may be taken as one more piece of evidence in support of our conjecture. 
\footnote{The function $T^{trial}_S$, and its fermionic counterpart clearly map to each other under duality.
In order to account for the nature of anyonic scattering, unitarity and the non-relativistic limit, we were 
forced to modify $T^{trial}_B$ and its fermionic counterpart by multiplicative and additive shifts. It is nontrivial that these shift functions, which were determined purely by consistency requirements in 
each theory, also turn out to transform into each other under duality. }

\subsection{A heuristic explanation for modified crossing symmetry} \label{expmd}

In this section we have conjectured that the naive crossing symmetry \eqref{pascat} are modified
in fundamental matter Chern-Simons theory; in the large N limit of interest to this paper, we have proposed 
that the second of \eqref{pascat} continues to apply, while the first of \eqref{pascat} is replaced by 
\eqref{conject}. The arguments presented so far for this replacement have been 
entirely pragmatic; we guessed the modified crossing relation in order that the S-matrix in the 
$S$-channel obey various consistency conditions.

 In this subsection we will attempt to sketch a logical explanation 
for this modified crossing relation \eqref{ttrial}. Our explanation is heuristic in several respects, but 
we hope that its defects will be remedied by more careful studies in the future.

The starting point of our analysis is the argument for crossing symmetry in the bosonic theory 
in the limit ${\lambda_{B}} \to 0$, briefly alluded to in subsection \ref{ncd}. When ${\lambda_{B}}$ is set to zero, 
the bosonic theory effectively reduces to a theory of scalars with global $U(N)$ symmetry . In this 
theory the offshell correlator 
\begin{equation}\label{correl}
C=\langle \phi_i(x_1) {\bar \phi}^j (x_2) {\bar \phi}^k(x_3) \phi_m(x_4)  \rangle
\end{equation}
is a well-defined meromorphic function of its arguments. By $U(N)$ invariance this correlator
is given by 
\begin{equation}\label{cform}
C_{im}^{jk}(x_1, x_2, x_3, x_4)= A(x_1, x_2, x_3, x_4) \delta_i^j \delta_m^k + B(x_1, x_2, x_3, x_4)
 \delta_i ^k \delta_m^j
\end{equation}
where the coefficient functions $A$ and $B$ are functions of the insertion points $x^1 \ldots x^4$. 
crossing symmetry follows from the observation that distinct scattering amplitudes are simply distinct 
onshell limits of the same correlators. 

This statement is usually made precise in momentum space, but 
we will find it more convenient to work in position space. Consider an $S^2$ of size $R$, inscribed 
around the origin in Euclidean $R^3$ (we will eventually be interested in the limit $R \to \infty$). 
The S-matrices $S_{U_d}$ and $S_{S}$ may both be obtained from the correlator $A$ as follows. 
Consider free incoming particles of momentum $p_i$ and $p_m$ starting out at very early times and 
focussed so that their worldlines will both intersect the origin of $R^3$. These two world lines 
intersect the $S^2$ described above at easily determined locations $x_1$ and $x_4$ 
respectively. Similarly the coordinates $x_2$ and $x_3$ are chosen to be the intercepts of the 
world lines of particles with index $j$ and $k$, starting out from the origin of $R^3$ and
proceeding to the future along world lines of momentum $p_2$ and $p_3$ respectively. 
Having now chosen the insertion points of all operators as definite functions of momenta, the 
correlator $A(x_1, x_2, x_3, x_4)$ is now a function only of the relevant particle - particle  scattering 
data; the 
particle-particle S-matrix may infact be read off from this correlator in the limit $R \to 
\infty$ after we strip off factors pertaining to free propagation of our particles from the 
surface of the $S^2$ to the origin of $R^3$. Particle- antiparticle scattering may be obtained 
in an identical manner, by choosing $x_1$ and $x_2$ to lie along the trajectory of incoming particles or
antiparticles of momentum $p_1$ and $p_2$ respectively, while $x_3$ and $x_4$ lie along 
particle trajectories of outgoing particles and antiparticles of momentum $p_3$ and $p_4$ respectively.
Intuitively we expect that crossing symmetry  - the first of \eqref{pascat} - follows from the analyticity of the correlator $A$ as a function of $x_1$, $x_2$, $x_3$ and $x_4$ on the large $S^2$.

In the large $N$ limit $A$ may be obtained from the correlator $C_{im}^{jk}$ in \eqref{cform} from 
the identity 
\begin{equation}\label{Acont}
A= \frac{1}{N^2} C^{jk}_{im} \delta_j^i \delta_k^m
\end{equation}
At nonzero ${\lambda_{B}}$ the correlator $C_{im}^{jk}$ no longer makes sense as it is not gauge invariant. 
In order to construct an appropriate gauge invariant quantity let $W_{12}$ denote an open Wilson line, in 
the fundamental representation, starting at $x_1$, ending at $x_2$ and running entirely outside 
the $S^2$ one which the operators are inserted. In a similar manner let $W_{43}$ denote an open Wilson
starting at $x_4$ and ending at $x_3$, once again traversing a path that lies entirely outside the 
$S^2$ on which operators are inserted. Then the quantity
\begin{equation}\label{apd}
A'=C_{im}^{jk} (W_{12})^i_j (W_{43})^m_k
\end{equation}
is a rough analogue of $A$ in the gauged theory. The precise relationship is that 
$A'$ reduces to $A$ in the limit ${\lambda_{B}} \to 0$ in which gauge dynamics decouples 
from matter dynamics.  $A'$ is clearly gauge invariant at all ${\lambda_{B}}$; moreover there 
seems no reason to doubt that $A'$ is an analytic function of $x_1 \ldots x_4$.

We can now evaluate $A'$ in the same two onshell limits discussed in the paragraph above; 
as in the paragraph above this yields two functions of onshell momenta that are analytic continuations of each other.
 In the limit ${\lambda_{B}} \to 0$ these two functions are 
 simply the direct channel and singlet channel S-matrices. We will now address 
the following question: what is the interpretation of these two functions, obtained out of 
$A'$, at finite ${\lambda_{B}}$?

The path integral that evaluates the quantity $A'$ may conceptually be split up into three parts. 
The path integral inside the $S^2$ may be thought of as defining a  ket $|\psi_1 >$ 
of the field theory that lives on $S^2$.  The path integral outside the $S^2$ defines a 
bra of the field theory on $S^2$, lets call it $<\psi_2|$. And, finally, the path integral on 
$S^2$ evaluates $<\psi_2 | \psi_1>$. 

The key observation here is that the inner product 
occurs in the  direct product of the matter Hilbert space, and the pure gauge 
Hilbert Space. The pure gauge Hilbert space is the two dimensional Hilbert Space of conformal 
blocks of pure Chern-Simons theory on $S^2$ with two fundamental and two antifundamental 
Wilson line insertions. 

The inner product in the gauge sector depends only on the topology of the paths of matter particles 
inside the $S^2$. The distinct topological sectors are distinguished by a relative winding number of 
the two scattering particles around each other. In the large $N$ limit where the probability for reconnections in the Skein relations (see
Eq. $4.22$ of \cite{Witten:1988hf}) vanishes, the gauge theroy inner product in a sector of winding number 
$w$ difffers from the inner product in a sector of winding number zero merely by 
the relevant Aharonov-Bohm phase. This relative weighting is, of course, a very important part of 
the scattering amplitude of the theory, producing all the nontrivial behaviour. However the gauge 
theory inner product is nontrivial even at $w=0$. The details of this extra factor depend on the apparently
unphysical external Wilson lines. This extra factor is not present in the `S-matrix' computed in this paper
(as we had no external Wilson lines connecting the various particles). In order to compare with the 
S-matrices presented in this paper,  we must remove this overall inner product factor.

\begin{figure}[tbp]
  \begin{center}
  \subfigure[]{\includegraphics[scale=.6]{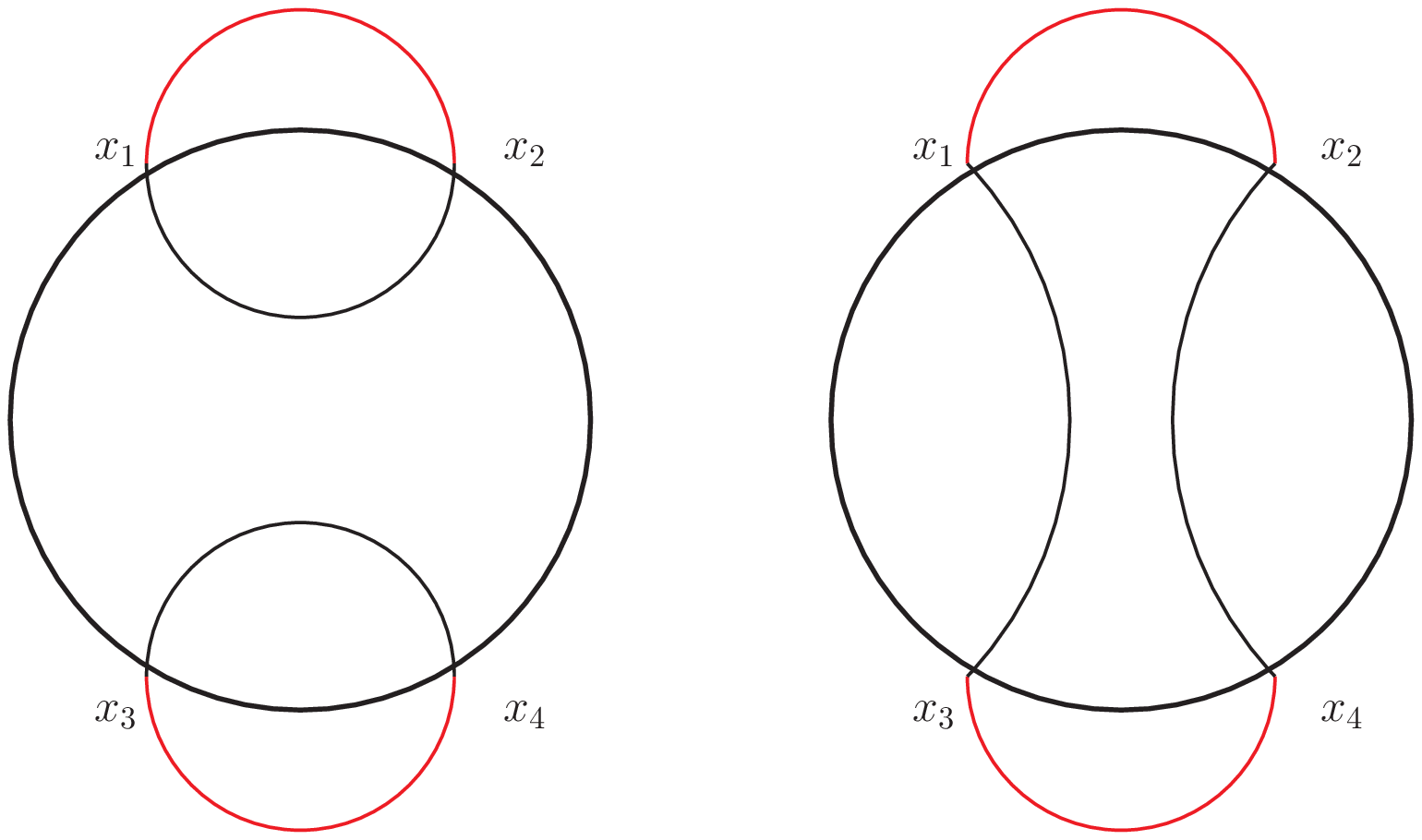}
  }\\
   \end{center}
  \vspace{-0.5cm}
  \caption{The full effective Wilson lines for $S$ and $U_d$ channels}
\label{SUd}
  \end{figure}

\begin{figure}[tbp]
  \begin{center}
  \subfigure[]{\includegraphics[scale=.6]{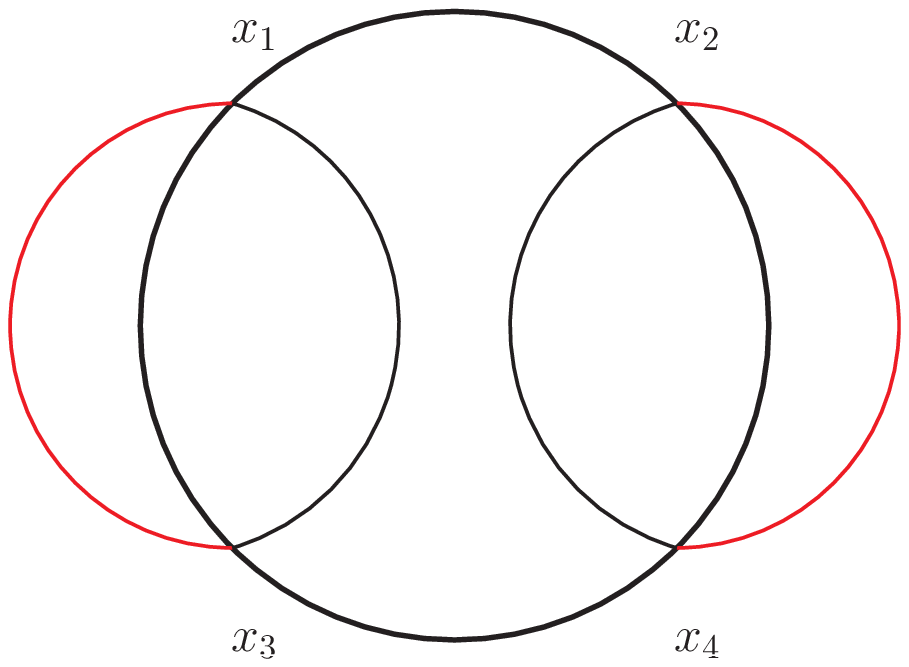}
  }\\
   \end{center}
  \vspace{-0.5cm}
  \caption{The full effective Wilson lines for $T$-channel}
\label{TWit}
  \end{figure}

The gauge inner product $<\psi_2^G| \psi_1^G>$ corresponding to identity matter scattering 
(i.e. the geodesic paths of the matter particles from prduction to annihilations) depends
on the scattering channel. Let us first 
study scattering in the  identity channel. The initial particle created at  $x_1$ connects up to the final 
particle at  $x_3$, while the particle created at $x_2$ connects up with the final particle at $x_4$. Combining with the external lines, the full effective Wilson line  is 
topologically a circle, see the second of Fig. \ref{SUd}.  On the other hand, in the case of particle-particle scattering, the dominant 
dynamical trajectories are from the initial insertion at $x_1$ to the final insertion at $x_2$ and from the 
initial insertion at $x_4$ to the final insertion at $x_3$. Including the external lines, the net effective Wilson line has the topology of two circles, see the first of Fig \ref{SUd}.

As the topology of the effective Wilson loops in the first and second of Fig. \ref{SUd} differs, it follows that 
the gauge theory inner product (even at zero winding) is different in the two sectors.
It was demonstrated by Witten in  \cite{Witten:1988hf} that the ratio of the  path integral with two circular Wilson lines 
to the path integral with a single circular Wilson
 line is infact given by 
$$ \frac{k \sin (\pi {\lambda_{B}})}{\pi}= N \frac{\sin (\pi {\lambda_{B}})}{\pi {\lambda_{B}}}$$
in the large $N$ limit. It follows that we should expect that 
\begin{equation}\label{expect}
T_S = \frac{k \sin (\pi {\lambda_{B}})}{\pi} T_{U_d}
\end{equation}
in perfect agreement with \eqref{ttrial} (the $\delta$ function piece in \eqref{ttrial} is presumably 
related to a contact term in the correlators described in this subsection).

A similar argument relates $T_{U_e}$ to $T_T$ without any relative factor, as in this case the 
closed Wilson lines described above has the topology of two circles in both cases.

\subsection{Direct evaluation of the S-matrix in the identity channel} \label{at}

The fact that we were able to solve the  integral equation that determines four particle scattering 
only for $q^\pm =0$ prevented us from evaluating the S-matrix in the identity channel by direct computation. For this reason we have been forced, in this section, to resort to guesswork and 
indirect arguments to conjecture a result for the S-matrix in the channel with identity exchange. It would, 
of course, be very satisfying to be able to verify our conjecture by direct computation. Unfortunately we 
have not succeeded in doing this. In this subsection we briefly report two potentially promising ideas 
for a direct evaluation.

\subsubsection{Double analytic continuation}

As we have already explained above, the planar graphs that evaluate $2 \rightarrow 2$ scattering 
may be summed by an integral equation. As a technical trick to solve the integral equation,
earlier in this paper we 
found it convenient to analytically continue momenta to Euclidean space according to the formula 
$p^0 = i p^0_E$. We then proceeded to solve the integral equation in Euclidean space. 
In order to evaluate $T$ and $U$-channel scattering we then analytically continued the final 
result back to Lorentzian space by setting $p^0_E=i p^0$. 

There is, however, a natural, inequivalent 
analytic continuation of the Euclidean space integral equation to Lorentzian space: the continuation
$$p^3= -i p^3_L$$
Under this continuation $x^3$ turns into a time like coordinate, while $x^\pm$ are complex coordinates 
$x^+ \sim z$, $x^- \sim {\bar z}$ that parameterize the spatial $R^2$. This at first strange sounding 
analytic continuation has been employed with great apparent success in several studies of the thermal 
partition function of large $N$ Chern-Simons theories \cite{Giombi:2011kc,Aharony:2011jz,Jain:2012qi,Maldacena:2011jn,Maldacena:2012sf, Aharony:2012ns,Jain:2013py,Takimi:2013zca,Aharony:2012nh}, a fact that suggests this analytic continuation should be taken seriously.

Under this analytic continuation a center of mass momentum with $q^\pm=0$ is timelike; indeed the 
condition $q^\pm=0$ is simply the assertion that the center of mass momentum points entirely in the 
time direction, so that in the $S$-channel we are studying scattering in the center of mass frame. \footnote{Recall that 
the 3 momentum $q^\mu$ had the interpretation of momentum transfer in the $T$ and the $U$-channels. 
As momentum transfer is necessarily spacelike for an onshell process, it follows that the $U$ and $T$ 
channel scattering processes are never onshell with this choice of Lorentzian continuation.}

In summary, it seems plausible that the double analytic continuation of the integral equation \eqref{sdeV}
at $q^\pm=0$ provides a direct computational handle on the S-matrix in the identity channel. 

The discussion of this subsection may seem, at first, to directly contradict \eqref{ttrial}; surely the 
solution of an analytically continued integral equation is simply the analytic continuation of
the solution of the original equation without any factors or additional singular terms? Infact this 
is not the case. It turns out that the integral equation after double analytic continuation has 
new singularities in the integral. These singularities - which are absent in the original equation - 
spoil naive analytic continuation. We illustrate this complicated set of affairs in Appendix \ref{dac}. 

If the central idea of this subsection is correct, then it should be possible to obtain the scattering 
cross section with identity exchange by solving the double analytic continued integral equation taking the new singular contributions into account. This appears to be a delicate task that we have not managed to implement.

As a warm up to the exercise suggested in this section it would be useful to rederive 
the ordinary non-relativistic Aharonov-Bohm equation by solving the Lippmann Schwinger equation, 
order by order in perturbation theory, in momentum space, perhaps at the value of the self 
adjoint extension parameter $w=1$ (see \cite{AmelinoCamelia:1994we} ) at which 
point the Aharonov-Bohm amplitude is an analytic function of $\nu$ so perturbation theory is 
well-defined. We suspect that this exercise will encounter all the subtle singularities discussed 
in this section, and it would be useful to learn how to carefully deal with these singularities 
in a context where the answer is known without doubt. We postpone further study of these ideas to future work.

\subsubsection{Schrodinger equation in lightfront quantization?}

It is striking that in the non-relativistic limit, the exact S-matrix was obtained rather easily by solving 
a Schrodinger equation in position space. One might wonder if the full non-relativistic S-matrix 
may similarly be obtained by solving an appropriate Schrodinger equation. 

An observation that supports this hope is the fact that genuine `particle creation' never occurs in the 
large $N$ limit. A Feynman diagram that describes virtual fundamental particles being created and destroyed 
during a scattering process has additional index loops and is suppressed in the large $N$ limit. It thus 
seems plausible that the scattering matrices of interest to us in this paper may be obtained by solving 
the relevant quantum mechanical problem.

Although we will not present the details here, we have succeeded in reproducing the effective scattering amplitude of the 
ungauged large $N$ $\phi^4$ theory by solving a two particle Schrodinger equation. The 
Schrodinger equation in question is obtained from a lightcone quantization of the quantum field theory.
It may well prove possible to extend this analysis to the gauged theory, and thereby extract the 
S-matrix from an effective Schrodinger equation; however we have not yet succeeded in 
implementing this idea. We leave further study of this idea to future work.

\section{Discussion}\label{disc}

In this paper we have presented computations and conjectures for the formulas for $2 \rightarrow 2$ scattering in large $N$ matter Chern-Simons theories at all orders in the 't Hooft coupling. 
All the computations presented in this paper were performed in the light cone gauge 
together with an assumption of involving the precise definition of the gauge propagator 
in this gauge. It would be useful to have checks of our results using different methods - perhaps working in a covariant gauge. It might be possible (and would be very interesting) to generalize 
the covariant computation of section \ref{landauonshell} to two loops. It would also be 
very interesting to study how (and whether) the unusual structural features predicted in our paper 
manifest themselves in a covariant computation.

Obvious extensions of our paper include the generalization of the computations presented 
here to the simplest $N=2$ supersymmetric matter Chern-Simons theories, and also 
to the large class of single boson-fermion theories studied in \cite{GurAri:2012is}.
It may also be possible to match the finite $b_4$ results of the bosonic computations 
in this paper with a generalized fermionic computation in which we include a 
$({\bar \psi} \psi)^2$ term in the fermionic Lagrangian.

Perhaps the most interesting formula presented in this paper is the formula for the scattering matrix 
in the $S$-channel.  (see \eqref{conject}, \eqref{ttrial}). This formula is manifestly unitary: it
includes an unusual rescaling of the identity piece in the S-matrix;  it agrees with the formula 
for Aharonov-Bohm scattering in the non-relativistic limit, and the formula for large $N$ 
$\phi^4$ scattering in the small ${\lambda_{B}}$ limit. It is also tightly related to scattering 
in the other channels via rescaled relations of crossing symmetry. In the case of the 
scalar theory, this S-matrix also has poles signalling the existence of a stable singlet
bound state of two particles in the singlet channel over a range of values of 
$b_4$.  Unfortunately the formula for $S$-channel scattering presented in this paper has not been derived but has simply 
been conjectured.  A very important problem for the future is to honestly derive the formula for $S$-channel scattering, perhaps along the lines sketched in subsection \ref{at}.

Another reason to understand scattering after the double analytic continuation described in 
subsection \ref{at} is to better understand the detailed connection between the Lorentzian 
results of our paper and the Euclidean results of earlier computations 
\cite{Giombi:2011kc,Aharony:2011jz,Jain:2012qi,Aharony:2012ns,Jain:2013py,Takimi:2013zca,Aharony:2012nh}.

The S-matrices derived in our paper have all been obtained for the scattering of massive particles. 
There is no barrier to taking the high energy (or equivalently zero mass) limit of our scattering amplitudes.
Interestingly, the scattering amplitudes develop no new infrared singlularities in this limit. This 
fact is probably an artifact of the large N limit that supresses the pair creation of fundamental particles;
it seems likely that $\frac{1}{N}$ corrections to the results presented in this paper will have new 
infrared singularities in the zero mass limit.

As we have explained, the formulas (and conjectures) presented in this paper imply that the 
usual rules of crossing symmetry are modified in matter Chern-Simons theories. In this paper we 
have presented a conjecture for the nature of that modification in the 't Hooft large $N$ limit. 
It would be interesting to prove this rule analogue of crossing symmetry (perhaps using a refinement of the arguments in subsection \ref{expmd}). 

A simplifying feature of the 't Hooft large $N$ is that scattering was truly anyonic (i.e. was characterized
by a nonzero anionic phase) only in the singlet channel of Particle-antiparticle scattering. 
In particular the anyonic phase vanishes in particle-particle scattering (see subsection \ref{abs}) 
so that we were never forced in this paper to address issues having to do with the generalization 
of Bose or Fermi statistics. At finite $N$ and $k$ this situation will change, presumably leading 
to nontrivial phases between particle-particle scattering in the direct and exchange channels. 
These considerations suggest that the crossing symmetry structure of scattering amplitudes will 
be very rich at finite $N$ and $k$; it would be fascinating to have even a well motivated 
conjecture for this structure. It is conceivable that the S-matrix presented in this paper
and its generalization to the finite $N$ and $k$ case may have useful applications in the condensed matter problems and also in the area 
of the topological quantum computation \cite{RevModPhys.80.1083}.
 \footnote{Perhaps there is a sense in which the finite $N$ and $k$ result is 
`quantum',  and results in the 't Hooft limit are obtained from the `classical limit' of the corresponding
`quantum structure'.}

If the unusual structural properties conjectured in this paper withstand further scrutiny, then they 
are likely to be general features of all matter Chern-Simons theories. We should, in particular, be able to probe
 these features in the scattering of maximally supersymmetric Chern-Simons theories 
(ABJ theories). In this connection it is interesting to note that there is an unresolved paradox in the 
study of scattering amplitudes in ABJM and ABJ theory. In this theory the $2 \rightarrow 2$ scattering amplitude has been argued 
to vanish at one loop \cite{Agarwal:2008pu, Bargheer:2012cp, Bianchi:2011fc},
 but to be non vanishing at two loops \cite{Chen:2011vv, Bianchi:2011dg, Bianchi:2011fc}. The paradox 
arises because although the proposed two loop formula for four particle scattering in ABJM theory
has cuts \cite{Chen:2011vv}, there do not seem to exist any candidate intermediate processes to saturate these 
cuts.  \footnote{There appear to be only two candidates for the processes that could produce
these cuts. The first is by sewing together two $2 \rightarrow 3$ tree level amplitudes, but there 
are no such amplitudes in ABJM theory. The second is by sewing together a tree level 
$2 \rightarrow 2$ amplitude with a one loop $2 \rightarrow 2 $ amplitude, but as we have remarked, 
the latter have been argued to vanish.}  While scattering amplitudes in ABJ theory are more 
confusing than those considered in this paper because they receive infrared divergences from 
intermediate massless scalar and fermion propagation, it is at least  conceivable that the 
resolution to this apparent unitarity paradox lies along the lines sketched in this paper.
The results of our paper should generalize, in the most straightforward fashion, to scattering in 
$U(M)\times U(N)$ theory when $\frac{M}{N}<< 1$ (in this limit the ABJ  theory begins to closely 
resemble a theory with a single gauge group and only fundamental matter, like the theories 
studied in this paper).   The analysis of our paper suggests that 
the $2 \rightarrow 2 $ scattering amplitude does not completely vanish at one loop: it should at least have 
a $\delta$ function localized singular piece. The contribution of this piece in a one loop sub diagram 
to two loop graphs could then, additionally, modify the scattering amplitudes as well as the usual 
rules of crossing symmetry. It would be fascinating to verify these expectations via a direct analysis 
of scattering amplitudes in the supersymmetric theories\footnote{It is interesting to note that, in \cite{Bianchi:2014iia} it was 
argued that in the case of ABJM, three loop amplitude is non zero. However, again they missed the existance of delta function. It would be 
interesting to see whether higher point functions also shows some nontrivial analytical structure. For a discussion of higher point function
in ABJM theory, we refer reader to \cite{Bianchi:2012cq}.}. 

A significant check of all the computations and conjectures presented in our paper is that they
are all consistent with the recently conjectured level-rank duality between bosonic and fermionic 
Chern-Simons theories. This works in a rather remarkable way. The bosonic S-matrices have nontrivial analytic structure 
(e.g. two particle cuts) at all values of $\lambda_B$
 (including $\lambda_B=0$ where the cuts come from the four boson contact interaction) 
 provided $|\lambda_B| \neq 1$. Precisely at $\lambda_B=1$, however, the bosonic S-matrix 
 collapses into precisely the analytically trivial constant that one predicts from fermionic 
 tree level scattering. Indeed the agreement between bosonic and fermionic S-matrices works 
 at all values of $\lambda_B$, not just at extreme ends. 

 Indeed the results of our paper shed some additional light 
on the working of this duality. The first point, as we have already emphasized in the 
introduction, is that our S-matrix is effectively anyonic in the singlet channel. The effective 
anyonic phase can be estimated very simply in the non-relativistic limit, and the duality map 
from $\lambda_B$ to $\lambda_F$ can simply be deduced by demanding that the dual theories
have equal anyonic phases. 

In the $U$-channel, on the other hand, the anyonic phase is trivial. Bosonic and fermionic S
matrices map to each other only after we transpose the exchange representations. 
As we have explained in more detail in the introduction, this suggests that, for scattering purposes, there exists a  map between asymptotic multi bosonic states that transform in representation $R$ of $U(N_B)$ and multi- fermionic states that transform in representation $R^T$ of $U(N_F)$. 

There is an obvious puzzle about the identification suggested above; namely the number of 
states on the two sides do not match (this is true even if we restrict to the simplest representation, 
namely the fundamental, simply because $N_B \neq N_F$). It seems possible that 
the duality between the bosonic and fermionic theories really works only 
on compact manifolds (and so, effectively, only in the singlet sector on $R^2$). If this turns out to 
be the correct eventual statement of the duality, then the perfect match under duality of the 
scattering amplidues in non singlet sectors may eventually find its explaination in the match of 
factorized subsectors in higher point scattering in the singlet channel. For instance one could consider 
the scattering of two particles, and simultaneously the 
scatering of two antiparticles  very far away, with colour indices 
chosen so that the full four particle initial state is a singlet and so duality invariant. Presumably the scattering amplitudes factor into the scattering amplitude for particle - particle scattering and the scatterng amplitude 
for antiparticle-antiparticle scattering, implying the duality invariance of these more basic 2 particle 
scattering amplitudes, even though they do not occur in a gauge singlet sector, explaining the 
results obtained in this paper. It would certainly be nice to understand this better.

In summary, the  results and conjectures presented in this paper have  several  unexpected  
features, have intriguing implications, and throw up several puzzles. If our results stand up to further 
scrutiny they suggest several fascinating new directions of investigation.

\acknowledgments

We would like to thank S. Datta for collaboration in the initial stages of this project. 
We would like especially to thank  S. Caron Huot, J. Maldacena,  N. Seiberg, A. Shivaji, E. Witten, Y. Tachikawa
 and R. Yacoby  for very useful discussions.  We would also like to thank 
O. Aharony, N. Arkani Hamed, D.Bak, D. Bardhan, J. Bhattacharya,  A.Dhar, R. Gopakumar,  
T. Faulkner, 
S. Giombi, R. Loganayagam, G.Mandal, S.Raju, M. Rangamani,  A. Sen,
 D. Son,  S.Trivedi, Y.-T. Huang and S. Zhibodeov  for useful discussions. We would also like to thank O. Aharony, T. Bargheer
S. Caron Hout, R. Gopakumar and  S. Giombi,  for comments on a preliminary version of this manuscript. 
S.J. would like to thank HRI, ICTS, IISc and IOP  for hospitality while this work was in progress.  S. M. would like to thank the Nishina Memorial Foundation,
 IPMU, Kyoto University and the University of Chicago for hospitality while this  work was in progress. We would all also like to  
acknowledge our debt to the people of India for their generous and steady support to research in the basic sciences.

\appendix

\section{The identity S-matrix as a function of $s$, $t$, $u$} \label{idt}
As explained in subsection \ref{ik}, the identity S-matrix has a simple form 
in the center of mass frame; it is given by 
$$ (2\pi)^3 \delta^3(p_1+p_2-p_3-p_4) 8 \pi \sqrt{s} \delta(\theta)$$
As we will see below, the expression $\delta(\theta)$ is slightly singular when recast in terms of invariants, 
so we will find it convenient to regulate this expression as
 $$ (2\pi)^3 \delta^3(p_1+p_2-p_3-p_4) 4 \pi \sqrt{s} \lim_{\epsilon \to 0}
\left(  \delta(\theta - \epsilon) + \delta(\theta +\epsilon) \right). $$
Using \eqref{stucm}, 
this expression may be recast in invariant form
\begin{equation}\label{mdm}
\delta\left(\sqrt{\frac{4 t}{t+u}}-\epsilon\right)(2\pi)^3\delta^3(p_1+p_2-p_3-p_4)
\end{equation}
as we have already noted in \eqref{idm}.

In this Appendix we present a cumbersome but direct algebraic check that $I$ as defined 
in \eqref{mdm} coincides with $I$ defined in \eqref{identsm}. Our strategy is as follows. 
We start with the expression \eqref{mdm}, and express the arguments of the delta functions 
in \eqref{mdm} entirely in terms of the 8 variables
 $p_1^x, p_1^y, p_2^x, p_2^y, p_3^x, p_3^y, p_4^x, p_4^y$ (the energies of the ingoing 
 and outgoing particles are solved for using the on shell condition). We choose to 
 view the resultant expression as follows. We think of 
 $p_1^x, p_1^y, p_2^x, p_2^y$ as fixed initial data and the remaining quantities 
$p_3^x, p_3^y, p_4^x, p_4^y$ as variable scattering data. The four delta functions in 
\eqref{mdm} thus determine $p_3^x, p_3^y, p_4^x, p_4^y$ as functions of $p_1^x, p_1^y, p_2^x, p_2^y$. At leading order in $\epsilon$  is not difficult to explicitly determine the values for $p_3^x, p_3^y, p_4^x, p_4^y$ obtained in this manner.  We find 
\be\label{peps} p_{3,x}=p_{1,x} \pm \epsilon a_{3,x},~ p_{4,x}=p_{2,x} \pm \epsilon a_{4,x},~ p_{3,y}=p_{1,y} \pm \epsilon a_{3,y},~ p_{4,y}=p_{2,y} \pm \epsilon a_{4,y}.\ee
where the four $a$ variables are obtained by solving four linear equations (the $\pm$ above
corresponds to the two possibilities $\theta= \epsilon$ or $\theta= -\epsilon$ in the centre-
of-mass frame). In what follows below we will not need the explicit form of the solutions for the $a$ variables, 
but will only need certain identities obeyed by these solutions. These identities turn out, 
in fact, to be three of the four equations that the $a$ variables obey. The relevant three
equations are  
\be\begin{split}\label{asoll}
&a_{4,x}=-a_{3,x},~~a_{4,y}=-a_{3,y},~~a_{3,x}=B a_{3,y},
\\
\text{where} \quad &B=\frac{p_{2,y} \sqrt{m^2+p_{1,x}^2+p_{1,y}^2}-p_{1,y} \sqrt{m^2+p_{2,x}^2+p_{2,y}^2}}{p_{1,x}
   \sqrt{m^2+p_{2,x}^2+p_{2,y}^2}-p_{2,x} \sqrt{m^2+p_{1,x}^2+p_{1,y}^2}} .
   \end{split}
\ee
Let us now return to our task of rewriting the delta function in \eqref{mdm} in terms of 
delta functions linear in $p_3^x, p_3^y, p_4^x, p_4^y$. It follows from the usual rules for 
manipulating delta functions that 
\be\label{delta32}\begin{split}
&\delta\left(\sqrt{\frac{4 t}{t+u}}-\epsilon\right)\delta^3(p_1+p_2-p_3-p_4)\\
&= J_1\delta^2(\overrightarrow{p}_3 -\overrightarrow{p}_1 +\epsilon a_3)\delta^2(\overrightarrow{p}_4 -\overrightarrow{p}_2 + \epsilon a_4) + J_2\delta^2(\overrightarrow{p}_3 - \overrightarrow{p}_1 -\epsilon a_3)\delta^2(\overrightarrow{p}_4-\overrightarrow{p}_2 - \epsilon a_4) 
\end{split}\ee 
where $J_1$ and $J_2$ are the relevant Jacobians. It remains to compute these Jacobians. 

The reader might naively expect that the Jacobians are independent of $a_3$ and $a_4$ in the limit $\epsilon \to 0$, but that is not the case. 
 It is not difficult verify that, in the $\epsilon \to 0$ limit the  derivatives $\frac{\partial \sqrt{ \frac{ 4 t}{t+u}} }{\partial p^x_3}$ 
and $\frac{\partial \sqrt{ \frac{ 4 t}{t+u}} }{\partial p^y_3}$ (which enter the expression for the Jacobians) are of the form $\frac{ A}{B}$ where $A$ and $B$ are both expressions of unit
homogeneity in $a_3$ and $a_4$. The ratio $\frac{A}{B}$ does not depend on the overall scale of $a_{3,x},a_{3,y}$ and $a_{4,x},a_{4,y}$, 
but does depend on their relative magnitudes.
 It turns out that the equations \eqref{asoll} are 
sufficient to unambiguously determine the ratio $\frac{A}{B}$  (which turns out to be the same
for the two solutions corresponding to the $\pm $ signs so that $J_1=J_2=J$) ; we find 
\be\label{JsEE}
J=\sqrt{s}\frac{1}{E_1 E_2}
\ee
 where
$E_i=\sqrt{m^2+p_{i,x}^2+p_{i,y}^2}$ and
\be s=\sqrt{2}\sqrt{\sqrt{m^2+p_{1,x}^2+p_{1,y}^2} \sqrt{m^2+p_{2,x}^2+p_{2,y}^2}+m^2-p_{1,x} p_{2,x}-p_{1,y}
   p_{2,y}}.\ee
Collecting factors, it follows that the RHS of \eqref{mdm} coincides with the RHS of \eqref{identsm}
in the limit $\epsilon \to 0$.

\section{Tree level S-matrix}\label{treeap}
The bosonic effective action is
\begin{equation}
T_B = \frac{1}{2}\int\frac{d^3p}{(2\pi)^3}\frac{d^3k}{(2\pi)^3}\frac{d^3q}{(2\pi)^3}V(p,k,q)\phi_i(p+q)\bar{\phi}^j(-k-q)\bar{\phi}^i(-p)\phi_j(k),
\end{equation}
where at tree level
\begin{equation}
V(p,k,q) = 8\pi i \lambda\epsilon_{\mu\nu\rho}\frac{q^{\mu}p^{\nu}k^{\rho}}{(k-p)^2}.
\end{equation}
And the fermionic effective action is
\begin{equation}
T_F = \frac{1}{2}\int\frac{d^3p}{(2\pi)^3}\frac{d^3k}{(2\pi)^3}\frac{d^3q}{(2\pi)^3}V^{\alpha\gamma}_{\beta\delta}(p,k,q)\psi_{i,\alpha}(p+q)\bar{\psi}^{j,\beta}(-k-q)\bar{\psi}^{i,\delta}(-p)\psi_{j,\gamma}(k),
\end{equation}
where at tree level
\begin{equation}
V^{\alpha\gamma}_{\beta\delta}(p,k,q) = 2i\pi\lambda\epsilon_{\mu\nu\rho}\frac{(\gamma^{\mu})^\alpha_\beta (\gamma^{\nu})^\gamma_\delta (k-p)^{\rho}}{(k-p)^2}.
\end{equation}
The gauge field propagator that we work with in this section is
\be
\langle A_{\mu}(p)A_{\nu}(-q)\rangle=(2\pi)^3\delta^3(p-q)\frac{4\pi}{p^2}\epsilon_{\mu\nu\rho}p^{\rho}. 
\ee
\subsection{Particle-particle scattering}
According to the momentum assignments in \eqref{pps}, The bosonic S-matrix is given by
\begin{equation}
\begin{split}
&S_{B}(p_1,p_2,p_3,p_4) \\
=&\langle out|1+iT_B|in\rangle \\
=& \langle0|a_n(p_4)a_m(p_3)a^{b\dagger}(p_2)a^{a\dagger}(p_1)|0\rangle \\
&+\frac{i}{2}\int\frac{d^3p}{(2\pi)^3}\frac{d^3k}{(2\pi)^3}\frac{d^3q}{(2\pi)^3}\Biggl[V(p,k,q)
\\
& ~~~~~~~~\times \langle 0|a_n(p_4)a_m(p_3)
\left(\phi_i(p+q)\bar{\phi}^j(-k-q)\bar{\phi}^i(-p)\phi_j(k)\right)
a^{b\dagger}(p_2)a^{a\dagger}(p_1)|0\rangle \Biggr].
\end{split}
\end{equation}
Using appropriate contractions and commutation relations, we find
\begin{equation}\begin{split}
S_B(p_1,p_2,p_3,p_4)&=\delta^a_m\delta^b_n\left(I(p_1,p_2,p_3,p_4)+iV(-p_3,p_2,p_1+p_3)\right)\\
&+\delta^a_n\delta^b_m\left(I(p_1,p_2,p_4,p_3)+iV(-p_4,p_2,-p_3-p_2)\right)
\end{split}\end{equation}
The first term is for the $U_d$ channel while the other is for the $U_e$ channel.\\
Whereas the fermionic S-matrix is
\begin{equation}
\begin{split}
&S_F(p_1,p_2,p_3,p_4)
\\
=&\langle out|1+iT_F|in\rangle 
\\
=& \langle0|a_n(p_4)a_m(p_3)a^{b\dagger}(p_2)a^{a\dagger}(p_1)|0\rangle \\
&+\frac{i}{2}\int\frac{d^3p}{(2\pi)^3}\frac{d^3k}{(2\pi)^3}\frac{d^3q}{(2\pi)^3}\Biggl[V^{\alpha\gamma}_{\beta\delta}(p,k,q)\\
&~~~~\times \langle 0|a_n(p_4)a_m(p_3)
\left(\psi_{i,\alpha}(p+q)\bar{\psi}^{j,\beta}(-(k+q))\bar{\psi}^{i,\delta}(-p)\psi_{j,\gamma}(k)\right)a^{b\dagger}(p_2)a^{a\dagger}(p_1)|0\rangle\Biggr]
\end{split}
\end{equation}
Using appropriate contractions and anticommutation relations,
\begin{equation}
\begin{split}
&S_F(p_1,p_2,p_3,p_4)
\\
=&-\delta^a_m\delta^b_n I(p_1,p_2,p_3,p_4)
\\
&-i\delta^a_m\delta^b_n V^{\alpha\gamma}_{\beta\delta}(-p_3,p_2,p_1+p_3)\bar{u}^\beta(-p_4)\bar{u}^\delta(-p_3)u_\alpha(p_1)u_\gamma(p_2)
\\
&+\delta^a_n\delta^b_m
I(p_1,p_2,p_4,p_3)
\\
&+i \delta^a_n\delta^b_m
V^{\alpha\gamma}_{\beta\delta}(-p_4,p_2,-p_3-p_2)\bar{u}^\beta(-p_3)\bar{u}^\delta(-p_4)u_\alpha(p_1)u_\gamma(p_2)
\end{split}
\end{equation}
Again, the first term is for the $U_d$ channel while the other is for the $U_e$ channel.\\

\subsection{Particle-antiparticle scattering}
According to the momentum assignments in \eqref{pas}, The bosonic S-matrix is given by
\begin{equation}
\begin{split}
&S_B(p_1,p_2,p_3,p_4)
\\
=&\langle out|1+iT_B|in\rangle 
\\
=& \langle0|b^n(p_4)a_m(p_3)b_b^{\dagger}(p_2)a^{a\dagger}(p_1)|0\rangle \\
&+\frac{i}{2}\int\frac{d^3p}{(2\pi)^3}\frac{d^3k}{(2\pi)^3}\frac{d^3q}{(2\pi)^3}\Biggl[ V(p,k,q)
\\
& ~~~~~ \times \langle0|b^n(p_4)a_m(p_3)\left(\phi_i(p+q)
\bar{\phi}^j(-k-q)\bar{\phi}^i(-p)\phi_j(k)\right)
b_b^{\dagger}(p_2)a^{a\dagger}(p_1)|0\rangle
\Biggr]
\end{split}
\end{equation}
Using appropriate contractions and commutation relations, we find
\begin{equation}\begin{split}
S_B(p_1,p_2,p_3,p_4)&=\left(\delta^a_m\delta_b^n-\frac{\delta^n_m\delta_b^a}{N}\right)\left(I(p_1,p_2,p_3,p_4)+iV(-p_3,p_4,p_1+p_3)\right)\\
&+\frac{\delta^n_m\delta_b^a}{N}\left(I(p_1,p_2,p_3,p_4)+iV(-p_2,p_4,p_1+p_2)\right)
\end{split}\end{equation}
The first term is for the $T$-channel while the other is for the $S$-channel.\\
Whereas the fermionic S-matrix is
\begin{equation}
\begin{split}
&S_F(p_1,p_2,p_3,p_4)=\langle out|1+iT_F|in\rangle = \langle0|b^n(p_4)a_m(p_3)b_b^{\dagger}(p_2)a^{a\dagger}(p_1)|0\rangle \\
&+\frac{i}{2}\int\frac{d^3p}{(2\pi)^3}\frac{d^3k}{(2\pi)^3}\frac{d^3q}{(2\pi)^3}V^{\alpha\gamma}_{\beta\delta}(p,k,q)\langle0|b^n(p_4)a_m(p_3)\\
&\Bigg(\psi_{i,\alpha}(p+q)\bar{\psi}^{j,\beta}(-k-q)\bar{\psi}^{i,\delta}(-p)\psi_{j,\gamma}(k)\Bigg)b_b^{\dagger}(p_2)a^{a\dagger}(p_1)|0\rangle
\end{split}
\end{equation}
Using appropriate contractions and anticommutation relations, we find
\begin{equation}
\begin{split}
&S_F(p_1,p_2,p_3,p_4)\\
&=-\left(\delta^a_m\delta_b^n-\frac{\delta^n_m\delta_b^a}{N}\right)\left(I(p_1,p_2,p_3,p_4)-iV^{\alpha\gamma}_{\beta\delta}(-p_3,p_4,p_1+p_3)\bar{u}^\beta(-p_3)\bar{v}^\delta(p_2)u_\alpha(p_1)v_\gamma(-p_4)\right)
\\&-\frac{\delta^n_m\delta_b^a}{N}\left(I(p_1,p_2,p_3,p_4)+iV^{\alpha\gamma}_{\beta\delta}(-p_2,p_4,p_1+p_2)\bar{v}^\beta(p_2)\bar{u}^\delta(-p_3)u_\alpha(p_1)v_\gamma(-p_4)\right)
\end{split}
\end{equation}
Again, the first term is for the $T$-channel while the other is for the $S$-channel.\\

\subsection{Explicit tree level computation}

Now we substitute for the $V$s for the respective channels in bosonic case, and obtain

While the fermionic expressions for $S$, $T$, $U_d$ and $U_e$ channels are (with respect to the identity) respectively,
\begin{equation}\label{stuchan}
\begin{split}
&T_{S}=\frac{2i\pi}{k_F(p_2+p_4)^2}\epsilon_{\mu\nu\rho}\left(\bar{u}(-p_3)\gamma^\mu u(p_1)\right)\left(\bar{v}(p_2)\gamma^\mu v(-p_4)\right)(p_2+p_4)^\rho\\
&T_{T}=-\frac{2i\pi}{k_F(p_3-p_4)^2}\epsilon_{\mu\nu\rho}\left(\bar{v}(p_2)\gamma^\mu u(p_1)\right)\left(\bar{u}(-p_3)\gamma^\mu v(-p_4)\right)(p_3+p_4)^\rho\\
&T_{U_d}=\frac{2i\pi}{k_F(p_2+p_3)^2}\epsilon_{\mu\nu\rho}\left(\bar{u}(-p_4)\gamma^\mu u(p_1)\right)\left(\bar{u}(-p_3)\gamma^\mu u(p_2)\right)(p_2+p_3)^\rho\\
&T_{U_d}=\frac{2i\pi\lambda_F}{(p_2+p_4)^2}\epsilon_{\mu\nu\rho}\left(\bar{u}(-p_3)\gamma^\mu u(p_1)\right)\left(\bar{u}(-p_4)\gamma^\mu u(p_2)\right)(p_2+p_4)^\rho
\end{split}
\end{equation}

These expressions can be manipulated conveniently using the Gordon Identities which are derived below:\\
The Dirac equation satisfied by $u(p),{\bar u(p)},v(p),{\bar v(p)}$ are given by
\be\label{diraceq12}\begin{split}
&\left(i\gamma^{\mu}p_{\mu}+m\right)u(p)=0,~~{\bar u(p)}\left(i\gamma^{\mu}p_{\mu}+m\right)=0,\\
&\left(-i\gamma^{\mu}p_{\mu}+m\right)v(p)=0,~~{\bar v(p)}\left(-i\gamma^{\mu}p_{\mu}+m\right)=0.
\end{split}\ee
The gamma matrices are given by 

\begin{equation}
\begin{split}
\gamma^0 = \left(\begin{array}{cc}
                 0 & 1 \\ -1 & 0
                 \end{array} \right), \\
\gamma^1 = \left(\begin{array}{cc} 0 & 1 \\ 1 & 0 \end{array}\right), \\
\gamma^2 = \left(\begin{array}{cc} 1 & 0 \\0 & -1 \end{array}\right).\\
\end{split}
\end{equation}
They satisfy
\be
\gamma^{\mu}\gamma^{\nu}=g^{\mu\nu}-\epsilon^{\mu\nu\rho}\gamma_{\rho}.
\ee
Now, using Dirac equation \eqref{diraceq12}, it is easy derive the Gordon identities
\be\label{Grdnid}\begin{split}
-{\bar u}(p_1)\gamma^{\mu}u(p_2)&=i\left({\bar u}(p_1)\frac{(p_1+p_2)^{\mu}}{2m}u(p_2)-\epsilon^{\mu\nu\rho}\frac{(-p_1+p_2)_{\nu}}{2m}{\bar u}(p_1)\gamma_{\rho}u(p_2)\right)\\
-{\bar u}(p_1)\gamma^{\mu}v(p_2)&=i\left({\bar u}(p_1)\frac{(p_1-p_2)^{\mu}}{2m}v(p_2)+\epsilon^{\mu\nu\rho}\frac{(p_1+p_2)_{\nu}}{2m}{\bar u}(p_1)\gamma_{\rho}v(p_2)\right)\\
-{\bar v}(p_1)\gamma^{\mu}u(p_2)&=i\left({\bar v}(p_1)\frac{(-p_1+p_2)^{\mu}}{2m}u(p_2)-\epsilon^{\mu\nu\rho}\frac{(p_1+p_2)_{\nu}}{2m}{\bar v}(p_1)\gamma_{\rho}u(p_2)\right)\\
-{\bar v}(p_1)\gamma^{\mu}v(p_2)&=i\left(-{\bar v}(p_1)\frac{(p_1+p_2)^{\mu}}{2m}v(p_2)+\epsilon^{\mu\nu\rho}\frac{(p_1+p_2)_{\nu}}{2m}{\bar v}(p_1)\gamma_{\rho}v(p_2)\right)
\end{split}\ee

Using this, it is easy to show that
\be\label{Grdnid1}\begin{split}
{\bar u}(p_1)\gamma^{\mu}u(p_2)&=\frac{1}{1+(\frac{p_1-p_2}{2m})^2}\left(-i\frac{(p_1+p_2)^{\mu}}{2m}-\frac{1}{2m^2}\epsilon^{\mu\nu\rho}(p_1)_{\nu}(p_2)_{\rho}\right){\bar u}(p_1)u(p_2)\\
{\bar v}(p_1)\gamma^{\mu}u(p_2)&=\frac{1}{1+(\frac{p_1+p_2}{2m})^2}\left(-i\frac{(-p_1+p_2)^{\mu}}{2m}+\frac{1}{2m^2}\epsilon^{\mu\nu\rho}(p_1)_{\nu}(p_2)_{\rho}\right){\bar v}(p_1)u(p_2)\\
{\bar u}(p_1)\gamma^{\mu}v(p_2)&=\frac{1}{1+(\frac{p_1+p_2}{2m})^2}\left(-i\frac{(p_1-p_2)^{\mu}}{2m}+\frac{1}{2m^2}\epsilon^{\mu\nu\rho}(p_1)_{\nu}(p_2)_{\rho}\right){\bar u}(p_1)v(p_2)\\
{\bar v}(p_1)\gamma^{\mu}v(p_2)&=\frac{1}{1+(\frac{p_1-p_2}{2m})^2}\left(i\frac{(p_1+p_2)^{\mu}}{2m}-\frac{1}{2m^2}\epsilon^{\mu\nu\rho}(p_1)_{\nu}(p_2)_{\rho}\right){\bar v}(p_1)v(p_2)\\
\end{split}\ee
The only thing that is remaining is to compute the  quantities, ${\bar u}(p')u(p),{\bar v}(p')v(p),{\bar u}(p')v(p),{\bar v}(p')u(p).$
For this, we explicitly construct the solution for $V$ and $u$ starting boosting the rest frame results which are easily computable to 
a frame where the momenta is $p.$ In the rest frame, equation satisfied by the $u$ and $v$ is given by,
\be
\left(-i\gamma^{0}+I\right)u(0)=0,~~~\left(i\gamma^{0}+I\right)v(0)=0,
\ee and for ${\bar u}$ and ${\bar v}$
\be
{\bar u}(0)\left(-i\gamma^{0}+I\right)=0,~~~{\bar v}(0)\left(i\gamma^{0}+I\right)=0,
\ee
where $I$ denotes, the $2\times2$ identity matrix. The solutions are
\be\label{sol0}
u(0)=\sqrt{m}\left( 1, -i\right),~~ v(0)=\sqrt{m}\left( 1, i\right),~~{\bar u}(0)=\sqrt{m}\left( 1, i\right),~~{\bar v}(0)=\sqrt{m}\left( -1, i\right).
\ee
Suppose we are now interested in solution for $u$ and $v$ at momenta $p,$ given by
\be
p_{\mu}=\left(-m \cosh(\alpha),m \sinh(\alpha) \cos(\theta),m \sinh(\alpha) \sin(\theta)\right).
\ee
The solutions are given by
\be\begin{split}
u(p)&=\left(\cosh(\frac{\alpha}{2})I-\sinh(\frac{\alpha}{2})\left(\cos(\theta)\gamma^{2}-\sin(\theta)\gamma^{1}\right)  \right)u(0)\\
{\bar u}(p)&={\bar u}(0)\left(\cosh(\frac{\alpha}{2})I+\sinh(\frac{\alpha}{2})\left(\cos(\theta)\gamma^{2}-\sin(\theta)\gamma^{1}\right)  \right)\\
v(p)&=\left(\cosh(\frac{\alpha}{2})I-\sinh(\frac{\alpha}{2})\left(\cos(\theta)\gamma^{2}-\sin(\theta)\gamma^{1}\right)  \right)v(0)\\
{\bar v}(p)&={\bar v}(0)\left(\cosh(\frac{\alpha}{2})I+\sinh(\frac{\alpha}{2})\left(\cos(\theta)\gamma^{2}-\sin(\theta)\gamma^{1}\right)  \right).
\end{split}\ee
It is now easy to compute ${\bar u}(p')u(p),{\bar v}(p')v(p),{\bar u}(p')v(p),{\bar v}(p')u(p).$ Results are given by
\be\begin{split}\label{dtpro}
{\bar u}(p_1)u(p_2)=&e^{i\tan^{-1}\frac{\sin(\theta_2-\theta_1)}{\cos\left(\theta_2-\theta_1\right)-\coth(\alpha_1)\coth(\alpha_2)}}\sqrt{\left(2m^2-2~ p_1\cdot p_2\right)},\\
{\bar v}(p_1)v(p_2)=&e^{i\tan^{-1}\frac{\sin(\theta_1-\theta_2)}{\cos\left(\theta_1-\theta_2\right)-\coth(\alpha_1)\coth(\alpha_2)}}\sqrt{\left(2m^2-2~ p_1\cdot p_2\right)},\\
{\bar v}(p_1)u(p_2)=&e^{i\tan^{-1}\frac{\sinh \left(\frac{\alpha _1}{2}\right) \cosh \left(\frac{\alpha _2}{2}\right) \sin \left(\theta _1\right)-\sinh \left(\frac{\alpha _2}{2}\right)
   \cosh \left(\frac{\alpha _1}{2}\right) \sin \left(\theta _2\right)}{\sinh \left(\frac{\alpha _1}{2}\right) \cosh \left(\frac{\alpha _2}{2}\right) \cos
   \left(\theta _1\right)-\sinh \left(\frac{\alpha _2}{2}\right) \cosh \left(\frac{\alpha _1}{2}\right) \cos \left(\theta _2\right)}}
\\
& \times \sqrt{\left(-2m^2-2~ p_1\cdot p_2\right)},\\
{\bar u}(p_1)v(p_2)=&e^{i\tan^{-1}\frac{\sinh \left(\frac{\alpha _2}{2}\right) \cosh \left(\frac{\alpha _1}{2}\right) \sin \left(\theta _2\right)-\sinh \left(\frac{\alpha _1}{2}\right)
   \cosh \left(\frac{\alpha _2}{2}\right) \sin \left(\theta _1\right)}{\sinh \left(\frac{\alpha _1}{2}\right) \cosh \left(\frac{\alpha _2}{2}\right) \cos
   \left(\theta _1\right)-\sinh \left(\frac{\alpha _2}{2}\right) \cosh \left(\frac{\alpha _1}{2}\right) \cos \left(\theta _2\right)}}
\\
& \times \sqrt{\left(-2m^2-2~ p_1\cdot p_2\right)},\\
\end{split}\ee
As a final ingredient to compute the tree level scattering is
\be\begin{split}\label{uchng}
&\epsilon_{\mu\nu\rho}{\bar u}(p_1)\gamma^{\mu}u(p_2){\bar u}(p_3)\gamma^{\nu}u(p_4)p_5^{\rho}\\
=&\frac{\left({\bar u}(p_1)u(p_2)\right)\left({\bar u}(p_3)u(p_4)\right)}{\left(1+\frac{(p_1-p_2)^2}{4m^2}\right)\left(1+\frac{(p_3-p_4)^2}{4m^2}\right)}
\\
&~ \times 
\Big[-\frac{1}{4m^2}\epsilon_{\mu\nu\rho}(p_1+p_2)^{\mu}(p_3+p_4)^{\nu}p_5^{\rho} 
\\
&~~~~~~+\frac{1}{4m^4}\left((p_1\cdot p_5) 
\epsilon_{\mu\nu\rho}p_{2}^{\mu}p_3^{\nu}p_4^{\rho}-
(p_2\cdot p_5) \epsilon_{\mu\nu\rho}p_{1}^{\mu}p_3^{\nu}p_4^{\rho}\right)\\
&
~~~~~~+\frac{i}{4m^3}
\biggl((p_4\cdot p_5) 
\left(p_3\cdot (p_1+p_2)\right)
-(p_3\cdot p_5) \left(p_4\cdot (p_1+p_2)\right) \biggr)
\\
&
~~~~~~+\frac{i}{4m^3}
\biggl(
(p_1\cdot p_5) \left(p_2\cdot (p_3+p_4)\right)
-(p_2\cdot p_5) \left(p_1\cdot (p_3+p_4)\right)\biggr)\Big],
\end{split}\ee where $p\cdot p'=p_\mu p'^{\mu}.$
Now just by few interchange of signs, as it follows from \eqref{Grdnid1}, one can compute tree level with any appropriate combinatios of $u's$ and
$v's$ using \eqref{dtpro}. For example,
 \be\begin{split}\label{tchng}
&\epsilon_{\mu\nu\rho}{\bar v}(p_1)\gamma^{\mu}v(p_2){\bar u}(p_3)\gamma^{\nu}u(p_4)p_5^{\rho}\\
=&\frac{\left({\bar v}(p_1)v(p_2)\right)\left({\bar u}(p_3)u(p_4)\right)}{\left(1+\frac{(p_1-p_2)^2}{4m^2}\right)\left(1+\frac{(p_3-p_4)^2}{4m^2}\right)}
\\
&~ \times \Big[\frac{1}{4m^2}\epsilon_{\mu\nu\rho}(p_1+p_2)^{\mu}(p_3+p_4)^{\nu}p_5^{\rho} \\
&
~~~~~~+\frac{1}{4m^4}\left((p_1\cdot p_5) 
\epsilon_{\mu\nu\rho}p_{2}^{\mu}p_3^{\nu}p_4^{\rho}-
(p_2\cdot p_5) \epsilon_{\mu\nu\rho}p_{1}^{\mu}p_3^{\nu}p_4^{\rho}\right)\\
&~~~~~~
+\frac{i}{4m^3}\left(-(p_4\cdot p_5) \left(p_3\cdot (p_1+p_2)\right)
+(p_3\cdot p_5) \left(p_4\cdot (p_1+p_2)\right) \right)
\\
&~~~~~~+\frac{i}{4m^3}\left(
(p_1\cdot p_5) \left(p_2\cdot (p_3+p_4)\right)
-(p_2\cdot p_5) \left(p_1\cdot (p_3+p_4)\right)\right)\Big].
\end{split}\ee
Using formulas presented in \eqref{uchng}, \eqref{tchng} we find
\be\begin{split}
{\bf S}_{F,U_d} &=  - I(p_1, p_2, p_3, p_4) -e^{i\alpha_1}\frac{8\pi}{k_F} \left( \frac{\epsilon_{\mu\nu\rho}p_1^\mu p_2^{\nu}p_3^{\rho}}{(p_2+p_3)^2}-2 i m_F\right) (2 \pi)^3 \delta(p_1+p_2+p_3+p_4),\\
{\bf S}_{F,U_e}&= I(p_1, p_2, p_4, p_3)- e^{i\alpha_2}\frac{8\pi}{k_F} \left( \frac{\epsilon_{\mu\nu\rho}p_1^\mu p_2^{\nu}p_3^{\rho}}{(p_2+p_4)^2}+2 i m_F\right) (2 \pi)^3 \delta(p_1+p_2+p_3+p_4),\\   
{\bf S}_{F,T} &= -I(p_1, p_2, p_3, p_4) +e^{i\alpha_3}\frac{8\pi}{k_F}\left(  \frac{\epsilon_{\mu\nu\rho}p_1^\mu p_2^{\nu}p_3^{\rho}}{(p_4+p_3)^2}+2im_F\right) (2 \pi)^3 \delta(p_1+p_2+p_3+p_4),
\\{\bf S}_{F,S}&= -I(p_1, p_2, p_3, p_4)+ e^{i\alpha_4}8\pi\lambda_F \left( \frac{\epsilon_{\mu\nu\rho}p_1^\mu p_2^{\nu}p_3^{\rho}}{(p_2+p_4)^2}-2 i m_F\right) (2 \pi)^3 \delta(p_1+p_2+p_3+p_4),  
\end{split}\ee
where $\alpha_1$ to $\alpha_4$ are some complicated physically irelevant phase factors. They obey interchange symmetry and for equal momneta (for 
example, in \eqref{dtpro}, $p_1=p_2$) phase vanishes. In particular, this implies that the phase factor in $U_d$ and $U_e$ channel are the same.
Although, these phases has no physical relevance, we present the results in the C.M. frame. Let the incoming momenta be $p_1,p_2$ and out going momenta 
are $-p_3,-p_4$ and the angle between $p_1$ and $-p_3$ is given by $\theta$ then we find $\alpha_1=\alpha_2=\alpha_3=\alpha_4=-\theta.$ Note that,
inparticular this has the property that, near identity, 
phase factors has no contribution, this is what we expect also from physical ground. So the answers obey the duality with the Bosonic answers in the respective channels.
\\
For completeness, we also write answers for bosonic case.
\be 
\begin{split}
{\bf S}_{B,U_d} &=   I(p_1, p_2, p_3, p_4)  -\frac{8\pi}{k_B}  \frac{\epsilon_{\mu\nu\rho}p_1^\mu p_2^{\nu}p_3^{\rho}}{(p_2+p_3)^2} (2 \pi)^3 \delta(p_1+p_2+p_3+p_4)\\
{\bf S}_{B,U_e}&= I(p_1, p_2, p_4, p_3)+ \frac{8\pi}{k_B}  \frac{\epsilon_{\mu\nu\rho}p_1^\mu p_2^{\nu}p_3^{\rho}}{(p_2+p_4)^2} (2 \pi)^3 \delta(p_1+p_2+p_3+p_4)\\   
{\bf S}_{B,T} &=   I(p_1, p_2, p_3, p_4)  +\frac{8\pi}{k_B}  \frac{\epsilon_{\mu\nu\rho}p_1^\mu p_2^{\nu}p_3^{\rho}}{(p_4+p_3)^2} (2 \pi)^3 \delta(p_1+p_2+p_3+p_4)\\
{\bf S}_{B,S}&=I(p_1, p_2, p_3, p_4)- 8\pi\lambda_B  \frac{\epsilon_{\mu\nu\rho}p_1^\mu p_2^{\nu}p_3^{\rho}}{(p_2+p_4)^2} (2 \pi)^3 \delta(p_1+p_2+p_3+p_4).\\   
\end{split}
\ee 
\section{Aharonov-Bohm scattering} \label{ab}

In this section we will review the classic computation, first performed by 
Aharonov and Bohm, of the scattering of a charged non-relativistic particle off 
a flux tube; see \cite{Aharonov:1959fk, Ruijsenaars:1981fp, Jackiw:1989qp,Bak:1994dj,Bak:1994zz,AmelinoCamelia:1994we} for relevant references. 
We assume that the flux tube is oriented in the $z$ direction, and sits 
at the origin of the transverse two dimensional space. We focus on states that also 
preserve translational invariance along the $z$ direction, so our problem is effectively 
two (spatial) dimensional. We assume that the integrated flux of the flux tube 
equals $2 \pi \nu$ so that the phase associated with the charge particle circling the 
flux tube is  $2 \pi i \nu$ (the particle is assumed to carry unit charge and mass $m$). 
Throughout this appendix we assume $|\nu|<1$.

\subsection{Derivation of the scattering wave function}

We will find scattering state solutions at energy $E= \frac{k^2}{2m}$ of the Schrodinger equation for this particle; intuitively $k$ is the momentum of the particle incident on the flux.

The time independent  Schrodinger equation that governs our system is 
\begin{equation} \label{Schro} 
 \left(-\frac{1}{2m}\left( {\bf \nabla} + 2 \pi i \nu {\bf G }  \right)^2 - 
\frac{k^2}{2 m}  \right)  { \psi} =0
\end{equation} 
where 
\begin{equation} \label{gfd} 
G_i= \frac{\epsilon_{i j}}{2 \pi}  \partial_j \ln r
\end{equation}
In polar coordinates the one form $G$ is given by 
$$G = \frac{d \phi }{2 \pi} .$$ 

Following Aharonov and Bohm we adopt `regular' boundary conditions at the origin of 
our space, i.e. we demand that the wave function at the origin remain finite. As we will see below
this requirement forces the wave function to vanish at the origin like $r^|\nu|$ in the $s$ wave 
channel. The appearance of $|\nu|$ in this boundary condition results in a scattering amplitude
that is non-analytic as a function of $\nu$ and $\nu=0$. \footnote{ See \cite{AmelinoCamelia:1994we} for a fascinating 
one parameter self adjoint relaxation of this boundary condition (which infact yields analytic S-matrices at $w=1$) .}

The most general solution to the Schrodinger equation consistent with the boundary 
conditions described above is given by 
\begin{equation}\label{pwe}
 \psi(r,\theta)= \sum_{n >0} a_{n} e^{i n \theta}  J_{n + \nu}(k r) 
+ \sum_{n>0} a_{-n} e^{-in \theta} J_{n-\nu} + a_{0} J_{|\nu|}(kr) 
\end{equation}
Recall the  asymptotic expansion of Bessel functions at small and large values of the argument 
\begin{equation}\label{bfae}
J_\alpha(x)= \frac{ \left( \frac{x}{2} \right)^\alpha}{\Gamma(\alpha+1)} + \ldots, ~~~
=\frac{1}{\sqrt{2 \pi x}} \left( e^{ix - i \frac{\pi}{4} 
-i \frac{\alpha \pi}{2} } + e^{-ix + i \frac{\pi}{4} 
+i \frac{\alpha \pi}{2} } \right) 
\end{equation}
and the expansion of the plane wave in terms of  Bessel functions 
\begin{equation}\label{pwbw}
e^{ikx}= \sum_{n}i^n J_n(kr) e^{i n \theta}
\end{equation}
and the large $r$ expansion of this plane wave 
(obtained by substituting \eqref{pwe} into \eqref{pwbw})
\begin{equation} 
\begin{split} \label{pwa} e^{i k x'}&= e^{i k r' \cos (\theta)}=
\sum_n i^n e^{i n \theta } J_n(kr)
\\
\sum_n i^n e^{i n \theta } J_n(kr) &\approx
\frac{1}{\sqrt{2 \pi k r}}
\sum_n i^n e^{i n \theta} \left( (e^{ikr -\frac{i\pi n }{2} -\frac{i \pi}{4}} 
+ e^{-ikr +\frac{i\pi n }{2} +\frac{i \pi}{4}} \right) 
\qquad (r \gg 1 )
\\
&= \frac{2\pi}{\sqrt{2 \pi k r}} \left( e^{\frac{-i \pi}{4}} e^{ikr} \delta(\theta)
+  e^{\frac{i \pi}{4}} e^{-ikr} \delta(\theta -\pi) \right).
\end{split}
\end{equation}
\footnote{
This formula is very picturesque; it describes an incoming wave from the 
negative $x$ axis (so at $\theta=-\pi$) and an outgoing wave along the 
positive $x$ axis (so at $\theta= 0$). In particular, the outgoing part of 
the incident wave is equivalent to a contribution to the scattering 
amplitude proportional to  $\delta(\theta)$ .}
It is easy to see that the unique solution  of the form \eqref{pwe} whose ingoing 
part  - i.e. part proportional to $e^{-ikr}$ - is identical to the plane wave \eqref{pwbw} is given 
by 
\begin{equation}\label{scatwf} 
\psi(r,\theta)= \sum_{n=1}^\infty i^n e^{-i \frac{\pi \nu }{2} } J_{n +\nu}(kr) e^{i n \theta}
+ 
\sum_{n=1}^\infty i^{n} e^{i \frac{\pi \nu }{2} } J_{n -\nu}(kr) e^{-i n \theta}  
+ e^{-i \frac{\pi |\nu|}{2}}  J_{|\nu|}(kr) 
\end{equation}

\subsection{The scattering amplitude}

 At large $r$ $\psi(r)$ reduces to 
\begin{equation} \label{sf} \frac{1}{\sqrt{ 2 \pi k r} } \left( 2 \pi e^{i \frac{\pi}{4} }  \delta(\theta - \pi) e^{-ikr} 
+ H(\theta) e^{-i \frac{\pi}{4} }  e^{ikr} \right) 
\end{equation}
where 
\begin{equation}\label{heq} 
H(\theta)=  e^{-i \pi |\nu|} + \sum_{n=1}^\infty \left( e^{-i \pi \nu} e^{i n \theta}  
+ e^{i \pi \nu} e^{-i n \theta} \right).
\end{equation}
Decomposing $H(\theta)$ up into its even and odd parts and then further processing we find  
\begin{equation}\label{hee}\begin{split} 
H (\theta)&= \left( \sum_{n=1}^\infty 2 \cos (\pi \nu) \cos (n \theta)  \right)  + e^{-i |\nu| \pi} 
+ \left( \sum_{n=1}^\infty  2 \sin (\pi \nu) \sin (n\theta) \right) \\
&= \left(  \cos (\nu \pi)  + \sum_{n=1}^\infty 2 \cos (\pi \nu) \cos (n \theta)  \right)  - i |\sin (\nu \pi)|
+ \left( \sum_{n=1}^\infty  2 \sin (\pi \nu) \sin (n\theta) \right)\\
&= 2 \pi \cos (\pi \nu) \delta(\theta) - i |\sin (\nu \pi)|
+ \left( \sum_{n=1}^\infty  2  \sin (\pi \nu) \sin (n\theta)  \right)\\
&= 2 \pi \cos (\pi \nu) \delta(\theta) + \sin (\pi \nu) {\rm Pv}
\left(\cot \left(\frac{\theta}{2}\right) \right) - i |\sin (\pi \nu)| \\
&= 2 \pi \cos (\pi\nu) \delta(\theta) + \sin (\pi \nu) {\rm Pv}  
\left( \frac{e^{-i \frac{\theta {\rm sgn} [\nu]}{2}}}{\sin \left(\frac{\theta}{2}\right) } \right). \\
\end{split}
\end{equation}
\footnote{In going from the third to the fourth line above we have used the formula 
\begin{equation}\label{cotexp}
{\rm Pv} \left( \cot \left(\frac{\theta}{2}\right) \right)= 2 \sum_{m=1}^\infty\sin (m \theta)
\end{equation}
This formula is equivalent to the assertion that 
\begin{equation}\label{mt}
\int \frac{d \theta}{2 \pi i} {\rm Pv} \cot \left(\frac{\theta}{2}\right) e^{im\theta} = {\rm sgn}(m)
\end{equation}
(the integral on the RHS of \eqref{mt} clearly vanishes when $m$=0 as ${\rm Pv} (\cot \left(\frac{\theta}{2}\right))$ is an odd function). The integral on the LHS of \eqref{mt} can be converted into a contour integral 
about the unit circle on the complex plane via the substitution $z=e^{i \theta}$. The contour integral 
in question is simply 
$$ \oint \frac{dz}{2 \pi i} {\rm Pv} \frac{z^{m-1}(z+1)}{z-1}$$
This integral is easily seen to evaluate to unity for $m\geq 1$ when it receives contributions only from 
the pole at unity. The substitution $z=\frac{1}{w}$ allows one to conclude as easily that the integral 
evaluates to $-1$ for $m \leq -1$, establishing \eqref{cotexp}.  }
It is conventional to write the wave function as a plane wave plus a scattered piece ; at large $r$ 
\begin{equation} \label{conwe}
\psi(r) = e^{ikx} + \frac{h(\theta) e^{-i \frac{\pi}{4} }  e^{ikr}}{\sqrt{ 2 \pi k r} } .
\end{equation}
Plugging \eqref{pwa} into \eqref{conwe} and comparing with \eqref{sf} we conclude that 
\begin{equation}\label{hto}
h(\theta)= H(\theta) - 2 \pi \delta(\theta)
\end{equation}
so that 
\begin{equation}\label{htff}
h(\theta)= 2 \pi  \left( \cos (\pi\nu) - 1 \right) \delta(\theta) + \sin (\pi \nu) {\rm Pv}  
\left( \frac{e^{-i \frac{\theta { \rm sgn} [\nu]}{2}}}{\sin \left(\frac{\theta}{2}\right) } \right) .
\end{equation}

\subsection{Physical interpretation of the $\delta$ function at forward scattering}

It is intuitively clear that the amplitude for propagation (path integral) for a particle 
starting out a large distance away from  the origin on the  negative real axis, to a position 
nearer the scattering center has enough information to compute the scattering S-matrix. 
\footnote{Let us explain how scattering data may be extracted in practice. 
Recall that the amplitude for a free particle to propagate from polar coordinates $r, \theta$ to 
polar coordinates $r', \theta'$ in time $t$ is given by 
\begin{equation}\label{freesoln}
 A_F(r, \theta, r', \theta', t)= \frac{1}{2 \pi  i t} e^{ i \left( 
\frac{r^2 +(r')^2 -2 r r' \cos (\theta -\theta')}{2 t}  \right) }
 \end{equation}
\begin{equation} \label{scatwf1}
\phi^F_k(r', \theta')= 2 \pi i \sqrt{t} e^{-i \frac{r^2}{2 t} } A_F(r, \theta, r', \theta', t)
\end{equation}
is the wave function at time $t$ of a particle, initially localized to a delta function located 
at $r, \theta$. In the limit 
\begin{equation}\label{limitwf}
r \to \infty, ~~~t \to \infty ~~~\frac{m r}{{\bar h} t}= k= {\rm fixed} , ~~~r', \theta'= {\rm fixed}
\end{equation}
we have 
\begin{equation}\label{freesol}
\phi^F_k= e^{i k x'} 
\end{equation}
i.e the wave function reduces to a plane wave. 
In the case of an interacting theory with interactions localized around the origin, 
let the amplitude for the particle to propagate from  $r, \theta$ to 
polar coordinates $r', \theta'$ in time $t$ be denoted by $A(r, \theta, r', \theta', t)$. 
It follows that the scattering wave function for our problem is given by 
\begin{equation} \label{scatwfi}
\phi_k(r', \theta')= 2 \pi i \sqrt{t} e^{-i \frac{r^2}{2 t} } A(r, \theta, r', \theta', t)
\end{equation}
in the limit \eqref{limitwf} as this path integral produces a wave function with an incoming 
piece that is indistinguishable from a plane wave near the origin. The scattering amplitude
 $h(\theta)$ is read off from the large $r'$ expansion of $\phi_k(r', \theta)$ in the usual manner. }
The amplitude for a particle to propagate from far to the left of the origin
to a point near the origin (lets say at angle $\theta \approx \pi$ for definiteness) receives 
contributions from path whose angular winding around the origin are approximately 
$...-3 \pi, -\pi, \pi, 3 \pi  ...$. Of these infinitely many paths those with winding approximately 
$\pi$ and $-\pi$ are special. These sectors consist of paths that go below the origin, and paths 
that go above the origin, but do not otherwise wind the origin. It may be shown that these paths 
are entirely responsible for the terms in $H(\theta)$ (see the previous subsection) proportional to 
$\delta(\theta)$. 

For a free plane wave $H(\theta) = 2 \pi \delta(\theta)$. In a `traditional' scattering problem 
$H(\theta)=2\pi \delta(\theta) + {\rm nonsingular}$ i.e. the incident wave goes through largely 
untouched, and in addition we have some scattering. In the problem with Aharonov-Bohm scattering, 
however, we have seen in the last subsection that
$H(\theta)=2 \pi \cos (\pi \nu) \delta(\theta)$. This fact is easily interpreted. The contribution of 
paths with winding $\pi$ and $-\pi$ in this problem is identical to the contribution of the same paths 
in the free theory except that the paths with winding $\pi$ are  weighted by an additional phase $e^{i \pi \nu}$ while the paths with winding $-\pi$ are  weighted by the additional phase $e^{-i \pi \nu}$. The two sectors are flipped by reflection and so otherwise contribute equally. This explains the modulation 
of the $\delta(\theta)$ part of  $H(\theta)$ by $\cos (\pi \nu)$, and the consequent appearance of 
the term $2 \pi(\cos (\pi \nu) -1) \delta(\theta)$ in $h(\theta)$.

\section{Details of the computation of the scalar S-matrix}

\subsection{Computation of the effective one particle exchange interaction} \label{die}

\begin{figure}[tbp]
  \begin{center}
  \subfigure[]{\includegraphics[scale=.3]{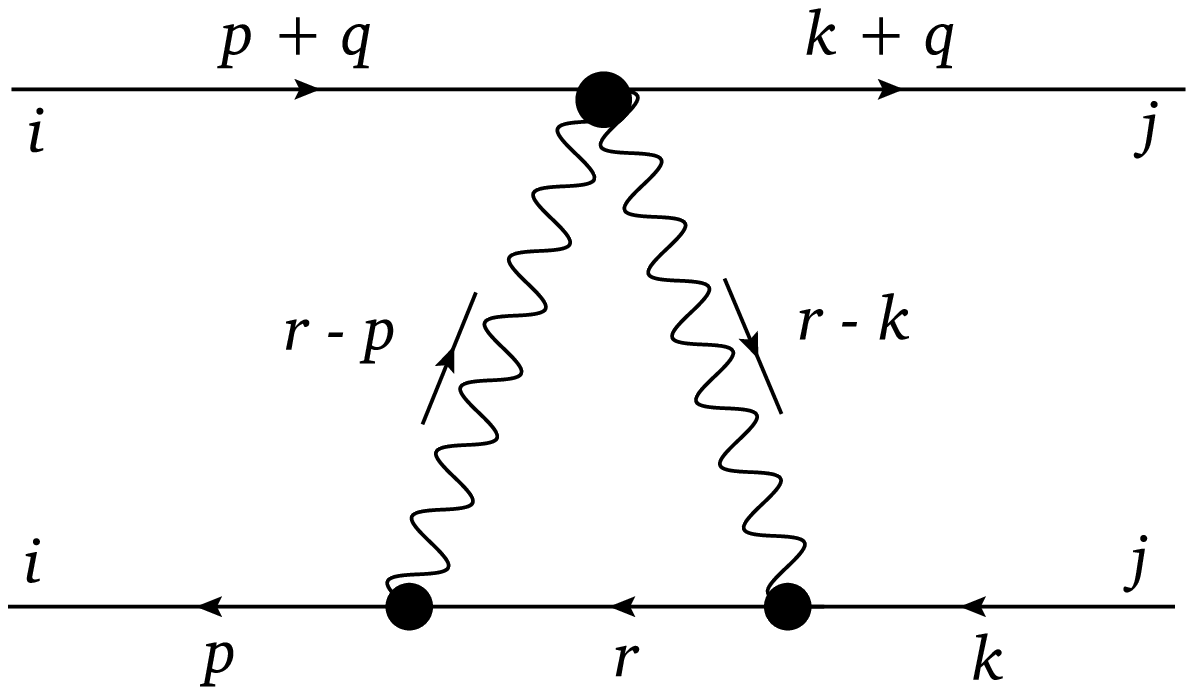}
\label{a}
  }
  \subfigure[]{\includegraphics[scale=.3]{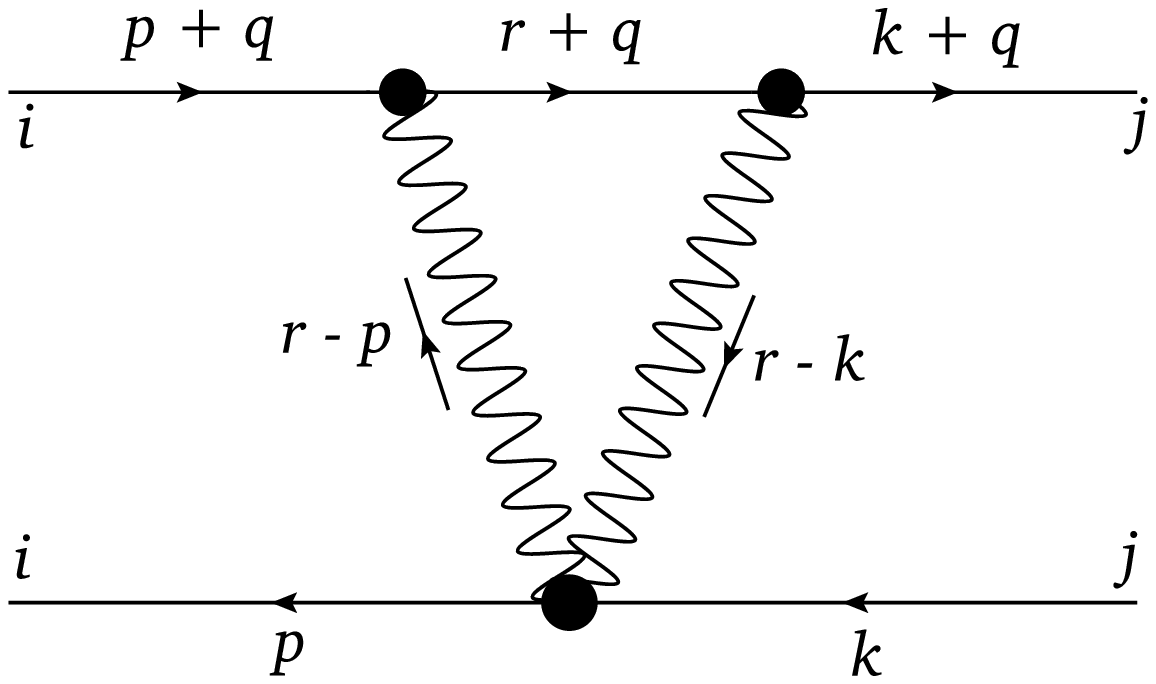}
\label{b}
  }
  \subfigure[]{\includegraphics[scale=.3]{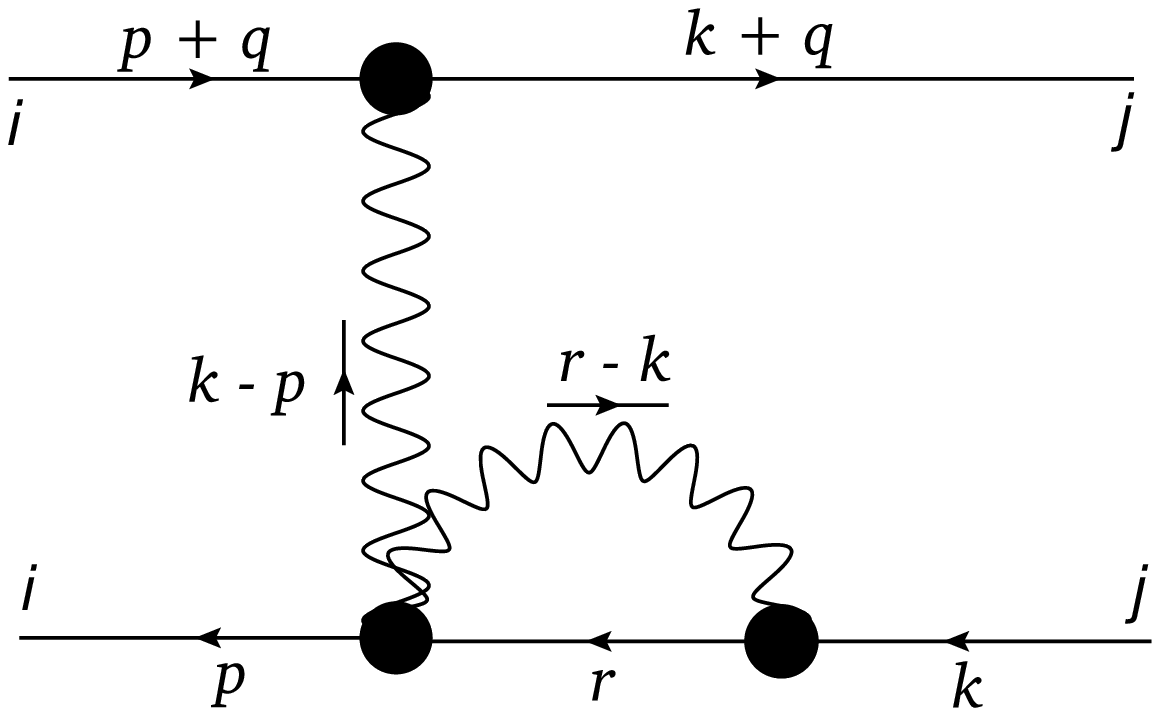}
\label{c}
  }\\
  \subfigure[]{\includegraphics[scale=.3]{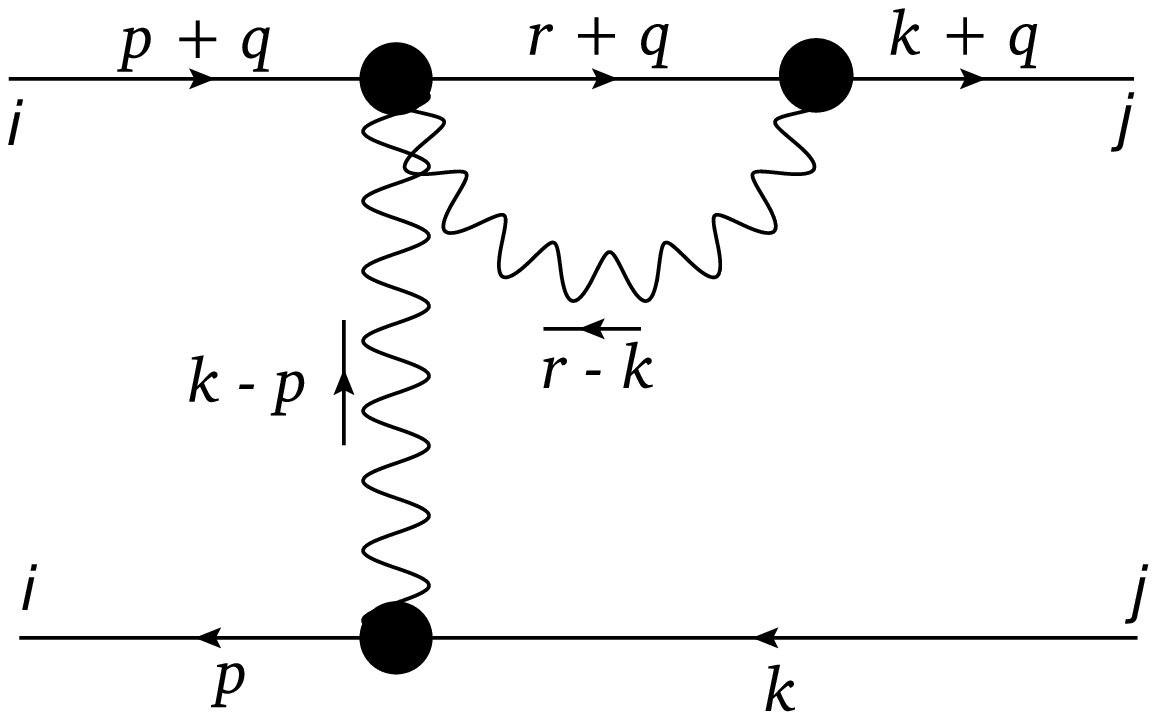}
\label{d}
  }
  \subfigure[]{\includegraphics[scale=.3]{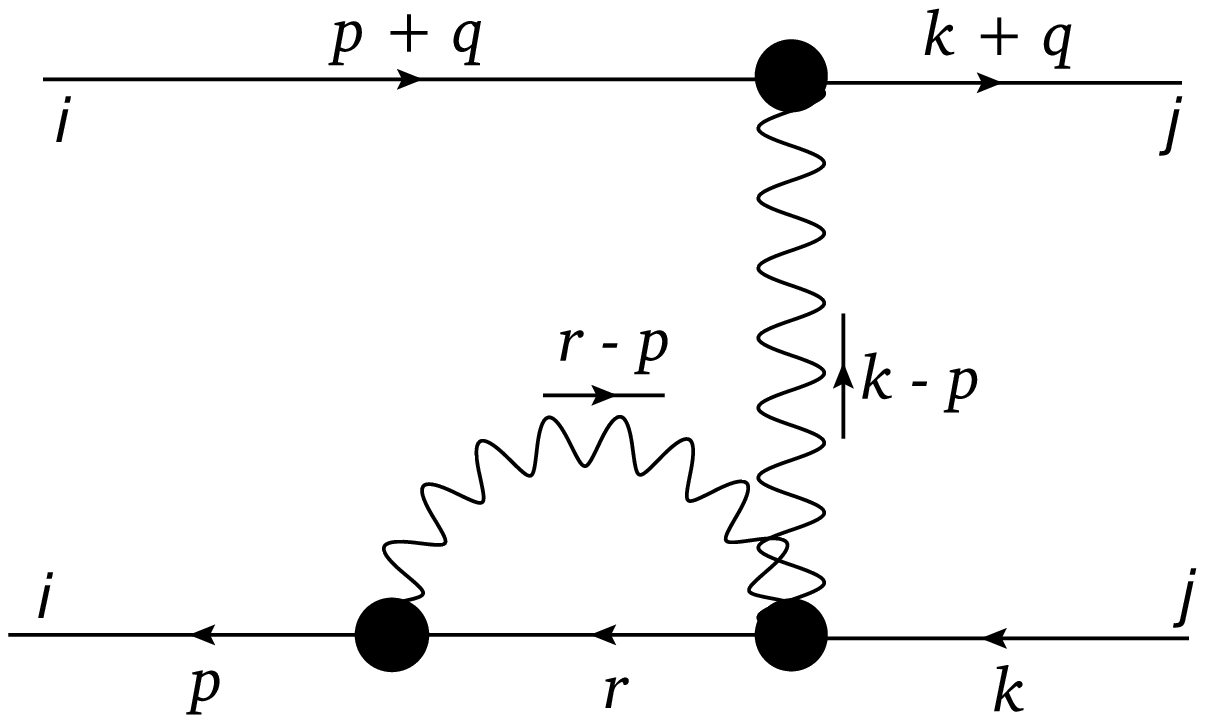}
\label{e}
  }
 \subfigure[]{\includegraphics[scale=.3]{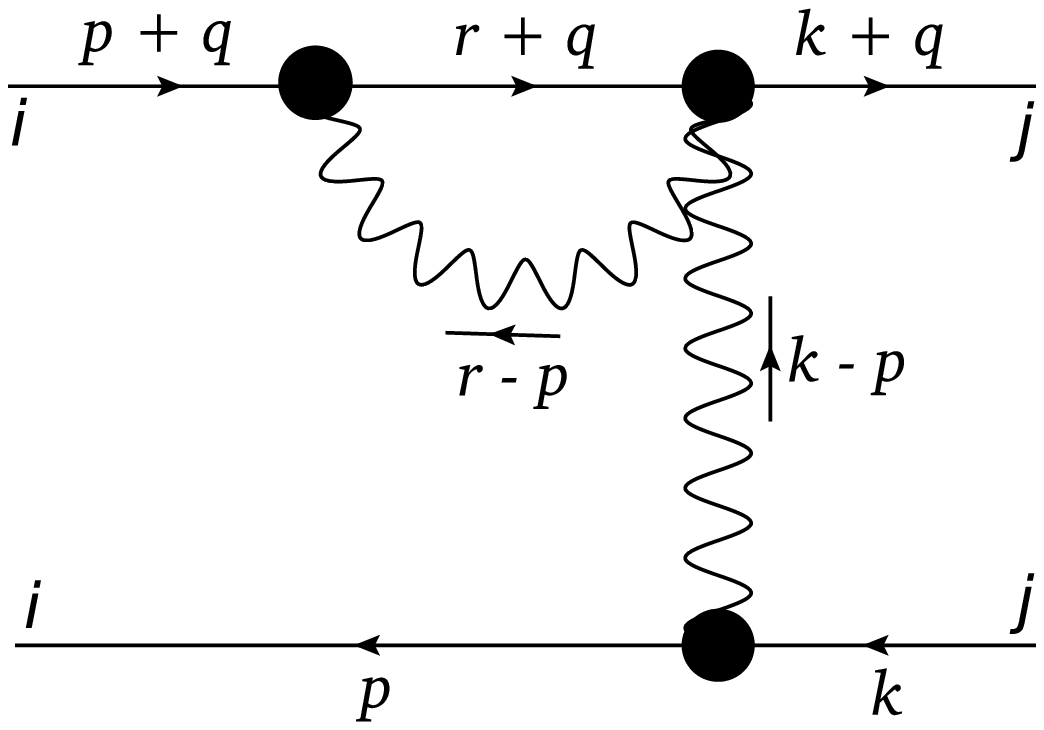}
\label{f}
  }
  \end{center}
  \vspace{-0.5cm}
\caption{The one loop diagrams that contribute to the unit represented by the triple line excluding the tree level diagram.
  Note that box diagram is not included here as it is one of the contributions from two units sewn together.}
  \label{loops}
  \end{figure}

In this subsection we explicitly compute the summation over the effective `one particle exchange' four 
point interactions depicted in Fig \ref{Unit}.  We perform our 
computation in Euclidean space and analytically continue our final result back to 
Euclidean space.  In Figure \ref{loops} we redraw the diagrams of 
Fig \ref{Unit}, this time including detailed momentum assignments for all legs. 

\footnote{All the graphs below have the common overall factor $-4 \pi^2 \lambda^2$, because they each 
have a single internal scalar propagator, two internal gauge propagators, and two $\phi \phi A$ 
3. The scalar propagators contribute with no factors. The gauge propagators are each proportional 
to $2 \pi i \lambda$. The triple vertices each contribute a factor of $i$. And finally we get an overall 
minus sign from the fact that we are computing the contribution to the Euclidean effective action which 
appears in the path integral as $e^{-S_E}$}. 

The graph in Fig. \ref{a} evaluates to 
\begin{align*}
NA_1 &= (-4\pi^2\lambda^2)\int-\frac{(r+p)_-}{(r-p)_-}\frac{(r+k)_-}{(r-k)_-}\frac{1}{r^2+c_{B}^2}\frac{d^3r}{(2\pi)^3}\\
&= \int-\left(1+2\frac{(p+k)_-}{(p-k)_-}\left(\frac{p_-}{(r-p)_-}-\frac{k_-}{(r-k)_-}\right)\right)\frac{1}{r^2+c_{B}^2}\frac{d^3r}{(2\pi)^3}
\end{align*}
Let $\theta$ denote the phase of the complex number $r_-$. Since $r^2$ doesn't have a $\theta$ dependence, performing the $\theta$ integration first,
\begin{equation}\label{na1}
NA_1 = (-4\pi^2\lambda^2)\int-
\left(1-2\frac{(p+k)_-}{(p-k)_-}\left(\theta(p_s-r_s)-\theta(k_s-r_s)\right)
\right)\frac{1}{r^2+c_{B}^2}\frac{dr_3r_sdr_s}{(2\pi)^2}
\end{equation}

The graph in Fig. \ref{b} evaluates to 
\begin{equation*}
NA_2 = (-4\pi^2\lambda^2)\int-\frac{(r+p+2q)_-}{(r-p)_-}\frac{(r+k+2q)_-}{(r-k)_-}\frac{1}{(r+q)^2+c_{B}^2}\frac{d^3r}{(2\pi)^3}
\end{equation*}
we can change the integration variable$r\rightarrow r-q$ and define variables $p'=p+q$,$k'=k+q$.
\begin{align*}
NA_2 &= (-4\pi^2\lambda^2)\int-\frac{(r+p')_-}{(r-p')_-}\frac{(r+k')_-}{(r-k')_-}\frac{1}{r^2+c_{B}^2}\frac{d^3r}{(2\pi)^3}\\
&= \int-\left(1+2\frac{(p'+k')_-}{(p-k)_-}\left(\frac{p'_-}{(r-p')_-}-\frac{k'_-}{(r-k')_-}\right)\right)\frac{1}{r^2+c_{B}^2}\frac{d^3r}{(2\pi)^3}
\end{align*}
Again, performing the $\theta$ integration first,
\begin{equation}\label{na2}
NA_2 = (4\pi^2\lambda^2)\int\left(1-2\frac{(p'+k')_-}{(p-k)_-}\left(\theta(p'_s-r_s)-\theta(k'_s-r_s)\right)\right)\frac{1}{r^2+c_{B}^2}\frac{dr_3r_sdr_s}{(2\pi)^2}
\end{equation}

Fig. \ref{c} evaluates to
\begin{equation}\label{na3}
 \begin{split}
NA_3 &= (-4\pi^2\lambda^2)\int-\frac{(p+k+2q)_-}{(p-k)_-}\frac{(r+k)_-}{(r-k)_-}\frac{1}{r^2+c_{B}^2}\frac{d^3r}{(2\pi)^3}\\
&= (-4\pi^2\lambda^2)\int-\frac{(p'+k')_-}{(p-k)_-}\left(1+2\frac{k_-}{(r-k)_-}\right)\frac{1}{r^2+c_{B}^2}\frac{d^3r}{(2\pi)^3}\\
&= (-4\pi^2\lambda^2)\int-\frac{(p'+k')_-}{(p-k)_-}\left(1-2\theta(k_s-r_s)\right)\frac{1}{r^2+c_{B}^2}\frac{dr_3r_sdr_s}{(2\pi)^2}
\end{split}
\end{equation}

Fig. \ref{d} evaluates to 
\begin{equation}\label{na4}
 \begin{split}
NA_4 &= (-4\pi^2\lambda^2)\int-\frac{(p+k)_-}{(p-k)_-}\frac{(r+k+2q)_-}{(r-k)_-}\frac{1}{(r+q)^2+c_{B}^2}\frac{d^3r}{(2\pi)^3}\\
&= (-4\pi^2\lambda^2)\int-\frac{(p+k)_-}{(p-k)_-}\frac{(r+k')_-}{(r-k')_-}\frac{1}{r^2+c_{B}^2}\frac{d^3r}{(2\pi)^3}\\
&= (-4\pi^2\lambda^2)\int-\frac{(p+k)_-}{(p-k)_-}\left(1+2\frac{k'_-}{(r-k')_-}\right)\frac{1}{r^2+c_{B}^2}\frac{d^3r}{(2\pi)^3}\\
&= (-4\pi^2\lambda^2)\int-\frac{(p+k)_-}{(p-k)_-}\left(1-2\theta(k'_s-r_s)\right)\frac{1}{r^2+c_{B}^2}\frac{dr_3r_sdr_s}{(2\pi)^2}
\end{split}
\end{equation}

Fig. \ref{e} evaluates to 
\begin{equation}\label{na5}
 \begin{split}
NA_5 &= (-4\pi^2\lambda^2)\int\frac{(p+k+2q)_-}{(p-k)_-}\frac{(r+p)_-}{(r-p)_-}\frac{1}{r^2+c_{B}^2}\frac{d^3r}{(2\pi)^3}\\
&= (-4\pi^2\lambda^2)\int\frac{(p'+k')_-}{(p-k)_-}\left(1+2\frac{p_-}{(r-p)_-}\right)\frac{1}{r^2+c_{B}^2}\frac{d^3r}{(2\pi)^3}\\
&= (-4\pi^2\lambda^2)\int\frac{(p'+k')_-}{(p-k)_-}\left(1-2\theta(p_s-r_s)\right)\frac{1}{r^2+c_{B}^2}\frac{dr_3r_sdr_s}{(2\pi)^2}
\end{split}
\end{equation}

Fig. \ref{f} evaluates to 
\begin{equation}\label{na6}
 \begin{split}
NA_6 &= (-4\pi^2\lambda^2)\int\frac{(p+k)_-}{(p-k)_-}\frac{(r+p+2q)_-}{(r-p)_-}\frac{1}{(r+q)^2+c_{B}^2}\frac{d^3r}{(2\pi)^3}\\
&= (-4\pi^2\lambda^2)\int-\frac{(p+k)_-}{(p-k)_-}\frac{(r+p')_-}{(r-p')_-}\frac{1}{r^2+c_{B}^2}\frac{d^3r}{(2\pi)^3}\\
&= (-4\pi^2\lambda^2)\int\frac{(p+k)_-}{(p-k)_-}\left(1+2\frac{p'_-}{(r-p')_-}\right)\frac{1}{r^2+c_{B}^2}\frac{d^3r}{(2\pi)^3}\\
&= (-4\pi^2\lambda^2)\int\frac{(p+k)_-}{(p-k)_-}\left(1-2\theta(p'_s-r_s)\right)\frac{1}{r^2+c_{B}^2}\frac{dr_3r_sdr_s}{(2\pi)^2}
\end{split}
\end{equation}

The total Amplitude is
\begin{equation}\label{natotdef}
NA_{tot} = \sum_{i=1}^6A_i
\end{equation}
Which gives
\begin{equation} \label{ampuu}
\begin{split}
NA_{tot}=\int
&\frac{dr_3r_sdr_s}{(2\pi)^2}\biggl[\frac{(-4\pi^2\lambda^2)}{r^2+c_{B}^2}
\\
&\times\left(-2 +\frac{ 4q_-}{(p-k)_-}\left[\theta(p'_s-r_s)-\theta(k'_s-r_s)+\theta(k_s-r_s)-\theta(p_s-r_s)\right]\right)
\biggr]
\end{split}
\end{equation}
Where we recall that 
$$p'=p+q, ~~~k'=k+q.$$
We are interested in the special case $q^\pm=0$. In this case the $p'_\pm= p_\pm$ and 
$k'_\pm=k_\pm$, and so $k's=k_s$ and $p'_s=p_s$. It follows that the $\theta$ functions in 
\eqref{ampuu} cancel in pairs and 
\begin{align}\label{natot}
NA_{tot}&=(-2)(-4\pi^2\lambda^2)\int\frac{1}{r^2+c_{B}^2}\frac{dr_3r_sdr_s}{(2\pi)^2} \nonumber \\
&=8\pi^2\lambda^2\int\frac{1}{r^2+c_{B}^2}\frac{d^3r}{(2\pi)^3}
\end{align}
We use dimensional regularization, which replaces the integral by ($\frac{m}{4\pi}$). So ultimately these diagrams give
\begin{equation}\label{naloop}
NA_{loop} = 2\pi\lambda^2 c_{B}
\end{equation}

\subsection{Euclidean rotation} \label{er}

The integral equation \eqref{sdee} may be used to solve for the function $V(p^0, {\vec p}, 
k^0, {\vec k}, q^3)$. In this subsection we will be interested only in the dependence of 
$V$ on $p^0$ and  $k^0$ and so use the notation $V=V(p^0, k^0)$. 

As is often the case in the study of relativistic scattering amplitudes, in this paper we will find it convenient 
determine $V$ by first computing its `Euclidean continuation'. In this brief subsection we pause to 
define the Euclidean continuation of $V$, and to determine the integral equation it obeys. 

Given the amplitude $V(p^0, k^0)$ we define the one parameter set of amplitudes, 
$V_\alpha(p^0, k^0)$ for $0 \leq \alpha \leq \frac{\pi}{2}$ as follows. Let us assume that $V(p^0, k^0)$ 
admits an analytic continuation to the function $V(z, w)$ for $ 0< Arg(z) < \frac{\pi}{2}$ and $0 < Arg(w) < \frac{\pi}{2}$.
 We also assume that this function can be defined to be free of singularities when $Arg(z)=Arg(w)$.  In terms of this analytic function, 
we define a one parameter extension, 
$V_\alpha$ of $V$ by 
$$V_\alpha(p^0, k^0)=  V(p^0 e^{i \alpha}, k^0 e^{i \alpha}).$$
It follows in particular that $V_\alpha$ is a smooth function of $\alpha$. 

The Euclidean continuation, $V_E$ of $V$ is defined by 
$$V_E(p^0, q^0) = V_{\frac{\pi}{2}}(p^0, k^0).$$ 
Note in particular that 
$$V_E(p^0, k^0)= V(i p^0, ik^0)$$
 \footnote{This equation that is sometimes summarized by the mnemonic 
$p^0_E=-i p^0_L, ~~~k^0_E= -ik^0_L$. }

In order to obtain the  integral equation obeyed by   $V_\alpha(p^0, q^0)$ one must, of course, make 
the replacement $p^0 \rightarrow e^{i \alpha} p^0$, $k^0 \rightarrow e^{i \alpha }k^0$ in \eqref{sde}. 
However this replacement must also be accompanied by a simultaneous change in the contour of 
integration of the variable $r^0$. If the $r^0$ contour is left unchanged then the pole
$$\frac{1}{(p-r)_+ (p-r)_- - i \epsilon} $$
in the integrand in the first of \eqref{sde} could cross the contour of integration at a particular 
value of $\alpha$, leading to a non-analyticity in $V_\alpha$ as a function of $\alpha$.
In order to define $V_\alpha$ as a smooth function with no singularities, we adopt the following procedure. For any given $p^0$ and $\alpha$ we  first deform the contour of integration over the variable $r^0$. This deformation is performed without crossing any singularities in the integrand, and so does not change 
the value of the integral. It is chosen in a manner that  ensures that the rotation $p^0 \rightarrow p^0 e^{i \alpha}$  can  be performed without the pole  crossing the contour of integration; for any fixed $p^0$ and 
$\alpha$ such a deformation may always be found. After the rotation on $p^0$ is now performed, the integration contour for $r^0$ is further modified to suit convenience.  It is convenient to choose the final contour for integration over $r^0$ to be the rotation of the initial contour counterclockwise by the angle $\alpha$, together with two arcs of angle  $\alpha$ at $\infty$. It is easily verified that the arcs at infinity do not contribute to the 
integral (because the integrand dies off fast enough at infinity) . 
 
 In summary, the  integral equation obeyed by the function $V_\alpha(p^0, k^0)$ is given 
 by making the replacements $p^0 \rightarrow e^{i \alpha} p^0, ~~k^0 \rightarrow e^{i \alpha} k^0, ~~
 r^0 \rightarrow e^{i \alpha} r^0$ in \eqref{sde} and then continuing to integrate 
 the new $r^0$ variable over its real axis.  The integral equation for $V_E$ is given by the special case 
 $\alpha= \frac{\pi}{2}$.

\subsection{Solution of the Euclidean integral equations} \label{seie}

In this subsection we determine the solution to the scalar Euclidean integral equation \eqref{sdee}.

Differentiating the first equation in \eqref{sdee} w.r.t. $p^3$ we conclude that $\partial_{p^3} V^E=0$. 
It follows that $V$ is independent of $k_3$ and $p_3.$
In a similar manner, from the second equation we conclude that $\partial_{k^3} V^E=0$.
The identity 
$$ \int_{-\infty}^\infty \frac{dx}{(x^2+a^2)((x+y)^2+a^2)}
= \frac{2 \pi}{|a|( y^2+4 a^2)}$$
may now be used to perform the integral over $r_3$ on the RHS of the first two equations in 
\eqref{sdee}. Defining 
$$a(p)= \sqrt{c_{B}^2 + {\vec p}^2} =\sqrt{c_{B}^2 + 2 p_+ p_-}$$ 
where the square root on the RHS is positive by definition, we find 
\begin{equation}\begin{split}\label{sdeei} 
V^E(p, k, q) &= V^E_0(p, k,  q) + \int \frac{d^2 r}{(2 \pi)^2} V_0^E(p, r, q_3) \frac{N}{a(r)(q_3^2+4 a^2(r) )} V^E(r, k, q_3) \\
V^E(p, k, q) &= V^E_0(p, k,  q) + \int \frac{d^2 r}{(2 \pi)^2} V^E(p, r, q_3) \frac{N}{a(r)(q_3^2+4 a^2(r) )} V^E_0(r, k, q_3) \\
NV^E_0(p, k, q_3) & = -4 \pi i \lambda q_3 \frac{(k+p)_-}{(k-p)_-}+ \wt b_4\\
\end{split} 
\end{equation}

Now if  $z=\frac{x+iy}{\sqrt{2}}$ then 
$$\partial_z= \frac{1}{2} \left( \partial_x - i \partial_y \right), ~~~\nabla^2= 2 \partial_z \partial_{{\bar z}}, ~~~ \partial_{\bar z} \frac{1}{z} = \partial_z \partial_{\bar z} \ln (z \bar{z} ) 
= \nabla^2 \ln r = 2 \pi \delta^2( {\vec r} ).$$
It follows from \eqref{sdeei} that 
\begin{equation}\label{sdp} \begin{split}
\partial_{p_+} \left( V-V_0 \right) 
= \frac{4 i \lambda  q_3 p_-}{a(p)(q_3^2+4 a^2(p))} V, \\
\partial_{k_+} \left( V-V_0 \right) 
= -\frac{4 i \lambda  q_3 k_-}{a(k)(q_3^2+4 a^2(k))} V.
\end{split}
\end{equation}
The equations \eqref{sdp} may be regarded as first order ordinary differential equations in the 
variables $p_+$ and $k_+$ respectively. These equations are easily solved. Using the identities
$$ \int \frac{dp_+ p_-}{a(p) (q_3^2 +4a(p)^2)}= \int \frac{d a}{q_3^2+4 a^2}
=\frac{1}{2 |q_3|} \tan^{-1} \left( \frac{2 a}{|q_3|} \right)$$
If we agree to choose a definition of $\tan^{-1}$ that makes it an odd function we can drop the modulus signs in this formula. Of course we would also like the $\tan^{-1}$ function to be 
continuous; these requirements together fix the branch choice 
$$ - \frac{\pi}{2} <\tan^{-1}(x) < \frac{\pi}{2}$$

It follows that \eqref{sdp} may be recast as 
\begin{equation} \begin{split} \label{pse}
\partial_{p_+} \left( e^{-2 i  \lambda \tan ^{-1}\left( \frac{2 a(p)}{q_3} \right)} V \right)
= \left( e^{-2 i  \lambda \tan ^{-1}\left( \frac{2 a(p)}{q_3} \right) } \right) \partial_{p_+} V_0, \\
\partial_{k_+} \left( e^{2 i  \lambda \tan ^{-1}\left( \frac{2 a(k)}{q_3} \right)} V \right)
= \left( e^{2 i  \lambda \tan ^{-1}\left( \frac{2 a(k)}{q_3} \right) } \right) \partial_{k_+} V_0.
\end{split}
\end{equation}
The equations \eqref{pse} are now easily solved by integration. It might at first seem that the integral 
of the RHS of these equations is complicated by the fact that the term multiplying $\partial_{p_+} V_0$
 in the first equation on the RHS of \eqref{pse} is actually a function of $p$. Recall, however, that  
$\partial_{p_+} V_0$ is proportional to the $\delta$ function; using the formula $f(x) \delta (x-a)= f(a) \delta(x-a)$ we can replace the argument of this prefactor by the corresponding function of $k_+$. 
Similar remarks apply to the second of \eqref{pse}. Integrating these two equations it follows that  
\begin{equation} \label{answer} \begin{split}
NV&= (4 \pi i \lambda  q_3 ) \frac{p_- + k_-}{p_- - k_-} e^{-2 i  \lambda \left( \tan ^{-1}\left( \frac{2 (a(k)}{q_3} \right) - \tan ^{-1}\left( \frac{2 (a(p)}{q_3} \right)\right)   } 
\\
& \qquad-
 e^{2 i \lambda \tan ^{-1}\left( \frac{2 a(p)}{q_3} \right) } h(k, p_-, q_3) \\
&= (4 \pi i \lambda  q_3 ) \frac{p_- + k_-}{p_- - k_-} e^{-2 i \lambda  \left( \tan ^{-1}\left( \frac{2 (a(k)}{q_3} \right) - 
 \tan ^{-1}\left( \frac{2 (a(p)}{q_3} \right)\right)   } 
\\
& \qquad -
 e^{-2 i \lambda \tan ^{-1}\left( \frac{2 a(k)}{q_3} \right) } \tilde{h}(k_{-}, p, q_3)
\end{split}
 \end{equation}
Comparing these two equations determines the $k_+$ dependence of $h$ and the $p_+$ dependence
of ${\tilde h}$, and we conclude
\begin{equation}\label{ansm}
NV(p,k,q_{3})=  e^{-2 i  \lambda \left( \tan ^{-1}\left( \frac{2 (a(k)}{q_3} \right) - 
 \tan ^{-1}\left( \frac{2 (a(p)}{q_3} \right)\right)   }  \left( 4 \pi i \lambda  q_3  \frac{p_- + k_-}{p_- - k_-}
+ j(k_-, p_-, q_3) \right)
\end{equation}
Now the function $j(k_-, p_-)$ above must be a function of charge zero, and so must be a function 
of $\frac{k_-}{p_-}$. It must also be singularity free (i.e. its derivative w.r.t both $p_+$ and $k_+$ must 
vanish. This seems impossible unless the function $j$ is a constant, so we conclude 
\begin{equation}\label{anon}
NV=  e^{-2 i \lambda   \left( \tan ^{-1}\left( \frac{2 (a(k)}{q_3} \right) - 
 \tan ^{-1}\left( \frac{2 (a(p)}{q_3} \right)\right)   }  \left( 4 \pi i \lambda  q_3  \frac{p_- + k_-}{p_- - k_-}
+ j(q_3)\right)
\end{equation}

In order to evaluate $j(q_3)$ we now plug the form \eqref{anon} back into  \eqref{sdeei}, explicitly 
perform the integral over ${\vec r}$ and compare both sides of the integral equation. The integral over 
${\vec r}$ may be evaluated in polar coordinates by integrating over the modulus $r$ and the angle $\theta$. 
We will find it convenient to perform the angular integral by contour methods. Let us define 
$z= e^{i \theta}$. Then $\int d \theta = \int_C \frac{dz}{2 \pi i z}$ where the contour $C$ runs counterclockwise over the unit circle on the complex plane.   The first of \eqref{sdeei} turns 
into 
\begin{equation} \begin{split} \label{intermed}
&e^{2 i \lambda  \tan^{-1} \left(\frac{2\sqrt{k^2+c_{B}^2}}{q_3} \right) }\left( 
NV(p, k, q) - NV_0(p, k, q) \right) \\
= &
\int dr \frac{ r e^{2 i \lambda \tan^{-1}\left( \frac{2 \sqrt{r^2+c_{B}^2} }{q_3} \right) } }
{\sqrt{c_{B}^2+r^2}(q_3^2+ 4(c_{B}^2+r^2)) } I(r) =\frac{1}{4 i \lambda q_3} \int dr \partial_r \left( e^{2 i \lambda \tan^{-1} \left(\frac{2\sqrt{r^2+c_{B}^2}}{q_3} \right) } \right) I(r)
\\
&I(r)=\int_C \frac{dz}{(2 \pi)^2 i z} \left( -4 \pi i \lambda q_3
\frac{rz + p_-}{rz -p_-} 
+ \wt b_4 \right) \left( - 4 \pi i \lambda q_3 
\frac{r z + k_-}{-rz + k_- }+ j(q_3) \right) 
\end{split}
\end{equation}
Where $z$ in $I(r)$ is integrated over the unit circle. We now proceed to evaluate I(r) using  Cauchy's theorem. We find 
\begin{equation}\label{ir} \begin{split}
 2 \pi I(r)& =(4 \pi i \lambda  q_3 + \wt b_4) (-4 \pi i \lambda q_3 + h)  \\
&-\theta(r - p) 8 \pi i \lambda q_3 \left( - 4 \pi i \lambda q_3 \frac{k_- + p_-}{k_- - p_-} + j(q_3) \right) \\
&+\theta(r - k) 8 \pi i \lambda q_3 \left( - 4 \pi i \lambda q_3 \frac{k_- + p_-}{k_- - p_-} + \wt b_4\right)
\end{split}
\end{equation}
where the first line is the contribution from the pole at $z=0$, the second line is the 
 contribution from the pole at $z=\frac{p_-}{r}$ and the third line is the contribution of the 
 pole at $z=\frac{k_-}{r}$.  Let us define 
 $$F(r)= e^{2 i \lambda \tan^{-1} \left(\frac{2\sqrt{r^2+c_{B}^2}}{q_3} \right)}$$
 It follows from \eqref{ir} and \eqref{intermed} that 
 \begin{equation}\begin{split} \label{alm}
& (8 \pi i \lambda q_3) e^{2 i \lambda  \tan^{-1} \left(\frac{2\sqrt{k^2+c_{B}^2}}{q_3} \right) }\left( NV(p, k, q) - NV_0(p, k, q) \right)\\
&=(4 \pi i \lambda q_3 + \wt b_4) (-4 \pi i q_3 + j(q_3)) (F(\infty)- F(0) ) \\
&- 8 \pi i \lambda q_3 \left( - 4 \pi i \lambda q_3 \frac{k_- + p_-}{k_- - p_-} + j(q_3) \right) 
(F(\infty)-F(p) )\\
&+  8 \pi i \lambda q_3 \left( - 4 \pi i \lambda q_3 \frac{k_- + p_-}{k_- - p_-} + \wt b_4\right) (F(\infty)-F(k) )
\end{split}
\end{equation}
Substituting in for $V$ and $V_0$, the LHS of this equation may be rewritten as 
$$(8\pi i \lambda q_3)
\left(  F(p) \left( 4 \pi i \lambda  q_3  \frac{p_- + k_-}{p_- - k_-}
+ j(k_-, p_-, q_3) \right) - F(k) \left( 4 \pi i \lambda  q_3  \frac{p_- + k_-}{p_- - k_-}
+ \wt b_4 \right) \right)$$
It follows LHS exactly cancels the terms proportional to $F(k)$ and $F(p)$, and \eqref{alm} 
may be rewritten as 
 $$(-4 \pi i \lambda  q_3 + \wt b_4) (+4 \pi i \lambda  q_3 + j(q_3)) F(\infty)
 =(4 \pi i \lambda  q_3 + \wt b_4) (-4 \pi i \lambda  q_3 + j(q_3)) F(0)$$
 This is a linear equation for $j(q_3)$ whose solution is given by 
 \begin{equation}\label{ansofse}
 j(q_3)=  {4 \pi i \lambda q_3} \left( \frac{ \left(4 \pi i \lambda q_3 + \wt b_4 \right) F(0) 
 +  \left(-4 \pi i \lambda  q_3 + \wt b_4 \right) F(\infty) } {\left(4 \pi i \lambda q_3 + \wt b_4 \right) F(0) 
 -  \left(-4 \pi i q_3 + \wt b_4 \right) F(\infty) } \right)
 \end{equation}
Using 
$$F(\infty)= e^{\pi i \lambda {\rm sgn} (q_3)}, ~~~F(0)=e^{2 i \lambda  \tan^{-1}\left( \frac{2c_{B}}{q_3} \right)}
$$
we have 
 \begin{equation}\label{ansofsef}
 j(q_3)=  {4 \pi i \lambda q_3} \left( \frac{ \left(4 \pi i \lambda q_3 +\wt b_4 \right) e^{2 i \lambda  \tan^{-1}\left( \frac{2c_{B}}{q_3} \right)}
  +  \left(-4 \pi i \lambda q_3 + \wt b_4 \right)  
e^{\pi i \lambda {\rm sgn} (q_3)}
 } {\left(4 \pi i \lambda q_3 + \wt b_4 \right) e^{2 i \lambda  \tan^{-1}\left( \frac{2c_{B}}{q_3} \right)}
 -  \left(-4 \pi i \lambda q_3 + \wt b_4 \right)  e^{\pi i \lambda {\rm sgn} (q_3)}
} \right)
 \end{equation}
In the limit $b_4 \to \infty$ we have 
 \begin{equation} \label{anssc1}\begin{split}
 j(q_3)&=-4 \pi i \lambda  q_3 \left( \frac{ e^{\pi i \lambda {\rm sgn} (q_3)}  + e^{2 i \lambda  \tan^{-1}\left( \frac{2c_{B}}{q_3} \right)} } { e^{\pi i \lambda {\rm sgn} (q_3)}  - e^{2 i \lambda  \tan^{-1}\left( \frac{2c_{B}}{q_3} \right)} }\right) \\
&= -4 \pi i \lambda  |q_3| \left( \frac{ 1  + e^{-2 i \lambda  \tan^{-1}\left( \frac{|q_3|}{2c_B} \right)} } { 1 - e^{-2 i \lambda  \tan^{-1}\left( \frac{|q_3|}{2c_B} \right)} }  
 \right) 
\end{split} 
 \end{equation}
In summary, the off shell Euclidean sum of the diagrams depicted in Fig \ref{ladder} is 
given by \eqref{anon} with $j(q_3)$ given by \eqref{ansofsef}.

\subsection{The one loop box diagram computed directly in Minkowski space} \label{olm}
\begin{figure}[tbp]
  \begin{center}
  \subfigure[]{\includegraphics[scale=.6]{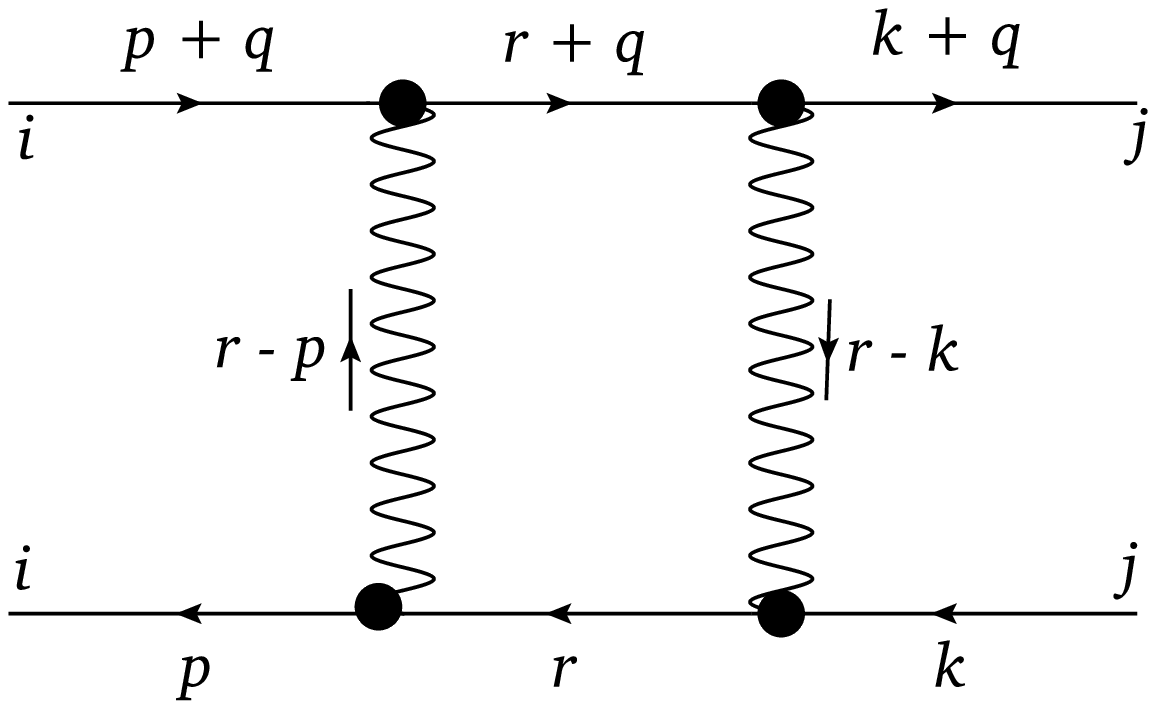}
  }\\
   \end{center}
  \vspace{-0.5cm}
  \caption{Box diagram in the light cone gauge.}
\label{olmfig}
  \end{figure}
In this subsection, 
by the direct calculation of the one loop box diagram in the Minkowski space,
we will show the cancellation of IR divergence 
of gauge propagator and that $P$ in \eqref{pamb} becomes unity $P = 1$.
In Minkowski space, the one loop box diagram (see Fig \ref{olmfig}) evaluates 
to 
\begin{equation}\label{olmin}
 \begin{split}
  I_{oneloop} =(4\pi\lambda q_3)^2 \int &\frac{d^3r}{(2\pi)^3}\frac{(r+p)_- (p-r)_+}{(p-r)_+(p-r)_- -i \epsilon_1}\frac{(r+k)_- (k-r)_+}{(k-r)_+(k-r)_- -i \epsilon_1}\\
&\frac{1}{2r_- r_+ +r_3^2+c_{B}^2-i \epsilon}\frac{1}{2(r+q)_- (r+q)_+ +(r+q)_3^2+c_{B}^2-i \epsilon}.
 \end{split}
\end{equation}
Although we are interested in the value of this integral at $q_\pm=0$, we have allowed $q_\pm \neq 0$ in the scalar propagators as a regulator; we will take the limit at the end of the computation. This manoever 
allows us to evaluate the integral in a particularly simple manner. 

Before embarking on the calculation, let us recall the issues involved. The term of 
${\cal O}(\lambda^2)$ in the expansion of the offshell amplitude \eqref{anonm} (we set $b_4=0$ for 
simplicity) is 
\begin{align}\label{expol}
V_2=& 8 \pi \lambda^2 q_3 \left( \tan^{-1} \left( \frac{2 a(k)}{q_3} \right)
-\tan^{-1} \left( \frac{2 a(p)}{q_3} \right) \right)
\frac{p_{-} + k_{-}}{p_{-}-k{-}}
\nonumber \\
&+ 16 \pi^2 q_{3}^2 \lambda^2 H(q_3) + 2 \pi c_{B} \lambda^2.
\end{align}
The last term in this equation is the contribution of the one loop diagrams in Fig. \ref{Unit} 
to $V$. Offshell, consequently, we expect \eqref{olmin} to evaluate to  
\begin{align}\label{expols}
-i I_{oneloop}
=& 8 \pi \lambda^2 q_3 \left( \tan^{-1} \left( \frac{2 a(k)}{q_3} \right)
-\tan^{-1} \left( \frac{2 a(p)}{q_3} \right) \right)
\frac{p_{-} + k_{-}}{p_{-}-k{-}}
\nonumber \\
&+ 16 \pi^2 q_{3}^2 \lambda^2 H(q_3) .
\end{align}
Here extra $-i$ factor comes from the analytic continuation
as we can check by the relationship between \eqref{sdeV} and \eqref{sdee}.
As mentioned at the beginning of this subsection,
the 
reason 
we are undertaking this whole exercise is that the first term in \eqref{expol}
is naively ambiguous onshell, 
and we aim to discover its true value via a careful evaluation of 
\eqref{olmin}.

In order to evaluate \eqref{olmin} 
we first evaluate the integral over $r_+$ integral using the methods 
of complex analysis. The integral over $r_+$ may be regarded as contour integral, where the 
contour runs from left to right along the real axis and then closes in a giant semi circle at infinity 
in the upper half plane. The integrand has four poles located at
\begin{equation} \label{poles}
 \begin{split}
  r_+&=p_++i\frac{\epsilon_1}{(r-p)_-},\\
r_+&=k_+ +i\frac{\epsilon_1}{(r-k)_-},\\
r_+&=-\frac{r_3^2+c_{B}^2}{2r_-}+i \frac{\epsilon}{2r_-},\\
r_+&=-q_+ -\frac{(r_3+q_3)^2+c_{B}^2}{2(r+q)_-}+i \frac{\epsilon}{2(r+q)_-}.\\
 \end{split}
\end{equation}

\subsubsection{Scalar poles}

From the point of view of IR divergences, the main point of interest in this section is the contribution from the first two poles in \eqref{poles}; the poles that have their origin in the gauge boson propagator. In order to be able to focus on the interesting part, however, it is useful to first get the `boring' part of the answer out 
of the way. 
(Irrelevant part for the subtraction between $\tan^{-1}$ functions.)
In this subsection we evaluate the contribution of the last two poles to the integral. In this 
subsection we assume for definiteness that the regulator $q_- <0$ (it is not difficult to see that the
final results do not depend on this assumption). 

If $r_-<0$ then neither of the third or fourth poles in \eqref{poles} lie in the upper half plane, and 
so these poles do not contribute to the $r_+$ integral.  On the other hand if $r_->-q_- > 0$, both poles 
contribute to the integral, and it is not difficult to verify that the contribution of the two poles infact
cancels. In other words the poles of interest contribute only in the range
$$ 0< r_- < -q_-.$$
When $r_-$ is in this window, we integrate over $r_+$ receives contributions only from the 
third pole in \eqref{poles}. Evaluating the residue of this pole redefining $r_-=-q_-x$, it is easily seen 
that 
\begin{equation}\label{olmin3}
 \begin{split}
  I^{3}_{oneloop} =\frac{i}{2}(4\pi\lambda q_3)^2 &\int_0^1\frac{dx}{2\pi} \int_{-\infty}^{\infty}\frac{dr_3}{2\pi}
\biggl[\left(1-2\frac{(p+k)_-}{(p-k)_-}\left(\frac{p_-}{p_-+q_-~x}-\frac{k_-}{k_-+q_-~x}\right)\right)\\
&~~~~~~~~~~~~\times\frac{1}{r_3^2+c_{B}^2+q_3^2~x+2q_3 r_3 x-i \epsilon -2 q_{-}q_{+}(x^2 -x)}\biggr].
 \end{split}
\end{equation}
In the limit $q_\pm \rightarrow 0$ ,
\begin{equation}\label{olmin31}
 \begin{split}
  I^{3}_{oneloop} =\frac{i}{2}(4\pi\lambda q_3)^2 \int_0^1\frac{dx}{2\pi} \int_{-\infty}^{\infty}\frac{dr_3}{2\pi}
\frac{1}{r_3^2+c_{B}^2+q_3^2~x+2q_3 r_3 x-i \epsilon},
 \end{split}
\end{equation}
(where $ I^{3}_{oneloop} $ is the contribution of the third and fourth poles to the integral 
\eqref{olmin}.)

We now evaluate the integral over $r_3$ by closing the contour in the upper half plane. 
The pole that contributes is at
$$r_3= -q_3 x + i \sqrt{c_{B}^2 +q_3^2 x-q_3^2 x^2-i \epsilon},$$
(note that $c_{B}^2+q_3^2 x-q_3^2 x^2 > 0$).
 We find
\begin{equation}\label{olmfnl1}
 \begin{split}
  I^3_{oneloop} &=\frac{i}{4}(4\pi\lambda q_3)^2 \int_0^1\frac{dx}{2\pi} 
\frac{1}{\sqrt{c_{B}^2+q_3^2 x-q_3^2 x^2}}\\
&=2\pi\lambda^2 \sqrt{q_3^2} \left( \log(2m+i \sqrt{q_3^2})-\log(2m-i \sqrt{q_3^2}) \right)\\
&= i(4\pi\lambda q_3)^2 H(q),
 \end{split}
\end{equation}
in precise agreement with the second term in \eqref{expols}.

As the sum of the third and fourth poles in \eqref{poles} yields the second term in \eqref{expols}, 
the sum of the first two poles must give rise to the first term in 
\eqref{expols}. We will now 
verify that that is indeed the case. 

\subsubsection{Contributions of the gauge boson poles off shell} \label{gbos}

The first two poles in 
\eqref{poles} 
are a consequence of our resolution of the singularity of the gauge 
boson propagator. Offshell, the contribution of these poles to the integral 
\eqref{olmin} 
is very simple: we pause to explain this fact. 
Consider the integral
$$ \int dl_+ dl_- \frac{l_+}{l_+l_- -i\epsilon_{1}} f(l_+, l_-),$$
where $f$ is any sufficiently smooth function.
The integrand has a pole at 
$$l_+= \frac{i \epsilon_{1}}{l_-}.$$
If we evaluate the $l_+$ integral by closing the contour with a giant semicircle in the upper 
half plane, this pole contributes only if $l_->0$. The contribution of this pole to the integral is 
$$ 2 \pi i  \int^{\infty}_{0} dl_-  \frac{i \epsilon_{1}}{l_-^2} \theta(l_-) f( 
\frac{i \epsilon_{1}}{l_-}, l_-).$$
Provided $f(l_+, l_-)$ has no singularities if either of its arguments vanish, then in the limit 
$\epsilon_{1} \to 0$ the integral over $l_-$ receives contributions only from $l_- \sim \epsilon_{1}$, i.e. 
at finite values of the variable $y= \frac{\epsilon_{1}}{l_-}$. 
Changing integration variables to $y$ we find 
that the contribution of this pole to the integral is given by
\begin{equation}\label{contftp} 
- 2 \pi \int_0^\infty dy f(iy, 0).
\end{equation}

Provided all external momenta are offshell the analysis of the paragraph above applies, and allows us 
to easily evaluate the contribution of the first two poles to \eqref{olmin}. Identifying the 
function $f$, applying the formula \eqref{contftp} and performing the integral over 
$r_3$ we find that the contribution of the first pole 
\begin{equation}\label{olmin-1}
 \begin{split}
  I^{1}_{oneloop} &=-(4\pi\lambda q_3)^2
\frac{(k+p)_- }{(k-p)_- }\\
&\times
\int_{0}^{\infty} \frac{d\wt y}{2\pi} \frac{1}{\sqrt{2p_- p_+ +c_{B}^2-i \epsilon+ i \wt y}}\frac{1}{4(2p_- p_+ +c_{B}^2+i\wt y -i \epsilon)+q_3^2},\\
 \end{split}
\end{equation}
where 
$$\tilde{y} = 2p_{-} y.$$
In \eqref{olmin-1}, we took $\epsilon_{1} \to 0$ limit already,
and we also take into account that $p_{-} < 0$ inside the lightcone.
Evaluating the integral we obtain
\begin{equation}\label{olmin-11}
 \begin{split}
  I^{1}_{oneloop} &=-(4\pi\lambda q_3)^2 
\frac{(k+p)_- }{(k-p)_- }\\
&\times\frac{i}{2\pi\sqrt{q_3^2}}\left(-\frac{\pi}{2}+\tan^{-1}\left(2\sqrt{\frac{2p_- p_++c_{B}^2- i\epsilon}{q_3^2}}\right)\right). \end{split}
\end{equation}
Similarly the contribution of the second term is 
\begin{equation}\label{olmin-22}
 \begin{split}
  I^{2}_{oneloop} &=(4\pi\lambda q_3)^2 
\frac{(k+p)_- }{(k-p)_- }\\
&\times \frac{i}{2\pi\sqrt{q_3^2}}\left(-\frac{\pi}{2}+\tan^{-1}\left(2\sqrt{\frac{2k_- k_++c_{B}^2- i\epsilon}{q_3^2}}\right)\right). \end{split}
\end{equation}
Summing these two contributions we find perfect agreement with the first term in 
\eqref{expols}.

All we have seen so far is that the one loop four point function in Minkowski space is, indeed, the 
continuation of its Euclidean counterpart. Of course we knew this had to be true on general 
grounds, so the agreements obtained so far have simply been internal consistency checks. 
In order to get new information we will now investigate the contribution of the first two poles 
in \eqref{poles} to the amplitude \eqref{olmin} 
when the external particles are all onshell. 
Recall that the continuation of the Euclidean answer - and the naive analysis of this subsection - 
yielded ambiguous answers for this quantity. Obtaining the correct result for this 
amplitude requires a more careful calculation which we now turn to .

\subsubsection{The onshell contribution of the gauge boson poles}
In this subsubsection finally we will show that
$P$ in \eqref{pamb} is unity $P = 1$.
In the previous two subsubsections,
we have seen that 
gauge boson poles contribute to the 
first term of \eqref{expols}
while the scalar boson poles contributes to
second term of the \eqref{expols}.
Therefore our concerning factor
\begin{equation}
\left( \tan^{-1} \left( \frac{2 a(k)}{q_3} \right)
-\tan^{-1} \left( \frac{2 a(p)}{q_3} \right) \right)
\to \left( \tan^{-1} \left( -i \right)
-\tan^{-1} \left( -i \right) \right)
\end{equation}
in the first term of \eqref{expols}
is given by the contribution of the on-shell gauge boson poles.

When the momenta $p$ and $k$ are onshell, the analysis of 
Appendix~\ref{gbos} yields an ambiguous 
result. This is because the analysis presented above applies only when the function $f$ of the previous 
section is sufficiently well behaved. This assumption is valid for generic values of $p$ and $k$. 
When the two external momenta are onshell, however, it turns out that the function $f(l_-, l_+)$ 
of the previous subsection blows up at $l_-=0$, invalidating the approximations used in the previous 
subsection. We will now present a more careful analysis of this special case. In this subsection we ignore 
the overall factor $(4 \pi \lambda q_3)^2$ in \eqref{olmin}; 
the factor is not important as the 
conclusion of this subsection is that the net contribution 
of the two gauge boson poles for the onshell 
4 point function actually vanishes.

The contribution of the first gauge boson pole to the 
$r_+$ integral in \eqref{olmin} is given as
\begin{align} 
  \int_{0}^\infty dy \int_{-\infty}^\infty  d r_3 &
\biggl[  -\frac{1}{2\pi}(1+\frac{\epsilon_1}{y})
\frac{1}{X}
\nonumber \\
& \times \left(
\frac{
(y(k_{-}+p_{-})
+ 2p_{-}\epsilon_{1}
)
(
2p_{-}(k_{+}-p_{+})-i y
)
}
{(y(k_{-}-p_{-})
- 2p_{-}\epsilon_{1}
)
(
2p_{-}(k_{+}-p_{+})-i y
)-i 2p_{-}y \epsilon_{1}
}\right)
\biggr]\label{fpc}
\end{align}
where we have made the variable redefinition 
\begin{equation}\label{vdo} 
r_-=p_-+2p_-\frac{\epsilon_1}{y}
\end{equation}
 and where 
\begin{equation}
\begin{split}
 X=&\left( r_3^2-p_3^2-2(p_3^2+c_{B}^2)\frac{\epsilon_1}{y}+i(y-\epsilon+2\epsilon_1)\right) 
\\
&\times
\left((r_3+q_3)^2-p_3^2-2(p_3^2+c_{B}^2)\frac{\epsilon_1}{y}+i(y-\epsilon+
2\epsilon_1) \right).
\end{split}
\end{equation}
We can obtain the second pole contribution by exchanging the 
momentum $k$ and $p$.

By noting that 
$$
p_{s}^2 = k_{s}^2, \qquad p_{3}^2 = k_{3}^2,
$$
when $p, k, (p+q), (k+q)$ are on-shell,
we can see that 
the first line of \eqref{fpc} is symmetric under the exchange.
We can see that $O(\epsilon_{1}^0)$ term of the 
second line of \eqref{fpc} is antisymmetric under the exchange
of $p$ and $k$ because its form is 
$$
\frac{k_{-}+p_{-}}{k_{-}-p_{-}}.
$$
Hence the sum of the contributions from 
first and second pole of \eqref{olmin} should be 
$O(\epsilon^1)$.
In the integrand of 
\eqref{fpc},
the variable $r_{3}$ appears only in factor $X$.
Therefore the sum of the 
first and second pole contribution becomes following form
\begin{equation} \label{apc}
\int_{y}^\infty d y 
\int_{-\infty}^\infty  d r_3  
\left(\epsilon_{1} \times \tilde{I}(y)\frac{1}{X(r_{3}, y)}\right).
\end{equation}
Because of this explicit factor, in order to establish 
that \eqref{apc} vanishes in the limit $\epsilon_1 \to 0$ 
it is sufficient to verify that the integral 
in \eqref{apc} has no compensating singularity as $\epsilon_1 \to 0$. 
To investigate it, it is important to note that 
\begin{equation}
(y(k_{-}-p_{-})
- 2p_{-}\epsilon_{1}
)
(
2p_{-}(k_{+}-p_{+})-i y
)-i 2p_{-}y \epsilon_{1}
= 0  
\Rightarrow y = \epsilon_{1} = 0.
\label{soluy}
\end{equation}
at the denominator of second line of 
\eqref{fpc} if $k_{-} \ne p_{-}$. 
\footnote{
In the case of $k_{-} - p_{-} = 0$ on-shell, 
LHS of \eqref{soluy} is always zero for any $y$, $\epsilon_{1}$.
This may intrigue to the delta function in the $S$-channel.}
It is also useful to expand
\begin{align}
-2\pi \tilde{I}(y) \sim&  
\frac{(k_{-}+p_{-})}{(k_{-}-p_{-})^2}
\left(\frac{2i p_{-}}{
2p_{-}(k_{+}-p_{+})-i y
}
+\frac{2i k_{-}}{
2k_{-}(p_{+}-k_{+})-i y
}\right)
\nn 
&+\frac{8p_{-}k_{-}}{y(k_{-}-p_{-})^2}
+{\cal O}(\epsilon_{1}).
\label{exp-I}
\end{align}

The integral over $r_3$ in factor $X$ is elementary, 
and may be explicitly performed; however the resultant expression is a slightly messy function of $y$ 
and we do not present the explicit form here. 

After performing the $r_3$ integral and further changing variables 
to $y_1 = \frac{y}{\epsilon_{1}}$, \eqref{apc} reduces to an expression of the schematic form 
\begin{equation}
 I=\int_0^\infty  dy_1 I(y_1).
\end{equation}
Naively $I(y_1)$ is of order $\epsilon_1^2$ (it picks up an additional factor of $\epsilon_1$ from 
the change of variables $y= \epsilon_1 y_1$). Infact the singular behavior that results in the ill 
definition of the naive expression modifies this estimate for $y_1$ of order unity or smaller. Nonetheless it is 
possible to demonstrate that $I(y_1)  \leq O(\epsilon_1)$ throughout its integration domain. 
In pariticular 
\footnote{For instance at 
$y \ll \epsilon_{1}$ namely $y_{1} \ll 1$ 
by performing the contour integral we get
\begin{equation}
\int dr_{3} \frac{1}{X} \sim 
\int dr_{3} \frac{1}{r_{3}^2 - 2 (p_{3}^2 + c_{B}^2) \frac{\epsilon}{y} 
+ i \tilde{\epsilon}}
\frac{1}{(r_{3}+q_{3})^2 - 2 (p_{3}^2 + c_{B}^2) \frac{\epsilon}{y} 
+ i \tilde{\epsilon}}
\sim \frac{\pi i y_{1}^{\frac{3}{2}}}{
4\sqrt{2}(p_{3}^2 +c_{B}^2)^{\frac{3}{2}}}.
\end{equation}
Then \eqref{fpc} behaves as
\begin{align}
\int dy  \biggl[\frac{\epsilon_{1}}{y}\int dr_{3}\frac{1}{X}
\left(\frac{
 2p_{-}\epsilon_{1}
(2p_{-}(k_{+}-p_{+}))
}
{- 2p_{-}\epsilon_{1}
(2p_{-}(k_{+}-p_{+}))
}
\right)
\biggr]
\sim 
\int dy_{1}  \epsilon_{1}
\frac{\sqrt{y_{1}}}{(p_{3}^2 +c_{B}^2)^{\frac{3}{2}}}.
\end{align}
At $y\gg p_{3}^2$, namely $y_{1} \gg \frac{p_{3}^2}{\epsilon_{1}}$, 
the integration over $r_{3}$ gives
\begin{equation}
\int dr_{3} \frac{1}{X} \sim 
\int dr_{3} \frac{1}{r_{3}^2 +i y}
\frac{1}{(r_{3}+q_{3})^2 +iy}
\sim \frac{-\pi e^{\frac{i\pi}{4}}}{2y^{\frac{3}{2}}}.
\end{equation}
Then from \eqref{exp-I} and $dy = \epsilon_{1}dy_{1}$,
we can see that \eqref{apc}
behaves as
\begin{equation}
 \int d y 
\int_{-\infty}^\infty  d r_3  
\left(\epsilon_{1} \times \tilde{I}(y)\frac{1}{X(r_{3}, y)}\right)
\sim 
 \int d y_{1} 
\epsilon_{1}^2 \frac{1}{y^{\frac{5}{2}}_{1}\epsilon_{1}^{\frac{5}{2}}}
\sim 
 \int d y_{1} \frac{1}{y^{\frac{5}{2}}_{1}\epsilon_{1}^{\frac{1}{2}}}.
\end{equation}
}
\begin{equation}
 I(y_1) \sim 
\begin{cases}
  \frac{ \epsilon_1\sqrt{y_{1}}}{(c_{B}^2+p_3^2)^{\frac{3}{2}}} &(y_1 \ll 1) \\
 \epsilon_1  & (y_1\sim 1)\\
 \epsilon_1^2 & (y_1\sim \frac{p_3^2}{\epsilon_1})\\
   \frac{1}{\sqrt{\epsilon_{1}}}\left(\frac{1}{y_1}\right)^{\frac{5}{2}} & 
(y_1 \gg \frac{p_3^2}{\epsilon_1})
\end{cases}.
\label{e-behavior}
\end{equation}
We can immediately see that
\eqref{apc} vanishes in the limit $\epsilon_1 \to 0$
in the first three cases in \eqref{e-behavior}. 
Actually also in the case $y_1 \gg \frac{p_3^2}{\epsilon_1}$,
we can check that it vanishes 
if we integrate over $y_{1}$
\begin{equation}
\left|\int^{\infty}_{\frac{p_{3}^2}{\epsilon_{1}}} dy_{1} 
\frac{1}{\sqrt{\epsilon_{1}}}\left(\frac{1}{y_1}\right)^{\frac{5}{2}}  
\right|
\sim  
\frac{1}{\sqrt{\epsilon_{1}}}\left(\epsilon_1\right)^{\frac{3}{2}}  \sim 
\epsilon_{1} \to 0.
\end{equation}
So the net contribution 
of two gauge boson poles for one loop 4 point function, 
namely contribution for the
first term of \eqref{olmin} vanishes.
This results that the subtraction of 
$\tan^{-1}$ function vanishes as
\begin{equation}
\begin{split}
0 =&  \left( \tan^{-1} \left( \frac{2 a(k)}{q_3} \right)
-\tan^{-1} \left( \frac{2 a(p)}{q_3} \right) \right)
\frac{p_{-} + k_{-}}{p_{-}-k{-}}
\\ 
\Rightarrow 0 = &\left( \tan^{-1} \left( \frac{2 a(k)}{q_3} \right)
-\tan^{-1} \left( \frac{2 a(p)}{q_3} \right) \right)
\end{split}
\end{equation}
in the on-shell $p$ and $k$.
Then finally we conclude that 
the $P$ in \eqref{pamb} becomes unity $P = 1$.

\section{Details of the one loop Landau gauge computation}\label{landau}
In this subsection we provide some details for our evaluation of the one loop scattering amplitude 
in the covariant Landau gauge. As we have explained in the main text, the evaluation consists of 
determining the integrand for each graph, and then following standard manipulations that allow 
one to re-express the integrand in a standard basis. In order to illustrate how this works, we first
present all steps in detail for the most complicated diagram (this is the box graph). For the 
remaining diagrams we content ourselves with a brief explanation or simply stating our results. 

\subsection{Simplification of the integrand of the box graph}
Straightforward use of the Feynman rules leads to an expression 
for the integrand of the box graph depicted in Fig. \ref{LgBox}
\begin{equation}\label{boxinto}
\begin{split}
\frac{1}{64\pi^2\lambda^2}I_{box}
=&
 \frac{ (\epsilon_{\nu_1\nu\beta}q_{\nu_1}p_{\nu}l_{\beta}
\epsilon_{\mu_1\mu\beta_1}q_{\mu_1}
(l+p)_{\mu}k_{\beta_1})}{l^2 ((l+p)^2+c_{B}^2)
(l+p-k)^2 ((l+p+q)^2 + c_{B}^2)}\\
=&\frac{k\cdot q 
\left[
2 \left((l\cdot k) \left(c_{B}^2-l\cdot p\right)
+(k\cdot p)( l\cdot (l+p))\right)-
(l\cdot q)( l\cdot (k+p))\right]}{l^2 ((l+p)^2+c_{B}^2)
(l+p-k)^2 ((l+p+q)^2 + c_{B}^2)}
\\
&+\frac{
(k\cdot q )
(q\cdot l) \left(k \cdot p+c_{B}^2\right)
+(k\cdot q)^2 
(l\cdot (-k+l+p))+(k\cdot p) (q\cdot l)^2}{l^2 ((l+p)^2+c_{B}^2)
(l+p-k)^2 ((l+p+q)^2 + c_{B}^2)}
\end{split}
\end{equation}
The denominator of the expression above is the product $E_1 E_2 E_3 E_4$ where
\begin{equation}\label{denomdef}
E_1 = c_{B}^2+(l+p)^2,~E_2=c_{B}^2+(p+q+l)^2,~E_3=l^2,~E_4 = (l+p-k)^2 .
\end{equation}
The terms in the numerator RHS of \eqref{boxinto} that involve the loop momentum $l$ can be 
re-expressed as functions of the denominators plus terms independent of $l$. 
For example 
\begin{align}
&l\cdot l=E_3, ~~~2 ~p \cdot l= E_1-E_3  , 
\nn
&2 ~q\cdot l=E_2-E_1,~~~2~ k\cdot l=E_1-E_4-2~c_{B}^2-2~k \cdot p ,
\nonumber
\end{align}

where we have used onshell conditions $$p^2+c_{B}^2=0,~~k^2+c_{B}^2=0,~~(p+q)^2+c_{B}^2=0,~~(k+q)^2+c_{B}^2=0.$$
Judiciously using these and similar identities, it is easy to show that the integrand in \eqref{boxinto} may be 
rewritten as
\begin{equation}\label{boxro}
\begin{split}
  &-\frac{(k\cdot k-k\cdot p) (k \cdot q) (k\cdot q+2 k \cdot k)}{E_1 E_2 E_3
   E_4}+\frac{k\cdot q 
\left(k\cdot q-2 c_{B}^2\right)}{2 E_1 E_2 E_3}+\frac{k \cdot q 
\left(k\cdot q-2 c_{B}^2\right)}{2 E_1 E_2 E_4}\\
&+\frac{k\cdot q \left(k\cdot p+c_{B}^2\right)}{E_2 E_3 E_4}+\frac{k\cdot q \left(k\cdot p+c_{B}^2\right)}{E_1 E_3 E_4}
-\frac{(k\cdot p) (q\cdot l)}{2 E_2 E_3 E_4}+\frac{
(k\cdot p) (q\cdot l)}{2 E_1 E_3 E_4}\\
&-\frac{k\cdot q}{2 E_1 E_2}+\frac{k\cdot q}{4 E_1 E_3}+\frac{k\cdot q}{4 E_1 E_4}+\frac{k\cdot q}{4 E_2 E_3}+\frac{k\cdot q}{4
   E_2 E_4}-\frac{k\cdot q}{2 E_3 E_4}
 \end{split}
\end{equation}
The expression in \eqref{boxro} includes a term with four denominators. As we have mentioned in the 
main text, under the integral sign it is always possible to reduce any such expression into a linear combination of expressions with three or fewer denominators (recall we work in 3 spacetime dimensions). 
This reduction may be 
achieved by the systematic procedure spelt out in \cite{vanNeerven:1983vr}.
Implementing this procedure in the case at hand we find the replacement rule
\begin{align} 
-\frac{(k\cdot k-k\cdot p) k\cdot q (k\cdot q+2 k\cdot k)}{E_1 E_2 E_3
   E_4}=\frac{k\cdot q \left(2 c_{B}^2-k\cdot q\right)}{2 E_1 E_2 E_3}+\frac{k\cdot q \left(2 c_{B}^2-k\cdot q\right)}{2 E_1 E_2 E_4}
\nn
-\frac{k\cdot q
   \left(k\cdot p+c_{B}^2\right)}{2 E_1 E_3 E_4}-\frac{k\cdot q \left(k\cdot p+c_{B}^2\right)}{2 E_2 E_3 E_4} \label {reprule}
\end{align}
Using \eqref{reprule}, the integrand for the box diagram reduces to 
\begin{equation}\label{boxrt}
\begin{split}
\frac{1}{64\pi^2\lambda^2}I_{box}&= 
\frac{k\cdot q \left(k\cdot p+c_{B}^2\right)}{2 E_1 E_3 E_4}
+\frac{k\cdot q \left(k\cdot p+c_{B}^2\right)}{2 E_2 E_3 E_4} 
-\frac{(k\cdot p) (q\cdot l)}{2 E_2 E_3 E_4}
+\frac{(k\cdot p) (q\cdot l)}{2 E_1 E_3 E_4}\\
&-\frac{k\cdot q}{2 E_1 E_2}
+\frac{k\cdot q}{4 E_1 E_3}
+\frac{k\cdot q}{4 E_1 E_4}
+\frac{k\cdot q}{4 E_2 E_3}
+\frac{k\cdot q}{4 E_2 E_4}
-\frac{k\cdot q}{2 E_3 E_4}
 \end{split}
\end{equation}
We now turn to a discussion of the relations between distinct scalar (and other) integrands. Expressing 
the corresponding integrals in terms of Feynman parameters, it is not difficult to demonstrate that, 
under the integral sign 
\begin{equation}
 \begin{split}
  &\frac{1}{E_1 E_2 E_3}=\frac{1}{E_1 E_2 E_4},~~~~~\frac{1}{E_1 E_3 E_4}=\frac{1}{E_2 E_3 E_4}\\
&\frac{q.l}{E_1 E_3 E_4}=-\frac{q.l}{E_2 E_3 E_4}\\
&\frac{1}{E_1 E_3}=\frac{1}{E_1 E_4}=\frac{1}{E_2 E_3}=\frac{1}{E_2 E_4}.
 \end{split}
\end{equation}
For instance
\begin{equation}
\begin{split}
\frac{1}{E_1E_3}=& \int_0^1 \frac{dx}{\left(x E_3 + (1-x)E_1\right)^2}
\\
=&\int_0^1 
\frac{dx}{\left(l^2 +2 (1-x)  l\cdot p \right)^2}\\
=&\int_0^1 \frac{dx}{\left({\tilde l}^2 +(1-x)^2 c_{B}^2\right)^2}.
\end{split}
\end{equation}
Similarly 
\begin{equation}
\begin{split}
\frac{1}{E_1 E_4} &= 
\int_0^1 \frac{dx}{\left(x E_4 + (1-x)E_1\right)^2}
\\
&=\int_0^1 \frac{dx}{\left(l^2+2 l\cdot p-2~p \cdot k x-2~
l\cdot k~x-2c_{B}^2~x\right)^2 }\\
&=\int_0^1 \frac{dx}{\left({\tilde l}^2 +(1-x)^2 c_{B}^2\right)^2}.
\end{split}
\end{equation}
Using these relations we may rewrite the integrand for the box diagram as 
\begin{equation}\label{boxrth}
\frac{1}{64 \pi^2 \lambda^2}
I_{box}=\frac{k\cdot q \left(k\cdot p+c_{B}^2\right)}{ E_1 E_3 E_4}
+\frac{(k\cdot p) (q\cdot l)}{ E_1 E_3 E_4}
-\frac{k\cdot q}{2 E_1 E_2}
+\frac{k\cdot q}{ E_1 E_3}
-\frac{k\cdot q}{2 E_3 E_4}.
\end{equation}
In order to complete our simplification, we must now re-express the term 
$$ \frac{(k\cdot p) (q\cdot l)}{E_1 E_3 E_4}$$
in terms of scalar integrals. 
The procedure for doing this is once again standard 
\cite{'tHooft:1978xw,Passarino:1978jh} 
and we find 
\begin{align}
\frac{(k\cdot p) (q\cdot l)}{E_1 E_3 E_4} 
\rightarrow &\frac{(k\cdot p) (k\cdot q)}{ 
\left(c_{B}^2-k\cdot p\right)}\frac{1}{E_1 ~E_4}-\frac{(k\cdot p) (k\cdot q)}{
c_{B}^2-k\cdot p}\frac{1}{E_3~E_4}
\nn
&+\frac{(k\cdot p) (k\cdot q) 
\left(k\cdot p+c_{B}^2\right)}{c_{B}^2-k\cdot p}\frac{1}{E_1~E_3~E_4}. 
\nonumber
\end{align}
Using this replacement rule the integrand for the box diagram finally 
reduces to 
\begin{align}
I_{box}=4\pi^2\lambda^2
\biggl(&-\frac{8 k\cdot q}{E_1~E_2}
-\frac{8(c_{B}^2+k\cdot p)k\cdot q}{c_{B}^2-k\cdot p}\frac{1}{E_3~E_4}
\nn
&+\frac{16~c_{B}^2 (k\cdot q)}{c_{B}^2-k\cdot p}\frac{1}{E_1~E_4}
+\frac{16~c_{B}^2(c_{B}^2+k\cdot p)k\cdot q}{c_{B}^2-k\cdot p}\frac{1}{E_1~E_3~E_4}\biggr).
\label{boxrthf}
\end{align}

\subsection{Simplification of the remaining integrands}

We are left with the task of evaluating and simplifying the integrand of the remaining one loop 
scattering diagrams.  These diagrams are listed in Fig. \ref{LgH}-\ref{Cancel}.  The simplification of the 
integrand follows a procedure that  similar to but much simpler than that adopted in the 
previous subsection.  The diagrams of \ref{LgH}-\ref{Cancel} are simpler than the box diagram considered in the 
previous subsection  because none of them involves more than 3 propagators, so we never 
have to employ a replacement rule analogous to \eqref{reprule}. 

We briefly illustrate how things work in the specially simple case of the $h$ graphs of Fig. \ref{LgH}.
Since all of the four h diagrams are interrelated by linear momentum redefinitions, we can evaluate 
any one of them and multiply the result by 4. 
 We consider first of these diagrams. 
Apart from come constant overall factor it gives
 \begin{align}
 &\int\frac{d^3l}{(2\pi)^3}
\frac{\epsilon^{\mu\nu\rho}(l+2p+2q)_{\nu}l_{\rho}g_{\mu\chi}\epsilon^{\chi\sigma\phi}(p+k)_{\sigma}(k-p)_{\phi}}{(k-p)^2l^2((l+p+q)^2+c_{B}^2)}\nn
 =&4\int\frac{d^3l}{(2\pi)^3}
\frac{\epsilon^{\mu\nu\rho}(p+q)_{\nu}l_{\rho}g_{\mu\chi}\epsilon^{\chi\sigma\phi}p_{\sigma}k_{\phi}}{(k-p)^2l^2((l+p+q)^2+c_{B}^2)}.
 \end{align}
Introducing Feynman parameter x and eliminating cross-terms including l in the denominator by usual drill we get
 \begin{equation}
4\int\frac{\epsilon^{\mu\nu\rho}(p+q)_{\nu}l_{\rho}g_{\mu\chi}\epsilon^{\chi\sigma\phi}p_{\sigma}k_{\phi}}{(k-p)^2(l^2+x(1-x)(p+q)^2+xc_{B}^2)^2}\frac{d^3l}{(2\pi)^3}
 \end{equation}
The integrand is odd in all components of l, hence the integration vanishes. It follows that 
$$I_{h}=0.  $$ 
In a similar manner we find that the  integrand for the sum of the two $V$ diagrams (see Fig. \ref{LgV}) is 
$$ I_{V}=4\pi^2\lambda^2\left(-\frac{2}{E_1}-\frac{8~c_{B}^2}{E_1~E_3}+\frac{6(c_{B}^2+k\cdot p)}{E_3~E_4}-\frac{8~c_{B}^2~(c_{B}^2+k\cdot p)}{E_1~E_3~E_4}\right). $$
The integrand for the sum of the two $Y$ diagrams (see Fig. \ref{LgY}) is 
\begin{equation}
\begin{split} 
\frac{1}{4\pi^2\lambda^2}I_{Y} = &
\frac{8 c_{B}^2 \left(k\cdot p+c_{B}^2\right) \left(-k\cdot p-2 k\cdot q+c_{B}^2\right)}{(c_{B}^2-k\cdot p)~E_1~E_3~E_4}
\\
&+\frac{8 c_{B}^2 \left(-k\cdot p-2 k\cdot q+c_{B}^2\right)}{(c_{B}^2-k\cdot p)~E_1~E_4}
-\frac{4 \left(k\cdot p+c_{B}^2\right) \left(-k\cdot p-2 k\cdot q+c_{B}^2\right)}{(c_{B}^2-k\cdot p)~E_3~E_4}. 
\end{split}
\end{equation}
The integrand for the sum of the eye diagrams (see Fig. \ref{LgEye}) is 
$$ I_{Eye}=4\pi^2\lambda^2\left(-\frac{2}{E_4}-\frac{2 \left(k\cdot p+c_{B}^2\right)}{E_3~E_4}\right)\cdot $$
Note that, contribution from lollipop diagrams (see Fig. \ref{LgLol}) vanishes. Similarly, one can show that two diagrams in Fig. \ref{Cancel} each other.
Summing all these contributions together, we find the remarkably simple integrand
\begin{equation}\label{finintgd}
I_{full}=4\pi^2\lambda^2\left(-\frac{2}{E_1}-\frac{2}{E_4}-\frac{8 k\cdot q}{E_1~E_2}\right).
\end{equation}
It follows that (modulo possible subtleties at special values of external momenta, see the main 
text) the full one loop four boson scattering 
amplitude is given by 
\begin{equation}\label{finansol}
S_{\rm one ~loop}
=2 \pi m \lambda^2+ 32 \pi^2 (k\cdot p) \lambda^2 H(q).
\end{equation}
Note, of course, that this result precisely matches  the ${\cal O}(\lambda^2)$ term in the Taylor expansion 
of the function $j(q)$ at $b_4=0$. 

\subsection{Absence of IR divergences}

Notice that our scattering amplitude 
is finite without regulation; in particular the amplitude has no IR divergences. This is satisfying. IR 
divergences in theories like QED result from the fact that the asymptotic electron states of the theory
are surrounded by a cloud of soft photons. The IR finiteness of our amplitude reflects the fact that 
Chern-Simons theories does not have massless gluonic states. Although the absence of IR divergences 
is physically very reasonable, at the technical level it appears to be a bit of a miracle, given the appearance
of the massless gauge boson propagator at intermediate steps in the computation. Integrands of the 
form, for instance
\begin{equation}
  \frac{1}{E_1 E_3 E_4},~~ \frac{1}{E_1  E_4},~~\frac{1}{E_1  E_3}
\end{equation}
that appear at intermediate steps in the computation, give rise to integrals 
that are IR divergent. The lack of IR divergences in our final result is a consequence of the cancellation 
of all these expressions in the final result for the integrand. For instance, the box diagram integrand Eq.\ref{boxrth}, first and second term are IR
finite where as third and fourth are IR divergent. However, one can show log divergence arising from both the third and fourth 
integrands cancel each other\footnote{  One simple way to check this is using the following trick (refer to Bern's paper \cite{Bern:1992em})
\begin{equation}\begin{split}
 \int \frac{d^3p}{2\pi^3} \frac{1}{E_1 E_3 E_4} 
&=-\frac{1}{ (k\cdot p+c_{B}^2)}(\int \frac{d^3p}{2\pi^3} \frac{1}{E_1 E_4 })+
 \frac{1}{2c_{B}^2}\int \frac{d^3p}{2\pi^3} \frac{1}{E_3 E_4 }\\
&-N\left(\frac{2}{(p-k)^2}+\frac{1}{2c_{B}^2}\right)\int \frac{d^5p}{2\pi^5} \frac{1}{E_1 E_3 E_4 }.
\end{split}\end{equation}
where in the last line $N$ is some number which is not important for our argument. Note that, third (last)  term in the last line is IR convergent as this is in the higher dimension. The second term in the 
last line also IR convergent where first term is not, 
 however, this IR divergence explicitly cancels the IR divergence coming from third term of   \eqref{boxrth}.}. Note that, the first line of box integral of Eq.\ref{boxinto} has no IR divergence (near $l\sim 0$), so final should 
also have no IR divergence.

\subsection{Absence of gauge boson cuts}\label{landaucut}

The imaginary part of any Feynman diagram may be determined using Cutkosky's rules. We pause to 
briefly review these rules  (we follow a presentation due to  't Hooft and Veltman \cite{'tHooft:1973pz}). 
Given a graph one divides the vertices of the graph into two groups; circled and uncircled vertices.
 Associated with a particular distribution of circles for vertices , one defines a `cut graph'.
The expression for the cut graph is obtained from a sequence of modifications on the expression 
for the usual (uncut) Feynman graph as we now describe. The factor of $i$ in each circled 
vertex is replaced by a factor of $-i$.  Propagators between two circled 
vertices are replaced by their complex conjugates. Every 
 factor of $\frac{1}{p^2+ c_{B}^2-i\epsilon}$ in a cut propagator: i.e. a propagator that runs between 
a circled and uncircled vertex -  is replaced by $\theta(p^0) \delta(p^2 +c_{B}^2 -i\epsilon )$
where $p^0$ is the energy running from the uncircled to the circled vertices. The sequence
of modifications described above gives the expression for the `cut graph' associated with 
a given distribution of circles for vertices. 

Cutkowski's rules state that the imaginary part of any  Feryman diagram is given by 
the sum of the expressions for cut graphs for all possible ways of distributing circles 
among the vertices of that graph subject to the restriction that at least one vertex in the 
graph is circled and at least one vertex is uncircled. Cutkowski's rules are the diagrammatic 
reflection of the unitarity of scattering amplitudes.

If we were to apply these rule to the one loop diagrams depicted in Fig. \ref{LgBox},\ref{LgH},\ref{LgV},\ref{LgY},\ref{LgEye}\ref{LgLol}, it would, at first 
appear that the imaginary part of the one loop graph would receive contributions from graphs in which 
two scalar propagators are cut and graphs in which two gauge boson propagators are cut. 
\footnote{A graph in which a one gauge boson and propagator is cut will contribute zero to the
imaginary part. All cut graphs may be regarded as the square of tree level processes. One of the 
tree process corresponding to such a cut would be decay of a single scalar to a scalar and  a gauge 
boson: this is kinematically forbidden and so does not contribute.} Our extremely simple final answer 
\eqref{finintm} and \eqref{vol} does have two scalar cuts, but has no cut contribution 
from two intermediate gauge boson lines. From a physical standpoint this is extremely satisfying; 
the Chern-Simons theory we study has no propagating gauge boson states, and so a two gauge boson
cut would likely have signalled a contradiction with unitarity.  From the purely technical point of view, 
however, the absence of two gauge boson cuts seems striking. Individual graphs in Fig. \ref{LgBox},\ref{LgH},\ref{LgV},\ref{LgY},\ref{LgEye}\ref{LgLol}
certainly have these cuts, which must, therefore cancel between graphs. In this subsubsection
we verify that this is indeed the case.

Two gauge boson cuts naively occur in the $T$-channel. In this channel the two external scalar lines at the 
top of the graphs in Fig. \ref{LgBox},\ref{LgH},\ref{LgV},\ref{LgY},\ref{LgEye}\ref{LgLol} represent initial states (one particle one antiparticle) while the 
two external lines at the bottom of the graph are final states. In order to focus on this channel 
we must take $p_0>0$, $k_0 <0$, $(p+q)_0>0$ and $(k+q)_0<0$. We find it useful to work in 
the `center of mass frame' in which the two incoming quanta approach each other along the $x$ axis.
Let the final scattering angle be $\alpha$. It follows that 
\begin{align}
& p=(p_0,p,0),\quad k=(-p_0,p,0),
\nn
&p+q=(p_0,p \cos(\alpha),p \sin(\alpha)),
\quad
k+q=(-p_0,p \cos(\alpha),p \sin(\alpha)).
\label{mominsc}
\end{align}

All two gauge boson cuts have a universal  factor that comes from delta functions that puts the gauge 
bosons on shell. This factor is given by
\begin{equation}\label{delta}
 \begin{split}
 & \int  \frac{d^3l}{(2\pi)^3}(-2\pi i)^2\delta (-l_0^2+l^2)\delta ((l+p-k)^2)\theta(-l_0)\theta(l_0+2 p_0)\\
=&\int  \frac{d^3l}{(2\pi)^3}(-2\pi i)^2\frac{1}{2|l_0|}\delta (l_0+l)\delta (-4l_0 p_0-4~p_0^2)\theta(-l_0)\theta(l_0+2 p_0)\\
=&\int  \frac{1}{(2\pi)^3}l dl_0 dl d\theta(-2\pi i)^2\frac{1}{8~p_0^2}\delta (l_0+l)\delta (l_0 +p_0)\theta(-l_0)\theta(l_0+2 p_0)\\
=&\int  \frac{1}{(2\pi)^3}p_0 dl_0 dl d\theta(-2\pi i)^2\frac{1}{8~p_0^2}\delta (-p_0+l)\delta (l_0 +p_0)\theta(-l_0)\theta(l_0+2 p_0)\\
=&-\frac{1}{16\pi p_0}\int   dl_0 dl d\theta\delta (-p_0+l)\delta (l_0 +p_0))
 \end{split}
\end{equation}
in the last line we have dropped the theta function because delta function clicks with in the theta function.
\footnote{The two delta functions in the final line of \eqref{delta} have a simple physical interpretation.
As the two gauge fields are on shell, the cut graph proceeds via two intermediate (tree level) scattering
processes, each of which take two scalar photons to two gauge bosons. The usual kinematical restrictions 
applied to these intermediate processes implies that the 3 momenta of the two intermediate gauge 
bosons - which, according to the labelling of 3 momenta in Fig \ref{LgBox} - is $-l$ and $l+p+k$ - 
$$(p_0, \pm p_0\cos (\alpha), \pm p_0\sin (\alpha)).$$ 
The $\delta$ functions in the last line of $\delta$ enforce this.}

 In addition to the universal factor evaluated in \eqref{delta} each diagram has its own particular factors 
that arise  from the vertex factors, from propagators between circles or 
between crosses, and from the numerator of the cut gauge boson propagators that we have not 
yet included in our analysis. For the various diagrams with two gauge bosons cuts, these factors 
are given by
\begin{equation}
 \begin{split}
 {\rm Eye}_{\rm ~diagram} =&  -4 p_0^2 \delta (-p_0+l)\delta (l_0 +p_0).\\
{\rm V}_{\rm ~diagram} =&4\left(2 p_0^2-l\cdot k-\frac{2c_{B}^2~p_0^2}{l\cdot p} 
\right)
\delta (-p_0+l)\delta (l_0 +p_0)\\
\frac{1}{16}\left({\rm Box}_{\rm ~diagram}\right) 
=&
\biggl[
-\frac{k\cdot q}{2}
+\frac{2 k\cdot q \left(k\cdot p+c_{B}^2\right)+(k\cdot p) (l\cdot q)}{4 l\cdot p}
\\
&+\frac{2 k\cdot q \left(k\cdot p+c_{B}^2\right)-(k\cdot p) (l\cdot q)}{
4 l\cdot (p+q)}
\\
&
-\frac{k\cdot q \left(k\cdot p+c_{B}^2\right) \left(2 c_{B}^2-k\cdot q\right)}{
4 (l\cdot p) ~~(l\cdot (p+q))}
\biggr] \times
\delta (l-p_0)\delta (l_0 +p_0)
\\
{\rm Y}_{\rm ~diagram}=&  
\biggl[8\left(-k\cdot p-k\cdot q-c_{B}^2+l\cdot p+q\cdot l\right)
\\ 
&~~+4\frac{c_{B}^2 \left(k\cdot p+c_{B}^2-2 q\cdot l\right)}{ l\cdot p} 
\biggr]
\times \delta (l-p_0)\delta (l_0 +p_0).
 \end{split}
\end{equation}

We must now sum these factors, multiply with the universal term in $\delta$ and then integrate the
result over the 3 momentum $l$. The delta functions in \eqref{delta} effectively turn this last integral into 
an integral over the angle of the spatial part of $l$. This angular integral is easily performed using 
\begin{equation}
 \begin{split}
&\int_c 
\frac{l\cdot q}{l\cdot p} =2\pi (\cos(\alpha) -1)-2\pi  (\cos(\alpha) -1) \frac{p_0}{m},~~
\\
&\int_c \frac{1}{l\cdot (p+q)~l\cdot p} =\frac{4\pi}{p_0 m (2c_{B}^2+p^2-p^2 \cos(\alpha))} ,
\\
 & \int_c 1 = 2\pi,~~
\int_c \frac{1}{l\cdot p} = \frac{2\pi}{m~p_0 },\qquad
\int_c l\cdot k =-\int_c l\cdot p =-2\pi p_0^2 , \qquad
\int_c l\cdot q =0
 \end{split}
\end{equation}
where the notation $\int_c$ is the angle integral or more formally
\begin{equation}
 \int_c= \int   dl_0 dl d\theta\delta (-p_0+l)\delta (l_0 +p_0).
\end{equation}
We find that the cut due to the various diagrams is given by  
$-\frac{1}{16\pi~ p_0}$ times 
\begin{equation}
 \begin{split}
 {\rm Box}_{\rm ~cut}
=& -\frac{1}{16\pi~ p_0} \times 
2 (-m + p_0) \sin^2\left(\frac{\alpha}{2}\right),
\\
{\rm Y}_{\rm ~cut}=&
-\frac{1}{16\pi~ p_0} \times
(p_0-m) \cos(\alpha),
\\ 
{\rm Eye}_{\rm ~cut}=&
-\frac{1}{16\pi~ p_0}
\frac{p_0}{2},
\\
{\rm V}_{\rm ~cut}=&
-\frac{1}{16\pi~ p_0}
(m-\frac{3 p_0}{2}).
 \end{split}
\end{equation}
It follows that 
\begin{equation}
( {\rm Box}_{\rm ~cut})
+ ({\rm Y}_{\rm ~cut})
+ ({\rm Eye}_{\rm ~cut})
+ ({\rm V}_{\rm ~cut}) = 0.
\end{equation}
\subsection{Potential subtlety at special values of external momenta}
We now turn to the discussion of an important subtlety that we have, so far, glossed over. As we have 
emphasized above, our determination of the integrand for the box diagram made crucial use of the 
replacement rule \eqref{reprule}. The derivation of this replacement rule works at generic values 
of the external momenta but turns out to fail when any two of the three independent external momenta are parallel
(in this case the Gram-determinant vanishes) to each other. As an example, consider the situation when $p^{\mu}\parallel k^{\mu}$ as appearing in
\ref{LgBox}. In this case, in the centre-of-mass frame, the angle of scattering, $\theta=0$ in $S$-channel.
Of course if the amplitude was an analytic function of external momenta then we could simply ignore
these exceptional momenta. The scattering amplitude at exceptional external momenta could be obtained
by analytic continuation from the generic case. However we have seen that, the scattering amplitude 
is {\it not} an analytic function of external momenta (in the $S$-channel, in the centre-of-mass frame we have a piece $\delta(\theta)$ and 
 this is precisely one of the points where the reduction that we discussed in \eqref{reprule} breaks down).
 The amplitude actually has singularities that 
are localized on the $s, t$ plane. Moreover these singularities play an important role in the discussion 
of unitarity in these theories, as we have already emphasized.

\section{Details of scattering in the fermionic theory}
\label{AP:fermion}

\subsection{Off shell four point function}
We now restrict our attention to the special case $q^\pm=0$. Plugging  \eqref{formv} into 
the Schwinger-Dyson equation \eqref{Masteq1}, performing the integral over the 
3 component of the momentum, and comparing coefficients of the different index structures 
on the two sides of this equation we find
\begin{equation}\small \label{werf}
 \begin{split}
 &f(p,k,q) = -\frac{\lambda}{2} G_{+3}(p-k)\\
&-4\pi i \lambda \int \frac{d^2p'}{(2\pi)^2} \frac{\left(p'_- f(p',k,q)(q_3 -2 i \Sigma_I(p') p'_s)+2 g(p',k,q) ((-1+\Sigma_I^2(p')){p'}_s^2-c_F^2)\right)G_{+3}(p'+p)}{\sqrt{{p'}_s^2+c_F^2}\left(q_3^2+4({p'}_s^2+c_F^2)\right)}\\
\\
\\&g(p,k,q) =\\
&-4\pi i \lambda \int \frac{d^2p'}{(2\pi)^2} \frac{p'_-}{\sqrt{{p'}_s^2+c_F^2}\left(q_3^2+4({p'}_s^2+c_F^2)\right)}\left(2 {p'}_- f(p',k,q)+ g(p',k,q) (q_3 +2 i \Sigma_I(p') {p'}_s)\right)G_{+3}(p'+p)\\
\end{split}
\end{equation}
\begin{equation}\small\label{werf1}\begin{split}
\\&f_1(p,k,q) =\\
&-4\pi i \lambda \int \frac{d^2p'}{(2\pi)^2} \frac{\left(p'_- f_1(p',k,q)(q_3 -2 i \Sigma_I(p') p'_s)+2 g_1(p',k,q) ((-1+\Sigma_I^2(p')){p'}_s^2-c_F^2)\right)G_{+3}(p'+p)}{\sqrt{{p'}_s^2+c_F^2}\left(q_3^2+4({p'}_s^2+c_F^2)\right)}\\
\\&g_1(p,k,q) = \frac{\lambda}{2} G_{+3}(p-k)\\
&-4\pi i \lambda \int \frac{d^2p'}{(2\pi)^2} \frac{p'_-}{\sqrt{{p'}_s^2+c_F^2}\left(q_3^2+4({p'}_s^2+c_F^2)\right)}\left(2 p'_- f_1(p',k,q)+ g_1(p',k,q) (q_3 +2 i \Sigma_I(p') p'_s)\right)G_{+3}(p'+p)\\
\end{split}
\end{equation}
We have played around with these equations and discovered that they admit a solution of the 
following structure 
\begin{equation}\label{ansatz}
 \begin{split}
  g(p,k,q)&=\frac{-p_-}{2(p-k)_-} W_0(y,x,q_3)+\frac{1}{2} W_1(y,x,q_3)\\
f(p,k,q)&=\frac{1}{2(p-k)_-} W_3(y,x,q_3)+ \frac{- p_+}{q_s^2} W_2(y,x,q_3)\\
g_1(p,k,q)&=\frac{k_{+}p_-}{2(p-k)_-} B_2(y,x,q_3)+\frac{1}{2(p-k)_-} B_3(y,x,q_3)\\
f_1(p,k,q)&=\frac{- p_{+}}{p_s^2(p-k)_-} B_0(y,x,q_3)+ \frac{-k_{+}}{2(p-k)_-} B_1(y,x,q_3)
 \end{split}
\end{equation}
where we use  $y=\frac{2}{q_3}\sqrt{k_s^2+c_F^2}$ and $x=\frac{2}{q_3}\sqrt{p_s^2+c_F^2}.$
Our ansatz completely specifies the dependence of $V$ on the argument of the complex 
variables $p^+$  and $k^+$, leaving undetermined
the dependence of $V$ on the modulus of these variables.
Plugging the above ansatz, it is possible to perform all angular integrals in  Eq. \ref{werf},\ref{werf1} using 
the formulae
\begin{equation}
 \begin{split}
  \int_0^{2\pi} \frac{d\theta}{2\pi}(p'_-)^2\frac{1}{p_{-}-p'_{-}}&=- p_{-}\theta(p'_s-p_s)\\
\int_0^{2\pi} \frac{d\theta}{2\pi}p'_-\frac{1}{p_{-}-p'_{-}}&=- \theta(p'_s-p_s)\\
\int_0^{2\pi} \frac{d\theta}{2\pi}\frac{1}{p'_{-}-p_{-}}&=-2 \frac{p_+}{p_s^2} \theta(p_s-p'_s)\\
\int_0^{2\pi} \frac{d\theta}{2\pi}\frac{1}{(p'_{-}-p_{-})(k-p')_-}&=\frac{2}{(k-p)_-}\left( \frac{k_+}{k_s^2} \theta(k_s-p'_s)-\frac{p_+}{p_s^2} \theta(p_s-p'_s)\right)\\
\int_0^{2\pi} \frac{d\theta}{2\pi}\frac{p'_- k_+}{(p'_{-}-p_{-})(k-p')_-}&=\frac{k_+}{(k-p)_-}\left( \theta(k_s-p'_s)- \theta(p_s-p'_s)\right)\\
\int_0^{2\pi} \frac{d\theta}{2\pi}\frac{(p'_-)^2 }{(p'_{-}-p_{-})(k-p')_-}&=-\frac{1}{(k-p)_-}\left(k_- \theta(-k_s+p'_s)- p_-\theta(-p_s+p'_s)\right)
\end{split}
\end{equation}
Equating the coefficients of the different functions of the arguments of $k^+$ and $p^+$
we obtain the following equations for the coefficient functions $W_1 \ldots W_4$ and
$B_1 \ldots B_4$.  
\begin{equation}\small \label{meqf}
 \begin{split}
 & W_1(y,x,q_3) = \frac{ i \lambda}{q_3} \int_{y}^{\infty} dx'\frac{X W_0(y,x',q_3)+2 W_3(y,x',q_3)}{(1+x'^2)}-\frac{ i \lambda}{q_3} \int_{x}^{\infty} dx'\frac{X W_1(y,x',q_3)+2 W_2(y,x',q_3)}{(1+x'^2)}\\
& W_0(y,x,q_3) =  \frac{ i \lambda}{q_3} \int_{y}^{x} dx'\frac{X W_0(y,x',q_3)+2 W_3(y,x',q_3)}{(1+x'^2)}\\
& W_3(y,x,q_3) = -\frac{ i \lambda}{q_3} \int_{x}^{y} dx'\frac{Y_1 W_0(y,x',q_3)+Y W_3(y,x',q_3)}{(1+x'^2)}-4\pi i \lambda\\
& W_2(y,x,q_3) = \frac{ i \lambda}{q_3} \int_{\frac{2|c_F|}{q_3}}^{x} dx'\frac{Y_1 W_1(y,x',q_3)+Y W_2(y,x',q_3)}{(1+x'^2)}\\
& B1(y,x,q_3)=\\
&-\frac{i\lambda}{q_3}\left(\frac{2}{(-c_F^2+q_3^2 \frac{y^2}{4})}    \int_{\frac{2|c_F|}{q_3}}^y \frac{Y B_0(y,x',q_3)+Y_1 B_3(y,x',q_3)}{x'^2+1} \,
   dx' + \int_x^y \frac{Y B_1(y,x',q_3)+Y_1
   B_2(y,x',q_3)}{x'^2+1} \, dx'\right)\\
\\& B_0(y,x,q_3)=\frac{i \lambda}{q_3}  \int_{\frac{2|c_F|}{q_3}}^x \frac{Y B_0(y,x',q_3)+Y_1 B_3(y,x',q_3)}{x'^2+1} \, dx'\\
& B_2(y,x,q_3)=-\frac{i \lambda }{q_3} \int_x^{\infty } \frac{2 B_1(y,x',q_3)+X B_2(y,x',q_3)}{x'^2+1} \, dx'\\
\\&B_3(y,x,q_3)=4\pi i ~\lambda
+\\& \frac{i\lambda}{q_3}\left(\frac{(-c_F^2+q_3^2 \frac{y^2}{4})}{2}\int_y^{\infty } \frac{2 B_1(y,x',q_3)+X B_2(y,x',q_3)}{x'^2+1} \, dx'
-\int_x^y \frac{2 B_0(y,x',q_3)+X B_3(y,x',q_3)}{x'^2+1} \, dx'\right)
\end{split}
\end{equation}
where 
\begin{equation}\begin{split}
a&=2\frac{|c_F|}{q_3},~~X=q_3 \left(1+i \left(\frac{2 m_f}{q_3}+\lambda  x\right)\right),\\
Y&=q_3 \left(1-i \left(\frac{2 m_f}{q_3}+\lambda  x\right)\right),~~Y_1=\frac{1}{2} q_3^2 \left(\left(\frac{2 m_f}{q_3}+\lambda  x\right){}^2-x^2\right)\\
x&=\frac{2}{q_3}\sqrt{p_s^2+c_F^2},~~y=\frac{2}{q_3}\sqrt{k_s^2+c_F^2}
\end{split}
\end{equation}
All of the equations above may be converted into differential equations by differentiating w.r.t. 
$x$. Notice that the first four equations in \eqref{meqf}  (equations for the $W$ variables) are decoupled from the last four variables (equations for the B variables). Furthermore the second and 
third of the equations above involve only the functions $W_0$ and $W_3$. These two equations 
are a set of linear first order differential equations for $W_0$ and $W_3$. These equations 
are given by 
\begin{equation}\label{}
\begin{split}
\partial_x W_0(y,x,q_3)&=I\frac{\lambda}{q_3}\frac{1}{1+x^2}\left(W_0(y,x,q_3)~X(x)+2 W_3(y,x,q_3)\right)\\
\partial_x W_3(y,x,q_3)&=I\frac{\lambda}{q_3}\frac{1}{1+x^2}\left(W_0(y,x,q_3)~Y_1(x)+Y(x)~W_3(y,x,q_3)\right)\\
\end{split}\end{equation}
It is not difficult to simultaneously solve these equations, using the observation that 
\be\partial_x \left(W_3(y,x,q_3)-\frac{Y(x)}{2}W_0(y,x,q_3)\right)=0.\ee
With this solution in hand, the first of  \eqref{meqf}  may then be used to solve for $W_1$ 
(we merely have to solve a linear first order differential equation) and the fourth of 
\eqref{meqf} may be solved for $W_2$. A very similar process may be employed to 
solve for $B_1$, $B_2$, $B_3$, and $B_4$. Of course the solution to the differential
equations so obtained have four integration `constants' (in the $Ws$) and four 
integration `constants' in the $B s$. These integration `constants' are really arbitrary 
functions of $y$. However their $y$ dependence may be determined either from the 
requirement of symmetry - or equivalently by setting up the analogue of the  \eqref{Masteq1}
`from the right' (this process yields a solutions to $Ws$ and $Bs$ upto unknown functions 
of $x$. Implementing these steps we find that our functions are given by 
\begin{equation}\small\label{intsolwb}\begin{split}
W_0(y,x,q_3)&=\frac{C_1(y)+C_2(y) e^{2 i \lambda  \tan ^{-1}(x)}}{q_3}\\
W_3(y,x,q_3)&=-C_1(y)+\frac{\left(C_1(y)+C_2(y) e^{2 i \lambda  \tan ^{-1}(x)}\right) \left(-2 i m_f-i \lambda  q_3 x+q_3\right)}{2 q_3}\\
W_1(y,x,q_3)&=\frac{D_1(y)+D_2(y) e^{2 i \lambda  \tan ^{-1}(x)}}{q_3}\\
W_2(y,x,q_3)&=-D_1(y)+\frac{\left(D_1(y)+D_2(y) e^{2 i \lambda  \tan ^{-1}(x)}\right) \left(-2 i m_f-i \lambda  q_3 x+q_3\right)}{2 q_3}\\
B_2(y,x,q_3)&=\frac{h_1(y)+h_2(y) e^{2 i \lambda  \tan ^{-1}(x)}}{q_3}\\
B_1(y,x,q_3)&=-h_1(y)+\frac{\left(-2 i m_f-i \lambda  q_3 x+q_3\right) \left(h_1(y)+h_2(y) e^{2 i \lambda  \tan ^{-1}(x)}\right)}{2 q_3}\\
B_3(y,x,q_3)&=\frac{h_3(y)+h_4(y) e^{2 i \lambda  \tan ^{-1}(x)}}{q_3}\\
B_0(y,x,q_3)&=-h_3(y)+\frac{\left(-2 i m_f-i \lambda  q_3 x+q_3\right) \left(h_3(y)+h_4(y) e^{2 i \lambda  \tan ^{-1}(x)}\right)}{2 q_3}.
\end{split}\end{equation}
The 8 undetermined constants in our solution are an artifact of the fact that we solved 
a set of integral equations by converting them into differential equations. In order to determine the 8 integration constants, we plug our solution back directly into the integral equations  \eqref{meqf}.
 It turns out that all integrals on the RHS of the equations \eqref{meqf} may be explicitly 
 performed. The undetermined constants are then easily obtained by comparing the 
 LHS and RHS of \eqref{meqf}. Implementing this procedure we obtain the final solution
 \begin{equation}\small \label{fsw}
 \begin{split}
W_0(y,x,q_3)&=-\frac{4 i \pi  \lambda  \left(-1+e^{2 i \lambda  \left(\tan ^{-1}(x)-\tan ^{-1}(y)\right)}\right)}{q_3}\\
W_1(y,x,q_3)&=\frac{4 i \pi  \lambda  \left(-1+e^{i \lambda  \left(\pi -2 \tan ^{-1}(y)\right)}\right) \left(e^{2 i \lambda  \tan ^{-1}(a)} (a \lambda +m_{f1}+i)-(a
   \lambda +m_{f1}-i) e^{2 i \lambda  \tan ^{-1}(x)}\right)}{q_3 \left(e^{i \pi  \lambda } (a \lambda +m_{f1}-i)-e^{2 i \lambda  \tan
   ^{-1}(a)} (a \lambda +m_{f1}+i)\right)}\\
W_2(y,x,q_3)&=\frac{2 \pi  \lambda  \left(-1+e^{i \lambda  \left(\pi -2 \tan ^{-1}(y)\right)}\right) }{e^{i \pi  \lambda } (a \lambda
   +m_{f1}-i)-e^{2 i \lambda  \tan ^{-1}(a)} (a \lambda +m_{f1}+i)}\\
&\left(e^{2 i \lambda  \tan ^{-1}(a)} (a \lambda +m_{f1}+i)
   (m_{f1}+\lambda  x-i)-(a \lambda +m_{f1}-i) (m_{f1}+\lambda  x+i) e^{2 i \lambda  \tan ^{-1}(x)}\right)\\
W_3(y,x,q_3)&=2 \pi  \lambda  \left(-(m_{f1}+\lambda  x+i) e^{2 i \lambda  \left(\tan ^{-1}(x)-\tan ^{-1}(y)\right)}+m_{f1}+\lambda  x-i\right).
\end{split}
\end{equation}
where
 $$m_{f1} =2 \frac{m_f}{q_3}.$$ 
The other components are
\begin{equation} \label{fsb}
 \begin{split}
&B_0(y,x,q_3)\\
&=\frac{\pi  \lambda  q_3 e^{-2 i \lambda  \tan ^{-1}(y)} \left(e^{i \pi  \lambda } (m_{f1}+\lambda  y-i)-(m_{f1}+\lambda  y+i) e^{2 i \lambda 
   \tan ^{-1}(y)}\right)}{e^{i \pi  \lambda } (a \lambda +m_{f1}-i)-e^{2 i \lambda  \tan ^{-1}(a)} (a \lambda +m_{f1}+i)}\\
&\left((i a \lambda +i m_{f1}+1) (m_{f1}+\lambda  x+i) e^{2 i \lambda  \tan ^{-1}(x)}-i e^{2 i \lambda  \tan ^{-1}(a)} (a \lambda +m_{f1}+i)
   (m_{f1}+\lambda  x-i)\right)\\
 &B_1(y,x,q_3)\\
 &=\frac{2 \pi  \lambda  q_3 e^{-2 i \lambda  \tan ^{-1}(y)} \left(e^{i \pi  \lambda } (m_{f1}+\lambda  x-i)-(m_{f1}+\lambda  x+i) e^{2 i
   \lambda  \tan ^{-1}(x)}\right)}{\left(\frac{y^2 q_3^2}{4}-c_F^2\right) \left(e^{i \pi  \lambda } (a \lambda +m_{f1}-i)-e^{2 i \lambda  \tan ^{-1}(a)} (a \lambda
   +m_{f1}+i)\right)}\\
 &\left(e^{2 i \lambda  \tan ^{-1}(a)} (a \lambda +m_{f1}+i) (i m_{f1}+i \lambda  y+1)-i (a \lambda +m_{f1}-i) (m_{f1}+\lambda  y+i) e^{2 i \lambda
    \tan ^{-1}(y)}\right)\\
 &B_2(y,x,q_3)=\\
 &\frac{4 \pi  \lambda  \left(e^{i \pi  \lambda }-e^{2 i \lambda  \tan ^{-1}(x)}\right) e^{-2 i \lambda  \tan ^{-1}(y)}}{\left(\frac{y^2 q_3^2}{4}-c_F^2\right) \left(e^{i \pi  \lambda
   } (a \lambda +m_{f1}-i)-e^{2 i \lambda  \tan ^{-1}(a)} (a \lambda +m_{f1}+i)\right)}\\
&\left((a \lambda +m_{f1}-i) (m_{f1}+\lambda  y+i) e^{2 i \lambda  \tan ^{-1}(y)}-e^{2 i \lambda  \tan ^{-1}(a)} (a \lambda +m_{f1}+i)
   (m_{f1}+\lambda  y-i)\right)\\
 &B_3(y,x,q_3)=\\
 &-\frac{2 \pi  \lambda  e^{-2 i \lambda  \tan ^{-1}(y)} \left(e^{2 i \lambda  \tan ^{-1}(a)} (a \lambda +m_{f1}+i)-(a \lambda +m_{f1}-i) e^{2 i
   \lambda  \tan ^{-1}(x)}\right)}{e^{i \pi  \lambda } (a \lambda +m_{f1}-i)-e^{2 i \lambda  \tan ^{-1}(a)} (a \lambda +m_{f1}+i)} \\
 &\left((m_{f1}+\lambda  y+i) e^{2 i \lambda  \tan ^{-1}(y)}-e^{i \pi  \lambda } (m_{f1}+\lambda  y-i)\right)
 \end{split}
\end{equation}
In summary, the offshell four point amplitude, defined in \eqref{defV}, takes the form \eqref{formv}, 
with the functions in this equation given by \eqref{ansatz} with the $W$ and $B$ functions given 
in \eqref{fsw} and \eqref{fsb} respectively.


\section{Preliminary analysis of the double analytic continuation}

\subsection{Analysis of the scalar integral equation after double analytic continuation} \label{dac}

In this appendix we initiate a very preliminary discussion of the bosonic integral equation after 
double analytic continuation discussed in subsection \ref{at} above. In subsection \ref{dco}
below we evaluate the one loop contribution to four boson scattering after double analytic 
continuation, and demonstrate that the computation includes a singular contribution, absent
from the naive analytic continuation of the U and $T$-channel results to the $S$-channel. Under 
certain assumptions this singular piece precisely reproduces the ${\cal O}(\lambda^2)$ term 
in the contact $\delta$ function part of the $S$-channel scattering amplitudes.  In subsection \ref{dct}
below we take a non-relativistic limit of the double analytic continued integral equation and 
demonstrate that it reduces to the non-relativistic Aharonov-Bohm equation with $\nu=\lambda$. 
\subsection{The oneloop box diagram after double analytic continuation } \label{dco}
Appendix \ref{olm} was devoted to a detailed study of the one loop diagram Fig. \ref{olmfig} at $q^{\pm}=0$ directly in usual Minkowski space. 
The conclusions of Appendix \ref{olm} may be summarized as follows. 
In the case that the momenta $p$ and $k$ both lie offshell, the Minkowskian one loop diagram agrees 
with the unambiguous analytic continuation of the Euclidean answer. In the case that the momenta 
$p$ and $k$ were both onshell, the continuation from Euclidean space was ambiguous, but 
the Minkowskian computation resolved the ambiguity.

In this Appendix we revisit the one loop diagram of Fig. \ref{olmfig} after performing the double analytic 
continuation described in subsection \ref{at}. We recompute the diagram, this time in the double 
analytically continued Minkowski space - the space in which the $3$ direction is taken to be time.
We address the following question: how does the answer of this computation compare with 
analytic continuation from usual Minkowski space (and the analytic continuation from Euclidean space, 
when this analytic continuation is unambiguous).

Although we will not present the detailed computation here we have indeed verified that when 
$p$ and $k$ are both offshell, the computation performed directly after the double analytic continuation 
agrees with the appropriate analytic continuations from usual Minkowski space as well as from Euclidean 
space. 

The situation is more delicate when $p$ and $k$ are both onshell. In this case though the 
Euclidean answer is ambiguous, the `usual' Minkowskian answer is not. We outline the 
computation of the double analytically continued result in this Appendix. In particular we  
show that the analytic continuation of this `usual' answer does {\it not} agree with the answer of the 
computation performed directly in double analytically continued Euclidean space. The details of the 
difference between these answer depends in a very unusual way on the relative smallness of the 
$i \epsilon$ in scalar propagators and $i \epsilon$ in the gauge propagators. In a natural limit 
(the one in which these two have the same degree of smallness), the difference between the 
two results agrees precisely with the difference between  $T_S^{trial}$ and $T_S^B$  (see \eqref{conject1}) 
lending some support to the conjecture \eqref{conject1}.

\subsubsection{Setting up the computation } \label{dcot}

Let $T(\alpha)$ denote the double analytic continuation of the one loop contribution to the $T$ matrix. 
$T(\alpha)$ is given by (see \eqref{olmin})  
\begin{equation}\small \label{moli}
\begin{split}
\frac{i T(\alpha)}{(4\pi\lambda q_3)^2} =
-\int \frac{d^3 r}{(2\pi)^3}\biggl[&\frac{2(r+p)_- (r-p)_+}{2(r-p)_- (r-p)_+ -i \epsilon}\frac{2(r+k)_- (r-k)_+}{2(r-k)_- (r-k)_+-i\epsilon}
\\
&\times \frac{1}{r_s^2-r_3^2+c_{B}^2-i\epsilon_1}\frac{1}{r_s^2-(r_3+q_3)^2+c_{B}^2-i\epsilon_1}\biggr]
\end{split}
\end{equation}

Note that after double analytic continuation $v_+$ is a complex number and $v_-$ is its complex 
conjugate for all $v_\pm$ (this was true also in Euclidean space). As in Euclidean space, we will find it convenient to work with the magnitude and phase of these complex numbers. Choosing axes so that 
$p_+$ is a real number we have
\begin{equation} \label{map}
 p_\pm=\frac{p_s}{\sqrt{2}},~~k_{\pm}=\frac{k_s}{\sqrt{2}} e^{\pm i\alpha},~~r_{\pm}=\frac{r_s}{\sqrt{2}} e^{\pm i\theta}.
\end{equation}
As we focus on the case of onshell scattering (and as $q^\pm =0$) we have
\begin{equation}\label{oscm}
p_s = k_s,~~q_3=-2p_3=-2k_3=2\sqrt{p_s^2+c_{B}^2}=\sqrt{s}
\end{equation}
Plugging \eqref{map} and \eqref{oscm} into \eqref{moli}  and using the fact that the scalar propagators 
are independent of $\theta$ and $\alpha$, while the gauge boson propagators are independent of 
$r_3$ we find
\begin{equation}\label{totint}
\frac{i T(\alpha)}{(4\pi\lambda q_3)^2} =\int_0^{\infty}\frac{r_sdr_s}{2\pi}\mathcal{I}_1(r_s,\alpha)\mathcal{I}_2(r_s)
\end{equation}
where $\mathcal{I}_2(r_s)$ is the integral of the product of the scalar propagators over the timelike coordinate $r_3$ 
\begin{equation}\label{I2}
\mathcal{I}_2(r_s) = \int_{-\infty}^{\infty}\frac{dr_3}{2\pi}\frac{1}{r_s^2-r_3^2+c_{B}^2-i\epsilon_1}\frac{1}{r_s^2-(r_3+q_3)^2+c_{B}^2-i\epsilon_1}
\end{equation}
and $\mathcal{I}_1(r_s,\alpha)$ is the integral of the product of the gauge boson propagators over the angle 
$\theta$
\begin{equation}\label{I1}
\mathcal{I}_1(r_s,\alpha) = -\int_0^{2\pi}\frac{d\theta}{2\pi}\frac{(r_se^{i\theta}-p_s)(r_se^{-i\theta}+p_s)}{(r_se^{i\theta}-p_s)(r_se^{-i\theta}-p_s)-i\epsilon}
\frac{(r_se^{i\theta}-p_se^{i\alpha})(r_se^{-i\theta}+p_se^{-i\alpha})}{(r_se^{i\theta}-p_se^{i\alpha})(r_se^{-i\theta}-p_se^{-i\alpha})-i\epsilon}
\end{equation}

The integral over $r_3$ in \eqref{I2} is easily evaluated by contour methods and we find 
\begin{equation}\label{I2f}
\begin{split}
\mathcal{I}_2(r_s) &= \frac{-i}{\sqrt{r_s^2+c_{B}^2}(q_3^2-4r_s^2-4c_{B}^2+4i\epsilon_1)}\\
&= \frac{-i}{4\sqrt{r_s^2+c_{B}^2}(p_s^2-r_s^2+i\epsilon_1)}
\end{split}
\end{equation}

The integral over $\theta$ in \eqref{I1} may also be evaluated by contour techniques. 
Let 
\begin{equation}
z = e^{i\theta},~~~w = e^{i\alpha}
\end{equation}
so that
\begin{equation} \label{cI}
\mathcal{I}_1(r_s,\alpha) = -\oint_{|z|=1}\frac{dz}{2\pi iz}\frac{(z+\frac{r_s}{p_s})(z-\frac{p_s}{r_s})}{(z-\frac{r_s}{p_s})(z-\frac{p_s}{r_s})+\frac{i\epsilon z}{r_sp_s}}
\frac{(z+w\frac{r_s}{p_s})(z-w\frac{p_s}{r_s})}{(z-w\frac{r_s}{p_s})(z-w\frac{p_s}{r_s})+\frac{i\epsilon zw}{r_sp_s}}
\end{equation}
where the integration contour in \eqref{cI} runs over the unit circle. 

\subsubsection{The contribution of the pole at zero}

The integrand in \eqref{cI} is a meromorphic function of $z$ with 5 poles. The simplest of these poles is 
at $z=0$. The contribution of this pole to $\mathcal{I}_1(r_s, \alpha) $ is simply $-1$ ; 
plugging this together with \eqref{I2f} into \eqref{totint} we find that the contribution of the pole at zero 
to ${i T}$ is given by 
\begin{equation}\label{cpz}
iT=i (4 \pi \lambda q_3)^2  H(q)
\end{equation}
in perfect agreement with the analytic continuation of \eqref{olmfnl1}. 
As the contribution of the pole at zero has already reproduced the analytic continuation of the `real' 
Minkowski scattering amplitude, It follow that the contribution of the remaining 4 poles in \eqref{cI} is 
simply the difference between this analytic continuation, and the result directly computed after double 
analytic continuation

\subsubsection{The contribution of the remaining four poles}

Let us retreat from the onshell limit for a moment, i.e. allow $p_s$ and $k_s$ to be different. A naive 
evaluation of the contribution of the remaining four poles in \eqref{cI} in the limit of vanishing $\epsilon_1$ 
yields and answer proportional to 
$$ \theta(p_s -r_s) - \theta(k_s-r_s)$$
This quantity vanishes when $p_s=k_s$ suggesting that the contribution of the remaining four poles 
to the angle integral should vanish in the onshell limit. \footnote{This is indeed how things worked in our
derivation of the Euclidean integral equation for $V$.} However this reasoning is a bit too quick 
for the following reason. Suppose $p_s -k_s= a$ where $a$ is a very small number and $k_s$ is the onshell
value of spatial momentum. Then $r_s$ is indeed constrained vary over a very small range. 
However this is not sufficient to guarantee that 
the integral over $r_s$ will vanish. The reason for this is that this small interval is concentrated around 
precisely the value of $r_s$ at which \eqref{I2f} is singular, and a singular integrand may well 
integrate to a finite quantity over a vanishing small integration domain. Cautioned by these considerations 
we now turn to a careful and honest evaluation of the contribution of the remaining 4 poles in \eqref{cI} to 
$\mathcal{I}_1(r_s, \alpha)$ 

The remaining four poles in \eqref{cI} are located at $z_\pm$ and $w z_\pm$ where
\begin{equation}
z_\pm = \frac{1}{2}\left(\frac{r_s}{p_s}+\frac{p_s}{r_s}-i\epsilon\pm\sqrt{(\frac{r_s}{p_s}+\frac{p_s}{r_s}-i\epsilon)^2-4}\right),~~~w = e^{i \alpha}
\end{equation}
where the square root function is defined to have a branch cut along the negative real axis. 
It is easily verified that 
\begin{equation}\label{prop}
z_+z_- = 1,~~~z_++z_-= \frac{r_s}{p_s}+\frac{p_s}{r_s},~~~z_+-z_-=\sqrt{\left(\frac{r_s}{p_s}-\frac{p_s}{r_s}\right)^2-2i\epsilon\left(\frac{r_s}{p_s}+\frac{p_s}{r_s}\right)}
\end{equation}
It may also be verified that  $|z_+|>1$, so $|z_-|<1$. The  two poles enclosed by the unit contour in 
\eqref{cI}  are located at  $z_-$ and $wz_-$ (the remaining two poles lie outside the contour 
and do not contribute to the integral). The contribution of these two poles to \eqref{cI} is given by 
\begin{equation}\label{fio}
\mathcal{I}_1(r_s,\alpha)=-\frac{(z_-+\frac{r_s}{p_s})(z_--\frac{p_s}{r_s})}{z_-^2(w-1)(z_+-z_-)}
\left(\frac{(z_-+w\frac{r_s}{p_s})(z_--w\frac{p_s}{r_s})}{z_--wz_+}-\frac{(wz_-+\frac{r_s}{p_s})(wz_--\frac{p_s}{r_s})}{wz_--z_+}\right)
\end{equation}
Using \eqref{prop} several times, \eqref{fio} may be simplified to
\begin{equation} \label{fion}
\begin{split}
\mathcal{I}_1(r_s,\alpha)&=\frac{-w(z_-+\frac{r_s}{p_s})(z_--\frac{p_s}{r_s})(z_--z_+-\frac{r_s}{p_s}+\frac{p_s}{r_s})}{z_-(wz_+-z_-)(wz_--z_+)}
\frac{z_++z_-}{z_+-z_-}\\
&=\frac{-w(z_--z_++\frac{r_s}{p_s}-\frac{p_s}{r_s})(z_--z_+-\frac{r_s}{p_s}+\frac{p_s}{r_s})}{(wz_+-z_-)(wz_--z_+)}
\frac{z_++z_-}{z_+-z_-}\\
&=\frac{2i\epsilon w(r_s^2+p_s^2)^2}{r_s^3p_s^3(z_+-z_-)(wz_+-z_-)(wz_--z_+)}\\
\end{split}
\end{equation}
Note that \eqref{fion} $\epsilon$ in apparent vindication of the intuition that suggests that these poles contribute vanishingly to the integral. Let us anyway proceed to complete our careful evaluation:
we conclude that the contribution of these poles to \eqref{totint} is given by 
\begin{equation} \label{sin}
i T(\alpha) = \frac{4\epsilon\pi\lambda^2q_3^2}{p_s^3}\int_0^\infty\frac{dr_sw(r_s^2+p_s^2)^2}{r_s^2(z_+-z_-)(w-\frac{z_-}{z_+})(w-\frac{z_+}{z_-})}
\frac{1}{\sqrt{r_s^2+c_{B}^2}(p_s^2-r_s^2+i\epsilon_1)}\end{equation}

In the limit $\epsilon \to 0$, the RHS in \eqref{sin} vanishes unless the integral in that equation develops 
a singularity. The integrand in \eqref{sin} does have a singularity that approaches the integration contour 
at $r_s= p_s$.  If $w \neq 1$, however, no other singularity in the integrand approaches the 
integration contour  $r_s=(0, \infty)$. A single singularity approaching an integration contour does not 
give rise to a singular contribution to the integral (because the integration contour can always be deformed 
to avoid the singularity). Provided $w \neq 1$ it follows that the integral on the RHS of \eqref{sin}
is nonsingular, and so the RHS of \eqref{sin} vanishes in the limit $\epsilon \to 0$. 

The situation is different, however, if $w$ tends to unity. In this case the singularities caused by 
the factors  $(w-\frac{z_-}{z_+})$,  $(w-\frac{z_+}{z_-})$ and  $(p_s^2-r_s^2+i\epsilon_1)$ all approach the same contour point, namely $r_s=p_s$ as $w \to 1$ and $\epsilon_1, \epsilon \to 0$. In this case 
the integral on the RHS conceivably develops a pinch singularity, and the RHS of \eqref{sin} 
does not necessarily vanish in this case.

In summary we have concluded that $i T(\alpha)$ vanishes for nonzero $\alpha$, but not necessarily 
at $\alpha=0$. In order to better understand the behaviour of $iT(\alpha)$ near $\alpha=0$ we now 
evaluate the integral of this quantity over $\alpha$. This integral may be affected by contour techniques 
and we find 
\begin{equation}
\begin{split}
&\int_0^{2\pi}d\alpha~i T(\alpha) 
\\
=& \oint_{|w|=1}\frac{dw}{iw}\frac{4\epsilon\pi\lambda^2q_3^2}{p_s^3}\int_0^\infty\frac{dr_sw(r_s^2+p_s^2)^2}{r_s^2(z_+-z_-)(w-\frac{z_-}{z_+})(w-\frac{z_+}{z_-})}
\frac{1}{\sqrt{r_s^2+c_{B}^2}(p_s^2-r_s^2+i\epsilon_1)}
\end{split}
\end{equation}
The integral runs counterclockwise over the unit circle in the $w$ plane. This contour encloses a single pole, 
at  $ w=\frac{z_-}{z_+}$. Evaluating the residue of this pole we find 
\begin{equation}\label{has}
\int_0^{2\pi}d\alpha~i T(\alpha) = -\frac{8\pi^2 \epsilon \lambda^2q_3^2}{p_s^2}\int_0^\infty\frac{dr_s(r_s^2+p_s^2)}{r_s(z_+-z_-)^2}
\frac{1}{\sqrt{r_s^2+c_{B}^2}(p_s^2-r_s^2+i\epsilon_1)}
\end{equation}

Because of the overall factor of $\epsilon$, it is clear that \eqref{has} receives contributions - if at all -  only from $r_s$ in the neighborhood of $p_s$.  It is not too difficult to convince oneself that the dominant 
contribution is from $r_s \sim \sqrt{\epsilon} $. In order to see this we make the variable change 
$r_s= \sqrt{\epsilon} x$. To leading order in $\sqrt{\epsilon}$ we find
\begin{equation}\label{bent}
\int_0^{2\pi}d\alpha~iI(\alpha) =-\frac{16\pi^2\lambda^2p_sq_3^2}{\sqrt{p_s^2+c_{B}^2}}\int_{-\infty}^\infty \frac{\sqrt{\epsilon}dx}{(i\epsilon_1-2xp_s\sqrt{\epsilon})(x^2-i)}
\end{equation}
(to obtain \eqref{bent} we have used here that $(z_+-z_-)^2 = 2\epsilon (x^2-i)$ at leading order in $\epsilon$)

Let us now assume that $\epsilon_1 \ll \sqrt{\epsilon}$ (this would in particular have been the case if 
$\epsilon_1= \epsilon$). In this case \eqref{bent} simplifies to 
\begin{equation} \label{fent}
\int_0^{2\pi}d\alpha~iI(\alpha) =4\pi^2\lambda^2\sqrt{s}\int_{-\infty}^\infty \frac{dx}{(x-ib)(x^2-i)}
\end{equation}
Where b is a positive infinitesimal. The integral on the RHS of \eqref{fent} evaluates (by a straightforward 
application of contour techniques) to $-\pi$.  We conclude that 
\begin{equation}\label{fof}
i T= -4 \pi^3 \lambda^2 \sqrt{s} \delta(\alpha)
\end{equation}
This is in perfect agreement with the expectation 
$$T= - 8 \pi i \sqrt{s} \left( \cos (\pi \lambda)-1 \right) \delta(\alpha) = 4 i  \pi^3 \lambda^2 \sqrt{s} 
\delta(\alpha) + {\cal O}(\lambda^4)$$

\subsection{Solutions of the Dirac equation at $q^\pm=0$ after double analytic continuation. } 

In order to compute S matrices in he Fermionic theory after double analytic continuation 
we need solutions to the relevant Dirac equations. We present the relevant solutions 
in this Appendix.

After a double analytic continuation 
$k_0=i k_3$ and the gamma matrix convention is 
$\gamma^{0}=-i\gamma^{3}.$
The Dirac equation is give by
\begin{equation}
\bar{\psi}(-p)\big( i\left( p_0\gamma^0 + p_-\gamma^- + p_+\gamma^+\left( 1 + g(p_s)\right)\right) + f(p_s)p_s\big) \psi(p) = 0.
\end{equation}
Where
\begin{equation}
p_s^2 = p_1^2+p_2^2.
\end{equation}
Our gamma matrix convention is
\begin{align}
\gamma^0 = \left(\begin{array}{cc}
                 -i & 0 \\ 0 & i
                 \end{array} \right) \\
\gamma^+ = \left(\begin{array}{cc} 0 & \sqrt{2} \\ 0 & 0 \end{array}\right) \\
\gamma^- = \left(\begin{array}{cc} 0 & 0 \\ \sqrt{2} & 0 \end{array}\right)
\end{align}

So now the Dirac equation is
\begin{equation}
\bar{\psi}(-p)\left(\begin{array}{cc} p_0 + f(p_s)p_s & i\sqrt{2}p_+( 1 + g(p_s)) \\ i\sqrt{2}p_- & -p_0 + f(p_s)p_s 
		    \end{array}\right)\psi(p) = 0
\end{equation}
Now we use the on-shell condition
\begin{equation}
p_0 = \pm E_{\vec{p}} 
\end{equation}
Where
\begin{equation}
E_{\vec{p}} = \sqrt{p_1^2+p_2^2+C_f^2}
\end{equation}
$C_f$ is the fermion pole mass.\\
The solution with $p_0 = -E_{\vec{p}}$ is particle solution $u(\vec{p})$ while the solution with $p_0 = E_{\vec{p}}$ is
the antiparticle solution $v(-\vec{p})$.\\

Now we need to solve
\begin{equation}
\bar{u}(\vec{p})\left(\begin{array}{cc} -E_{\vec{p}} + f(p_s)p_s & i\sqrt{2} p_+( 1 + g(p_s)) \\ i\sqrt{2}p_- & E_{\vec{p}} + f(p_s)p_s 
		    \end{array}\right)u(\vec{p}) = 0
\end{equation}
Which on solving on right and on left gives respectively,
\begin{align}
u(\vec{p}) = \frac{1}{\sqrt{E_{\vec{p}}+f(p_s)p_s}}\left(\begin{array}{c} E_{\vec{p}}+f(p_s)p_s \\~~ -i\sqrt{2}p_- \end{array}\right)\\
\bar{u}(\vec{p}) = \frac{1}{\sqrt{E_{\vec{p}}+f(p_s)p_s}}\left(\begin{array}{cc} E_{\vec{p}}+f(p_s)p_s & -i\sqrt{2} p_+( 1 + g(p_s)) \end{array}\right)
\end{align}
Where normalization is set to be $\bar{u}(\vec{p})u(\vec{p}) = 2f(p_s)p_s$.

We also need to solve
\begin{equation}
\bar{v}(\vec{p})\left(\begin{array}{cc} -E_{\vec{p}} - f(p_s)p_s & i\sqrt{2} p_+( 1 + g(p_s)) \\ i\sqrt{2}p_- & E_{\vec{p}} - f(p_s)p_s 
		    \end{array}\right)v(\vec{p}) = 0
\end{equation}
Which on solving on right and on left gives respectively,
\begin{align}
v(\vec{p}) = \frac{1}{\sqrt{E_{\vec{p}}-f(p_s)p_s}}\left(\begin{array}{c} E_{\vec{p}}-f(p_s)p_s \\ -i\sqrt{2}p_- \end{array}\right)\\
\bar{v}(\vec{p}) = \frac{1}{\sqrt{E_{\vec{p}}-f(p_s)p_s}}\left(\begin{array}{cc} E_{\vec{p}}-f(p_s)p_s & ~~-i\sqrt{2} p_+( 1 + g(p_s)) \end{array}\right)
\end{align}
Where normalization is set to be $\bar{v}(\vec{p})v(\vec{p}) = -2f(p_s)p_s$.

\subsection{Aharonov-Bohm in the non-relativistic limit} \label{dct}

After double analytic continuation, the four boson four point function satisfies the integral equation 
\begin{equation}\label{integral}
V(\vec{p},\vec{k}) = V_0(\vec{p},\vec{k}) + \int \frac{(i)^2V_0(\vec{p},\vec{l})V(\vec{l},\vec{k})\frac{d^3l}{(2\pi)^3}}{\left(-l_0^2+l_s^2+c_{B}^2-i\epsilon\right)\left(-(l_0+q_0)^2+l_s^2+c_{B}^2-i\epsilon\right)}
\end{equation}
where
\begin{equation}
V_0(\vec{p},\vec{k}) = 4\pi i\lambda q_0\frac{(k+p)_-}{(k-p)_-} 
- 2i\pi\lambda^2 c_{B}
\end{equation}
Since both $V_0$ and $V$ depend only on the spatial components of momenta, we can perform $l_0$ integral in (\ref{integral}) to get
\begin{equation}\label{spatial}
V(\vec{p},\vec{k}) = V_0(\vec{p},\vec{k}) +i \int \frac{V_0(\vec{p},\vec{l})V(\vec{l},\vec{k})}{\sqrt{l_s^2+c_{B}^2}\left(q_0^2-4l_s^2-4c_{B}^2+i\epsilon\right)}\frac{d^2l}{(2\pi)^2}
\end{equation}
Let us focus on the special case in which $k$ and $k+q$ are taken to be onshell, i.e. $q_0 = -2k_0 = -2\sqrt{k_s^2+c_{B}^2}$ while $p$ and $p+q$ are generically offshell. Let us define
\begin{equation} \label{wf}
\psi(\vec{p}) = (2\pi)^2\delta^2(\vec{p}-\vec{k}) +i \frac{V(\vec{p},\vec{k})}{4\sqrt{p_s^2+c_{B}^2}(k_s^2-p_s^2+i\epsilon)}
\end{equation}
Where $k$ is onshell. Then (\ref{spatial}) can be written as
\begin{equation}\label{fourschr}
-4i\sqrt{p_s^2+c_{B}^2}\left(k_s^2-p_s^2\right)\psi(\vec{p}) = \int V_0(\vec{p},\vec{l})\psi(\vec{l})\frac{d^2l}{(2\pi)^2}
\end{equation}
In the non-relativistic limit 
\begin{align*}
\sqrt{p_s^2+c_{B}^2} &=  c_{B}\\
q_0 &= -2 c_{B}
\end{align*}
and so (\ref{fourschr}) becomes
\begin{equation} \label{sche}
\left(k_s^2-p_s^2\right)\psi(\vec{p}) = \int\left(2\pi\lambda \frac{(l+p)_-}{(l-p)_-} + \frac{\pi\lambda^2}{2}\right)\psi(\vec{l})\frac{d^2l}{(2\pi)^2}
\end{equation}
\eqref{sche} takes the form of a non-relativistic Schrodinger equation of a particle propagating in a 
potential whose nature we will soon identify. \eqref{wf} is the assertion that the wave function $\psi(r)$
that obeys this Schrodinger equation takes the Lippmann Schwinger scattering form, with a scattering 
function (roughly $h(\theta)$) proportional to $V(k, p)$ once $p$ is set onshell. Restated, the non 
relativistic limit of the integral equation \eqref{integral} is simply the Lippmann Schwinger equation 
for the scattering matrix of a non-relativistic quantum mechanical problem, whose precise nature we now 
investigate.
 
In order to better understand the Schrodinger equation \eqref{sche} we transform it to position space. 
Multiplying \eqref{sche} by $\frac{e^{i px}} {(2 \pi)^2}$ and integrating over $p$ we find 
\begin{equation} \label{schep}
\int\left(k_s^2-p_s^2\right)\psi(\vec{p})e^{ip.x}\frac{d^2p}{(2\pi)^2} = \int\left(2\pi\lambda \frac{(l+p)_-}{(l-p)_-} + \frac{\pi\lambda^2}{2}\right)\psi(\vec{l})\frac{d^2l}{(2\pi)^2}e^{ip.x}\frac{d^2p}{(2\pi)^2}
\end{equation}
Let us define the position space wave function 
$$\psi(x) = \int\frac{d^3 p}{(2 \pi)^3} e^{i p.x} \psi(p)$$
Changing the integration variable on the RHS of \eqref{sche} as  $p\rightarrow p+l$, and recalling $z=x^+=\frac{x^1+ix^2}{\sqrt{2}}$ and $\bar{z}=x^-=\frac{x^1-ix^2}{\sqrt{2}}$, \eqref{sche} may 
be rewritten as 
\begin{equation}
\left(2\partial_z\partial_{\bar{z}}+k_s^2\right)\psi(z,\bar{z}) =\int\left(-4\pi\lambda \frac{l_-}{p_-} + \frac{\pi\lambda^2}{2}-2\pi\lambda\right)\psi(\vec{l})\frac{d^2l}{(2\pi)^2}e^{ip.x}e^{il.x}\frac{d^2p}{(2\pi)^2}
\end{equation}
The first term on RHS of \eqref{schep} is 
\begin{align}
-4\pi\lambda\int\frac{e^{ip.x}}{p_-}\frac{d^2p}{(2\pi)^2}\int l_-\psi(\vec{l})e^{il.x}\frac{d^2l}{(2\pi)^2}&=-4\pi\lambda \left(\frac{i}{2\pi z}\right)\left(-i\partial_{\bar{z}}\psi(z,\bar{z})\right)\\
&=\frac{-2\lambda}{z}\psi(z,\bar{z})
\end{align}
While the rest of the RHS of \eqref{schep} is 
\begin{equation}
\left(\frac{\pi\lambda^2}{2}-2\pi\lambda\right)\int\psi(\vec{l})e^{il.x}\frac{d^2l}{(2\pi)^2}\int e^{ip.x}\frac{d^2p}{(2\pi)^2}=\left(\frac{\pi\lambda^2}{2}-2\pi\lambda\right)\psi(z,\bar{z})\delta^2(z)
\end{equation}
It follows that \eqref{schep} may be recast as
\begin{equation}
\left(\partial_z\partial_{\bar{z}}+\frac{k_s^2}{2}\right)\psi(z,\bar{z}) = \frac{-\lambda}{z}\partial_{\bar{z}}\psi(z,\bar{z})+\left(\frac{\pi\lambda^2}{4}-\pi\lambda\right)\psi(z,\bar{z})\delta^2(z)
\end{equation}
Let us now define a gauge covariant derivative as
\begin{equation}\label{fschep}
\begin{split}
D_z &= \partial_z+iA_z\\
A_z &= \frac{-i\lambda}{z}\\
D_{\bar{z}} &= \partial_{\bar{z}}
\end{split}
\end{equation}

in terms of which \eqref{schep} reduces to 
\begin{equation} \label{sscc}
\left(D_zD_{\bar{z}}+\frac{k_s^2}{2}\right)\psi(z,\bar{z}) = -\left(\frac{\pi\lambda^2}{4}+\pi\lambda\right)\psi(z,\bar{z})\delta^2(z)
\end{equation}

How is the gauge potential $A_z$ in \eqref{fschep} to be interpreted? Firstly, clearly this potential is 
pure gauge away from $z=0$, as the antiholomorphic derivative of $A_z$ vanishes away from $z=0$. 
In other words $A_z$ is the gauge potential of a localized point flux.  The magnitude of this 
flux is given by the contour integral $\int A_z dz$ over the unit circle and so is $ 2 \pi^2 \lambda$. 
In other words \eqref{sscc} is the Schrodinger equation for the Aharonov-Bohm problem with $\nu=\lambda$ (plus  delta function contact interaction), in  an unusual complex gauge. The contact interaction plausibly 
makes do difference to scattering computations if the Schrodinger equation is studied with boundary
conditions (like those adopted by Aharonov and Bohm) that force $\psi(r)$ to vanish at the origin.

\bibliographystyle{JHEP}
\bibliography{CS}
\end{document}